\documentclass[printer]{aa}
\usepackage{graphics,epsfig,lscape,txfonts,color}
\usepackage{natbib}
\bibpunct{(}{)}{;}{a}{}{,}
\usepackage[english]{babel}
\usepackage{longtable}
\usepackage{lscape}
\usepackage{url}            
\usepackage{subfigure}

\def\ie{\textit{i.\,e.}}                                      % i.e. (kursiv) \ie
\def\eg{\textit{e.\,g.}}                                      % e.g. (kursiv) \eg
\def\cf{\textit{cf.}}

\def\cm-2{cm$^{-2}$}

\def\HII{\hbox{H\,{\sc ii}}}

\def\ros{{\it ROSAT}}
\def\ein{{\it Einstein}}
\def\chandra{{\it Chandra}}

\def\xmm{{XMM-Newton}}

\def\m31{\object{M~31}}

\def\la{\mathrel{\mathchoice {\vcenter{\offinterlineskip\halign{\hfil
$\displaystyle##$\hfil\cr<\cr\sim\cr}}}
{\vcenter{\offinterlineskip\halign{\hfil$\textstyle##$\hfil\cr
<\cr\sim\cr}}}
{\vcenter{\offinterlineskip\halign{\hfil$\scriptstyle##$\hfil\cr
<\cr\sim\cr}}}
{\vcenter{\offinterlineskip\halign{\hfil$\scriptscriptstyle##$\hfil\cr
<\cr\sim\cr}}}}}                                            % ungef�r kleiner \la
\def\ga{\mathrel{\mathchoice {\vcenter{\offinterlineskip\halign{\hfil
$\displaystyle##$\hfil\cr>\cr\sim\cr}}}
{\vcenter{\offinterlineskip\halign{\hfil$\textstyle##$\hfil\cr
>\cr\sim\cr}}}
{\vcenter{\offinterlineskip\halign{\hfil$\scriptstyle##$\hfil\cr
>\cr\sim\cr}}}
{\vcenter{\offinterlineskip\halign{\hfil$\scriptscriptstyle##$\hfil\cr
>\cr\sim\cr}}}}}                                            % ungef�r gr�er \ga

\def\subsun{\mbox{$_{\odot}$}}

\def\arcmin{\hbox{$^\prime$}}
\def\arcsec{\hbox{$^{\prime\prime}$}}

\def\farcm{\hbox{$.\mkern-4mu^\prime$}}
\def\farcs{\hbox{$.\!\!^{\prime\prime}$}}

\def\XLPt{XMM\ LP-total}

\def\num{\hbox{N$^{\underline{o}}$}}

\newcommand{\ergcm}[1]{$\times 10^{#1}$ \hbox{erg cm$^{-2}$ s$^{-1}$}}
\newcommand{\oergcm}[1]{$10^{#1}$ erg cm$^{-2}$ s$^{-1}$}
\newcommand{\ergs}[1]{$\times 10^{#1}$ \hbox{erg s$^{-1}$}}
\newcommand{\oergs}[1]{$10^{#1}$ erg s$^{-1}$}
\newcommand{\hcm}[1]{$\times 10^{#1}$ cm$^{-2}$}

\newcommand{\oexpo}[1]{$10^{#1}$}

\def\cm-2{cm$^{-2}$}

\newcommand{\nh}{\hbox{$N_{\rm H}$}}

\newcommand{\mr}{\mathrm}
\newcommand{\lb}{\left}
\newcommand{\rb}{\right}
\begin{document}
\originalTeX

   \title{The deep XMM-Newton Survey of M~31
   \thanks{Based on 
   observations obtained with XMM-Newton, an ESA science mission with 
   instruments and contributions directly funded by ESA Member States 
   and NASA.}\fnmsep
   \thanks{Tables~5 and~8 are only available in electronic form
    at the CDS via anonymous ftp to cdsarc.u-strasbg.fr (130.79.128.5)
    or via http://cdsweb.u-strasbg.fr/cgi-bin/qcat?J/A+A/ }
}

   \author{H.~Stiele\inst{1} \and
           W.~Pietsch\inst{1} \and 
           F.~Haberl\inst{1} \and
	   D.~Hatzidimitriou\inst{2,}\inst{3} 
	   R.~Barnard\inst{4,}\inst{5} \and
	   B.~F.~Williams\inst{6} \and
	   A.~K.~H.~Kong\inst{7} \and 
	   U.~Kolb\inst{4} 
        }
\institute{Max-Planck-Institut f\"ur extraterrestrische Physik, Giessenbachstra{\ss}e,
           85748 Garching, Germany 
\and Department of Astrophysics, Astronomy and Mechanics, Faculty of
Physics, University of Athens, Panepistimiopolis, 15784 Zografos,
Athens, Greece
\and IESL, Foundation for Research and Technology, 71110 Heraklion, Greece
\and Department of Physics and Astronomy, The Open University, Walton Hall, Milton Keynes, MK7 6AA, UK
\and Harvard-Smithsonian Center for Astrophysics, 60 Garden Street, Cambridge, MA 02138, USA
\and Department of Astronomy, Box 351580, University of Washington, Seattle, WA 98195, USA
\and Institute of Astronomy and Department of Physics, National Tsing Hua University, Hsinchu 30013, Taiwan
}

     \offprints{H.~Stiele, \email{holger.stiele@brera.inaf.it}}

   \date{Received  / Accepted }
   \titlerunning{The deep XMM-Newton Survey of M~31}

   	\abstract{}
        {The largest Local Group spiral galaxy, \m31, has been completely imaged for the first time obtaining a luminosity lower limit $\sim$\oergs{35} in the 0.2--4.5\,keV band.  Our \xmm\ EPIC survey combines archival observations along the major axis, from June 2000 to July 2004, with observations taken between June 2006 and February 2008 that cover the remainder of the $\mr{D}_{25}$ ellipse. The main goal of the paper is a study of the X-ray source population of \m31.}
        {An X-ray  catalogue of 1\,897  sources was created; 914 were detected for the first time. Source classification and identification were based on X-ray hardness ratios, spatial extent of the sources, and by cross correlating with catalogues in the X-ray, optical, infrared and radio wavelengths. We also analysed the long-term variability of the  X-ray sources; this variability allows us to distinguish between X-ray binaries and active galactic nuclei (AGN).  Furthermore, supernova remnant classifications of previous studies that did not use long-term variability as a classification criterion, could be validated. Inclusion of previous \chandra\ and \ros\ observations in the long-term variability study allowed us to detect additional transient or at least highly variable sources, which are good candidate X-ray binaries.}
        {Fourteen of the 30 supersoft source (SSS) candidates represent supersoft emission of optical novae. Many of the 25 supernova remnants (SNRs) and 31 SNR candidates lie within the 10\,kpc dust ring and other star forming regions in \m31;  this connection between SNRs and star forming regions implies that most of the remnants originate in type II supernovae. The brightest sources in X-rays in \m31\ belong to the class of X-ray binaries (XRBs). Ten low mass XRBs (LMXBs) and 26 LMXB candidates were identified based on their temporal variability. In addition 36 LMXBs and 17 LMXB candidates were identified due to correlations with globular clusters and globular cluster candidates. From optical and X-ray colour-colour diagrams, possible high mass XRB (HMXB) candidates were selected. Two of these candidates have an X-ray spectrum as is expected for an HMXB containing a neutron star primary.}
        {While our survey has greatly improved our understanding of the X-ray source populations in \m31, at this point 65\% of the sources can still only be classified as ``hard" sources, \ie\ it is not possible to decide whether these sources are X-ray binaries or Crab-like supernova remnants in \m31, or X-ray sources in the background. Deeper observations in X-ray and at other wavelengths would help classify these sources.}
	
 
\maketitle

\section{Introduction}
\label{Sec:Intro}
Our nearest neighbouring large spiral galaxy, the Andromeda galaxy, also known as \m31\ or \object{NGC~224}, is an ideal target for an X-ray source population study of a galaxy similar to the Milky Way. Its proximity \citep[distance 780 kpc,][]{1998AJ....115.1916H,1998ApJ...503L.131S} and the moderate Galactic foreground absorption \citep[\nh = 7\hcm{20}, ][]{1992ApJS...79...77S} allow a detailed study of source populations and individual sources.

After early detections  of \m31\ with X-ray detectors mounted on rockets \citep[\eg\ ][]{1974ApJ...190..285B} and the {\it Uhuru} satellite \citep[][]{1974ApJS...27...37G}, the imaging X-ray optics flown on the \ein\ X-ray observatory permitted the resolution of individual X-ray sources in \m31 for the first time. In the entire set of \ein\ imaging observations of \m31, \citet[][hereafter TF91]{1991ApJ...382...82T} found 108 individual X-ray sources brighter than $\sim6.4$\ergs{36}, of which 16 sources showed variability \citep{1979ApJ...234L..45V,1990ApJ...356..119C}.

In July 1990, the bulge region of \m31\ was observed with the \ros\ High Resolution Imager (HRI) for $\sim 48$\,ks.\@ \citet[][hereafter PFJ93]{1993ApJ...410..615P} reported 86 sources brighter than $\sim1.8$\ergs{36} in this observation. Of the \ros\ HRI sources located within 7\farcm5 of the nucleus, 18 sources were found to vary when compared to previous \ein\ observations and about three of the sources may be ``transients".
Two deep PSPC (Position Sensitive Proportional Counter) surveys of \m31\ were performed with \ros, the first in July 1991 \citep[][hereafter SHP97]{1997A&A...317..328S}, the second in July/August 1992 \citep[][hereafter SHL2001]{2001A&A...373...63S}. In total 560 X-ray sources were detected in the field of \m31; of these, 491 sources were not detected in previous \ein\ observations. In addition, a comparison with the results of the \ein\ survey revealed long term variability in 18 sources, including 7 possible transients. Comparing the two \ros\ surveys, 34 long term variable sources and 8 transient candidates were detected. The derived luminosities of the detected \m31\ sources ranged from 5\ergs{35} to 5\ergs{38}.\@ Another important result obtained with \ros\ was the establishment of supersoft sources (SSSs) as a new class of \m31\ X-ray sources \citep[\cf\ ][]{1999A&A...344..459K} and the identification of the first SSS with an optical nova in \m31\ \citep{2002A&A...389..439N}. 

\citet{2000ApJ...537L..23G} reported on first observations of the nuclear region of \m31\ with \chandra. They found that the nuclear source has an unusual X-ray spectrum compared to the other point sources in \m31. \citet{2002ApJ...577..738K} report on eight \chandra\ ACIS-I observations taken between 1999 and 2001, which cover the central $\sim 17\arcmin\!\times\!17\arcmin$ region of \m31. They detected 204 sources, of which $\sim$50\% are variable on timescales of months and 13 sources were classified as transients. \citet{2002ApJ...578..114K} detected 142 point sources ($L_X=2\!\times\!10^{35}$ to 2\ergs{38} in the 0.1--10\,keV band) in a 47\,ks \chandra/HRC observation of the central region of \m31. A comparison with a \ros\ observation taken 11\,yr earlier, showed that 46$\pm$26\% of the sources with $L_X>5$\ergs{36} are variable. Three different \m31\ disc fields, consisting of different stellar population mixtures, were observed by \chandra. \citet{2002ApJ...570..618D} investigated bright X-ray binaries (XRBs) in these fields, while \citet{2004ApJ...610..247D} examined the populations of supersoft sources (SSSs) and quasisoft sources (QSSs), including observations of the central field. Using \chandra\ HRC observations, \citet{2004ApJ...609..735W} measured the mean fluxes and long-term time variability of 166 sources detected in these data. \citet{2007A&A...468...49V} used \chandra\ data to examine the low mass X-ray binaries (LMXBs) in the bulge of \m31. Good candidates for LMXBs are the so-called transient sources. Studies of transient sources in \m31\ are presented in numerous papers, e.\,g.~\citet{2006ApJ...643..356W}, \citet[][ hereafter TPC06]{2006ApJ...645..277T}, \citet{2005ApJ...632.1086W}, \citet[][ hereafter WGM06]{2006ApJ...637..479W}, and \citet{2008A&A...489..707V}. 

Using \xmm\ and \chandra\ data, \citet{2004ApJ...616..821T} detected 43 X-ray sources coincident with globular cluster candidates from various optical surveys. They studied their spectral properties, time variability and log\,N-log\,S relations. 

\citet{2001A&A...378..800O} used \xmm\ Performance Verification observations to study the variability of X-ray sources in the central region of \m31. They found 116 sources brighter than a limiting luminosity of 6\ergs{35} and examined the $\sim60$ brightest sources for periodic and non-periodic variability. At least 15\% of these sources appear to be variable on a time scale of several months. \citet{2003A&A...411..553B} used \xmm\ to study the X-ray binary RX J0042.6+4115 and suggested it as a Z-source. \citet{2006ApJ...643..844O} studied the population of SSSs and QSSs with \xmm. 
Recently, \citet{2008ApJ...676.1218T} reported the discovery of 217s pulsations in the bright persistent SSS XMMU~J004252.5+411540.\@ \citet[][hereafter SBK2009]{2009A&A...495..733S} presented the results of a complete spectral survey of the 335 X-ray point sources they detected in five \xmm\ observations located along the major axis of \m31.\@ They obtained background subtracted spectra and lightcurves for each of the 335 X-ray sources. Sources with more than 50 source counts were individually spectrally fitted. In addition, they selected 18 HMXB candidates, based on a power law photon index of $0.8\!\le\!\Gamma\!\le\!1.2$.

\citet[][ hereafter PFH2005]{2005A&A...434..483P} prepared a catalogue of \m31\
point-like X-ray sources analysing all observations available at that time in the \xmm\ archive which overlap at least in part with the optical $\mr{D}_{25}$ extent of the galaxy. 
In total, they detected 856 sources. The central part of the galaxy was covered four times with a separation of the observations of about half a year starting in June 2000. PFH2005 only gave source properties derived from an analysis of the combined observations of the central region. Source identification and classification were based on hardness ratios, and correlations with sources in other wavelength regimes.  In follow-up work, (i) \citet[][]{2005A&A...430L..45P} searched for X-ray burst sources in globular cluster (GlC) sources and candidates and identified two X-ray bursters and a few more candidates, while (ii) \citet[][ hereafter PFF2005]{2005A&A...442..879P} searched for correlations with optical novae. They identified 7 SSSs and 1 symbiotic star from the catalogue of PFH2005 with optical novae, and identified anadditional \xmm\ source with an optical nova. This work was continued and extended on archival \chandra\ HRC-I and ACIS-I observations by \citet[][ hereafter PHS2007]{2007A&A...465..375P}. 

\citet[][hereafter SPH2008]{2008A&A...480..599S} presented a time variability analysis of all of the \m31\ central sources. 
They detected 39 sources not reported at all in PFH2005. 21 sources were detected in the July 2004 monitoring observations of the low mass X-ray binary RX J0042.6+4115 (PI Barnard), which became available in the meantime. Six sources, which were classified as ``hard" sources by PFH2005, show distinct time variability and hence are classified as XRB candidates in SPH2008. The SNR classifications of three other sources from PFH2005 had to be rejected due to the distinct time variability found by SPH2008. \citet{2009A&A...500..769H} reported on the first two SSSs ever discovered in the \m31\ globular cluster system, and \citet{2009A&A...498L..13H} discussed the very short supersoft X-ray state of the classical nova M31N 2007-11a. A comparative study of supersoft sources detected with \ros, \chandra\ and \xmm, examining their long-term variability, was presented by \citet{2010AN....331..212S}.

An investigation of the log\,N-log\,S relation of sources detected in the 2.0--10.0\,keV range will be presented in a forthcoming paper (Stiele et al. 2011 in prep.). In this work the contribution of background objects and the spatial dependence of the log\,N-log\,S relations for sources of \m31\ is studied.

In this paper we report on the large \xmm\ survey of \m31, which covers the entire $\mr{D}_{25}$ ellipse of \m31, for the first time, down to a limiting luminosity of $\sim$\oergs{35} in the 0.2--4.5\,keV band. In Sect.\,\ref{sec:obsana} information about the observations used is provided. The analysis of the data is presented in Sect.\,\ref{Sec:analys}. Section \,\ref{Sec:coim} presents the combined colour image of all observations used. The source catalogue of the deep \xmm\ survey of \m31\ is described in Sect.\,\ref{Sec:srccat}. 
The results of the temporal variability analysis are discussed in Sect.\,\ref{Sec:var}. Cross-correlations with other \m31\ X-ray catalogues are discussed in Sect.\,\ref{Sec:CrossX-ray}, while Sect.\,\ref{SEC:CCow} discusses cross-correlations with catalogues at other wavelengths. Our results related to foreground stars and background sources in the field of \m31\ are presented in Sect.\,\ref{Sec:fgback}. 
Individual source classes belonging to M31 are discussed in Sect.\,\ref{Sec:Srcsm31}. 
We draw our conclusions in Sect.\,\ref{Sec:Concl}.

\begin{table*}
\scriptsize
\begin{center}
\caption{A selection of important X-ray surveys of \m31.}
\begin{tabular}{lcrcll}
\hline\noalign{\smallskip}
\hline\noalign{\smallskip}
Paper & S$^{+}$ & \#ofSrc$^{*}$ & \multicolumn{1}{c}{L$_{X}^{\dagger}$} & field & comments \\
& & & erg cm$^{-2}$ s$^{-1}$& \\
\noalign{\smallskip}
\hline\noalign{\smallskip}
\protect{\citet[][TF91]{1991ApJ...382...82T}} & E & 108 & $6.4\!\times\!10^{36}$--$1.3\!\times\!10^{38}$ & entire set of \ein\ & 16 sources showed variability\\
& & & (0.2--4\,keV) &imaging observations & \\
\protect{\citet[][PFJ93]{1993ApJ...410..615P}} & R (HRI) & 86 & $\ga1.8\!\times\!10^{36}$ & bulge region & 18 sources variable; $\sim$3 transients\\
& & & (0.2--4\,keV) & &\\
\protect{\citet{1997A&A...317..328S,2001A&A...373...63S}} & R (PSPC) & 560 & $5\!\times\!10^{35}$--$5\!\times\!10^{38}$ &
whole galaxy & two deep surveys\\
(SPH97, SHL2001) & & & (0.1--2.4\,keV) & & 491 sources not detected with \ein \\
& & & & & 11 sources variable, 7 transients compared to \ein \\
& & & & & 34 sources variable, 8 transients between \ros\ surveys \\
\protect{\citet{2001A&A...378..800O}} & X & 116 & $\ga6\!\times\!10^{35}$ & centre & examined the $\sim60$ brightest sources for variability\\
& & & (0.3--12\,keV) & & \\
\protect{\citet{2002ApJ...577..738K}} & C (ACIS-I) & 204 & $\ga2\!\times\!10^{35}$ & central $\sim 17\arcmin\!\times\!17\arcmin$ & observations between 1999 and 2001\\
& & & (0.3--7\,keV) & & $\sim$50\% of the sources are variable, 13 transients\\
\protect{\citet{2002ApJ...578..114K}} & C (HRC) & 142 & $2\!\times\!10^{35}$--$2\!\times\!10^{38}$ & centre & one 47\,ks observation; 46$\pm$26\% of the sources \\
& & & (0.1--10\,keV) & & with $L_X>5$\ergs{36} are variable\\
\protect{\citet{2002ApJ...570..618D}} & C (ACIS-I/S) & 28 & $5\!\times\!10^{35}$--$3\!\times\!10^{38}$ & 3 disc fields & bright X-ray binaries\\
& & & (0.3--7\,keV) & & \\
\protect{\citet{2004ApJ...610..247D}} & C (ACIS-S S3) & 33 & & 3 disc fields + centre & supersoft sources and quasisoft sources\\
\protect{\citet{2004ApJ...609..735W}} & C (HRC) & 166 & $1.4\!\times\!10^{36}$--$5\!\times\!10^{38}$ & major axis + centre & $\ga$25\% showed significant variability\\
& & & (0.1--10\,keV) & & \\
\protect{\citet{2004ApJ...616..821T}} & C, X & 43 & $\sim10^{35}$--$\sim10^{39}$ & major axis + centre & globular cluster study \\
& & & (0.3--10\,keV) & & \\
\protect{\citet[][PFH2005]{2005A&A...434..483P}} & X & 856 & $4.4\!\times\!10^{34}$--$2.8\!\times\!10^{38}$ & major axis + centre & source catalogue\\
& & & (0.2--4.5\,keV) & & \\
\protect{\citet[][PFF2005]{2005A&A...442..879P}} & C, R, X & 21 & $\sim10^{35}$--$\sim10^{38}$ & centre & correlations with optical novae\\
& & & (0.2--1\,keV) & & \\
\protect{\citet{2006ApJ...643..844O}} & C, X & 42 & $6\!\times\!10^{35}$--$\sim10^{39}$ & major axis + centre & supersoft sources and quasisoft sources\\
& & & (0.2--2\,keV) & & \\
& & & (0.3--10\,keV) & & \\
\protect{\citet[][PHS2007]{2007A&A...465..375P}} & C, X & 46 & $\sim10^{35}$--$\sim10^{38}$ & centre & correlations with optical novae\\
& & & (0.2--1\,keV) & & \\
\protect{\citet{2007A&A...468...49V}} & C & 263 & $5\!\times\!10^{33}$--$1.5\!\times\!10^{38}$ & bulge region & low mass X-ray binary study\\
& & & (0.5--8\,keV) & & \\
\protect{\citet[][SPH2008]{2008A&A...480..599S}} & X & 39 & $7\!\times\!10^{34}$--$6\!\times\!10^{37}$ & centre & re-analysis of archival and new 2004 observations\\
& & 300 & $4.4\!\times\!10^{34}$--$2.8\!\times\!10^{38}$ & & time variability analysis; 149 sources with a significance\\
& & & (0.2--4.5\,keV) & & for variability $>$3; 6 new X-ray binary candidates,\\
& & & & & 3 supernova remnant classifications were rejected\\
\protect{\citet[][SBK2009]{2009A&A...495..733S}} & X & 335 & $\sim10^{34}$--$\sim10^{39}$ & 5 fields along & background subtracted spectra and lightcurves for\\
& & & (0.3--10\,keV) & major axis & each source; 18 HMXB candidates, selected from their\\
& & & & & power law photon index\\
\protect{\citet{2010AN....331..212S}} & X & 40 & & whole galaxy & supersoft sources; comparing \ros, \chandra\ and\\
& & & & & \xmm\ catalogues\\
\noalign{\smallskip}
\hline
\noalign{\smallskip}
\end{tabular}
\label{Tab:VarSNRs1}
\end{center}
Notes:\\
$^{ +~}$: X-ray satellite(s) on which the study is based: E for \ein, R for \ros, C for \chandra, and X for \xmm\ (EPIC)\\
$^{ *~}$: Number of sources\\
$^{ \dagger~}$: observed luminosity range in the indicated energy band, assuming a distance of 780\,kpc to \m31
\normalsize
\end{table*}

\section{Observations}
\label{sec:obsana}
\begin{figure}
\resizebox{\hsize}{!}{\includegraphics[clip]{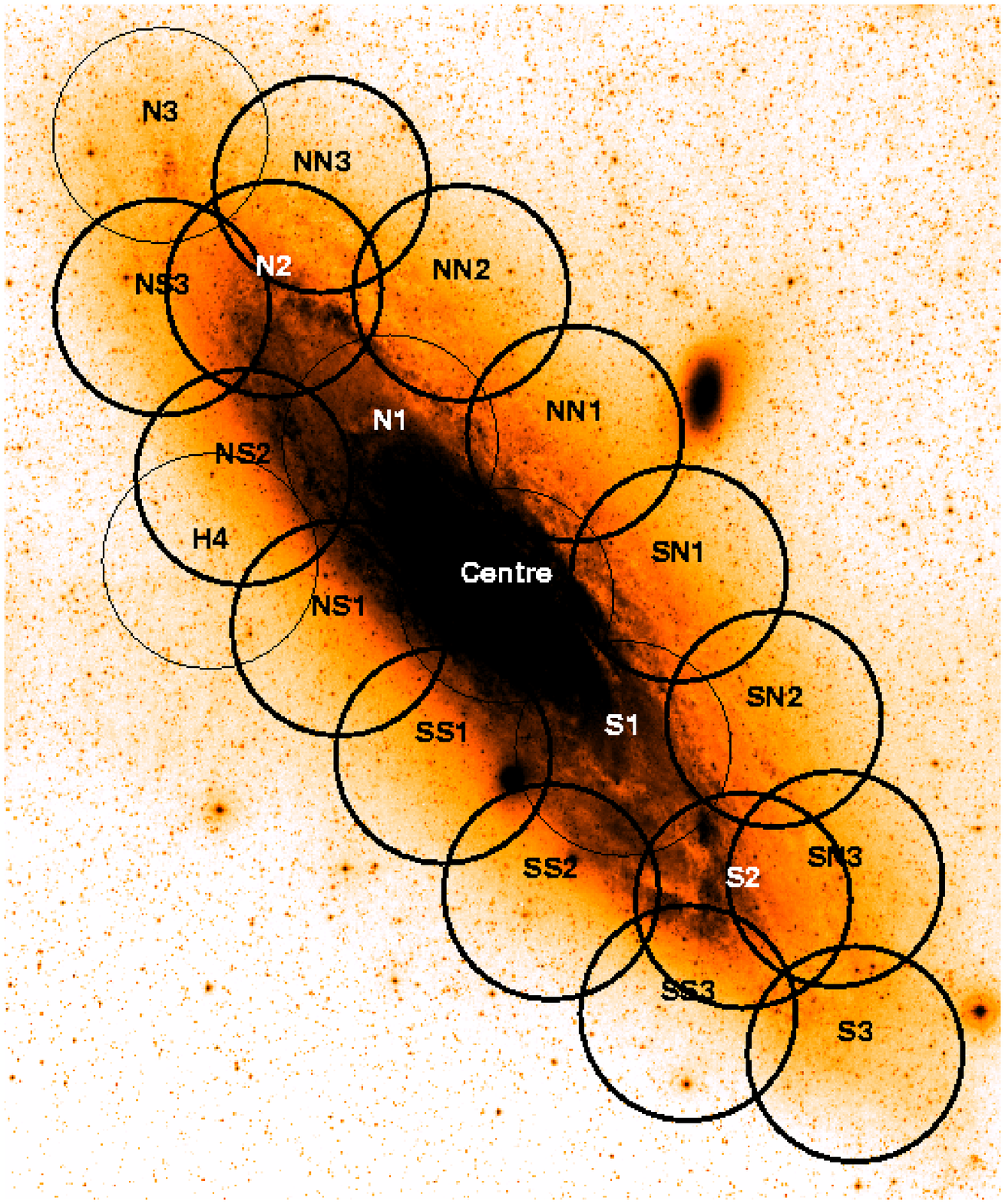}}
\caption{A deep optical image of \m31\ (private communication: V.~Burwitz) overplotted with the \xmm\ fields of the survey. The area covered by individual EPIC observations is approximated by circles with 14 arcmin radius. Fields observed in the ``Deep \xmm\ Survey of \m31'' are marked with thicker lines. For presentation purposes, the ToO observation and the observations of RX J0042.6+4115 are omitted.}
\label{fig:deepsurveyfields}
\end{figure}

Figure~\ref{fig:deepsurveyfields} shows the layout of the individual \xmm\ observations over the field of \m31. The observations of the ``Deep \xmm\ Survey of \m31'' (PI Pietsch) mainly point at the outer parts of \m31, while the area along the major axis is covered by archival \xmm\ observations (PIs Watson, Mason, Di Stefano). 
To treat all data in the same way, we re-analysed all archival \xmm\ observations of \m31, which were used in \citet{2005A&A...434..483P}.\@ In addition we included an \xmm\ target of opportunity (ToO) observation of source CXOM31~J004059.2+411551 and the four observations of source RX J0042.6+4115 (PI Barnard).

All observations of the ``Deep \xmm\ Survey of \m31'' and the ToO observation were taken between June 2006 and February 2008.\@ All other observations were available via the \xmm\ Data Archive\footnote{\url{http://xmm.esac.esa.int/xsa/}} and were taken between June 2000 and July 2004.    

The journal of observations is given in Table~\ref{tab:observations}. It includes the \m31\ field name (Column~1), the identification number (2) and date (3) of the observation and the pointing direction (4, 5), while col.~6 contains the systematic offset (see Sect.\,\ref{SubSec:AstCorr}). For each EPIC camera the filter used and the exposure time after screening for high background is given (see Sect.\,\ref{sec:Screening}).

\begin{table*}
\scriptsize
\begin{center}
\caption[]{\xmm\ log of the {\em Deep Survey} and archival \m31\ observation overlapping with the optical $D_{25}$ ellipse.\label{tab:observations}}
\begin{tabular}{llllrrrlrlrlr}
\hline\noalign{\smallskip}
\hline\noalign{\smallskip}
\multicolumn{2}{c}{M 31 field} & \multicolumn{1}{c}{Obs. id.} &\multicolumn{1}{c}{Obs. dates} &
\multicolumn{2}{c}{Pointing direction} & \multicolumn{1}{c}{Offset~$^*$} & \multicolumn{2}{c}{EPIC PN} & 
\multicolumn{2}{c}{EPIC MOS1} & \multicolumn{2}{c}{EPIC MOS2}  \\ 
\noalign{\smallskip}
& & & & \multicolumn{2}{c}{RA/Dec (J2000)} & \multicolumn{1}{c}{} 
& \multicolumn{1}{c}{Filter$^{+}$}  & \multicolumn{1}{c}{$T_{exp}^{\dagger}$}
& \multicolumn{1}{c}{Filter$^{+}$}  & \multicolumn{1}{c}{$T_{exp}^{\dagger}$}
& \multicolumn{1}{c}{Filter$^{+}$}  & \multicolumn{1}{c}{$T_{exp}^{\dagger}$}\\
\noalign{\smallskip}
\multicolumn{2}{c}{(1)} & \multicolumn{1}{c}{(2)} & \multicolumn{1}{c}{(3)} & 
\multicolumn{1}{c}{(4)} & \multicolumn{1}{c}{(5)} & \multicolumn{1}{c}{(6)} & 
\multicolumn{1}{c}{(7)} & \multicolumn{1}{c}{(8)} & \multicolumn{1}{c}{(9)} & 
\multicolumn{1}{c}{(10)} & \multicolumn{1}{c}{(11)} & \multicolumn{1}{c}{(12)}\\
\noalign{\smallskip}\hline\noalign{\smallskip}
Centre 1 & (c1) & 0112570401 & 2000-06-25    & 0:42:36.2 & 41:16:58 & $-1.9,+0.1$ & medium  & 23.48(23.48) & medium  & 29.64(29.64) &medium  & 29.64(29.64) \\
Centre 2 & (c2) & 0112570601 & 2000-12-28    & 0:42:49.8 & 41:14:37 & $-2.1,+0.2$ & medium  &  5.82( 5.82) & medium  & 6.42( 6.42) &medium  & 6.42( 6.42) \\
Centre 3 & (c3) & 0109270101 & 2001-06-29    & 0:42:36.3 & 41:16:54 & $-3.2,-1.7$ & medium  & 21.71(21.71) & medium  & 23.85(23.85) &medium  & 23.86(23.86) \\
N1 &  (n1) & 0109270701 & 2002-01-05    & 0:44:08.2 & 41:34:56 & $-0.3,+0.7$ & medium  & 48.31(48.31) & medium  & 55.68(55.68) &medium  & 55.67(55.67) \\
Centre 4 & (c4) & 0112570101 & 2002-01-06/07 & 0:42:50.4 & 41:14:46 & $-1.0,-0.8$ & thin	& 47.85(47.85) & thin	 & 52.87(52.87) &thin	 & 52.86(52.86) \\
S1 & (s1) & 0112570201 & 2002-01-12/13 & 0:41:32.7 & 40:54:38 & $-2.1,-1.7$ & thin	& 46.75(46.75) & thin	 & 51.83(51.83) &thin	 & 51.84(51.84) \\
S2 & (s2) & 0112570301 & 2002-01-24/25 & 0:40:06.0 & 40:35:24 & $-1.1,-0.3$ & thin	& 22.23(22.23) & thin	 & 24.23(24.23) &thin	 & 24.24(24.24) \\
N2 & (n2) & 0109270301 & 2002-01-26/27 & 0:45:20.0 & 41:56:09 & $-0.3,-1.5$ & medium	& 22.73(22.73) & medium	 & 25.22(25.22) &medium	 & 25.28(25.28) \\
N3 & (n3) & 0109270401 & 2002-06-29/30 & 0:46:38.0 & 42:16:20 & $-2.3,-1.7$ & medium	& 39.34(39.34) & medium	 & 43.50(43.50) &medium	 & 43.63(43.63) \\
H4  & (h4) & 0151580401 & 2003-02-06 & 0:46:07.0 & 41:20:58 & $+0.3,+0.0$ & medium	& 10.14(10.14) & medium	 & 12.76(12.76) &medium	 & 12.76(12.76) \\
RX~1 & (b1)$^{\ddagger}$ & 0202230201 & 2004-07-16 & 0:42:38.6 & 41:16:04 & $-1.3,-1.2$ & medium	& 16.32(16.32) & medium & 19.21(19.21) &medium	 & 19.21(19.21) \\
RX~2 & (b2) & 0202230301 & 2004-07-17 & 0:42:38.6 & 41:16:04 & $-1.0,-0.9$ & medium	& 0.0(0.0) & medium & 0.0(0.0) &medium & 0.0(0.0) \\           
RX~3 & (b3)$^{\ddagger}$ & 0202230401 & 2004-07-18 & 0:42:38.6 & 41:16:04 & $-1.7,-1.5$ & medium  & 12.30(12.30) & medium  & 17.64(17.64) &medium	 & 17.68(17.68) \\
RX~4 & (b4)$^{\ddagger}$ & 0202230501 & 2004-07-19 & 0:42:38.6 & 41:16:04 & $-1.4,-1.8$ & medium  & 7.94(7.94) & medium  & 10.12(10.12) &medium	  & 10.13(10.13) \\
S3   & (s3) & 0402560101 & 2006-06-28 & 0:38:52.8 & 40:15:00 & $-3.1,-3.0$ & thin	&  4.99(4.99) & medium	 &  6.96(6.96) &medium	 &   6.97(6.97)\\
SS1 & (ss1) & 0402560201 & 2006-06-30 & 0:43:28.8 & 40:55:12 & $-4.4,-3.7$ & thin	& 14.07(9.57) & medium	 & 24.56(10.65) &medium	 & 24.58(10.66) \\
SN1 & (sn1) & 0402560301 & 2006-07-01 & 0:40:43.2 & 41:17:60 & $-2.7,-1.5$ & thin	& 41.23(35.42) & medium	 & 47.60(39.40) &medium	 & 47.64(39.44) \\
SS2 & (ss2) & 0402560401 & 2006-07-08 & 0:42:16.8 & 40:37:12 & $-1.2,-1.3$ & thin	& 21.64(9.92) & medium	 & 25.59(11.04) &medium	 & 25.64(11.05) \\
SN2 & (sn2) & 0402560501 & 2006-07-20 & 0:39:40.8 & 40:58:48 & $-0.8,-0.7$ & thin	& 48.79(21.45) & medium	 & 56.13(23.85) &medium	 & 56.17(23.86) \\
SN3 & (sn3) & 0402560701 & 2006-07-23 & 0:39:02.4 & 40:37:48 & $-0.9,-2.0$ & thin	& 23.80(15.43) & medium	 & 28.02(17.16) &medium	 & 28.04(17.17) \\
SS3 & (ss3) & 0402560601 & 2006-07-28 & 0:40:45.6 & 40:21:00 & $-1.8,-1.7$ & thin	& 27.77(20.22) & medium	 & 31.92(22.49) &medium	 & 31.94(22.5) \\
S2~& (s21) & 0402560801 & 2006-12-25 & 0:40:06.0 & 40:35:24 & $-1.6,-0.7$ & thin	& 39.12(39.12) & medium	 & 45.19(45.19) &medium	 & 45.21(45.21) \\
NN1 & (nn1) & 0402560901 & 2006-12-26 & 0:41:52.8 & 41:36:36 & $-1.5,-1.5$ & thin	& 37.9(37.9) & medium	 & 43.08(43.08) &medium	 & 43.1(43.1) \\
NS1 & (ns1) & 0402561001 & 2006-12-30 & 0:44:38.4 & 41:12:00 & $-1.0,-1.3$ & thin	& 45.11(45.11) & medium	 & 50.9(50.9) &medium	 & 50.93(50.93) \\
NN2 & (nn2) & 0402561101 & 2007-01-01 & 0:43:09.6 & 41:55:12 & $-0.0,-1.2$ & thin	& 41.73(41.73) & medium	 & 46.45(46.45) &medium	 & 46.47(46.47) \\
NS2 & (ns2) & 0402561201 & 2007-01-02 & 0:45:43.2 & 41:31:48 & $-2.3,-1.7$ & thin	& 34.96(34.96) & medium	 & 40.55(40.55) &medium	 & 40.58(40.58) \\
NN3 & (nn3) & 0402561301 & 2007-01-03 & 0:44:45.6 & 42:09:36 & $-1.4,-0.7$ & thin	& 31.04(31.04) & medium	 & 34.81(34.81) &medium	 & 34.81(34.81) \\
NS3 & (ns3) & 0402561401 & 2007-01-04 & 0:46:38.4 & 41:53:60 & $-2.1,+0.3$ & thin	& 39.41(39.41) & medium	 & 45.50(45.50) &medium	 & 45.52(45.52) \\
N2~& (n21) & 0402561501 & 2007-01-05 & 0:45:20.0 & 41:56:09 & $-2.6,-1.3$ & thin	& 37.18(37.18) & medium	 & 41.98(41.98) &medium	 & 42.03(42.03) \\
SS1 & (ss11) & 0505760201 & 2007-07-22 & 0:43:28.8 & 40:55:12 & $-2.5,-2.6$ & thin	& 30.07(23.90) & medium	 & 34.01(26.70) &medium	 & 34.02(26.72) \\
S3~& (s31) & 0505760101 & 2007-07-24 & 0:38:52.8 & 40:15:00 & $-1.8,-1.0$ & thin	& 21.86(15.74) & medium	 & 24.74(17.65) &medium	 & 24.74(17.65) \\
CXOM31& (sn11)$^{\diamond}$ & 0410582001 & 2007-07-25 & 0:40:59.2 & 41:15:51 & $-1.2,-0.3$ & thin	& 11.27(11.27) & medium	 & 14.01(14.01) &medium	 & 14.02(14.02) \\
SS3 & (ss31) & 0505760401 & 2007-12-25 & 0:40:45.6 & 40:21:00 & $-1.0,+0.1$ & thin	& 23.56(22.82) & medium	 & 28.18(25.8) &medium	 & 28.2(25.82) \\
SS2 & (ss21) & 0505760301 & 2007-12-28 & 0:42:16.8 & 40:37:12 & $+1.3,-0.1$ & thin	& 35.28(35.28) & medium	 & 40.00(40.00) &medium	 & 40.01(40.01) \\
SN3 & (sn31) & 0505760501 & 2007-12-31 & 0:39:02.4 & 40:37:48 & $-1.6,-1.3$ & thin	& 24.26(24.26) & medium	 & 28.77(28.77) &medium	 & 28.78(28.78) \\
S3 &  (s32) & 0511380101 & 2008-01-02 & 0:38:52.8 & 40:15:00 & $-1.7,-3.3$ & thin	& 38.31(38.31) & medium	 & 44.92(44.92) &medium	 & 44.95(44.95) \\
SS1 & (ss12) & 0511380201 & 2008-01-05 & 0:43:28.8 & 40:55:12 & $-0.9,-1.4$ & thin	&  8.85( 8.85) & medium	 & 11.28(11.28) &medium		 & 11.29(11.29) \\
SN2 & (sn21) & 0511380301 & 2008-01-06 & 0:39:40.8 & 40:58:48 & $-0.2,-0.4$ & thin	& 24.79(24.79) & medium	 & 29.28(29.28) &medium	 & 29.29(29.29) \\
SS1 & (ss13) & 0511380601 & 2008-02-09 & 0:43:28.8 & 40:55:12 & $-0.8,-1.8$ & thin	& 13.35(13.35) & medium	 & 15.07(15.07) &medium	 &  15.08(15.08)\\
\noalign{\smallskip}
\hline
\noalign{\smallskip}
\end{tabular}
\end{center}
\scriptsize{Notes:\\
$^{ *~}$: Systematic offset in RA and Dec in arcsec determined from correlations with 2MASS, USNO-B1, LGGS and \chandra\ catalogues \\
$^{ +~}$: All observations in full frame imaging mode\\
$^{ {\dagger}~}$: Exposure time in units of ks after screening for high
background used for detection, for colour image in brackets\\
$^{ {\ddagger}~}$: Combination of the three observations is called b (see text), RX denotes RX J0042.6+4115 \\
$^{ {\diamond}~}$: CXOM31 denotes CXOM31~J004059.2+411}
\normalsize
\end{table*}

\section{Data analysis}
\label{Sec:analys}
In this section, the basic concepts of the X-ray data reduction and source detection processes are described.

\subsection{Screening for high background}
\label{sec:Screening}
The first step was to exclude times of increased background, due to soft proton flares. Most of these times are located at the start or end of an orbit window. We selected good time intervals (GTIs) -- intervals where the intensity was lower than a certain threshold -- using 7--15\,keV light curves constructed from source-free regions of each observation. 
The GTIs with PN and MOS data were determined from the higher statistic PN light curves. Outside the PN time coverage, GTIs were determined from the combined MOS light curves. For each observation, the limiting thresholds for the count rate were adjusted individually; this way we avoided cutting out  short periods (up to a few hundred seconds) of marginally increased background. Short periods of low background, which were embedded within longer periods of high background, were omitted. For most observations, the PN count rate thresholds were 2--8\,cts\,ks$^{-1}$\,arcmin$^{-2}$.

As many of the observations were affected by strong background flares, the net exposure which can be used for our analysis was strongly reduced. 
The GTIs of the various observations ranged over 6--56\,ks, 
apart from observation b2 (ObsID 0202230301) which had to be rejected, because it showed high background throughout the observation. The exposures for all three EPIC instruments are given in Cols. 8, 10 and 12 of Table~\ref{tab:observations}.\@ The observations obtained during the summer visibility window of \m31\ were affected more strongly by background radiation than those taken during the winter window. The most affected observations of the deep survey were reobserved.

After screening for times of enhanced particle background, the second step was to examine the influence of solar wind charge exchange. This was done by producing soft energy \linebreak($<\!2$\,keV) background light curves. These lightcurves varied only for 10 observations, for which additional screening was necessary.
The screening of enhanced background due to solar wind charge exchange was applied to the observations only for the creation of colour images, in order to avoid that these observations will appear in the mosaic image with a tinge of red. The screening was not used for source detection.

The third and last step includes the study of the background due to detector noise. The processing chains take into account all known bad or hot pixels and columns and flag the affected pixels in the event lists. We selected data with {\tt (FLAG \& 0xfa0000)=0}, excluded rows and columns near edges, and searched by eye for additional warm or hot pixels and columns in each observation.  
To avoid background variability over the PN images, we omitted the energy range from 7.2--9.2\,keV where strong fluorescence lines cause higher background in the outer detector area \citep{2004SPIE.5165..112F}. 

An additional background component can occur during the EPIC PN offset map calculation.  
If this period is affected by high particle background, the offset calculation will lead to a slight underestimate of the offset in some pixels which can then  result in blocks of pixels ($\approx 4\!\times\!4$) with enhanced low energy signal.\footnote{See also \url{http://xmm2.esac.esa.int/docs/documents/CAL-TN-0050-1-0.ps.gz}} These blocks will be found by the {\tt SAS} detection tools and appear as sources with extremely soft spectrum (so called supersoft sources). To reduce the number of false detections in this source class, we decided to include the task {\tt epreject} in {\tt epchain}, which locates the pixels with a slight underestimate of the offset and corrects this underestimate. To ensure that {\tt epreject} produces reliable results, difference images of the event lists obtained with and without {\tt epreject}, were created. Only events with energies above 200\,eV were used. We checked whether {\tt epreject} removed all pixels with an enhanced low energy signal. Only in observation ns1 the difference image still shows a block of pixels with enhanced signal. As this block is also visible at higher energies (PHA$>30$) it cannot be corrected with {\tt epreject}. Additionally, we ascertained that almost all pixels not affected during the offset map calculation have a value consistent with zero in the difference images, with two exceptions discussed in Sect.\,\ref{Sec:srccat}.

\subsection{Images}
\label{Sec:Images}
For each observation, the data were split into five energy bands: (0.2--0.5)\,keV, (0.5--1.0)\,keV, (1.0--2.0)\,keV, (2.0--4.5)\,keV, and (4.5--12)\,keV. For the PN data, we used only single-pixel events (PATTERN\,$=$\,0) in the first energy band, while for the other bands, single-pixel and double-pixel events were selected (PATTERN\,$\le$\,4). In the MOS cameras, single-pixel to quadruple-pixel events (PATTERN\,$\le$\,12) were used. We created images, background images and exposure maps (with and without vignetting correction) for PN, MOS\,1 and MOS\,2 in each of the five energy bands and masked them for the acceptable detector area. The image bin size is 2\arcsec. The same procedure was applied in our previous \m31 and M~33 studies \citep[PFH2005 and][]{2004A&A...426...11P}.

To create background images, the {\tt SAS} task {\tt eboxdetect} was run in local mode, in which it determines the background from the surrounding pixels of a sliding box, with box sizes of $5\times5$, $10\times10$ and $20\times20$ pixels (10\arcsec$\times$10\arcsec, 20\arcsec$\times$20\arcsec and 40\arcsec$\times$40\arcsec). The detection threshold is set to {\tt likemin\,=\,15}, which is a good compromise between cutting out most of the sources and leaving sufficient area to derive the appropriate background. For the background calculation, a two dimensional spline is fitted to a rebinned and exposure corrected image (task {\tt esplinemap}). The number of bins used for rebinning is controlled by the parameter {\tt nsplinenodes}, which is set to 16 for all but the observations of the central region, where it was set to 20 (maximum value). For PN, the background maps contain the contribution from the ``out of time (OoT)" events. 

\subsection{Source detection}
\label{Sec:SrcDet}
For each observation, source detection was performed simultaneously on 5 energy bands for each EPIC camera, using the XMM-{\tt SAS} detection tasks {\tt eboxdetect} and {\tt emldetect}, as such fitting provides the most statistically robust measurements of the source positions by including all of the data. This method was also used to generate the 2XMM catalog \citep[cf][]{2009A&A...493..339W}. In the following we describe the detection procedure used. 

The source detection procedure consists of two consecutive detection steps. An initial source list is created with the task {\tt eboxdetect} (\cf\ Sect.\,\ref{Sec:Images}). To select source candidates down to a low statistical significance level, a low likelihood threshold of four was used at this stage. The background was estimated from the previously created background images (see Sect.\,\ref{Sec:Images}). 

This list is the starting point for the XMM-{\tt SAS} task {\tt emldetect} (v.~4.60.1).\@ The {\tt emldetect} task performs a Maximum Likelihood fit of the distribution of source counts \citep[based on Cash C-statistics approach;][]{1979ApJ...228..939C}, using a point spread function model obtained from ray tracing calculations. If $P$ is the probability that a Poissonian fluctuation in the background is detected as a spurious source, the likelihood of the detection is then defined as $\mathcal{L}=-\ln\lb( P \rb)$.\footnote{This is a simplified description as {\tt emldetect} transforms the derived likelihoods to equivalent likelihoods, corresponding to the case of two free parameters. This allows comparison between detection runs with different numbers of free parameters.} The fit is performed simultaneously in all energy bands for all three cameras by summing the likelihood contribution of each band and each camera. Sources exceeding the detection likelihood threshold in the full band (combination of the 15 bands) are regarded as detections; the catalogue is thus full band selected. 

The detection threshold used is 7, as in PFH2005. Some other parameters differ from the values used in PFH2005, as in this work a parameter setting optimised for the detection of extended sources was used (G.~Lamer; private communication). The parameters in question are the event cut-out ({\tt ecut\,=\,30.0}) and the source selection radius ({\tt scut\,=\,0.9})  for multi-source fitting, the maximum number of sources into which one input source can be split ({\tt nmulsou\,=\,2}), and the maximum number of sources that can be fitted simultaneously ({\tt nmaxfit\,=\,2}).\@ Multi-PSF fitting was performed in a two stage process for objects with a detection likelihood larger than ten. All of the sources were also fitted with a convolution of a $\beta$-model cluster brightness profile \citep[][]{1976A&A....49..137C} with the \xmm\ point spread function, in order to detect any possible extension in the detected signal. Sources which have a core radius significantly larger than the PSF are flagged as extended. The free parameters of the fit were the source location, the source extent and the source counts in each energy band of each telescope.

To derive the X-ray flux of a source from its measured count rate, one uses the so-called energy conversion factors (ECF): 
\begin{equation}
\mr{Flux}=\frac{\mr{Rate}}{\mr{ECF}}
\end{equation}
These factors were calculated using the detector response, and depended on the used filter, the energy band in question, and the spectrum of the source. As we wanted to apply the conversion factors to all sources found in the survey, we assumed a power law model with photon index $\Gamma\!=\!1.7$ and the Galactic foreground absorption of $N_{\mr{H}}\!=\!7\times10^{20}$\,cm$^{-2}$ \citep[][see also PFH2005]{1992ApJS...79...77S} to be the universal source spectrum for the ECF calculation.

The ECFs (see Table~\ref{tab:ECFvalues}) were derived with {\tt XSPEC}\footnote{\url{http://heasarc.gsfc.gov/docs/xanadu/xspec}}(v~11.3.2) using response matrices (V.7.1) available from the \xmm\ calibration homepage\footnote{\url{http://xmm2.esac.esa.int/external/xmm_sw_cal/calib/epic_files.shtml}}. As all necessary corrections of the source parameters (\eg\ vignetting corrections) were included in the image creation and source detection procedure\footnote{especially in the {\tt emldetect} task}, the \emph{on axis} ECF values were derived \citep[\cf\ ][]{2009A&A...493..339W}. The fluxes determined with the ECFs given in Table~\ref{tab:ECFvalues} are absorbed (\ie\ observed) fluxes and hence correspond to the observed count rates, which are derived in the {\tt emldetect} task.

During the mission lifetime, the MOS energy distribution behaviour has changed. Near the nominal boresight positions, where most of the detected photons hit the detectors, there has been a decrease in the low energy response of the MOS cameras \citep{2006ESASP.604..925R}. To take this effect into account, different response matrices for observations obtained before and after the year 2005 were used (see Table~\ref{tab:ECFvalues}). 

\begin{table}
\begin{center}
\caption{Count rate to energy conversion factors. 
The ECFs used for observations obtained before revolution 534 are marked with ``OLD".}
\begin{tabular}{lrrrrrr}
\hline\noalign{\smallskip}
\hline\noalign{\smallskip}
\multicolumn{1}{c}{Detector} & \multicolumn{1}{c}{Filter} & \multicolumn{1}{c}{B1} & \multicolumn{1}{c}{B2} & \multicolumn{1}{c}{B3} & 
\multicolumn{1}{c}{B4} & \multicolumn{1}{c}{B5}  \\ 
\noalign{\smallskip}
& & \multicolumn{5}{c}{$(10^{11}\mr{cts\,cm^2\,erg^{-1}})$} \\
\noalign{\smallskip}\hline\noalign{\smallskip}
EPIC PN & thin & $11.33$ & $8.44$ & $5.97$ & $1.94$ & $0.58$ \\
& medium & $10.05$ & $8.19$ & $5.79$ & $1.94$ & $0.58$ \\
EPIC MOS\,1 & thin & $2.25$ & $1.94$ & $2.06$ & $0.76$ & $0.14$ \\
& medium & $2.07$ & $1.90$ & $2.07$ & $0.75$ & $0.15$ \\
EPIC MOS\,2 & thin & $2.29$ & $1.98$ & $2.09$ & $0.78$ & $0.15$ \\
& medium & $2.06$ & $1.90$ & $2.04$ & $0.75$ & $0.15$ \\
EPIC MOS\,1 & thin & $2.59$ & $2.04$ & $2.12$ & $0.76$ & $0.15$ \\
OLD & medium & $2.33$ & $1.98$ & $2.09$ & $0.76$ & $0.15$ \\
EPIC MOS\,2 & thin & $2.58$ & $2.04$ & $2.13$ & $0.76$ & $0.15$ \\
OLD & medium & $2.38$ & $1.99$ & $2.09$ & $0.75$ & $0.16$ \\
\noalign{\smallskip}
\hline
\noalign{\smallskip}
\end{tabular}
\label{tab:ECFvalues}
\end{center}
\normalsize
\end{table}

For most sources, band 5 just adds noise to the total count rate. If converted to flux, this noise often dominates the total flux due to the small ECF.\@ To avoid this problem we calculated count rates and fluxes for detected sources in the ``XID" (0.2--4.5)\,keV band (bands 1 to 4 combined).  While for most sources this is a good solution, for extremely hard or soft sources there may still be bands
just adding noise. This, then, may lead to rate and flux errors that seem to falsely indicate a lower source significance. A similar effect occurs in the combined rates and fluxes, if a source is detected primarily by one instrument (\eg\ soft sources in PN).

Sources are entered in the \XLPt\ catalogue from the observation in which the highest source detection likelihood is obtained (either combined or single observations). For variable sources this means that the source properties given in the \XLPt\ catalogue (see Sect.\,\ref{Sec:srccat} and Table~5) are those observed during their brightest state.

We rejected spurious detections in the vicinity of bright sources. In regions with a highly structured background, the {\tt SAS} detection task {\tt emldetect} registered some extended sources. We also rejected these ``sources" as spurious detections. In an additional step we checked whether an object had visible contours in at least one image out of the five energy bands. The point-like or extended nature, which was determined with {\tt emldetect}, was taken into account. In this way, ``sources" that are fluctuations in the background, but which were not fully modelled in the background images, were detected. In addition, objects located on hot pixels, or bright pixels at the rim or in the corners of the individual CCD chips (which were missed during the background screening) were recognised and excluded from the source catalogue, especially if they were detected with a likelihood larger than six in one detector only.

To allow for a statistical analysis, the source catalogue only contains sources detected by the {\tt SAS} tasks {\tt eboxdetect} and {\tt emldetect} as described above, \ie\ the few sources that were not detected by the analysis program, despite being visible on the X-ray images, have not been added by hand as it was done in previous studies (SPH2008; PFH2005).

To classify the source spectra, we computed four hardness ratios. The hardness ratios and errors are defined as:
\begin{equation}
\mr{HR}i = \frac{B_{i+1} - B_{i}}{B_{i+1} + B_{i}}\; \mbox{and}\;\; \mr{EHR}i = 2  \frac{\sqrt{(B_{i+1} EB_{i})^2 + (B_{i} EB_{i+1})^2}}{(B_{i+1} + B_{i})^2},
\label{Eq:hardr}
\end{equation}
for {\it i} = 1 to 4, where $B_{i}$ and $EB_{i}$ denote count rates and corresponding errors in energy band {\it i}.

\subsection{Astrometrical corrections}
\label{SubSec:AstCorr}
To obtain astrometrically-corrected positions for the sources of the five central fields we used the {\tt SAS}-task {\tt eposcorr} with \chandra\ source lists \citep{2002ApJ...577..738K,2002ApJ...578..114K,2004ApJ...609..735W}.\@
For the other fields we selected sources from the USNO-B1 \citep{2003AJ....125..984M}, 2MASS \citep{2006AJ....131.1163S} and Local Group Galaxy Survey \citep[LGGS; ][]{2006AJ....131.2478M} catalogues\footnote{For the remainder of the subsection we will call all three catalogues ``optical catalogues" for easier readability, although the 2MASS catalogue is an infrared catalogue.}.

\subsubsection{Astrometry of optical/infrared catalogues}
In a first step, we examined the agreement between the positions given by the various optical catalogues.\footnote{From the LGGS catalogue only sources brighter than 21\,mag were used in order to be comparable to the brightness limit of the USNO-B1 catalogue.} A close examination of the shifts obtained, showed significant differences between the positions given in the individual catalogues. In summary, between the USNO-B1 and LGGS catalogues we found an offset of: $-$0\farcs197 in R.A.\ and 0\farcs067 in Dec\footnote{the offset in declination is negligible}; and between the USNO-B1 and 2MASS catalogues we found an offset of: $-$0\farcs108 in R.A.\ and 0\farcs204 in Dec. We chose the USNO-B1 catalogue as a reference, since it covers the entire field observed in the Deep \xmm\ survey, and in addition it provides values for the proper motion of the optical sources.

Since the optical catalogues, as well as the Deep \xmm\ catalogue, are composed of individual observations of sub-fields of \m31, we searched for systematic drifts in the positional zero points from region to region. However no systematic offsets were found.

Finally, we applied the corrections found to the sources in the LGGS and 2MASS catalogues, to bring all catalogues to the USNO-B1 reference frame.

The offsets found between the USNO-B1 and 2MASS catalogues can be explained by the independent determination of the astrometric solutions for these catalogues. Given that the positions provided in the LGGS catalogue are corrected with respect to the USNO-B1 catalogue \citep[see][]{2006AJ....131.2478M}, the offset found in right ascension was totally unexpected and cannot be explained. 

\subsubsection{Corrections of the X-ray observations}
From the positionally corrected catalogues, we selected sources which either correlate with globular clusters from the Revised Bologna Catalogue \citep[V.3.4, January 2008; ][]{2004A&A...416..917G,2005A&A...436..535G,2006A&A...456..985G,2007A&A...471..127G} or with foreground stars, characterised by their optical to X-ray flux ratio \citep{1988ApJ...326..680M} and their hardness ratio \citep[see source selection criteria given in Table~\ref{Tab:class} and][]{2008A&A...480..599S}. For sources selected from the USNO-B1 catalogue, we used the proper motion corrected positions. We then used the {\tt SAS}-task {\tt eposcorr} to derive the offset of the X-ray aspect solution. Four observations did not have enough optical counterparts to apply this method. The lack of counterparts is due to the very short exposure times resulting after the screening for high background (obs.~s3, ss12, ss13) and the location of the observation (obs.~sn11). In these cases, we used bright persistent X-ray sources, which we correlated with another observation of the same field. We checked for any residual systematic uncertainty in the source positions and found it to be well characterised by a conservative $1\sigma$ value of 0\,\farcs5.\@ This uncertainty is due to positional errors of the optical sources as well as inaccuracy in the process of the determination of the offset between optical and X-ray sources, and is called systematic positional error. The appropriate offset, given in Col.~6 of Table~\ref{tab:observations}, was applied to the event file of each pointing, and images and exposure maps were then reproduced with the corrected astrometry.\\

Fields that were observed at least twice are treated in a special way, which is described in the following section.

\subsection{Multiple observations of selected fields}
The fields that were observed more than once were the central field, the fields pointing on RX J0042.6+4115\footnote{The combination of observations b1, b3 and b4 is called b.}, two fields located on the major axis of \m31\  (S2, N2) and all fields of the ``Large Survey" located in the southern part of the galaxy (SS1, SS2, SS3, S3, SN3, SN2, SN1).\@ To reach higher detection sensitivity we merged the images, background images and exposure maps of observations which have the same pointing direction and were obtained with the same filter setting. Subsequently, source detection, as described in Sect.~\ref{Sec:SrcDet}, was repeated on the merged data. For the S2 field, there are two observations with different filter settings. In this case, source detection was performed simultaneously on all 15 bands of both observations, \ie\ on 30 bands simultaneously. The N2 field was treated in the same way. For the central field images, background images and exposure maps of observations c1, c2 and c3 were merged. These merged data were used together with the data of observation c4  to search for sources simultaneously; in this way it was possible to take into account the different ECFs for the different filters. One field was observed twice with slightly different pointing direction in observations sn1 and sn11; simultaneous source detection was used for these observations also.

\subsection{Variability calculation}
\label{Sec:DefVar}
To examine the time variability of each source listed in the total source catalogue, we determined the XID flux at the source position in each observation or at least an upper limit for the XID flux. We used the task {\tt emldetect}  
with fixed source positions when calculating the total flux. To get fluxes and upper limits for all sources in the input list we set the detection likelihood threshold to 0.
    
A starting list was created from the full source catalogue, which only contains the identification number and position of each source located in the field examined. To give correct results, the task {\tt emldetect} has to process the sources from the brightest one to the faintest one. We, therefore, had to first order the sources in each observation by the detection likelihood. For sources not visible in the observation in question we set the detection likelihood to 0. This list was used as input for a first {\tt emldetect} run. In this way we achieved an output list in which a detection likelihood was allocated to every source. For a final examination of the sources in order of detection likelihood, a second {\tt emldetect} run was necessary. 

We only accepted XID fluxes for detections $\ge$ 3 $\sigma$; otherwise we used a 3 $\sigma$ upper limit. 
To compare the XID fluxes between the different observations, we calculated the significance of the difference 
\begin{equation}
S=\frac{F_{\mr{max}}- F_{\mr{min}}}{\sqrt{\sigma_{\mr{max}}^2+\sigma_{\mr{min}}^2}}
\end{equation}
and the ratio of the XID fluxes $V=F_{\mr{max}}/F_{\mr{min}}$, where $F_{\mr{max}}$ and $F_{\mr{min}}$ are the maximum and minimum (or upper limit) source XID flux, and $\sigma_{\mr{max}}$ and $\sigma_{\mr{min}}$ are the errors of the maximum and minimum flux, respectively. This calculation was not performed whenever $F_{\mr{max}}$ was an upper limit. Finally, the largest XID flux of each source was derived, excluding upper limits.   

\subsection{Spectral analysis}
To extract the X-ray spectrum of individual sources, we selected an extraction region and a corresponding background region which  was at least as large as the source region, was located on the same CCD at a similar off axis angle as the source, and did not contain any point sources or extended emission. For EPIC PN, we only accepted single-pixel events for the spectra of supersoft sources, while for all other spectra single and double-pixel events were used. For the EPIC-MOS detectors, single-pixel through to quadruple-pixel events were always used. Additionally, we only kept events with FLAG\,$=$\,0 for all three detectors. For each extraction region, we produced the corresponding response matrix files and ancillary response files. 

For each source, the spectral fit was obtained by fitting all three EPIC spectra simultaneously, using the tool {\tt XSPEC}.\@ For the absorption, we used the {\tt TBabs} model, with abundances from \citet{2000ApJ...542..914W} and photoelectric absorption cross-sections from \citet{1992ApJ...400..699B} with a new He cross-section based on \citet{1998ApJ...496.1044Y}.\@

\subsection{Cross correlations}
\label{Sec:CrossCorr_Tech}
Sources were regarded as correlating if their positions overlapped within their 3$\sigma$ (99.73\%) positional errors, defined as \citep{2009A&A...493..339W}:
\begin{equation}  
\Delta\mr{pos}\le3.44\times\sqrt{\sigma_{\mr{stat}}^2 + \sigma_{\mr{syst}}^2}+3\times\sigma_{\mr{ccat}}
\label{Eq:Cor}
\end{equation}
where $\sigma_{\mr{stat}}$ is the statistical and $\sigma_{\mr{syst}}$ the systematic error of the X-ray sources detected in the present study. The statistical error was derived by {\tt emldetect}.\@ The determination of the systematic error is described in Sect.\,\ref{SubSec:AstCorr}. We use a value of 0\,\farcs5, for all sources. The positional error of the sources in the catalogue used for cross-correlation is given by $\sigma_{\mr{ccat}}$. The values of $\sigma_{\mr{ccat}}$ (68\% error) used for the different X-ray catalogues can be found in Table~\ref{Tab:XrayRefCat}.\@ Exceptions to Eq.~\ref{Eq:Cor} are sources that are listed in more than one catalogue or that are resolved into multiple sources with \chandra.\@ The first case is restricted to catalogues with comparable spatial resolution and hence positional uncertainty. 

To identify the X-ray sources in the field of \m31\ we searched for correlations with catalogues in other wavelength regimes. The \xmm\ source catalogue was correlated with the following catalogues and public data bases:
\begin{description} 
\item [Globular Clusters:] Bologna Catalogue \citep[V.3.5, March 2008; ][$\sigma_{\mr{ccat}}=0.\arcsec2$; RBV~3.5]{2004A&A...416..917G,2005A&A...436..535G,2006A&A...456..985G,2007A&A...471..127G,2009A&A...508.1285G},  \citet[][$\sigma_{\mr{ccat}}=0.\arcsec2$]{2009AJ....137...94C}, \citet[][$\sigma_{\mr{ccat}}=0.\arcsec5$]{2009AJ....138..770H}, \citet[][$\sigma_{\mr{ccat}}=0.\arcsec2$]{2008PASP..120....1K}, \citet[][$\sigma_{\mr{ccat}}=0.\arcsec2$]{2007PASP..119....7K}, \citet[][]{2005PASP..117.1236F}, \citet[][$\sigma_{\mr{ccat}}=1\arcsec$]{1993PhDT........41M}
\item [Novae:] Nova list of the \m31\ Nova Monitoring Project\footnote{\url{http://www.mpe.mpg.de/~m31novae/opt/m31/M31_table.html}} ($\sigma_{\mr{ccat}}$ is given for each individual source), PHS2007, \citet{2010AN....331..187P}
\item [Supernova Remnants:] \citet[][19 srcs]{1980A&AS...40...67D}, \citet[][967 srcs]{1992A&AS...92..625W} and \citet[][58 srcs]{1993A&AS...98..327B}, \citet[][233 srcs]{1995A&AS..114..215M}; An X-ray source is considered as correlating with a SNR, if the X-ray source position (including 3$\sigma$ error) lies within the extent given for the SNR.
\item [Radio Catalogues:] \citet[][$\sigma_{\mr{ccat}}$ is given for each individual source]{2005ApJS..159..242G},  \citet[][$\sigma_{\mr{ccat}}$ is given for each individual source]{2004ApJS..155...89G}, \citet[][$\sigma_{\mr{ccat}}=3\arcsec$]{2008AJ....136..684K}, \citet[][$\sigma_{\mr{ccat}}$ is given for each individual source]{1990ApJS...72..761B}, NVSS \citep[NRAO/VLA Sky Survey\footnote{\url{http://www.cv.nrao.edu/nvss/NVSSlist.shtml}};][$\sigma_{\mr{ccat}}$ is given for each individual source]{1998AJ....115.1693C}
\item [H {\small II} Regions, H $\alpha$ Catalogue:] \citet[][$\sigma_{\mr{ccat}}$ is given for each individual source]{1992A&AS...92..625W}, \citet[][$\sigma_{\mr{ccat}}=0.\arcsec2$]{2007AJ....134.2474M}
\item [Optical Catalogues:] USNO-B1 \citep[][$\sigma_{\mr{ccat}}$ is given for each individual source]{2003AJ....125..984M}, Local Group Survey \citep[LSG; ][$\sigma_{\mr{ccat}}=0.\arcsec2$]{2006AJ....131.2478M}
\item [Infrared catalogues:] 2MASS \citep[][$\sigma_{\mr{ccat}}$ is given for each individual source]{2006AJ....131.1163S}, \citet[][$\sigma_{\mr{ccat}}=0.\arcsec8$, for Table~2: $\sigma_{\mr{ccat}}=0.\arcsec5$]{2008ApJ...687..230M}
\item [Data bases:] the SIMBAD catalogue\footnote{\url{http://simbad.u-strasbg.fr/simbad}} (Centre de Donn\'ees astronomiques de Strasbourg; hereafter SIMBAD) , the NASA Extragalactic Database\footnote{\url{http://nedwww.ipac.caltech.edu}} (hereafter NED)
\end{description}

\begin{table}
\begin{center}
\caption{X-ray source catalogues used for cross-correlation and the used positional errors}
\begin{tabular}{lrlr}
\hline\noalign{\smallskip}
\hline\noalign{\smallskip}
\multicolumn{1}{l}{X-ray catalogue$^{\ddagger}$} & \multicolumn{1}{l}{$\sigma_{\mr{ccat}}^{\dagger}$} & \multicolumn{1}{l}{X-ray catalogue$^{\ddagger}$} & \multicolumn{1}{l}{$\sigma_{\mr{ccat}}^{\dagger}$}\\
\hline\noalign{\smallskip}
PFH2005 & $*$ & DKG2004 & 0\,\farcs3 \\
SPH2008 & $*$ & WNG2006 & 0\,\farcs3 \\
SHP97 & $*$ & VG2007 & 0\,\farcs4 \\
SHL2001 & $*$ & OBT2001 & 3\arcsec \\
PFJ93 & $*$ & O2006 & 1\arcsec \\
TF91 & $*$ & SBK2009 & 3\arcsec$^{+}$ \\
Ka2002 & 0\,\farcs3 & D2002 & 0\,\farcs5 \\
KGP2002 & $*$ & TP2004 & 1\arcsec \\
WGK2004 & 1\arcsec$^{+}$ & ONB2010 & 1\arcsec \\
\noalign{\smallskip}
\hline
\noalign{\smallskip}
\end{tabular}
\label{Tab:XrayRefCat}
\end{center}
Notes:\\
$^{ {\dagger}~}$: $*$ indicates that the catalogue provides $\sigma_{\mr{ccat}}$ values for each source individually\\
$^{ +~}$: value taken from indicated paper\\
$^{ {\ddagger}~}$: TF91: \citet{1991ApJ...382...82T}, PFJ93: \citet{1993ApJ...410..615P}, SHP97: \citet{1997A&A...317..328S}, SHL2001: \citet{2001A&A...373...63S}, OBT2001: \citet{2001A&A...378..800O}, D2002: \citet{2002ApJ...570..618D}, KGP2002: \citet{2002ApJ...577..738K}, Ka2002: \citet{2002ApJ...578..114K}, WGK2004: \citet{2004ApJ...609..735W}, DKG2004: \citet{2004ApJ...610..247D}, TP2004: \citet{2004ApJ...616..821T}, PFH2005: \citet{2005A&A...434..483P}, O2006: \citet{2006ApJ...643..844O}, WNG2006: \citet{2006ApJ...643..356W}, VG2007: \citet{2007A&A...468...49V}, SPH2008: \citet{2008A&A...480..599S}, SBK2009: \citet{2009A&A...495..733S}, ONB2010: \citet{2010ApJ...717..739O}
\normalsize
\end{table}

\section{Colour image}
\label{Sec:coim}
\begin{figure*}
\sidecaption
\includegraphics[width=12cm]{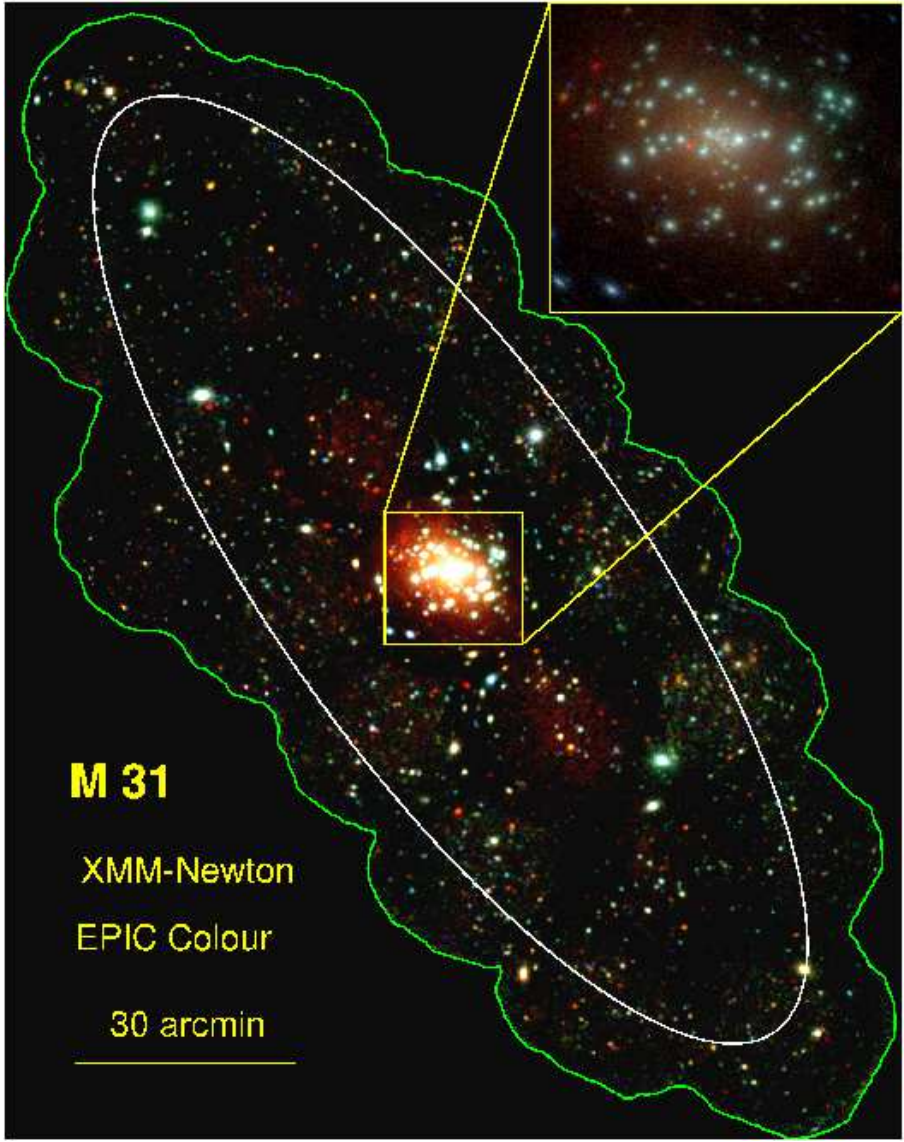}
\caption{Combined EPIC PN, MOS\,1 and MOS\,2 RGB image of the Deep \m31\ Survey including archival data. The optical extent of \m31\ is indicated by the $\mathrm{D_{25}}$ ellipse and the boundary of the observed field is given by the green contour. The central region, marked with the yellow square, is shown in higher resolution in the upper right corner. For more details see Sect.\,\ref{Sec:coim}.
\label{Fig:cimage}}
\end{figure*}

Figure~\ref{Fig:cimage} shows the combined, exposure corrected EPIC PN, MOS\,1 and MOS\,2 RGB (red-green-blue) mosaic image of the Deep Survey and archival data. The colours represent the X-ray energies as follows: red: 0.2--1.0\,keV, green: 1.0--2.0\,keV and blue: 2.0--12\,keV. The optical extent of \m31\ is indicated by the $\mathrm{D_{25}}$ ellipse and the boundary of the observed field is given by the green contour. The image is smoothed with a 2D-Gaussian of 20\arcsec\ FWHM. In some observations, individual noisy MOS\,1 and MOS\,2 CCDs are omitted. 
The images have not been corrected for the background of the detector or for vignetting.\\ 
The colour of the sources reflects their class. Supersoft sources appear in red. Thermal SNRs and foreground stars are orange to yellow. ``Hard" sources (background objects, mainly AGN, and X-ray binaries or Crab-like SNRs) are blue to white. 

Logarithmically scaled \xmm\ EPIC low background images made up of the combined images from the PN, MOS\,1 and MOS\,2 cameras in the (0.2--4.5) keV XID band for each \m31\ observation can be found in the Appendix. The images also show X-ray contours, and the sources from the \XLPt\ catalogue are marked with boxes. 

\section{Source catalogue (\XLPt)}
\label{Sec:srccat}
The source catalogue of the Deep \xmm\ survey of \m31\ (hereafter \XLPt\ catalogue) contains 1\,897 X-ray sources. Of these sources 914 are detected for the first time in X-rays. 

The source parameters are summarised in Table~5, which gives the source number (Col.~1), detection field from which the source was entered into the catalogue (2), source position (3 to 9) with $3\sigma$ (99.73\%) uncertainty radius (10), likelihood of existence (11), integrated PN, MOS\,1 and MOS\,2 count rate and error (12,13) and flux and error (14,15) in the (0.2--4.5) keV XID band, and hardness ratios and errors (16--23). Hardness ratios are calculated only for sources for which at least one of the two band count rates has a significance greater than $2\sigma$. Errors are the properly combined statistical errors in each band and can extend beyond the range of allowed values of hardness ratios as defined previously (--1.0 to 1.0; Eq.~\ref{Eq:hardr}). The ``Val'' parameter (Col 24) indicates whether the source is  within the field of view  (true or false, ``T'' or ``F'') in the PN, MOS\,1 and MOS\,2 detectors respectively.

Table~5 also gives the exposure time (25), source existence likelihood (26), the count rate and error (27, 28) and the flux and error (29, 30) in the (0.2--4.5)\,keV XID band, and hardness ratios and errors (31--38) for the EPIC PN. Columns 39 to 52 and 53 to 66 give the same information corresponding to Cols.\ 25 to 38, but for the EPIC MOS\,1 and MOS\,2 instruments. Hardness ratios for the individual instruments were again screened as described above. From the comparison between the hardness ratios derived from the integrated PN, MOS\,1 and MOS\,2 count rates (Cols. 16--23) and the hardness ratios from the individual instruments (Cols. 31--38, 45--52 and 59--66), it is clear that the combined count rates from all instruments yielded a significantly larger fraction of hardness ratios above the chosen significance threshold.

Column~67 shows cross correlations with published \m31\ X-ray catalogues (\cf\ Sect.\,\ref{Sec:CrossCorr_Tech}). We discuss the results of the cross correlations in Sects.\,\ref{Sec:fgback} and \ref{Sec:Srcsm31}. 

In the remaining columns of Table~5,  
we give information extracted from the USNO-B1, 2MASS and LGGS catalogues (\cf\ Sect.\,\ref{Sec:CrossCorr_Tech}).  The information from the USNO-B1 catalogue (name, number of objects within search area, distance, B2, R2 and I magnitude of the brightest\footnote{in B2 magnitude} object) is given in Cols.~68 to 73.\@ The 2MASS source name, number of objects within search area, and the distance  can be found in Cols.~74 to 76. Similar information from the LGGS catalogue is given in Cols.~77 to 82 (name, number of objects within search area, distance, V magnitude, V-R and B-V colours of the brightest\footnote{in B magnitude} object).\@ To improve the reliability of source classifications we used the USNO-B1 B2 and R2 magnitudes to calculate 
\begin{equation}
\log\lb(\frac{f_{\mr{x}}}{f_{\mr{opt}}}\rb) = \log\lb(f_{\mr{x}}\rb) + \frac{m_{\mr{B2}} + m_{\mr{R2}}}{2\times2.5} + 5.37,
\label{Eq:fxopt}
\end{equation}
and the LGGS V magnitude to calculate
\begin{equation}
\log\lb(\frac{f_{\mr{x}}}{f_{\mr{opt}}}\rb) = \log\lb(f_{\mr{x}}\rb) + \frac{m_{\mr{V}}}{2.5} + 5.37,
\label{Eq:fxvopt}
\end{equation}
following \citet[][ see Cols. 83--86]{1988ApJ...326..680M}.

The X-ray sources in the \XLPt\ catalogue are identified or classified based on properties in X-rays (HRs, variability, extent) and of the correlated objects in other wavelength regimes (Cols.\, 87 and 88 in Table~5).\@ For classified sources the class name is given in angled brackets. Identification and classification criteria are summarised in Table~\ref{Tab:class}, which provides, for each source class (Col.\,1), the classification criteria (2), and the numbers of identified (3) and classified (4) sources.
The hardness ratio criteria are based on model spectra. Details on the definition of these criteria can be found in Sect.\,6 of PFH2005. As we have no clear hardness ratio criteria to discriminate between XRBs, Crab-like supernova remnants (SNRs) or AGN we introduced a $<$hard$>$ class for those sources. If such a source shows strong variability (i.\,e.\ V$\ge$10) on the examined time scales it is likely to be an XRB. Compared with SPH2008  
the HR2 selection criterion for SNRs was tightened (from HR2$<\!-0.2$ to HR2$+$EHR2$<\!-0.2$) to exclude questionable SNR candidates from the class of SNRs. If we applied  the former criterion to the survey data, $\sim$35 sources would be classified as SNRs in addition to those listed in Table~\ref{Tab:class}. Most of the 35 sources are located outside the D$_{25}$ ellipse, and none of them correlates with an optically identified SNR, a radio source, or an H{\small II} region. In addition, the errors in HR2 are of the same order as the HR2 values. It is therefore very likely that these sources do belong to other classes, since the strip between $-0.3\!<$HR2$<$0 is populated by foreground stars, XRBs, background objects, and candidates for these three classes.
Outcomes of the identification and classification processes are discussed in detail in Sects.\,\ref{Sec:fgback} and \ref{Sec:Srcsm31}. 

The last column (89) of Table~5 contains the \xmm\ source name as registered to the IAU Registry. Source names consist of the acronym XMMM31 and the source position as follows: XMMM31~Jhhmmss.s+ddmmss, where the right ascension is given in hours~(hh), minutes~(mm) and seconds~(ss.s) truncated to decimal seconds and the declination is given in degrees~(dd), arc minutes~(mm) and arc seconds~(ss) truncated to arc seconds, for equinox 2000. In the following, we refer to individual sources by their source number (Col.\,1 of Table~5), which is marked with a ``\num" at the front of the number.

Of the 1\,897 sources, 1\,247 can only be classified as $<$hard$>$ sources, while 123 sources remain without classification. Two of them (\num\ 482, \num\ 768) are highly affected by optical loading; both ``X-ray sources" coincide spatially with very bright optical foreground stars (USNO-B1 R2 magnitudes of 6.76 and 6.74 respectively). The spectrum of source \num\ 482 is dominated by optical loading. This becomes evident from the hardness ratios which indicate an SSS. For \num\ 768 the hardness ratios would allow a foreground star classification. The obtained count rates and fluxes of both sources are affected by the usage of {\tt epreject}, which neutralises the corrections applied for optical loading. Therefore residuals are visible in the difference images created from event lists obtained with and without {\tt epreject}. As we cannot exclude the possibility that some of the detected photons are true X-rays -- especially for source \num\ 768 --, we decided to include them in the \XLPt\ catalogue, but without a classification.

\begin{table*}
\addtocounter{table}{+1}
\scriptsize
\begin{center}
\caption{Summary of identifications and classifications.}
\begin{tabular}{llrr}
\hline\noalign{\smallskip}
\hline\noalign{\smallskip}
\multicolumn{1}{c}{Source class} & 
\multicolumn{1}{c}{Selection criteria} &
\multicolumn{1}{c}{identified} &
\multicolumn{1}{c}{classified}  \\ 
\noalign{\smallskip}\hline\noalign{\smallskip}
fg Star & ${\rm log}({{f}_{\rm x} \over {f}_{\rm opt}})\!<\!-1.0$ and HR2$-$EHR$2\!<\!0.3$ and HR3$-$EHR$3\!<\!-0.4$ or not defined & 40 & 223 \\
AGN  &  Radio source and not classification as SNR from HR2 or optical/radio & 11 & 49 \\
Gal  &  optical id with galaxy  & 4 & 19 \\
GCl  &  X-ray extent and/or spectrum & 1 & 5\\
SSS  &  HR$1\!<\!0.0$, HR2$-$EHR$2\!<\!-0.96$ or HR2 not defined, HR3, HR4 not defined &   & 30 \\
SNR  &  HR$1\!>\!-0.1$ and HR2$+$EHR$2\!<\!-0.2$ and not a fg Star, or id with optical/radio SNR   & 25 & 31 \\
GlC  &  optical id & 36 & 16\\
XRB  &  optical id or X-ray variability  & 10 & 26 \\
hard &  HR2$-$EHR$2\!>\!-0.2$ or only HR3 and/or HR4 defined, and no other classification&  & 1\,247 \\
\noalign{\smallskip}
\hline
\noalign{\smallskip}
\end{tabular}
\normalsize
\label{Tab:class}
\end{center}
\end{table*}

\subsection{Flux distribution}
\label{Sec:flux_dist}
The faintest source (\num\ 526) has an XID band flux of 5.8\ergcm{-16}.\@ The source with the highest XID Flux (\num\ 966, XID band flux of 3.75\ergcm{-12}) is located in the centre of \m31\ and identified as a Z-source LMXB \citep{2003A&A...411..553B}. This source has a mean  absorbed XID luminosity of 2.74\ergs{38}.\@ 

\begin{figure}
\resizebox{\hsize}{!}{\includegraphics[clip]{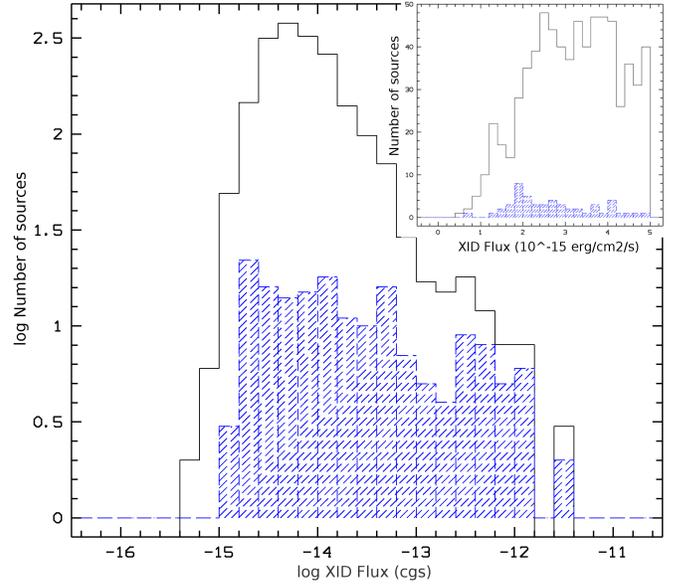}}
     \caption{Distribution of the source fluxes in the 0.2--4.5\,keV (XID) band. The diagrams show the number of sources at each flux bin, plotted versus the flux, using logarithmic scales. The inlay shows the number of sources for XID fluxes smaller than 5\ergcm{-15}, on linear scales.
The blue histogram gives the distribution of sources classified or identified as either SSSs, SNRs, XRBs or GlCs. 
}
    \label{Fig:XIDfluxdist} 
\end{figure}

Figure \ref{Fig:XIDfluxdist} shows the distribution of the XID (0.2--4.5\,keV) source fluxes. Plotted are the number of sources in a certain flux bin. We see from the inlay that the number of sources starts to decrease in the bin from 2.4 to 2.6\ergcm{-15}. This XID flux roughly determines the completeness limit of the survey and corresponds to an absorbed 0.2--4.5\,keV limiting luminosity of $\sim\!2$\ergs{35}. 

Previous X-ray studies \citep[][and references therein]{2004ApJ...609..735W} noted a lack of bright sources ($L_{\mr{X}}\!\ga$\oergs{37}; 0.1--10\,keV) in the northern half of the disc compared to the southern half. This finding is not supported in the present study. Excluding the pointings to the centre of \m31, we found in the remaining observations 13 sources in each hemisphere that were brighter than $L_{\mr{X\,abs}}\!\ga$\oergs{37}.\footnote{The luminosity is based on XID Fluxes. Using the total 0.2--12\,keV band the result does not change (23 in the northern half and 24 in the southern half).}\@ The reason our survey does not support the old results is that we found several bright sources in the outer regions of the northern half of the disk, which have not been covered  in \citet[][and references therein]{2004ApJ...609..735W}. In the central field of \m31, a total of 41 sources brighter than $L_{\mr{X}}\!\ga$\oergs{37} (0.2--4.5\,keV) were found.

\begin{figure}
 \resizebox{\hsize}{!}{\includegraphics[clip]{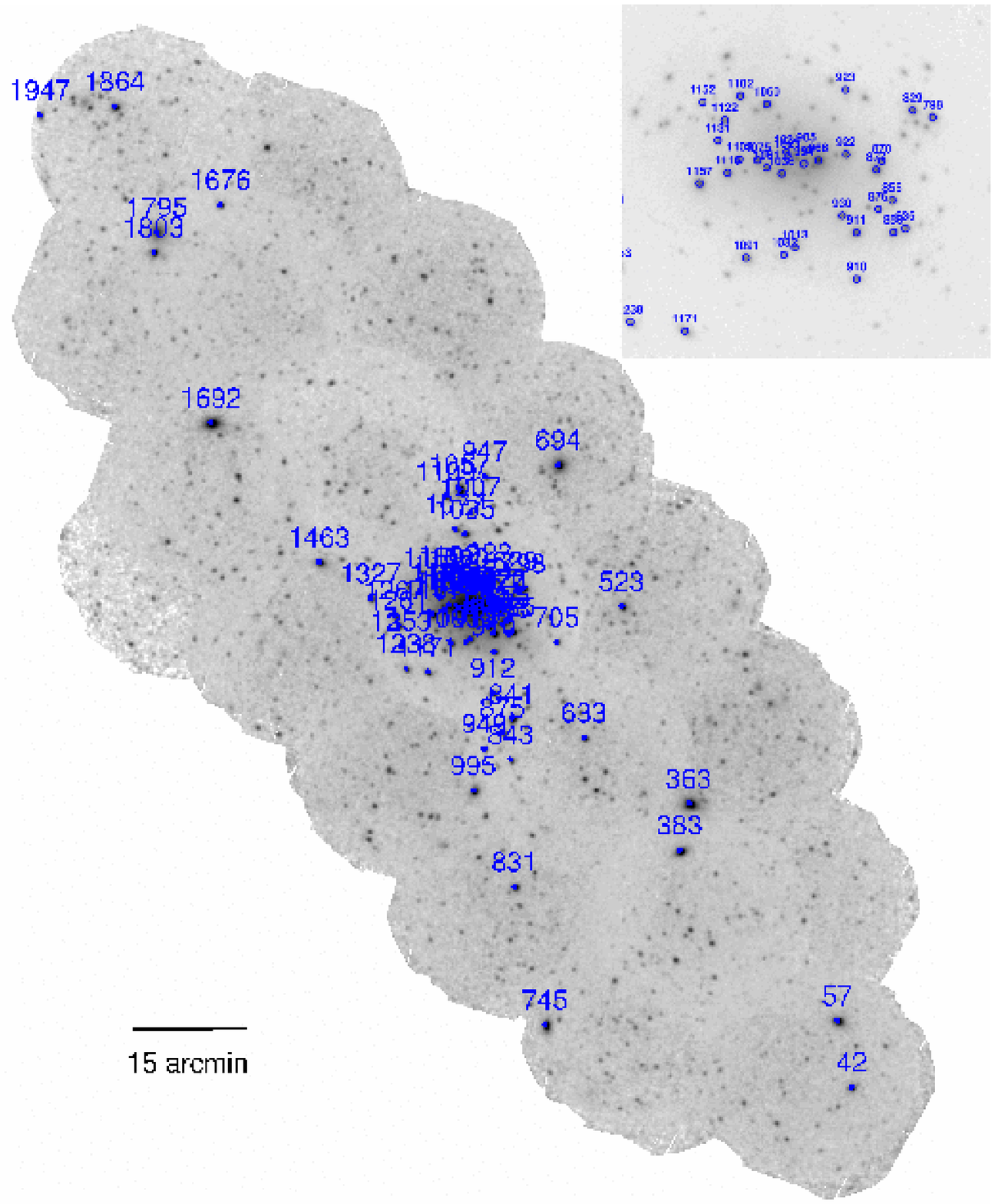}}
 \caption{\xmm\ Deep Survey image over plotted with sources that have an absorbed 0.2--4.5\,keV luminosity larger than \oergs{37}. Striking features are the two patches located north and south of the centre. The central region (same as in Fig.\,\ref{Fig:cimage}) is shown with higher resolution in the upper right corner.}
\label{Fig:brightS} 
\end{figure}

Figure~\ref{Fig:brightS} shows the spatial distribution of the bright sources. Striking features are the two patches located north and south of the centre. The southern one seems to point roughly in the direction of M~32 (\num\ 995), while the northern one ends in the globular cluster B\,116 (\num\ 947). However there is no association to any known spatial structure of \m31, like \eg\ the spiral arms. 

\subsection{Exposure map}
\label{Sec:ExpMap}
\begin{figure}
 \resizebox{\hsize}{!}{\includegraphics[clip]{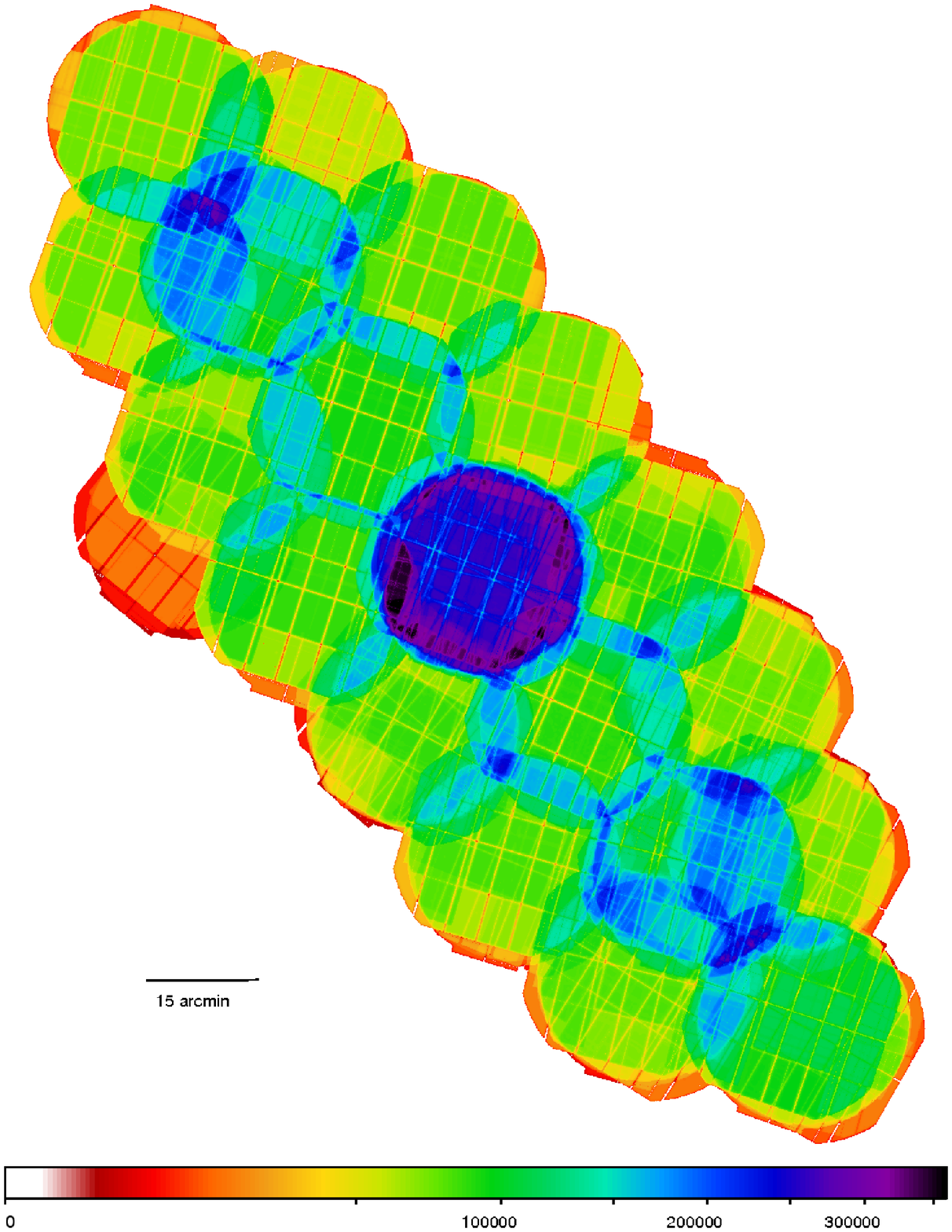}}
 \caption{Exposure map of all fields of the \XLPt\ catalogue. For details see Sect.\,\ref{Sec:ExpMap}.}
\label{Fig:ExpMap} 
\end{figure}

Figure~\ref{Fig:ExpMap} shows the exposure map  
used to create the colour image of all \xmm\ Large Survey and archival observations (Fig.\,\ref{Fig:cimage}). The combined MOS exposure was weighted by a factor of 0.4, before being added to the PN exposure. However, this map does not quite represent the exposures used in source detection; overlapping regions were not combined during source detection. 

From Fig.\,\ref{Fig:ExpMap} we see that the exposure for most of the surveyed area is rather homogeneous. Exceptions are the central area, overlapping regions and observation h4.

\subsection{Hardness ratio diagrams}
\begin{figure}
 \resizebox{\hsize}{!}{\includegraphics[clip]{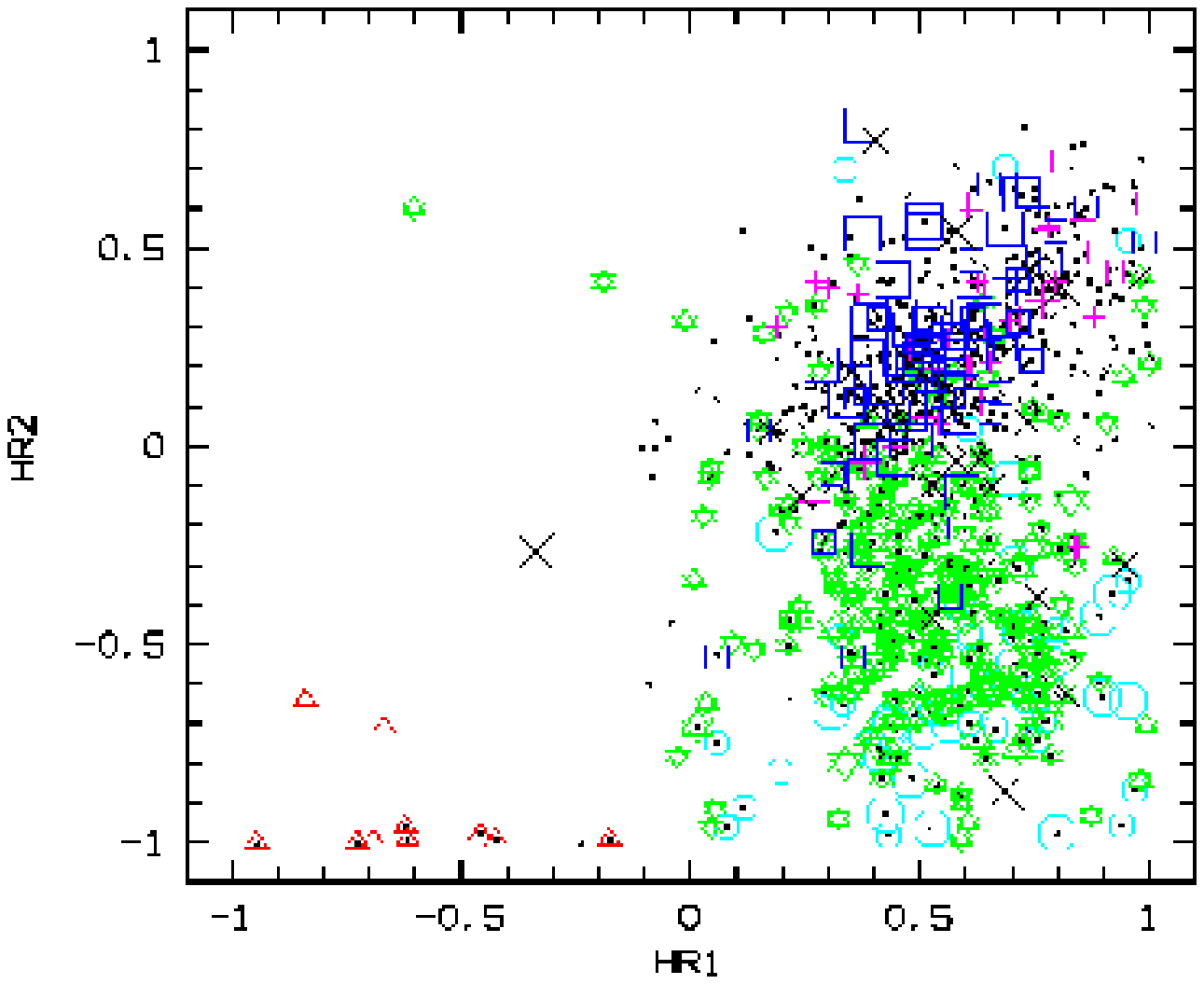}}
 \resizebox{\hsize}{!}{\includegraphics[clip]{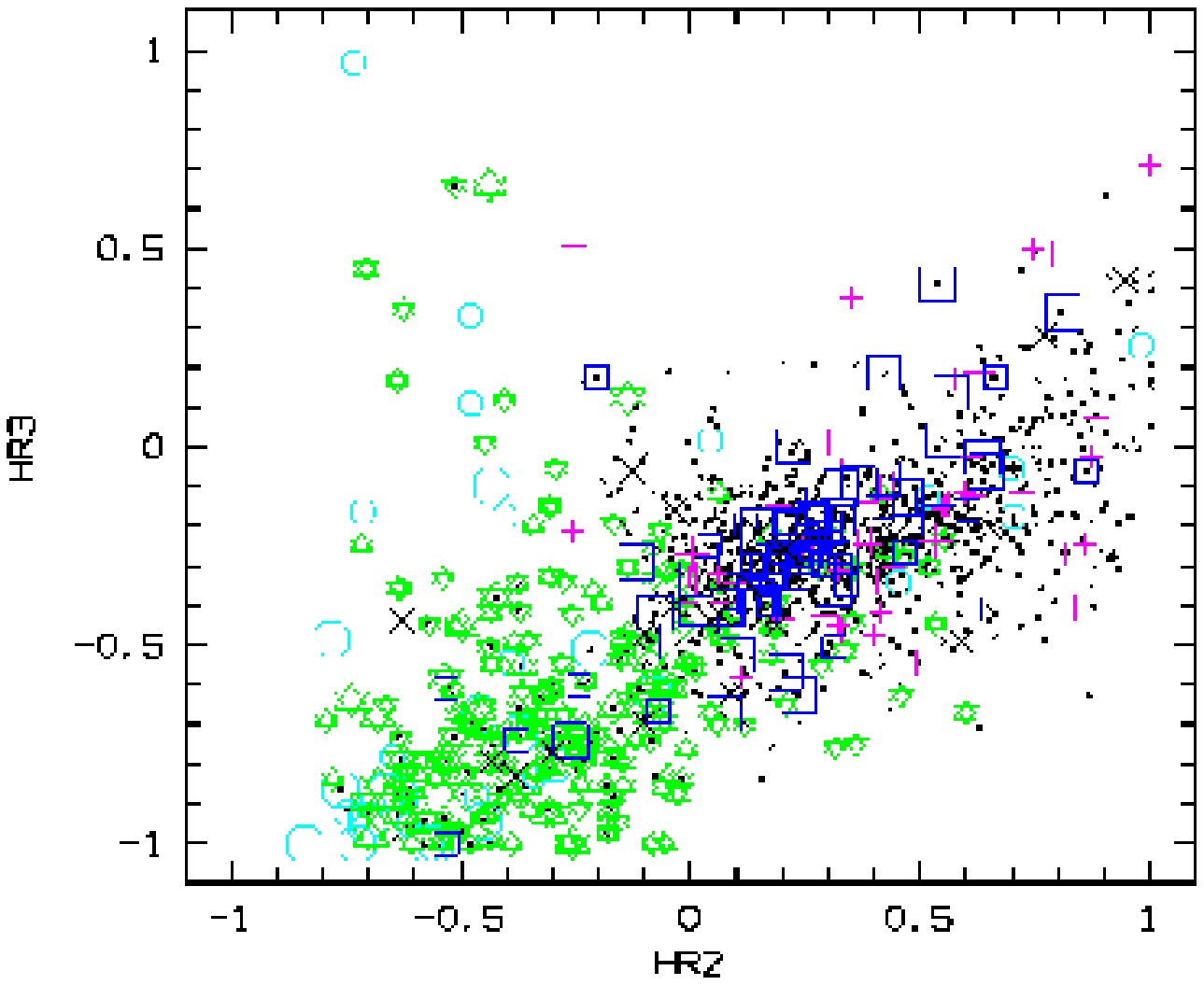}}
 \resizebox{\hsize}{!}{\includegraphics[clip]{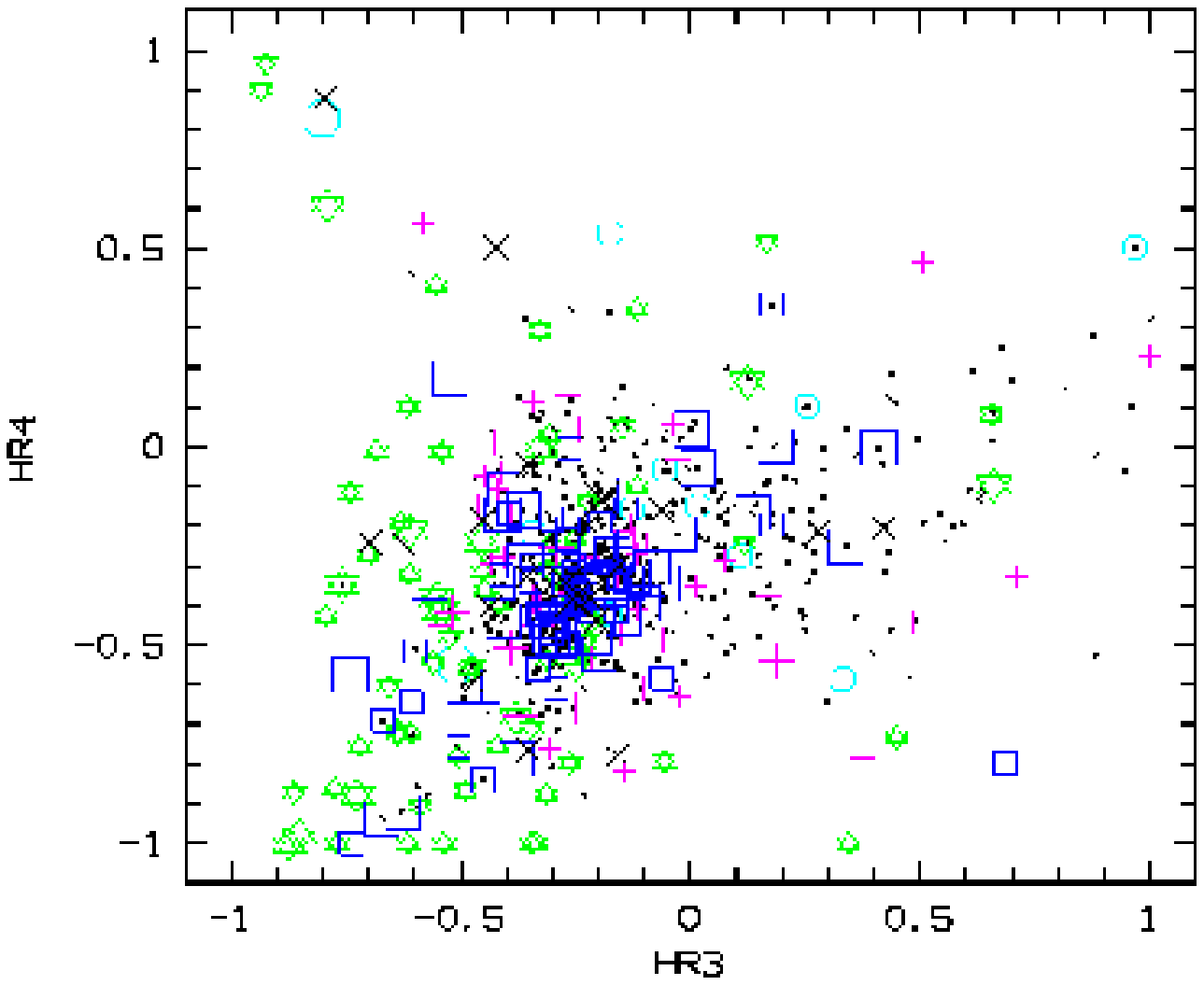}}
 \caption{Hardness ratios of sources detected by \xmm\ EPIC.\@ Sources with HR errors smaller then 0.20 on both HR$(i)$ and HR$(i+1)$ are shown as dots. Foreground stars and candidates are marked as big and small stars, AGN and candidates as big and small crosses, background galaxies and galaxy clusters as big ``X" and their candidates as small ``X", SSS candidates as triangles, SNRs and candidates as big and small octagons, GlCs and XRBs as big squares and their candidates as small squares.}
\label{Fig:HR_diagrams} 
\end{figure}

We plot X-ray colour/colour diagrams based on the HRs (see Fig.\,\ref{Fig:HR_diagrams}). Sources are plotted as dots if the error in both contributing HRs is below 0.2. Classified and identified sources are plotted as symbols in all cases. Symbols including a dot therefore mark the well-defined HRs of a class.

From the HR1-HR2 diagram (upper panel in Fig.\,\ref{Fig:HR_diagrams}) we note that the class of SSSs is the only one that can be defined based on hardness ratios alone. In the part of the HR1-HR2 diagram that is populated by SNRs, most of the foreground stars and some background objects and XRBs  are also found. 

Foreground star candidates can be selected from the HR2-HR3 diagram (middle panel in Fig.\,\ref{Fig:HR_diagrams}), where most of them are located in the lower left corner. The HR3-HR4 diagram (lower panel in Fig.\,\ref{Fig:HR_diagrams}) does not help to disentangle the different source classes. Thus, we need additional information from correlations with sources in other wavelengths or on the source variability or extent to be able to classify the sources.

\subsection{Extended sources}
\label{Sec:ExtSrcs}
The \XLPt\ catalogue contains 12 sources which are fitted as extended sources with a likelihood of extension larger than 15.\@ This value was chosen so as to minimise the number of spurious detections of extended sources (H.~Brunner; private communication), as well as keeping all sources that can clearly be seen as extended sources in the X-ray images. A convolution of a $\beta$-model cluster brightness profile \citep{1976A&A....49..137C} with the \xmm\ point spread function was used to determine the extent of the sources (\cf\ Sect.\,\ref{Sec:SrcDet}). This model describes the brightness profile of galaxy clusters, as
\begin{equation}
f\lb(x,y\rb)=\lb(1+\frac{\lb(x-x_0\rb)^2+\lb(y-y_0\rb)^2}{r_{\rm{c}}^2}\rb)^{-3/2},
\end{equation}
where $r_{\rm{c}}$ denotes the core radius; this is also the extent parameter given by {\tt emldetect}.

Table~\ref{Tab:ExtSrcs} gives the source number (Col.~1), likelihood of detection (2), the extent found (3) and its associated error (4) in arcsec, the likelihood of extension (5), and the classification of the source (6, see Sect.\,\ref{SubSec:Gal_GCl_AGN}) for each of the 12 extended sources. Additional comments taken from Table~5 are provided in the last column.

\begin{table*}
\scriptsize
\begin{center}
\caption{Extended sources in the \XLPt\ catalogue}
\begin{tabular}{rrrrrrrcl}
\hline\noalign{\smallskip}
\hline\noalign{\smallskip}
\multicolumn{1}{l}{SRC} & \multicolumn{1}{l}{DET\_ML}& \multicolumn{1}{l}{EXT$^{+}$}& \multicolumn{1}{l}{EEXT$^{+}$}& \multicolumn{1}{l}{EXT\_ML}& \multicolumn{1}{l}{XFLUX$^{*}$}& \multicolumn{1}{l}{XEFLUX$^{*}$}& \multicolumn{1}{l}{class}& \multicolumn{1}{l}{comment$^{\dagger}$}\\
\multicolumn{1}{c}{(1)} & \multicolumn{1}{c}{(2)}& \multicolumn{1}{c}{(3)}& \multicolumn{1}{c}{(4)}& \multicolumn{1}{c}{(5)}& \multicolumn{1}{c}{(6)}& \multicolumn{1}{c}{(7)}& \multicolumn{1}{c}{(8)}& \multicolumn{1}{c}{(9)}\\
\hline\noalign{\smallskip}
 141 &    65.08 & 11.22 & 1.29 &   23.68 &  1.45 & 0.20 & $<$GCl$>$  & GLG127(Gal), 37W 025A (IR, RadioS; NED) \\
 199 &   275.16 & 17.33 & 1.05 &  174.73 &  4.31 & 0.29 & $<$hard$>$ &  \\
 252 &   222.05 & 14.64 & 1.12 &   81.60 &  4.40 & 0.49 & $<$GCl$>$ & 5 optical objects in error box \\
 304 &   299.75 & 15.10 & 0.92 &  133.62 &  2.20 & 0.18 & $<$GCl$>$  & B242 [CHM09]; RBC3.5: $<$GlC$>$ \\
 442 &    33.76 & 11.60 & 1.71 &   15.44 &  1.62 & 0.28 & $<$hard$>$ &  \\
 618 &   271.08 &  6.20 & 0.73 &   42.86 &  3.15 & 0.21 & $<$hard$>$ &  \\
 718 &    77.75 &  7.18 & 1.23 &   21.47 &  0.58 & 0.07 & Gal        & B052 [CHM09], RBC3.5 \\
1\,130 &   168.31 & 10.80 & 0.97 &   44.23 &  3.27 & 0.31 & $<$hard$>$ &  \\
1\,543 &    70.49 & 11.87 & 1.37 &   28.63 &  1.51 & 0.19 & $<$GCl$>$ & [MLA93] 1076 PN (SIM,NED) \\
1\,795 & 11\,416.36 & 18.79 & 0.29 & 4\,169.74 & 98.87 & 1.43 & GCl   & GLG253 (Gal), [B90] 473, z=0.3 [KTV2006] \\
1\,859 &   107.09 & 13.73 & 1.40 &   43.89 &  1.23 & 0.19 & $<$hard$>$ &  \\
1\,912 &   332.06 & 23.03 & 1.23 &  213.90 &  5.43 & 0.37 & $<$GCl$>$  & cluster of galaxies candidate \\
\noalign{\smallskip}
\hline
\noalign{\smallskip}
\end{tabular}
\label{Tab:ExtSrcs}
\end{center}
Notes:\\
$^{ +~}$: Extent and error of extent in units of 1\arcsec; 1\arcsec\ corresponds to 3.8\,pc at the assumed distance of \m31 \\
$^{ *~}$: XID Flux and flux error in units of 1\ergcm{-14} \\
$^{ \dagger~}$: Taken from Table~5
\normalsize
\end{table*}

\begin{figure}
 \resizebox{\hsize}{!}{\includegraphics[clip]{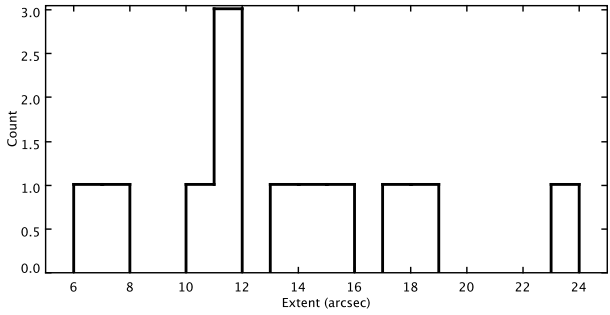}}
 \caption{Distribution of extent parameter.}
\label{Fig:extdist} 
\end{figure}

The extent parameter found for the sources ranges from 6\farcs2 
 to 23\farcs03 (see Fig.\,\ref{Fig:extdist}). The brightest source (\num\ 1\,795), which has the highest likelihood of extension and the second largest extent, was identified from its X-ray properties as a galaxy cluster located behind \m31\ \citep{2006ApJ...641..756K}.\@ The iron emission lines in the X-ray spectrum yield a cluster redshift of $z\!=\!0.29$. For further discussion see Sect.\,\ref{SubSec:Gal_GCl_AGN}.

\section{Variability between \textit{XMM-Newton} observations}
\label{Sec:var}
To examine the long-term time variability of each source, we determined the XID flux at the source position in each observation or at least an upper limit for the XID flux. The XID fluxes were used to derive the variability factor and the significance of variability (\cf\ Sect.\,\ref{Sec:DefVar}).

The sources are taken from the \XLPt\ catalogue (Table~5). Table~8 contains all information necessary to examine time variability. Sources are only included in the table if they are observed at least twice. Column 1 gives the source number. Columns 2 and 3 contain the flux and the corresponding error in the (0.2--4.5) keV XID band. The hardness ratios and errors are given in columns 4 to 11. Column 12 gives the type of the source. All this information was taken from Table~5. 

The subsequent 140 columns provide information related to individual observations in which the position of the source was observed. Column 13 gives the name of one of these observations, which we will call observation 1. The EPIC instruments contributing to the source detection in observation 1, are indicated by three characters  in the ``obs1\_val" parameter (Col. 14, first character for PN, second MOS\,1, third MOS\,2), each one being either a ``T" if the source is inside the FoV, or ``F" if it lies outside the FoV.\@ Then the count rate and error (15,16) and flux and error (17,18) in the (0.2--4.5) keV XID band, and hardness ratios and error (19--26) of observation 1 are given. Corresponding information is given for the remaining observations which cover the position of the source: obs.~2 (cols.~27--40), obs.~3 (41--54), obs.~4 (55--68), obs.~5 (69--82), obs.~6 (83--96), obs.~7 (97--110), obs.~8 (111--124), obs.~9 (125--138), obs.~10 (139--152). Whether the columns corresponding to obs.~3 -- obs.~10 are filled in or not, depends on the number of observations in which the source has been covered in the combined EPIC FoV.\@ This number is indicated in column 153. The maximum significance of variation and the maximum flux ratio (fvar\_max) are given in columns 154 and 155.\@ As described in Sect.\,\ref{Sec:DefVar}, only detections with a significance greater than 3$\sigma$ were used, otherwise the 3$\sigma$ upper limit was adopted. Column 156 indicates the number of observations that provide only an upper limit. The maximum flux (fmax) and its error are given in columns 157 and 158. 

In a few cases a maximum flux value could not be derived, because each observation only yielded an upper limit. There can be two reasons for this: The first reason is that faint sources detected in merged observations may not be detected in the individual observations at the 3$\sigma$ limit. The second reason is that in cases where the significance of detection was not much above the 3$\sigma$ limit, it can become smaller than the 3$\sigma$ limit when the source position is fixed to the adopted final mean value from all observations. 

\addtocounter{table}{+2}

\begin{table*} 
\scriptsize
\begin{center}
\caption{Sources with maximum flux larger than 8\ergcm{-13}, a statistical significance of variability larger than 10 and a flux variability smaller than 5, ordered by flux.} 
\begin{tabular}{llrrrcl}
\hline\noalign{\smallskip}
\hline\noalign{\smallskip}
\multicolumn{1}{c}{Source} &
\multicolumn{1}{c}{fvar} & 
\multicolumn{1}{c}{svar} & 
\multicolumn{1}{c}{fmax$^{\ddagger}$} &
\multicolumn{1}{c}{efmax$^{\ddagger}$} &
\multicolumn{1}{c}{class$^{+}$} &
\multicolumn{1}{c}{Comment$^{\dagger}$}  \\
\noalign{\smallskip}\hline\noalign{\smallskip}
966   &   1.63 & 49.01 & 46.73 & 0.59 & XRB    & 1(sv,z), 2, 10(v), 12(v), 13, 14, 20, 22(v), 25(LMXB), 27, 28(1.56)\\
877   &   3.13  & 49.13 & 16.06 & 0.20 & $<$hard$>$ & 1(sv), 2, 10(v), 12(v), 13, 14, 20(v), 22(v), 27, 28(3.05)\\
745   &   2.43 & 26.89 & 12.65 & 0.18 & AGN    & 13, 14\\ 
1\,157  &   1.32 & 11.10 & 9.87  & 0.25 & GlC    & 1(sv), 2, 5, 10, 12, 13, 14, 20, 21, 22(v), 27, 28(1.37)\\ 
1\,060  &   2.13 & 30.00  & 9.04 & 0.14 & $<$XRB$>$  & 1(sv), 2, 10, 12, 13, 14, 20(v, NS-LMXB), 22(v), 27\\
1\,171  &   4.14 & 18.86 & 9.02  & 0.41 & GlC    & 1(d,sv), 2(t, 53.4), 5, 10, 12, 13, 14, 16, 20, 22, 27, 28(2.47)\\
1\,116  &   3.76 & 51.98 & 8.16 & 0.10 & GlC    & 1(sv), 2(t, 58.6), 3(t, 33), 5, 10, 12, 13, 14, 16, 20, 21, 22(v,t), 27\\
\noalign{\smallskip}
\hline
\noalign{\smallskip}
\end{tabular}
\label{Tab:varlist_bright}
\end{center}
Notes:\\
$^{ {\ddagger}~}$: maximum XID flux and error in units of 1\ergcm{-13} or maximum absorbed 0.2--4.5\,keV luminosity and error in units of 7.3\ergs{36}\\
$^{ {+}~}$: class according to Table~\ref{Tab:class}\\ 
$^{{\dagger}~}$: for comment column see Table~\ref{Tab:varlist}
\normalsize
\end{table*}

\begin{figure}
   \resizebox{\hsize}{!}{\includegraphics[clip,angle=-90,bb=52 90 355 460]{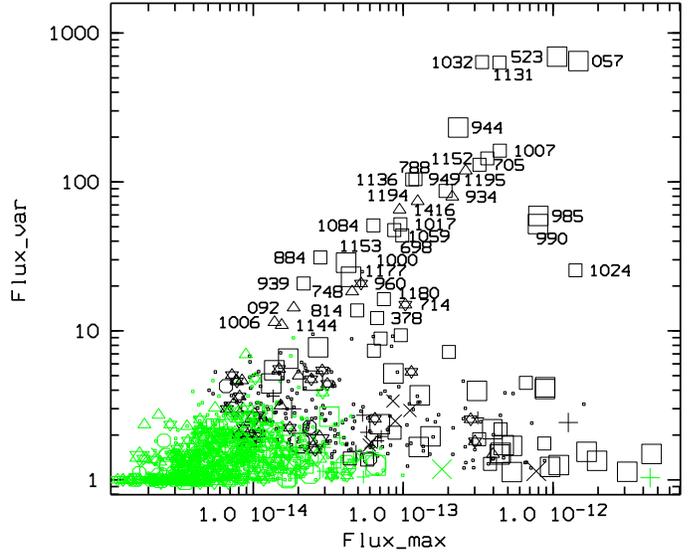}}
     \caption{
     Variability factor of sources from the \XLPt\ catalogue in the 0.2--4.5 keV band derived from average fluxes of the \xmm\ EPIC observations plotted versus maximum detected flux (\hbox{erg cm$^{-2}$ s$^{-1}$}). Source classification is indicated: Foreground stars and candidates are marked as big and small stars, AGN and candidates as big and small crosses, background galaxies and galaxy clusters as big ``X" and their candidates as small ``X", SSS candidates as triangles, SNRs and candidates as big and small octagons, GlCs and XRBs as big squares and their candidates as small squares. Sources with a statistical significance for the variability below 3 are marked in green. 
     }
    \label{Fig:var_fmax} 
\end{figure}

\begin{figure*}
   \resizebox{\hsize}{!}{\includegraphics[clip,angle=-90,bb=50 85 360 490]{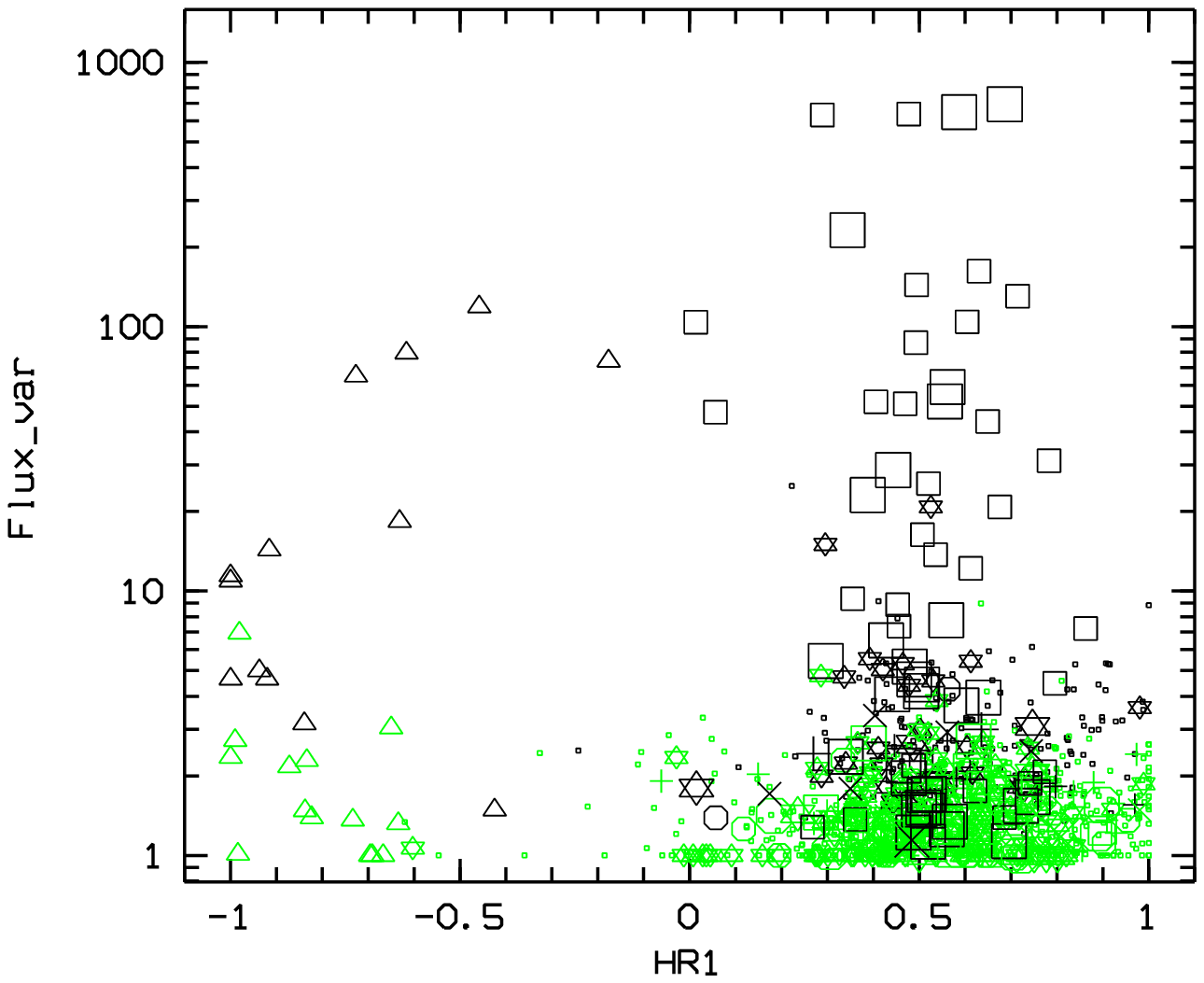}\hskip0.2cm\includegraphics[clip,angle=-90,bb=50 85 360 490]{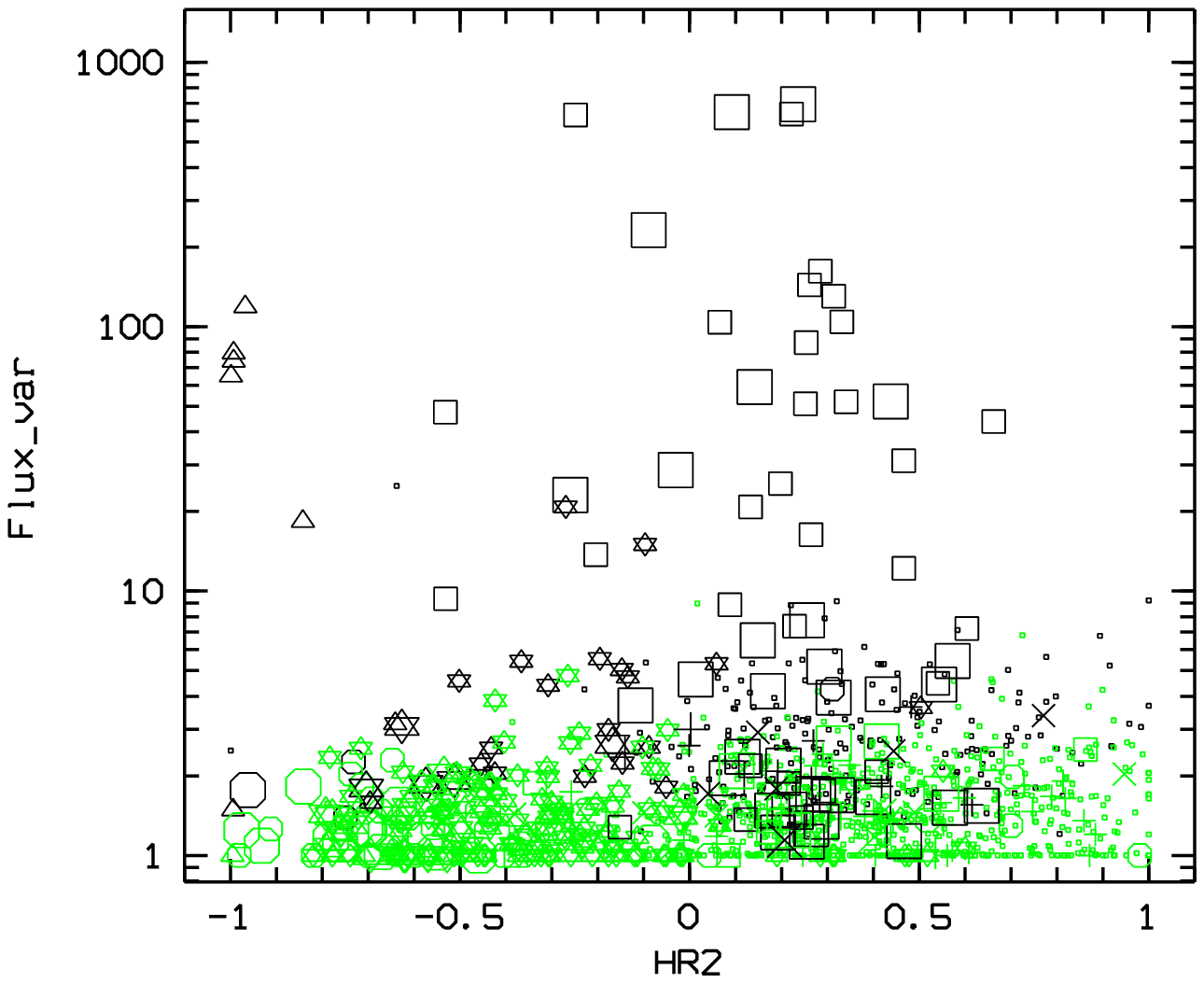}}
     \caption{
     Variability factor of sources from the \XLPt\ catalogue in the 0.2--4.5 keV band (derived from the average fluxes of the \xmm\ EPIC observations) plotted versus HR1 in the left panel and HR2 in the right panel. For source classification see Fig.\,\ref{Fig:var_fmax}. Sources with a statistical significance of the variability below 3 are marked in green. 
     }
    \label{Fig:var_hr} 
\end{figure*}

\vspace{5mm}
Figure~\ref{Fig:var_fmax} shows the variability factor plotted versus maximum detected XID flux. Apart from XRBs, or XRBs in GlCs, or candidates of these source classes, which were selected based on their variability, there are a few SSS candidates showing pronounced temporal variability. 
The sources classified or identified as AGN, background galaxies or galaxy clusters all show $F_{\mr{var}}\!<\!4$.\@ Most of the  foreground stars show $F_{\mr{var}}\!<\!4$. 

Out of the 1\,407 examined sources, we found 317 sources with a  variability significance $>\!3.0$, \ie\ 182 more than reported in SPH2008. 
For bright sources it is much easier to detect variability than for faint sources, because the difference between the maximum observed flux and the detection limit is larger. Therefore the significance of the variability declines with decreasing flux. This is illustrated by the distribution of the sources marked in green in Fig.\,\ref{Fig:var_fmax}.

Table~\ref{Tab:varlist} lists all sources with a variability factor larger than five. There are 69 such sources (34 in addition to SPH2008). The sources are sorted in descending order with respect to their variability factors. Table~\ref{Tab:varlist} gives the source number (Col.~1),  maxima of flux variability (2) and maxima of the significance parameter (3). 
The next columns (4, 5) indicate the maximum observed flux and its error. Column~6 contains the class of the source.  Sources with  $F_{\mr{var}}\!\ge\!10$ that were not already classified as SSS or foreground stars, were classified as XRB. 

Time variability can also be helpful to verify a SNR candidate classification. If there is significant variability, the SNR classification must be rejected, and if an optical counterpart is detected, the source has to be re-classified as foreground star candidate. Column 7 contains references to the individual sources in the literature. In some cases the reference provides information on the temporal behaviour and a more precise classification (see brackets). The numbers given in connection with \citet{2007A&A...468...49V} and \citet{2006ApJ...643..356W} are the variability factors obtained in these papers from \chandra\ data. From the 69 sources of Table~\ref{Tab:varlist}, ten show a flux variability larger than 100.\@ With a flux variability factor $>\!690$ source \num\ 523 is the most variable source in our sample. Source \num\ 57 has the largest significance of variability, with a value of $\approx 97$. The variability significance is below 10 for just 33 sources, 15 of which show significance values below 5. 
Thirty-five of the variable sources are classified as XRBs or XRB candidates, and eight of them are located in globular clusters. Nine of the variable sources are SSS candidates, while six variable sources are classified as foreground stars and foreground star candidates.

Table~\ref{Tab:varlist_bright} lists all ``bright" sources with a maximum flux larger than 8\ergcm{-13} and a flux variability smaller than five (the description of the columns is the same as in Table~\ref{Tab:varlist}). 
All seven sources listed in Table~\ref{Tab:varlist_bright} (three in addition to SPH2008) have a significance of variability $>\!10$.\@ Apart from two sources, they are XRBs (three in globular clusters) or XRB candidates. The most luminous source in our sample is source \num\ 966 with an absorbed 0.2--4.5\,keV luminosity of $\approx 3.3$\ergs{38} at maximum.

Figure~\ref{Fig:var_hr} shows the relationship between the variability factor and the hardness ratios HR1 and HR2, respectively. The hardness ratios are taken from Table~5. The HR1 plot shows that the sample of highly variable sources includes SSS and XRB candidates, which occupy two distinct regions in this plot \citep[see also ][ for the LMC]{1999A&A...344..521H}. The SSSs marked by triangles, appear on the left hand side, while the XRBs or XRB candidates have much harder spectra, and appear on the right. It seems that foreground stars, SSSs and XRBs can be separated, on the HR2 diagram, although there is some overlap between foreground stars and XRBs.

Individual sources are discussed in the Sects.\,\ref{Sec:fgback} and \ref{Sec:Srcsm31}.

\section{Cross-correlations with other \m31\ X-ray catalogues}
\label{Sec:CrossX-ray}
Cross-correlations were determined by applying Eq.\,\ref{Eq:Cor} to the sources of the \XLPt\ catalogue and to sources reported in earlier X-ray catalogues. The list of X-ray catalogues used is given in Table~\ref{Tab:XrayRefCat}.

\subsection{Previous \textit{XMM-Newton} catalogues}
\label{SubSec:prevXMM}
Previous source lists based on archival \xmm\ observations were presented in \citet{2001A&A...378..800O}, PFH2005, \citet{2006ApJ...643..844O},   
SPH2008, and SBK2009. Of these four studies, PFH2005 covers the largest area of \m31. Table \ref{Tab:CompXMM} lists all sources from previous \xmm\ studies that are not detected in the present investigation.

\begin{table*}
\scriptsize
\begin{center}
\caption{Sources from previous \xmm\ studies that are not listed in the \XLPt\ catalogue.}
\begin{tabular}{ll}
\hline\noalign{\smallskip}
\hline\noalign{\smallskip}
\multicolumn{2}{l}{PFH2005 856 sources}\\
\multicolumn{2}{l}{103 not detected}\\
6 not detected, LH$>$100: & 327 ($<$SNR$>$,LH$=$2140.0), 384 (XRB,667.0), 332 ($<$SNR$>$,654.0), \\
& 316 ($<$SNR$>$,259.0), 312 ($<$SNR$>$,241.0), 281($<$hard$>$,160.0)\\
10 not detected, 20$\le$LH$<$50: & 75 ($<$SSS$>$), 423 ($<$fg Star$>$), 120 ($<$hard$>$), 505 ($<$hard$>$), \\
& 220 ($<$SNR$>$), 304 ($<$fg Star$>$), 819 ($<$hard$>$), 799 ($<$SSS$>$), 413 ($<$SNR$>$), 830 ($<$hard$>$)\\
14 not detected, 15$\le$LH$<$20: & 427($<$hard$>$), 734 ($<$hard$>$), 424 ($<$hard$>$), 518 ($<$SSS$>$), \\
& 232 ($<$hard$>$), 339 ($<$hard$>$), 446 ($<$SSS$>$), 219 ($<$fg Star$>$), 567 ($<$hard$>$), 256 ($<$fg Star$>$), \\
& 356 ($<$hard$>$), 248 ($<$hard$>$), 160 ($<$hard$>$), 399 ()\\
21 not detected, 10$\le$LH$<$15: & 375 ($<$hard$>$), 17 ($<$hard$>$), 195 ($<$hard$>$), 417 ($<$SNR$>$), \\
& 783 ($<$hard$>$), 803 ($<$hard$>$), 829 ($<$hard$>$), 135 ($<$hard$>$), 151 ($<$hard$>$),  131 ($<$hard$>$), \\
& 426 ($<$hard$>$), 593 ($<$fg Star$>$), 526 ($<$hard$>$), 250 ($<$hard$>$), 62 ($<$hard$>$), 67 ($<$hard$>$), \\
& 188 ($<$hard$>$), 186 ($<$AGN$>$), 510 ($<$hard$>$), 529 ($<$hard$>$),  754 ($<$hard$>$)\\ 
52 not detected, LH$<$10: & 599 ($<$hard$>$), 439 ($<$hard$>$), 809 ($<$hard$>$), 14 ($<$SNR$>$), 743 ($<$hard$>$),\\
& 433 ($<$hard$>$), 5 (), 210 ($<$hard$>$), 97 ($<$hard$>$), 708 ($<$hard$>$), 476 (), 534 ($<$hard$>$), 501 (),\\
& 170 ($<$hard$>$), 146 (SNR), 769 (), 838 ($<$hard$>$), 571 ($<$hard$>$), 816 ($<$hard$>$, 554 (), 627 ($<$hard$>$),\\
& 464 ($<$fg Star$>$), 811 ($<$hard$>$), 655 ($<$hard$>$), 184 ($<$hard$>$), 447 ($<$hard$>$), 380 ($<$hard$>$),\\
& 566 ($<$hard$>$), 137 ($<$fg Star$>$), 63 (), 48 (), 152 ($<$fg Star$>$),  291 ($<$hard$>$), 559 ($<$hard$>$),\\
& 102 ($<$hard$>$), 740 ($<$hard$>$), 540 ($<$fg Star$>$), 240 ($<$hard$>$), 485 (), 668 ($<$hard$>$), 44 (),\\
& 560 ($<$hard$>$), 836 ($<$hard$>$), 436 ($<$hard$>$), 484 ($<$fg Star$>$), 216 ($<$hard$>$), 362 ($<$hard$>$), 527 ($<$$>$), 179 ($<$hard$>$),\\
& 834 ($<$hard$>$), 86 ($<$hard$>$), 455 ()\\
\noalign{\smallskip}
\hline
\noalign{\smallskip}
\multicolumn{2}{l}{SPH2008 39 sources}\\
\multicolumn{2}{l}{15 not detected}\\
3 not detected, 50$\le$LH$<$100: & 874 ($<$SNR$>$,LH$=$85.5), 895 ($<$hard$>$,75.9), 882 ( ,56.4)\\
6 not detected, 10$\le$LH$<$50: & 869 (), 885 ($<$SNR$>$), 863 ($<$hard$>$), 875 ($<$SSS$>$), 893  ($<$hard$>$), 866  ($<$hard$>$)\\
6 not detected, LH$<$10: & 870 ($<$SNR$>$), 891 ($<$hard$>$), 889 ($<$hard$>$), 872 ($<$SNR$>$), 867 ($<$hard$>$), 862 ($<$SNR$>$)\\
\noalign{\smallskip}
\hline
\noalign{\smallskip}
\multicolumn{2}{l}{SBK2009 335 sources}\\
\multicolumn{2}{l}{31 not detected}\\
& 4 ($<$hard$>$), 18 ($<$hard$>$), 29 ($<$hard$>$), 32 ($<$hard$>$), 34 ($<$hard$>$), 45 ($<$SSS$>$), 67 ($<$hard$>$),\\
& 102 ($<$hard$>$), 106 ($<$hard$>$), 117 ($<$hard$>$), 149 ($<$hard$>$), 152 ($<$hard$>$), 179 ($<$hard$>$),\\
& 183 ($<$hard$>$), 184 ($<$hard$>$), 188 ($<$hard$>$), 191 ($<$hard$>$), 192 ($<$AGN$>$), 202 ($<$hard$>$),\\
& 204 ($<$fg star$>$), 217 ($<$hard$>$), 249 ($<$hard$>$), 250 ($<$hard$>$), 260 ($<$hard$>$), 274 ($<$hard$>$),\\
& 279 ($<$hard$>$), 285 ($<$hard$>$), 295 ($<$hard$>$), 306 ($<$hard$>$), 325 ($<$hard$>$), 333 ($<$hard$>$)\\
\noalign{\smallskip}
\hline
\noalign{\smallskip}
\end{tabular}
\label{Tab:CompXMM}
\end{center}
\normalsize
\end{table*}

In the ten observations covering the major axis, and a field in the halo of \m31, PFH2005 detected 856 X-ray sources with a detection likelihood threshold of 7 (\cf\ Sect.\,\ref{Sec:SrcDet}). Of these 856 sources, only 753 sources are also present in the \XLPt\ catalogue, \ie\ 103 sources of PFH2005 were not detected. This can be due to:  
the search strategy; the parameter settings used in the {\tt emldetect} run; the determination of the extent of a source for the \XLPt\ catalogue; the more severe screening for GTIs for the \XLPt\ catalogue, which led to shorter final exposure times; the use of the {\tt epreject} task and last but not least due to the {\tt SAS} versions and calibration files applied. The search strategy of PFH2005 was optimised to detect sources located close to each other in crowded fields. This point especially explains the non-detection of the bright PFH2005 sources [PFH2005] 281, 312, 316, 327, 332, 384 ($\mathcal{L}>50$) in the present study, as four of them ([PFH2005] 312, 316, 327, 332) are located in the innermost central region of \m31\ where source detection is complicated by the bright diffuse X-ray emission, while [PFH2005] 281 and 384 lie in the immediate vicinity of two bright sources ([PFH2005] 280 and 381  at distances of 7.7\arcsec and 5.5\arcsec, respectively). The changes in the {\tt SAS} versions and the GTIs, in particular, affect sources with small detection likelihoods ($\mathcal{L}<10$). 

The improvements in the {\tt SAS} detection tools and calibration files should reduce the number of spurious detections, which increase with decreasing detection likelihood. However, this does not necessarily imply that \emph{all} undetected  sources with $\mathcal{L}<10$ of PFH2005 are spurious detections. The changes in the {\tt SAS} versions, calibration files and GTIs do not only affect the source detection tasks, but also can cause changes in the background images. These changes may increase the assumed background value at the position of a source, which would result in a lower detection likelihood. Going from {\tt mlmin\,=\,7} to {\tt mlmin\,=\,6}, but leaving everything else unchanged, we detected an additional nine sources of PFH2005. One of the previously undetected sources ([PFH2005] 75) was classified as $<$SSS$>$, but correlates with blocks of pixels with enhanced low energy signal in the PN offset map and was corrected by {\tt epreject}. Another source classified as $<$SSS$>$ ([PFH2005] 799) is only detected in the MOS\,1 camera, but not in MOS\,2. From an examination by eye, it seems that source [PFH2005] 799 is the detection of some noisy pixels at the rim of the MOS\,1 CCD\,6 and not a real X-ray source.\@

SPH2008 extended the source catalogue of PFH2005 by re-analysing the data of the central region of \m31\ and also including data of monitoring observations of LMXB RX J0042.6+4115.\@ Of the 39 new sources presented in SPH2008, 24 are also listed in the \XLPt\ catalogue, \ie\ 15 sources of SPH2008 were not detected. Differences between the two studies include the detection likelihood thresholds used for {\tt eboxdetect} (SPH2008: {\tt likemin}=5) and {\tt emldetect} (SPH2008: {\tt mlmin}=6), the lower limit for the likelihood of extention (SPH2008: {\tt dmlextmin}\,=\,4; \XLPt: 15), the screening for GTIs, the use of the {\tt epreject} task and the {\tt SAS} versions and calibration files used. Concerning the GTIs, images, background images and exposure maps SPH2008 followed the same procedures as in PFH2005. The arguments given above are therefore also valid here. From the 14 undetected sources, three sources were detected in SPH2008 with {\tt mlmin} $<$ 7.\@  
One source ([SPH2008] 882) was added by hand to the final source list, as SPH2008 could not find any reason why {\tt emldetect} did not automatically find it. The two extended sources ([SPH2008] 863, 869) detected with extent likelihoods between 4.7 and 5.1 in SPH2008, are neither detected as extended nor as pointlike sources in the present study, where the extent likelihood has to be larger than 15.\@

SBK2009 re-analysed the \xmm\ observations located along the major axis of \m31, ignoring all observations pointing to the centre of the galaxy. They used a detection likelihood threshold of ten. Of the 335 sources detected by SBK2009, 304 sources are also contained in the \XLPt\ catalogue, \ie\ 31 sources are not detected. Of the 304 re-detected sources, two ([SBK2009] 298, 233) are found with a detection likelihood below ten. Of the 31 undetected sources, 27 were also not detected in PFH2005. The remaining four sources correlate with PFH2005 sources, which were not detected in the present study. SBK2009 state that they find 34 sources not present in the source catalogue of PFH2005. A possible reason for this may be that SBK2009 used different energy bands for source detection. They also had five bands, but they combined bands 2 and 3 from PFH2005 into one band in the range 0.5--2\,keV, and on the other hand they split band 5 of PFH2005 into two bands from 4.5--7\,keV and from 7--12\,keV, respectively. This might also explain why most of the additional found sources were classified as $<$hard$>$.

\citet{2006ApJ...643..844O} addressed the population of SSSs and QSSs based on the same archival observations as PFH2005. \citet{2006ApJ...643..844O} detected 15 SSSs, 18 QSSs and 10 SNRs of which one ([O2006]~Table\,4, Src.\,3) is also listed as an SSS ([O2006]~Table\,2, Src.\,13). Of these sources two SSSs, four QSSs and two SNRs (among them is the source [O2006]~Table\,4, Src.\,3) are not contained in the \XLPt\ catalogue. 
These seven sources are also not present in the PFH2005 catalogue.

The nine bright variable sources from \citet{2001A&A...378..800O} were all detected.

\subsection{\textit{Chandra} catalogues}
\label{SubSec:Chcat}
The \chandra\ catalogues used for cross-correlations were presented in Sect.\,\ref{Sec:Intro} (see also Table~\ref{Tab:XrayRefCat}).

Details of the comparison between the \XLPt\ catalogue and the different \chandra\ catalogues can be found in Table~\ref{Tab:CompChan}.\@ Here, we only give a few general remarks. A non-negligible number of \chandra\ sources not reported in the \XLPt\ catalogue have already been classified as transient or variable sources. Thus, it is not surprising that these sources were not detected in the \xmm\ observations \citep[parts of:][DKG2004]{2007A&A...468...49V,2006ApJ...643..356W}. One \chandra\ source (n1-66) lies outside the field of \m31\ covered by the \xmm\ observations. For the innermost central region of \m31, the point spread function of \xmm\ causes source confusion and therefore only \chandra\ observations are able to resolve the individual sources, especially if they are faint compared to the diffuse emission or nearby bright sources  \citep{2002ApJ...577..738K,2002ApJ...578..114K,2004ApJ...609..735W,2004ApJ...610..247D,2006ApJ...643..356W,2007A&A...468...49V}. This explains why a certain number of these sources are not detected in \xmm\ observations.

\begin{table*}
\scriptsize
\begin{center}
\caption{Sources detected in previous \chandra\ studies that are not present in the \XLPt\ catalogue.}
\begin{tabular}{ll}
\hline\noalign{\smallskip}
\hline\noalign{\smallskip}
\multicolumn{2}{l}{\citet{2002ApJ...577..738K} 204 sources}\\
\multicolumn{2}{l}{58 not detected}\\
5 transient: & r3-46,r3-43,r2-28,r1-23,r1-19 \\
20 variable: & r3-53,r3-77,r3-106,r3-76,r2-52,r2-31,r2-23,r1-31,r2-20,r1-24,r1-28,r1-27,r1-33,r1-21,r1-20,r1-7,r2-15,r1-17,r1-16,r2-47\\
33 unclassified: & r3-102,r3-92,r3-51,r3-75,r3-91,r3-89,r3-101,r3-88,r2-44,r2-55,r2-54,r3-32,r2-53,r1-30,r3-99,r1-22,r1-26,r1-18,r3-26,\\
&r2-41,r2-40,r3-71,r2-50,r2-49,r2-38,r3-97,r2-46,r3-12,r3-66,r3-104,r3-82,r3-5,r3-4\\
\noalign{\smallskip}
\hline
\noalign{\smallskip}
\multicolumn{2}{l}{\citet{2002ApJ...578..114K} 142 sources}\\
\multicolumn{2}{l}{26 not detected}\\
3 transient: & J004217.0+411508,J004243.8+411604,J004245.9+411619\\
7 variable: & J004232.7+411311,J004242.0+411532,J004243.1+411640,J004244.3+411605,J004245.2+411611,J004245.5+411608,\\
&J004248.6+411624\\
16 unclassified: & J004207.3+410443,J004229.1+412857,J004239.5+411614,J004239.6+411700,J004242.5+411659,J004242.7+411503,\\
&J004243.1+411604,J004244.2+411614,J004245.0+411523,J004246.1+411543,J004247.4+411507,J004249.1+411742,\\
& J004251.2+411639,J004252.3+411734,J004252.5+411328,J004318.5+410950\\
\noalign{\smallskip}
\hline
\noalign{\smallskip}
\multicolumn{2}{l}{\citet{2004ApJ...609..735W} 166 sources}\\
\multicolumn{2}{l}{28 not detected}\\
12 transient: & s1-79,s1-80,s1-82,r3-46,r2-28,r1-23,r1-19,r2-69,r1-28,r1-35,r1-34,n1-85 \\
7 variable: & r2-31,r1-31,r1-24,r1-20,r1-7,r1-17,r1-16\\
9 unclassified: & s1-81,r2-68,s1-85,r1-30,r1-22,r1-26,r1-18,n1-77,n1-84\\
\noalign{\smallskip}
\hline
\noalign{\smallskip}
\multicolumn{2}{l}{\citet{2007A&A...468...49V} 261 sources}\\
\multicolumn{2}{l}{104 not detected}\\
11 transient: & 6,12,29,32,41,51,59,84,118,130,146 \\
15 variable: & 3,5,8,9,18,22,24,27,44,63,92,96,99,149,169\\
78 unclassified: & 4,19,21,25,26,30,37,39,40,42,48,49,53,56,57,58,60,62,65,70,73,75,76,77,80,82,84,86,87,89,91,94,97,98,104,109,114,\\
&115,117,119,122,124,129,133,138,141,143,144,145,150,152,158,162,164,167,171,173,182,183,188,189,191,193,194,\\
&197,202,205,206,210,213,217,219,220,225,256,257,263\\
\noalign{\smallskip}
\hline
\noalign{\smallskip}
\multicolumn{2}{l}{\citet{2006ApJ...643..356W} 45 sources}\\
\multicolumn{2}{l}{25 not detected}\\
25 transient: & n1-26,n1-85,n1-86,n1-88,n1-89,r1-19,r1-23,r1-28,r1-34,r1-35,r2-28,r2-61,r2-62,r2-66,r2-69,r2-72,r3-43,r3-46,s1-18,\\
& s1-27,s1-69,s1-79,s1-80,s1-82,s2-62 \\
\noalign{\smallskip}
\hline
\noalign{\smallskip}
\multicolumn{2}{l}{DKG2004 43 sources}\\
\multicolumn{2}{l}{15 not detected}\\
9 transient: & s2-62,s1-27,s1-69,s1-18,n1-26,r2-62,r1-35,r2-61,r2-66 \\
5 unclassified: & s2-27,s2-10,n1-29,n1-46,r2-54\\
1 not in FoV: & n1-66\\
\noalign{\smallskip}
\hline
\noalign{\smallskip}
\multicolumn{2}{l}{\citet{2002ApJ...570..618D} 28 sources}\\
\multicolumn{2}{l}{2 not detcted}\\
2 unclassified: & 17 ($\hat{=}$ r2-15), 28 ($\hat{=}$ r3-71) \\
\noalign{\smallskip}
\hline
\noalign{\smallskip}
\end{tabular}
\label{Tab:CompChan}
\end{center}
Notes:\\
Variability information (transient, variable) is taken from the papers. ``Unclassified" denotes sources which are not indicated as transient or variable sources in the papers.
\normalsize
\end{table*}

Of the 28 bright X-ray sources located in globular clusters \citep{2002ApJ...570..618D}, two were not found in the \xmm\ data (see Table~\ref{Tab:CompChan}). They are also not included in the source catalogue of PFH2005 and SPH2008. Hence, both objects are good candidates for being transient or at least highly variable sources (\cf\ Sect.\,\ref{SubSub:comp_GlC}). Another study of the globular cluster population of \m31\ is presented by \citet{2004ApJ...616..821T}.\@ Their work is based on \xmm\ and \chandra\ data and contains 43 X-ray sources. Of these sources three were not found in the present study. One of them ([TP2004] 1) is located well outside the field of \m31\ covered by the Deep \xmm\ Survey\footnote{The source was observed with \xmm\ on 11 January 2001. Obs.~id.: 0065770101}. The second source ([TP2004] 21) correlates with r3-71, which is discussed above (see \citet{2002ApJ...570..618D} in Table~\ref{Tab:CompChan}). The transient nature of the third source ([TP2004] 35), and the fact that it was not observed in any \xmm\ observation taken before 2004 was already reported by \citet{2004ApJ...616..821T}. The source was first detected with \xmm\ in the observation from 31 December 2006. 

\subsection{\textit{ROSAT} catalogues}
Of the 86 sources detected with \ros\ HRI in the central $\sim$34\arcmin\ of \m31 (PFJ93), all but eight sources ([PFJ93] 1,2,31,33,40,48,63,85) are detected in the \xmm\ observations. Six of these eight sources ([PFJ93] 1,2,31,33,63,85) have already been discussed in PFH2005 and classified as transients. Sources [PFJ93] 40 and 48 correlate with [PFH2005] 312 and 332, respectively, which are discussed in Sect.\,\ref{SubSec:prevXMM}. In addition to these eight sources, PFH2005 did not detect source [PFJ93] 51. This source was detected in the \xmm\ observations centred on RX J0042.6+4115 and was thus classified as a recurrent transient (see SPH2008).

In each of the two \ros\ PSPC surveys of \m31, 396 individual X-ray sources were detected (SHP97 and SHL2001). From the SHP97 catalogue 130 sources were not detected. Of these sources 48 are located outside the FoV of our \xmm\ \m31\ survey. From the SHL2001 catalogue, 93 sources are not detected, 60 of which lie outside the \xmm\ FoV.\@ For information on individual sources see Table~\ref{Tab:CompRos}. 
\begin{table*}
\scriptsize
\begin{center}
\caption{Sources from the \ros\ PSPC catalogues that are not present in the \XLPt\ catalogue.}
\begin{tabular}{ll}
\hline\noalign{\smallskip}
\hline\noalign{\smallskip}
\multicolumn{2}{l}{SHP97 396 sources}\\
\multicolumn{2}{l}{130 not detected}\\
48 outside FoV: & 1,2,3,4,5,7,8,14,31,41,72,91,98,104,120,125,159,202,209,271,276,285,286,290,300,312,314,\\
& 320,342,350,363,367,371,374,383,385,386,387,388,389,390,391,392,393,394,395,396 \\
1 transient: & 69 \\
21 not detected, LH$<$12: & 19,24,27,33,46,52,59,63,68,71,133,149,161,264,273,275,307,329,330,358,377\\
16 not detected, 12$\le$LH$<$15: & 12,15,49,82,93,113,114,128,196,230,262,283,334,364,372,376\\
44 not detected, LH$\ge$15: & 16(LH$=$26.6),32(30.2),43(18.2),45(51.2),60(20.1),66(36.2),67(4536.2),78(20.5),80(16.3),81(26.6),\\
& 88(33.7),95(548.0),102(16.4),126(217.3),141(843.3),145(46.9),146(673.7),166(17.4),167(90.0),\\
& 171(54.3),182(454.4),186(39.8),190(113.0),191(54.5),192(54.3),203(103.3),214(400.2),215(251.0),\\
& 232(104.4),245(26.0),260(54.6),263(38.1),265(24.6),268(54.3),270(40.4),277(15.6),309(81.8),\\
& 319(23.4),331(19.5),335(51.2),340(27.5),341(28.1),365(22.4),373(69.5)\\
\noalign{\smallskip}
\hline
\noalign{\smallskip}
\multicolumn{2}{l}{SHL2001 396 sources}\\
\multicolumn{2}{l}{93 not detected}\\
60 outside FoV: & 1,2,3,4,5,6,7,8,9,10,11,12,14,15,16,21,22,32,39,58,67,69,75,77,81,83,85,90,93,125,141,146,\\
& 160,164,192,202,243,260,282,296,298,302,325,326,328,355,371,372,378,379,383,388,389,390,\\
& 391,392,393,394,395,396 \\ 
4 not detected, LH$<$12: & 62,96,238,269\\
2 not detected, 12$\le$LH$<$15: & 231,361\\
27 not detected, LH$\ge$15: & 51(LH$=$28.4),104(901.2),121(94.1),126(46.2),143(34.7),168(131.9),171(43.0),173(317.8),190(215.8),\\
& 207(98.0),208(298.8),226(73.1),230(75.6),232(1165.6),240(218.4),246(39.9),248(219.6),256(60.0),\\
& 267(22.2),271(52.8), 322(2703.3),324(147.7),344(40.7),356(15.3),365(19.0),380(17.4),384(15.8)\\
\noalign{\smallskip}
\hline
\noalign{\smallskip}
\end{tabular}
\label{Tab:CompRos}
\end{center}
\normalsize
\end{table*}

Forty-four (out of 302) sources from SHP97 and 27 (out of 293) sources from SHL2001, have \ros\ detection likelihoods larger than 15, but are not listed in the \XLPt\ catalogue. These sources have to be regarded as transient or at least highly variable.

\subsection{\textit{Einstein} catalogue}

The list of \ein\ X-ray sources in the field of \m31\ reported by TF91 contains 108 sources, with 81 sources taken from the \ein\ HRI data with an assumed positional error of 3\arcsec\ \citep[reported by][]{1984ApJ...284..663C}, and 27 sources based on \ein\ IPC data with a 45\arcsec\ positional error. Applying the above mentioned correlation procedure to the \ein\ HRI sources, 64 of these sources are also detected in this work and listed in the \XLPt\ catalogue, \ie\ 17 sources are not detected ([TF91] 29, 31, 35, 39, 40, 43, 46, 50, 53, 54, 65, 66, 72, 75, 78, 93, 96). For the \ein\ IPC sources only the 1~$\sigma$ positional error was used to search for counterparts among the \xmm\ sources. Of the 27 \ein\ IPC sources six remain without a counterpart in our catalogue ([TF91] 15, 99, 100, 106, 107, 108), of which [TF91] 15 and 108 are located outside the field of \m31\ covered by the \XLPt\ catalogue. Sources [TF91] 50 and 54 correlate with [PFH2005] 312 and 316, respectively. Both sources were already discussed in Sect.\,\ref{SubSec:prevXMM}. Apart from [TF91] 106, which is suggested as a possible faint transient by SHL2001, the remaining 18 sources are also not detected by PFH2005. They classified those sources as transient. 
   
\section{Cross-correlations with catalogues at other wavelengths}
\label{SEC:CCow}
The \XLPt\ catalogue was correlated with the catalogues and public data bases given in Sect.\,\ref{Sec:CrossCorr_Tech}. Two sources (from the \XLPt\ and from the reference catalogues) were be considered as correlating, if their positions matched within the uncertainty (see Eq.~\ref{Eq:Cor}).

However, the correlation of an X-ray source with a source from the reference catalogue does not necessarily imply that the two sources are counterparts. To confirm this, additional information is needed, like corresponding temporal variability of both sources or corresponding spectral properties. We should also take into account the possibility that the counterpart of the examined X-ray source is not even listed in the reference catalogue used (due to faintness for example).

The whole correlation process will get even more challenging if an X-ray source correlates with more than one source from the reference catalogue. In this case we need a method to decide which of the correlating sources is the most likely to correspond to the X-ray source in question. Therefore, the method used should indicate how likely the correlation is with each one of the sources from the reference catalogue. Based on these likelihoods one can define criteria to accept a source from the reference catalogue as being the most likely source to correspond to the X-ray source. 

The simplest method uses the spatial distance between the X-ray source and the reference sources to derive the likelihoods. In other words, the source from the reference catalogue that is located closest to the X-ray source is regarded as the most likely source corresponding to the X-ray source. 

An improved method is a ``likelihood ratio" technique, were an additional source property (\eg\ an optical magnitude in deep field studies) is used to strengthen the correlation selection process. This technique was applied successfully to deep fields to find optical counterparts of X-ray sources \citep[\eg\ ][]{2007ApJS..172..353B}. A drawback of this method is that one a priori has to know the expected probability distribution of the optical magnitudes of the sources belonging to the studied object. In our case, this means that we have to know the distribution function for all optical sources of \m31\ that can have X-ray counterparts, \emph{without} including foreground and background sources. Apart from the fact that such distribution functions are unknown, an additional challenge would be the time dependence of the magnitude of the optical sources (\eg\ of novae) and of the connection between optical and X-ray sources (\eg\ optical novae and SSSs). Therefore it is not possible to apply this ``likelihood ratio" technique to the sources in the \XLPt\ survey. The whole correlation selection process becomes even more challenging if more than one reference catalogue is used.

To be able to take all available information into account, we decided not to automate the selection process, but to select the class and most likely correlations for each source by hand (as it was done \eg\ in PFH2005). Therefore the source classification, and thus the correlation selection process, is based on the cross correlations between the different reference catalogues, on the X-ray properties (hardness ratios, extent and time variability), and on the criteria given in Table~\ref{Tab:class}. For reasons of completeness we give for each X-ray source the number of correlations found in the USNO-B1, 2MASS and LGGS catalogues in Table~5.\@ The caveat of this method is that it cannot quantify the probability of the individual correlations.

\section{Foreground stars and background objects}
\label{Sec:fgback}
\subsection{Foreground stars}  
\label{Sec:fgStar}
X-ray emission has been detected from many late-type -- spectral types F, G, K, and M -- stars, as well as from hot OB stars \citep[see review by][]{2000RvMA...13..115S}. Hence, X-ray observations of nearby galaxies also reveal a significant fraction of Galactic stars. 
With typical absorption-corrected luminosities of $L_{\mathrm{0.2-10\,keV}}\!<$\oergs{31}, single stars in other galaxies are too faint to be detected with present instruments. However, concentrations of stars can be detected, but not resolved.

\begin{figure}
 \resizebox{\hsize}{!}{\includegraphics[clip,angle=0]{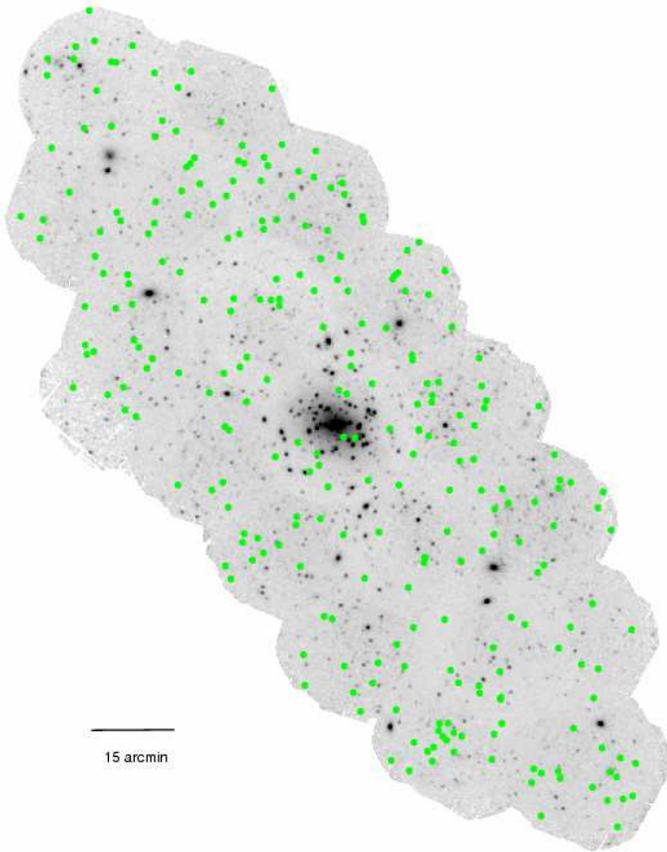}}
 \caption{The spatial distribution of foreground stars and candidates, classified in the \XLPt\ catalogue. The image shows the homogeneous distribution of the sources over the covered field (marked with green dots).}
 \label{Fig:fgS_spdist} 
\end{figure}

Foreground stars (fg Stars) are a class of X-ray sources which are homogeneously distributed over the field of \m31 (Fig.\,\ref{Fig:fgS_spdist}). The good positional accuracy of \xmm\ and the available catalogues USNO-B1, 2MASS and LGGS allow us to efficiently select this type of source. The selection criteria are given in Table~\ref{Tab:class}. The optical follow-up observations of \citet{2006A&A...451..835H} and \citet{2009A&A...507..705B} have confirmed the foreground star nature of bright foreground star candidates selected in PFH2005, based on the same selection criteria as used in this paper. 
Somewhat different criteria were applied for very red foreground stars, with an LGGS colour $\mr{V}-\mr{R}\!>\!1$ or USNO-B1 colour $\mr{B2}-\mr{R2}\!>\!1$. These are classified as foreground star candidates, if $f_{\mr{x}}/f_{\mr{opt}}\!<\!-0.65$ and $f_{\mr{x}}/f_{\mr{opt,R}}\!<\!-1.0$.\@ A misclassification of symbiotic systems in \m31 as foreground objects by this criterion can be excluded, as symbiotic systems  typically have X-ray luminosities below \oergs{33}, which is more than a factor 100 below the detection limit of our survey.

If the foreground star candidate lies within the field covered by the LGGS we checked its presence in the LGGS images (as the LGGS catalogue itself does not list bright stars, because of saturation problems). Otherwise DSS2 images were used. Correlations with bright optical sources from the USNO-B1 catalogue, with an $f_{\mr{x}}/f_{\mr{opt}}$ in the range expected for foreground stars, that were not visible in the optical images were rejected as spurious. We found 223 foreground star candidates. Fourty sources were identified as foreground stars, either because they are listed in the globular cluster catalogues as spectroscopically confirmed foreground stars or because they have a spectral type assigned to them in the literature \citep[][SIMBAD]{2009A&A...507..705B,2006A&A...451..835H}.

Two of the foreground star candidates close to the centre of \m31\ (\num\ 826, \num\ 1\,110) have no entry in the USNO-B1 and LGGS catalogues, and one has no entry in the USNO-B1 R2 and B2 columns (\num\ 976). However, they are clearly visible on LGGS images, they are 2MASS sources and they fulfil the X-ray hardness ratio selection criteria. Therefore, we also classify them as foreground stars.

The following 19 sources were selected as very red foreground star candidates: \num\ 54, \num\ 118, \num\ 384, \num\ 391, \num\ 393, \num\ 585, \num\ 646, \num\ 651, \num\ 711, \num\ 1\,038, \num\ 1\,119, \num\ 1\,330, \num\ 1\,396, \num\ 1\,429, \num\ 1\,506, \num\ 1\,605, \num\ 1\,695, \num\ 1\,713 and \num\ 1\,747. 
A further 10 sources (\num\ 210, \num\ 269, \num\ 278, \num\ 310, \num\ 484, \num\ 714, \num\ 978, \num\ 1\,591, \num\ 1\,908 and \num\ 1\,930) fulfil the hardness ratio criteria, but violate the $f_{\mr{x}}/f_{\mr{opt}}$ criteria and are therefore marked as ``foreground star candidates" in the comment column of Table~5.\@

\begin{figure*}
 \subfigure[\num\ 473]{\includegraphics[scale=0.3, angle=-90]{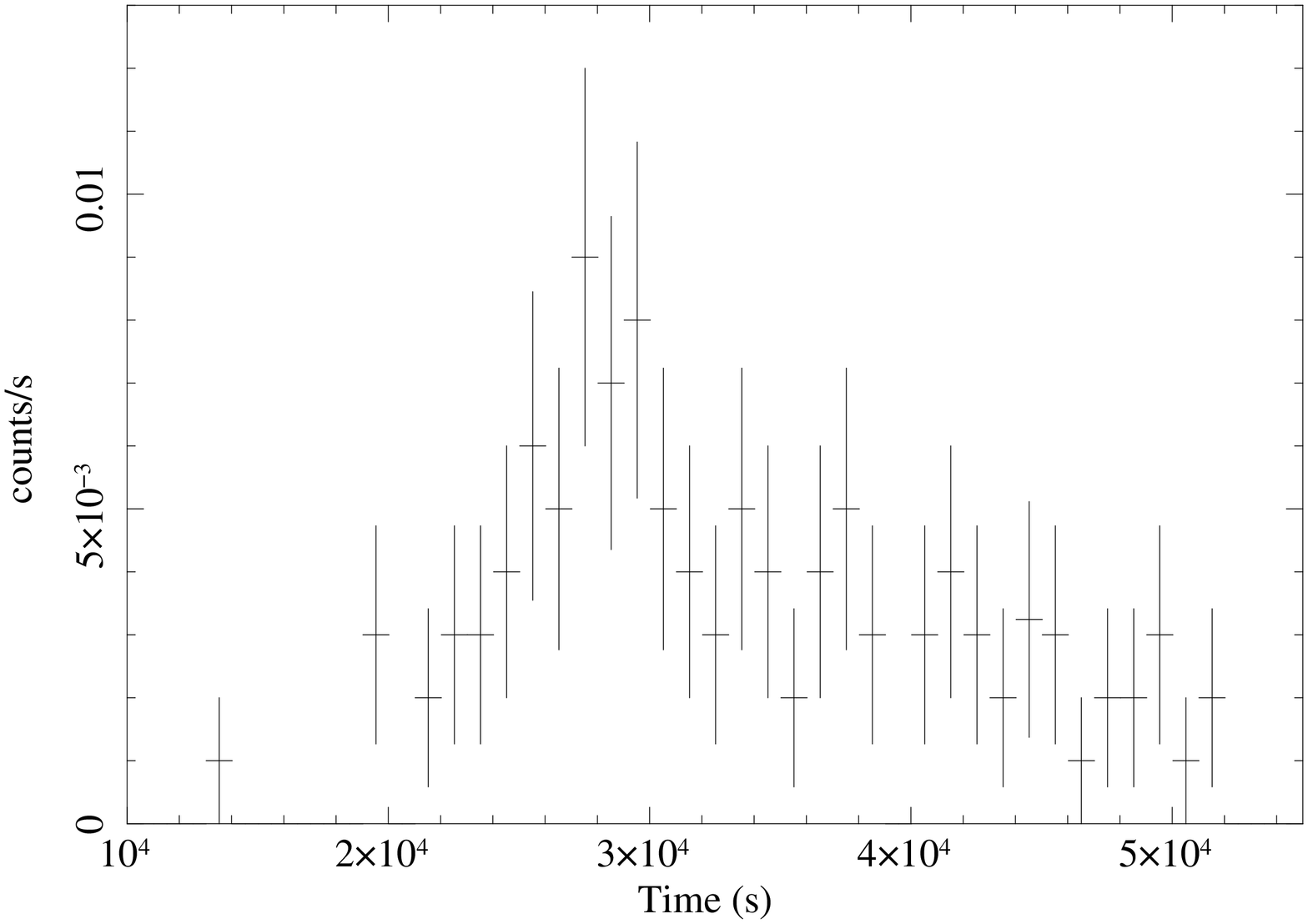}\label{SubFig:fgS_flare_1}}
 \subfigure[\num\ 780]{\includegraphics[scale=0.3, angle=-90]{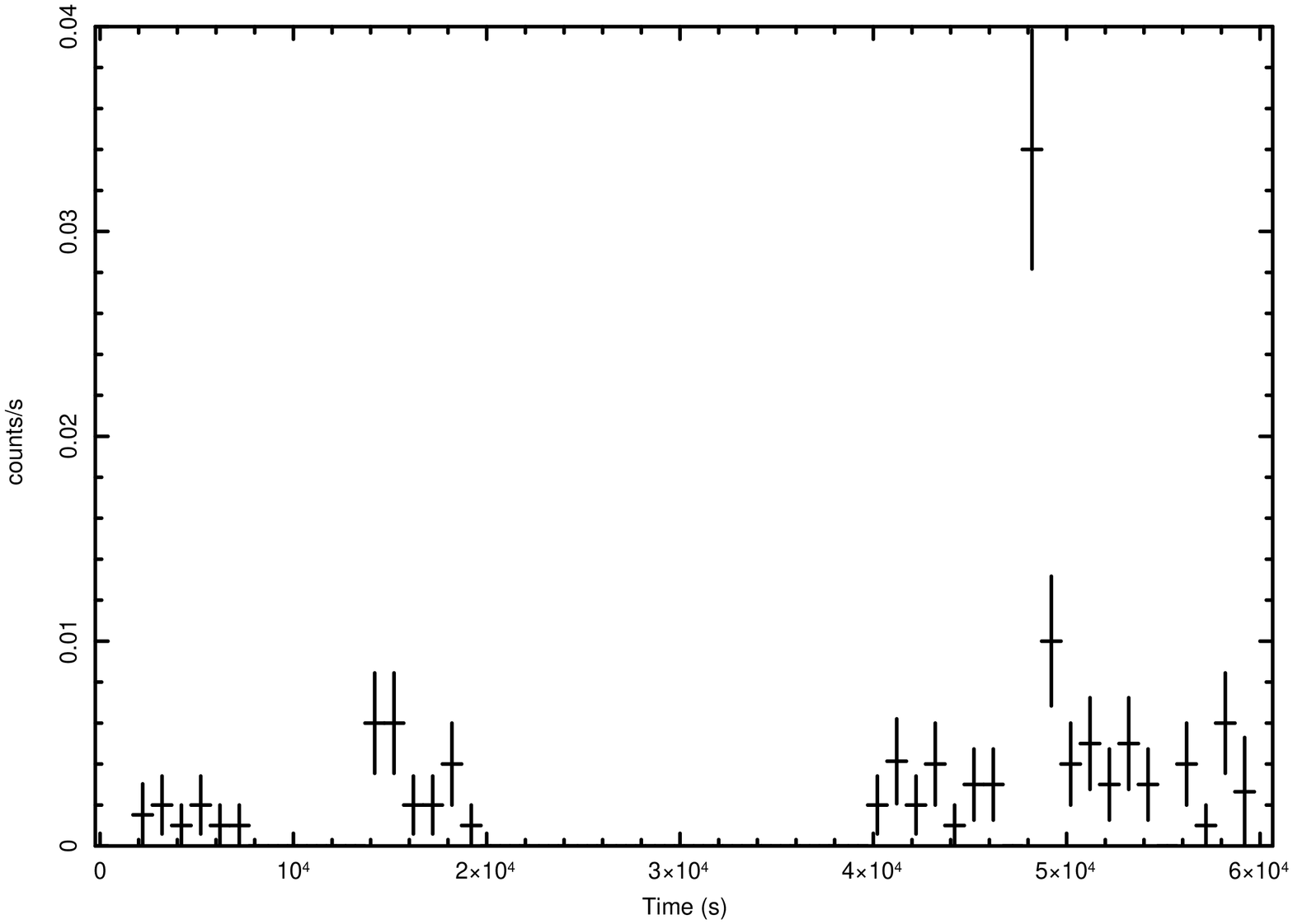}\label{SubFig:fgS_flare_2}}\\
 \subfigure[\num\ 1\,551]{\includegraphics[scale=0.3, angle=-90]{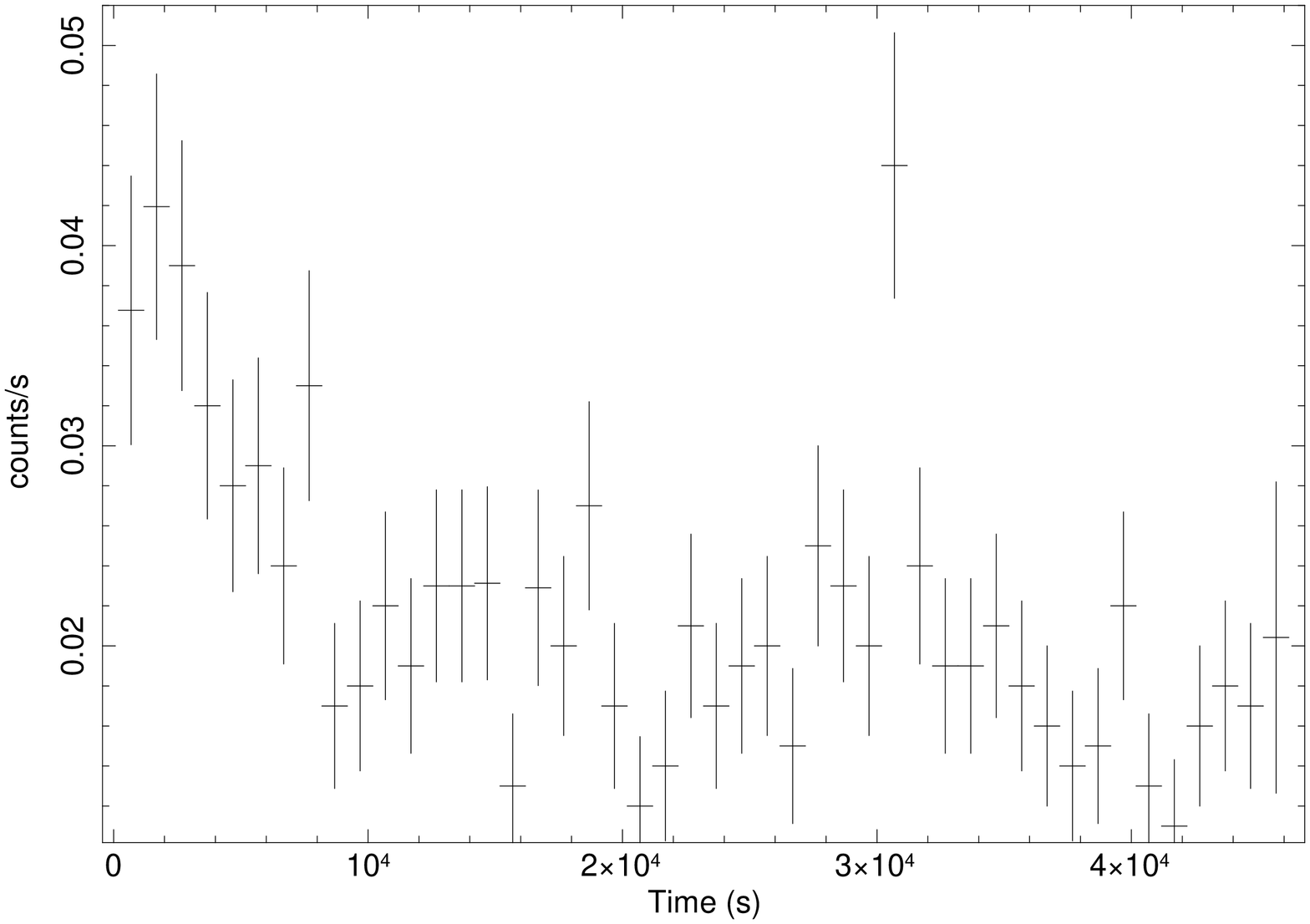}\label{SubFig:fgS_flare_3}}
 \subfigure[\num\ 1\,585]{\includegraphics[scale=0.3, angle=-90]{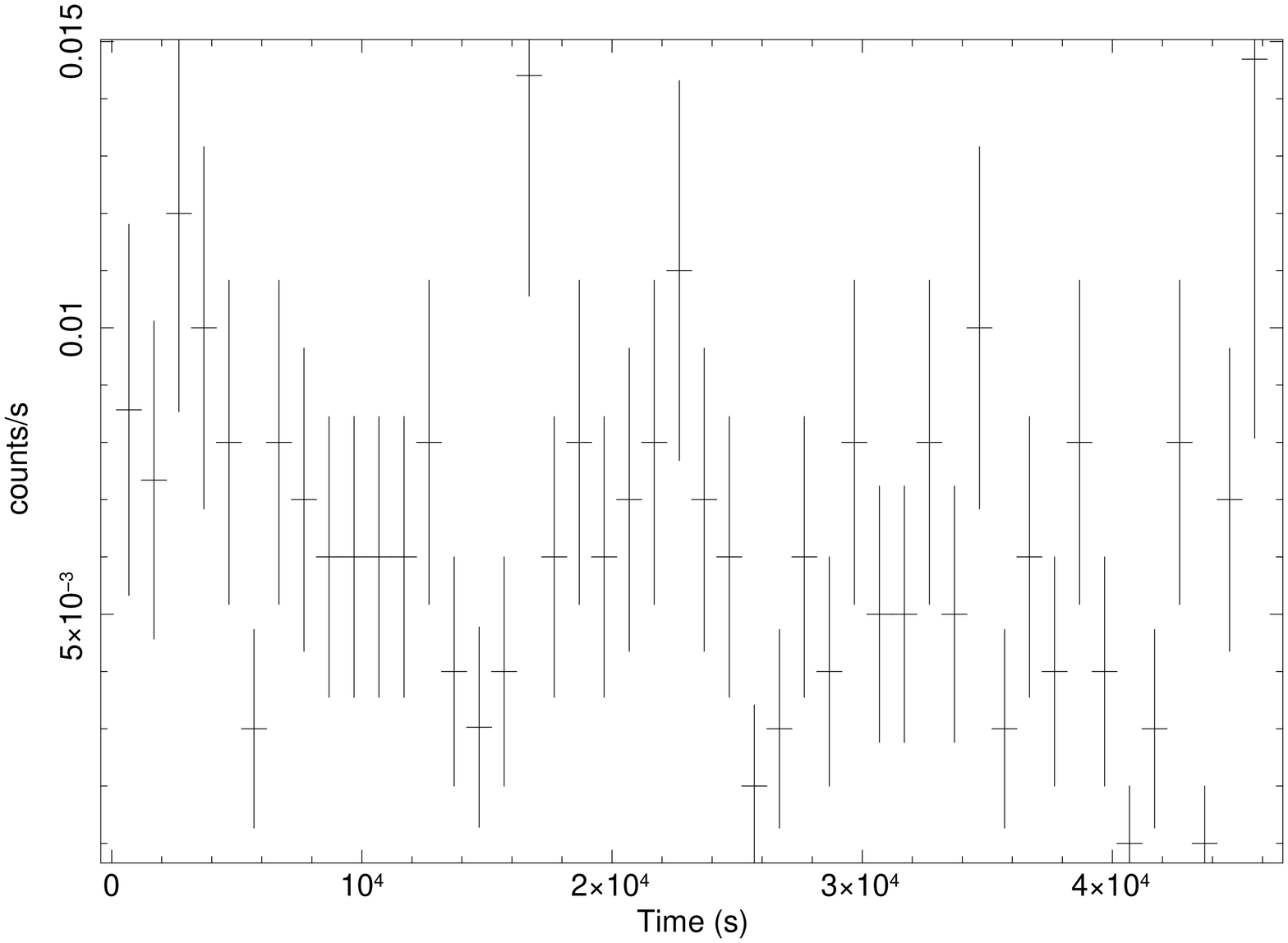}\label{SubFig:fgS_flare_4}}\\
 \subfigure[\num\ 1\,676]{\includegraphics[scale=0.3, angle=-90]{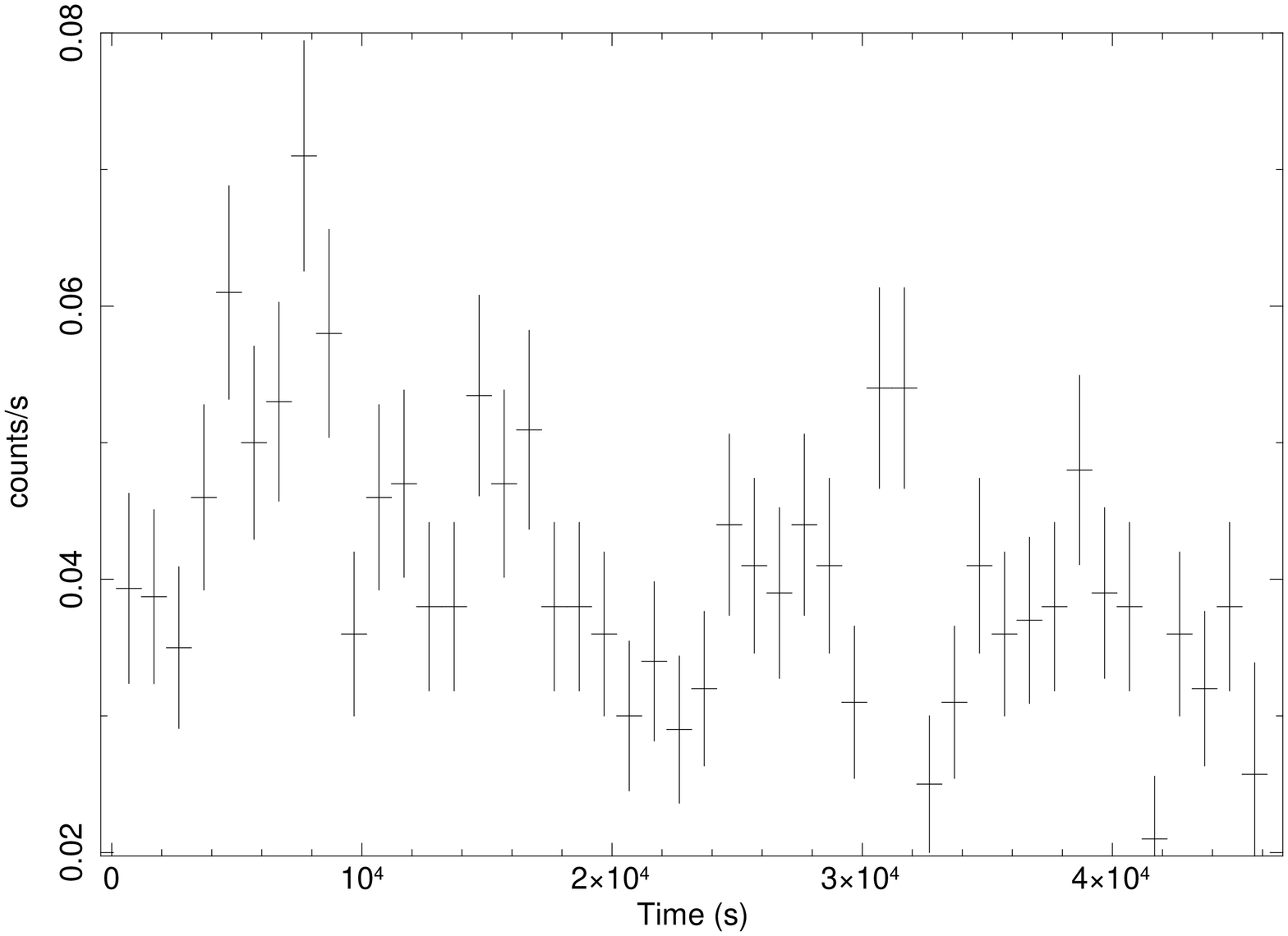}\label{SubFig:fgS_flare_5}}
 \subfigure[\num\ 1\,742]{\includegraphics[scale=0.3, angle=-90]{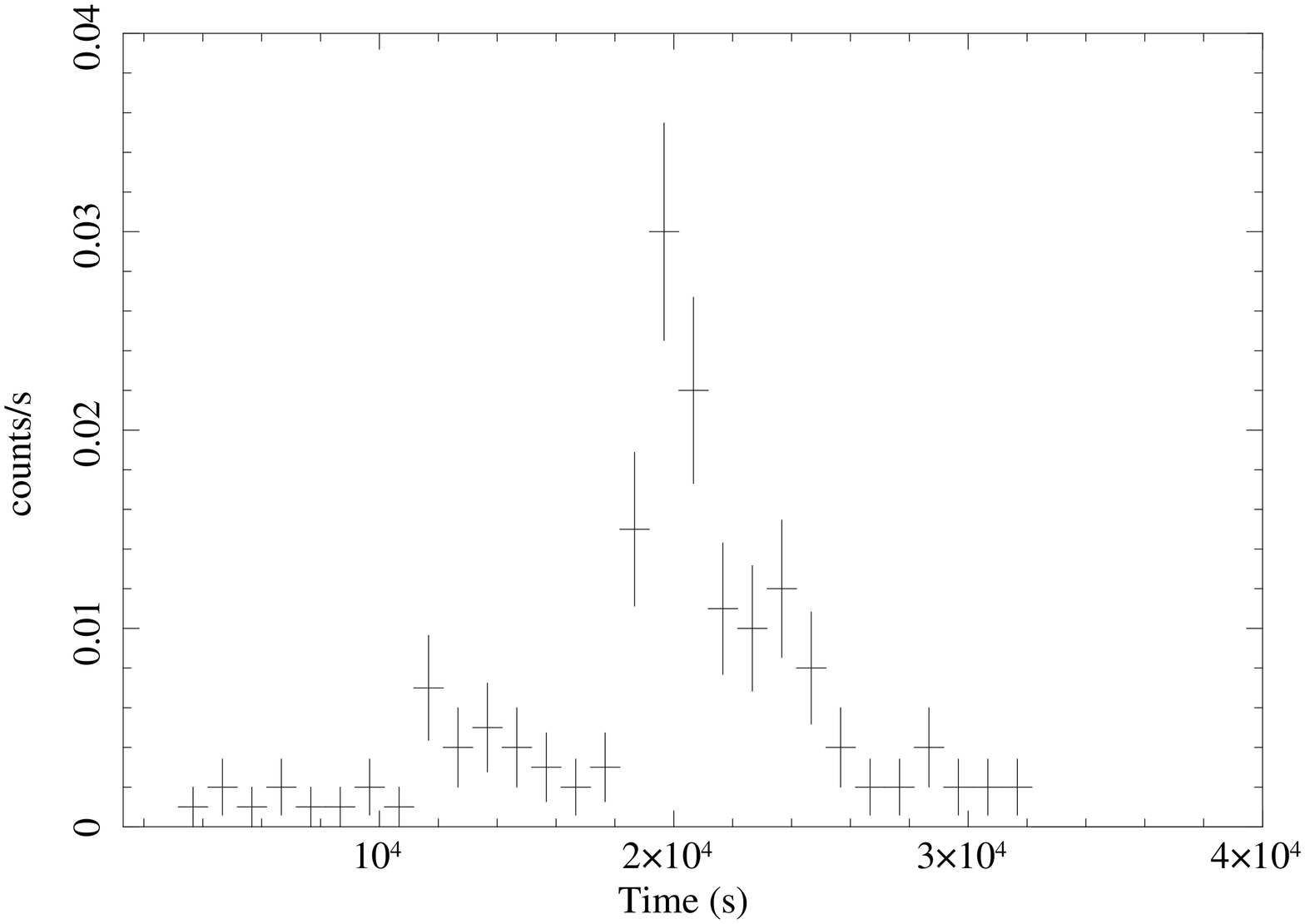}\label{SubFig:fgS_flare_6}}\\
 \subfigure[\num\ 714]{\includegraphics[scale=0.3, angle=-90]{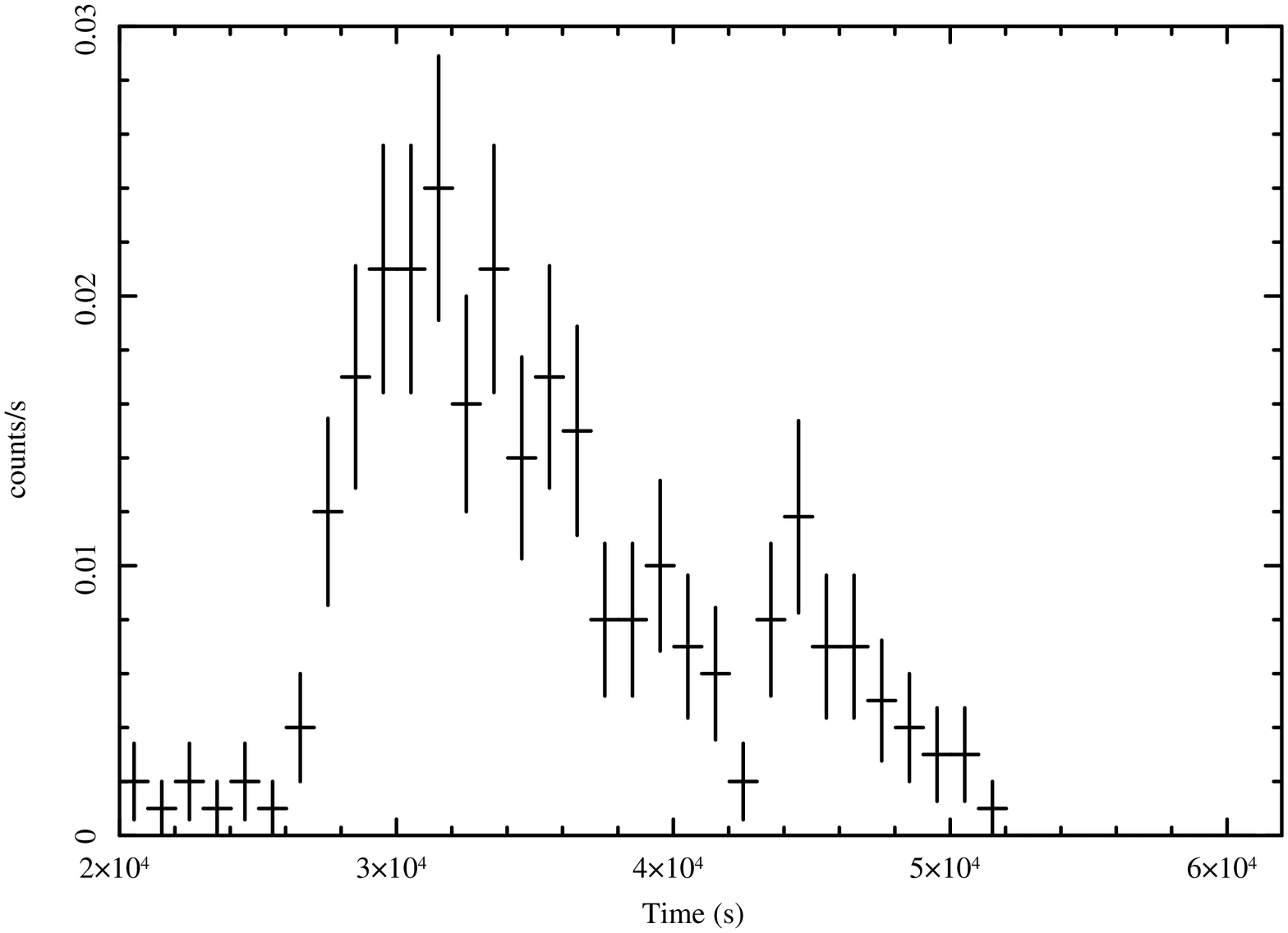}\label{SubFig:fgS_flare_8}}\\
 \caption{X-ray light curves of foreground stars and candidates that, with a binning of 1000\,s, show flares.}
 \label{Fig:fgS_flare} 
\end{figure*}

Six sources (\num\ 473, \num\ 780, \num\ 1\,551, \num\ 1\,585, \num\ 1\,676, \num\ 1\,742), classified as foreground star candidates, have X-ray light curves that in a binning of 1\,000\,s showed flares (see Fig.\,\ref{Fig:fgS_flare}). These observations strengthen the foreground star classification. A seventh source (\num\ 714) is classified as a foreground star candidate, since its hardness ratios and its $f_{\mr{x}}/f_{\mr{opt}}$ ratio in the quiescent state fulfil the selection criteria of foreground star candidates. In addition, the source shows a flare throughout observation ss3.\@ Hence, the $f_{\mr{x}}/f_{\mr{opt}}$ ratio for this observation, in which the source is brightest, is too high to be consistent with the range of values expected for foreground stars. 

\begin{table}
\begin{center}
\caption{Infrared colours and spectral types of foreground stars that show flares.}
\begin{tabular}{rrrrrr}
\hline\noalign{\smallskip}
\hline\noalign{\smallskip}
\multicolumn{1}{l}{\num} & \multicolumn{1}{c}{J mag} & \multicolumn{1}{l}{H mag} & \multicolumn{1}{c}{K mag} & \multicolumn{1}{c}{SpT$^{*}$}& \multicolumn{1}{c}{err$^{+}$}\\
\hline\noalign{\smallskip}
 473 & 12.984 & 12.681 & 12.558 & K0 & 0.4 \\
 714 & 14.310 & 13.618 & 13.458 & M0 & 0.2 \\
 780 & 14.251 & 13.595 & 13.351 & M3 & 0.1 \\
1551 & 12.666 & 12.009 & 11.806 & M2 & 0.1 \\
1585 & 13.488 & 12.899 & 12.650 & M2 & 0.1 \\
1676 & 10.460 &  9.878 &  9.798 & K1 & 0.2 \\
1742 & 13.722 & 13.138 & 12.896 & M1 & 0.2 \\
\noalign{\smallskip}
\hline
\noalign{\smallskip}
\end{tabular}
\label{Tab:fgStar_flare}
\end{center}
Notes:\\
$^{ *~}$: spectral type\\
$^{ +~}$: error (in subtypes)
\normalsize
\end{table}

Table~\ref{Tab:fgStar_flare} gives the J, H and K magnitudes taken from the 2MASS catalogue for each of the six flaring foreground stars. Using the standard calibration of spectral types for dwarf stars based on their near infrared colours (from the fourth edition of Allen's astrophysical quantities, edt. A.Cox, p.151) we derived the spectral classification for the objects, using both H-K and J-K. The spectral types (and "error") we give in Table~\ref{Tab:fgStar_flare} are derived from averaging the two classes (derived from the two colours). The spectral types are entirely consistent with those expected for flare stars (usually K and M types).

\begin{figure}
\resizebox{\hsize}{!}{\includegraphics[clip,angle=0]{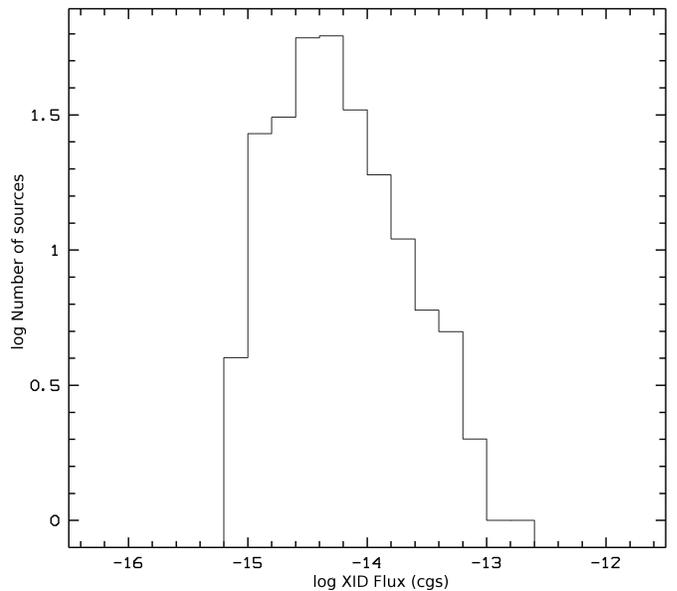}}
\caption{Distribution of the source fluxes in the 0.2--4.5\,keV (XID) band. The diagram shows a histogram of the number of foreground stars and candidates per flux bin, in logarithmic scales.} 
\label{Fig:fgS_fldist}
\end{figure}

Figure~\ref{Fig:fgS_fldist} shows the XID flux distribution for foreground stars and foreground star candidates, which ranges from 6.9\ergcm{-16} to 2.0\ergcm{-13}. Most of the foreground stars and candidates (257 sources) have fluxes below 5\ergcm{-14}.
 
\subsubsection{Comparing \textit{XMM-Newton}, \textit{Chandra} and \textit{ROSAT} catalogues}
In the combined \ros\ PSPC survey (SHP97, SHL2001) 55 sources were classified as foreground stars. Of these, 14 sources remain without counterparts in the present \xmm\ survey. Five of these 14 sources are located outside the field observed with \xmm. Forty-one \ros\ foreground star candidates have counterparts in the \XLPt\ catalogue. Of these counterparts, 16 were classified as foreground star candidates and four were identified as foreground stars \citep[spectral type from][or SIMBAD]{2009A&A...507..705B,2006A&A...451..835H}.\@ In addition 12 sources were listed as $<$hard$>$, two as AGN candidates and one as a globular cluster candidate in the \XLPt\ catalogue. The counterparts of three \ros\ sources remain without classification in the \XLPt\ catalogue. 

Another three \ros\ sources have more than one counterpart in the \xmm\ data. Source [SHP97]~109 correlates with sources \num\ 597, \num\ 604, \num\ 606, and \num\ 645. The former three are classified as $<$hard$>$, while source \num\ 645 is classified as a foreground star candidate. However source \num\ 645 has the largest distance from the position of [SHP97]~109 compared to the other three \xmm\ counterparts. Furthermore, this source had a flux below the \ros\ detection threshold (about a factor 2.6) in the \xmm\ observations and is about a factor 3--34 fainter than the three other possible \xmm\ counterparts. Thus it is very unlikely that [SHP97]~109 represents the X-ray emission of a foreground star. 

Source [SHL2001]~156 has two \xmm\ counterparts and is discussed in Sect.\,\ref{Sec:SSS_comp}.\@ The third source ([SHL2001]~374) correlates with sources \num\ 1\,922 and \num\ 1\,924. The two \xmm\ sources are classified as $<$hard$>$ and as a foreground star candidate, respectively. In the source catalogue of SHL2001 source [SHP97]~369 is listed as the counterpart of [SHL2001]~374. The source in the first \ros\ survey has a smaller positional error and only correlates with source \num\ 1\,924. Although this seems to indicate that source \num\ 1\,924 is the counterpart of [SHL2001]~374, we cannot exclude the possibility that [SHL2001]~374 is a blend of both \xmm\ sources, as these two sources have similar luminosities in the \xmm\ observations.

\citet{2002ApJ...577..738K} classified four sources as foreground stars. For two sources (\num\ 960$\hat{=}$r2-42 and \num\ 976$\hat{=}$r3-33) the classification is confirmed by our study. The third source (\num\ 1000$\hat{=}$r2-19) remained without classification in the \XLPt\ catalogue, as it is too soft to be classified as $<$hard$>$ and the optical counterpart found in the LGGS catalogue does not fulfil the $f_{\mr{x}}/f_{\mr{opt}}$ criteria. The fourth source (r2-46) was not detected in the \xmm\ observations.

The foreground star classification of three sources (s1-74, s1-45, n1-82) in \citet{2004ApJ...609..735W} is confirmed by the \XLPt\ study (\num\ 289, \num\  603, \num\ 1\,449). For source \num\ 289 the spectral type F0 was determined \citep{2006A&A...451..835H}.

The source list of DKG2004 contains six sources (s2-46, s2-29, s2-37, s1-45, s1-20, r3-122) that are classified as foreground stars. All six sources are confirmed as foreground star candidates by our \xmm\ study (\cf\ Table~5). For source \num\ 696 ($\hat{=}$s1-20) \citet{2006A&A...451..835H} obtained the spectral type G0.

Of the four sources listed as foreground stars in \citet{2007A&A...468...49V} only one source (\num\ 936$\hat{=}$ [VG2007]~168) was confirmed as a foreground star, based on the entry in the RBC\,V3.5 and \citet{2009AJ....137...94C}. The second source (\num\ 1\,118$\hat{=}$[VG2007]~180) is listed in the RBC\,V3.5 and \citet{2009AJ....137...94C} as a globular cluster. The third source (\num\ 829$\hat{=}$[VG2007]~181) does not have a counterpart in the USNO-B1, 2MASS or LGGS catalogues, nor does it fulfil the hardness ratio criteria for foreground stars. Hence, the source is classified as $<$hard$>$. The fourth source ([VG2007]~81) is not spatially resolved from its neighbouring source [VG2007]~79 in our \xmm\ observations (source \num\ 1\,078). Hence source \num\ 1\,078 is classified as $<$hard$>$.

\subsection{Galaxies, galaxy clusters and AGN}
\label{SubSec:Gal_GCl_AGN}
The majority of background sources belong to the class of active galactic nuclei (AGN). This was shown by the recent deepest available surveys of the X-ray background \citep[][]{2000Natur.404..459M,2001A&A...365L..45H,2005ARA&A..43..827B}. The class of AGN is divided into many sub-sets. The common factor in all the sub-sets is that their emission emanates from a small, spatially unresolved galactic core. The small size of the emitting region is implied by the X-ray flux variability observed in many AGN, which is on time scales as short as several minutes (to years). The observed X-ray luminosities range from \oexpo{39} to \oergs{46}, sometimes even exceeding \oergs{46}. Although AGN show many different properties, like the amount of radio emission or the emission line strengths and widths, they are believed to be only different facets of one underlying basic phenomenon \citep[\cf\ ][]{1995PASP..107..803U}: the accretion of galactic matter onto a supermassive black hole ($\sim\!10^{6}\!-\!10^{9}$\,M\subsun) in the centre of the galaxy. 

It is difficult and, to some extent, arbitrary to distinguish between active and normal galaxies, since most galaxies are believed to host a black hole at the position of their kinetic centre \citep{2005ApJ...631..280B}. In normal galaxies the accretion rate to the central supermassive BH is so low, that only weak activity can be detected -- if at all. The overall thermal emission of the nuclear region is due to bremsstrahlung from hot gas. The total X-ray luminosity of a normal galaxy can reach some \oergs{41}, at most. It consists of diffuse emission and emission of unresolved individual sources.

Galaxy clusters (GCls) are by far the largest and most massive virialised objects in the Universe. Their masses lie in the range of $10^{14}$--\,$10^{15}$M\subsun\ and they have sizes of a few megaparsecs (Mpc). A mass-to-light ratio of $M/L\!\simeq\!200\,$\,M\subsun/L\subsun\ indicates that galaxy clusters are clearly dominated by their dark matter content. Furthermore, galaxy clusters allow us to study the baryonic matter component, as they define the only large volumes in the Universe from which the majority of baryons emit detectable radiation. This baryonic gas, the {\em hot intracluster medium} (ICM), is extremely thin, with electron densities of $n_{\mathrm{e}}\!\simeq\!10^2$--10$^5$\,m$^{-3}$, and fills the entire cluster volume. Owing to the plasma temperatures of $k_{\mathrm{B}}\,T\!\simeq\!2$--10\,keV, the thermal ICM emission gives rise to X-ray luminosities of $L_{\mathrm{X}}\!\simeq\!10^{43}$--\,$3\!\times\!10^{45}$\,erg\,s$^{-1}$. Therefore galaxy clusters are the most X-ray luminous objects in the Universe next to AGN.

\begin{figure}
 \resizebox{\hsize}{!}{\includegraphics[clip,angle=0]{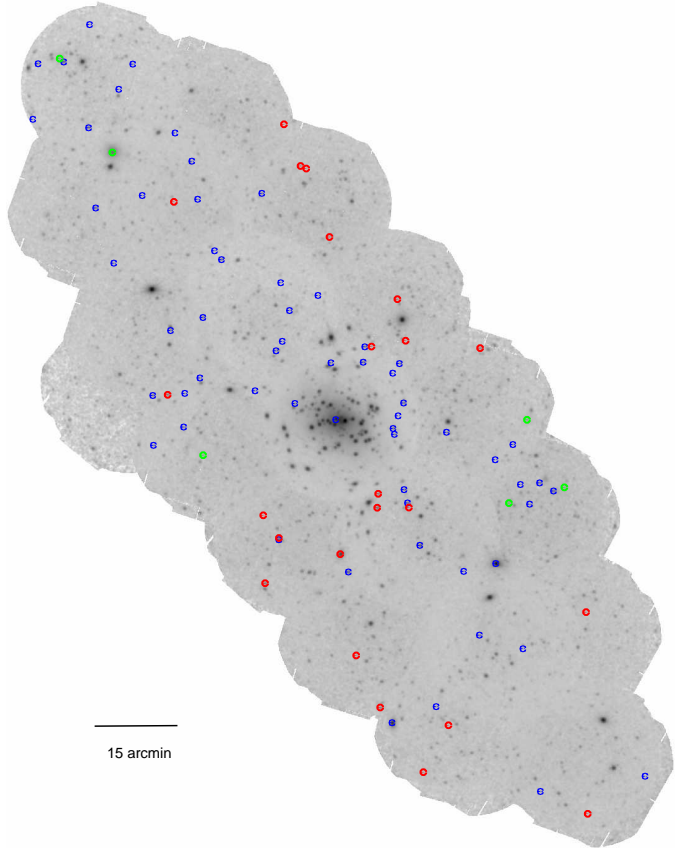}}
 \caption{The spatial distribution of background sources and candidates, classified in the \XLPt\ catalogue. AGN are marked with blue dots, ``normal" galaxies with red dots and galaxy clusters with green dots.}
 \label{Fig:BG_spdist} 
\end{figure}

We identified four sources as background galaxies and 11 as AGN, and classified 19 galaxy and 49 AGN candidates. The classification is based on SIMBAD and NED correlations and correlations with sources listed as background objects in the globular cluster catalogues \citep[RBC\,V3.5 and ][]{2009AJ....137...94C}. Sources are classified as AGN candidates, if they have a radio counterpart \citep[NVSS;][]{1990ApJS...72..761B,2004ApJS..155...89G} with the additional condition of being neither a SNR nor a SNR candidate from X-ray hardness ratios, as well as not being listed as a ``normal" background galaxy in \citet{2004ApJS..155...89G}. Most AGN will be classified as $<$hard$>$ ((HR2$-$EHR2)$>-$0.2, see Table~\ref{Tab:class}) because of their intrinsic power law component. Additional absorption in the line of sight by the interstellar medium of \m31\ will lead to an even higher HR2. Only the few AGN with a dominant component in the measured flux below 1\,keV may lead to a classification $<$SNR$>$ or $<$fg Star$>$ in our adapted scheme.

One (\num\ 995) of the four identified galaxies is M~32. An overview of previous X-ray observations of this galaxy is given in PFH2005. They also discuss the fact that \chandra\ resolved the X-ray emission of M~32 into several distinct point sources (maximum separation of the three central \chandra\ sources 8\farcs3). Although M~32 is located closer to the centre of the FoV in the observations of field SS1, than it was in the s1 observation used in PFH2005, \xmm\ still detects only one source. The remaining three sources (\num\ 88, \num\ 403, \num\ 718) are identified as galaxies, because they are listed as background galaxies in both the RBC\,V3.5 and \citet{2009AJ....137...94C}. For source \num\ 403 (B\,007) NED gives a redshift of $0.139692\pm0.000230$ \citep{2007AJ....134..706K}.

Eleven X-ray sources are identified as AGN. The first one (\num\ 363) correlates with a BL Lac object located behind \m31\ (NED, see also PFH2005). The second source (\num\ 745) correlates with a Seyfert 1 galaxy (5C~3.100), which has a redshift of $\approx 0.07$ (SIMBAD). The third source (\num\ 1\,559) correlates with a quasar (Sharov~21) that showed a single strong optical flare, during which its UV flux has increased by a factor of $\sim$20  \citep{2010A&A...512A...1M}. The remaining sources were spectroscopically confirmed (from our optical follow-up observations) to be AGN (D.~Hatzidimitriou, private communication; and Hatzidimitriou et al.~(2010) in prep.).  

\begin{table}
\begin{center}
\caption{Spectral fit parameters for extended sources}
\begin{tabular}{rcllc}
\hline\noalign{\smallskip}
\hline\noalign{\smallskip}
\multicolumn{1}{l}{Src ID} & \multicolumn{1}{c}{$N_{\rm H}$/10$^{21}$\,cm$^{-2}$} & \multicolumn{1}{l}{$k_{\mr{B}}T$/keV} & \multicolumn{1}{c}{redshift} & \multicolumn{1}{c}{$\chi^2$/dof}\\
\hline\noalign{\smallskip}
 141 & $1.19^{+1.63}_{-0.88}$ & $2.17^{+2.30}_{-0.68}$ & $0.24^{+1.24}_{-0.11}$ & 78.5/53\\
\noalign{\smallskip}
 252 & $0.61^{+1.16}_{-0.43}$ & $1.95^{+0.64}_{-0.29}$ & $0.22^{+0.15}_{-0.07}$ & 56.4/151\\
\noalign{\smallskip}
 304 & $2.68^{+2.64}_{-1.85}$ & $0.95^{+3.32}_{-1.95}$ & $0.12^{+0.07}_{-0.05}$ & 50.9/57\\
\noalign{\smallskip}
1543 & $2.74^{+6.91}_{-1.76}$ & $2.08^{+2.31}_{-1.11}$ & $0.61^{+1.11}_{-0.26}$ & 32.9/34\\
\noalign{\smallskip}
\hline
\noalign{\smallskip}
\end{tabular}
\label{Tab:spfit_ext}
\end{center}
\normalsize
\end{table}

In Sect.\,\ref{Sec:ExtSrcs} the 12 extended sources in the \XLPt\ catalogue were presented. \citet{2006ApJ...641..756K} showed that the brightest of these sources (\num\ 1795) is a galaxy cluster located at a redshift of $z\!=\!0.29$.
For the remaining 11 sources, X-ray spectra were created and fitted with the {\tt MEKAL} model in {\tt XSPEC}. Unfortunately, for most of the examined sources the spectral parameters (foreground absorption, temperature and redshift) are not very well constrained. Nevertheless four sources (\num\ 141, \num\ 252, \num\ 304, \num\ 1543) with temperatures in the range of $\sim\!1$--2\,keV and proposed redshifts between 0.1\,--\,0.6 were found (Table~\ref{Tab:spfit_ext}). Inspection of optical images (DSS\,2 images and if available LGGS images) revealed an agglomeration of optical sources at the positions of these four extended X-ray sources. Thus they are classified as galaxy cluster candidates. 

Although, B242 (the optical counterpart of source \num\ 304) is listed as a globular cluster candidate in the RBC3.5 catalogue, \citet{2009AJ....137...94C} classified this source as a background object. Our findings from the X-rays favour the background object classification. Hence a globular cluster classification for this source seems to be excluded. 

Source \num\ 1\,912 was already classified as a galaxy cluster candidate in PFH2005. The spectrum confirms this classification. The best fit parameters are \nh$=\!1.29^{+0.53}_{-0.41}$\hcm{21}, $T\!=\!2.8^{+0.8}_{-0.5}$\,keV and redshift of $0.06^{+0.03}_{-0.04}$.

A plot of the spatial distribution of the classified\,/\,identified background sources is given in Fig.\,\ref{Fig:BG_spdist}, which shows that these sources are rather homogeneously distributed over the observed field. However, in the fields located along the major axis of \m31\ we mainly see AGN, which are bright enough to be visible through \m31, while most of the galaxies and galaxy clusters are detected in the outer fields. 

\subsubsection{Comparing \textit{XMM-Newton}, \textit{Chandra} and \textit{ROSAT} catalogues}
Of the ten \ros\ PSPC survey sources classified as background galaxies one is located outside the field of the Deep \xmm\ Survey. The remaining objects are confirmed to be background sources and are classified or identified as galaxies or AGN. The only case which is worth discussing in more detail is the source pair [SHP97]~246 and [SHL2001]~252. From the \xmm\ observations it is evident that this source pair is not one source, as indicated in the combined \ros\ PSPC source catalogue (SHL2001), but consists of three individual sources (\num\ 1\,269, \num\ 1\,279 and \num\ 1\,280). [SHL2001]~252 correlates spatially with all three \xmm\ sources, while [SHP97]~246 correlates only with source \num\ 1\,269, which is identified as a foreground star of type K2 (SIMBAD). The two other \xmm\ counterparts of [SHL2001]~252 are classified as a galaxy candidate and an AGN candidate, respectively. In summary, [SHL2001]~252 is most likely a blend of both background sources and maybe even a blend of all three \xmm\ sources, while [SHP97]~246 seems to be the X-ray counterpart of the foreground star mentioned above. 

\citet{2002ApJ...577..738K} classified source r3-83 (\num\ 1\,132) as an extragalactic object, as it is listed in SIMBAD and NED as an emission line object. Following PFH2005, we classified source \num\ 1\,132 as $<$hard$>$. The BL Lac object (\num\ 363) was also detected in \chandra\ observations \citep{2004ApJ...609..735W}.

\section{M~31 sources}
\label{Sec:Srcsm31}
\subsection{Supersoft sources}
Supersoft source (SSS) classification is assigned to sources showing extremely soft spectra with equivalent blackbody temperatures of $\sim$15--80\,eV. The associated bolometric luminosities are in the range of \oexpo{36}--\oergs{38} \citep[][]{1997ARA&A..35...69K}.

Because of the phenomenological definition, this class is likely to include objects of several types. The favoured model for these sources is that they are close binary systems with a white dwarf (WD) primary, burning hydrogen on the surface \citep[\cf\ ][]{1997ARA&A..35...69K}. Close binary SSSs include post-outburst, recurrent, and classical novae, the hottest symbiotic stars, and other LMXBs containing a WD (cataclysmic variables, CVs).\@ Symbiotic systems, which contain a WD in a wide binary system, may  also be observed as SSSs \citep[][]{1997ARA&A..35...69K}. Because matter that is burned can be retained by the WD, some SSS binaries may be progenitors of type-Ia supernovae \citep[\cf\ ][]{1992A&A...262...97V}.

The \XLPt\ catalogue contains 30 SSS candidates that were selected on the basis of their hardness ratios (see Fig.\,\ref{Fig:HR_diagrams} and Table~\ref{Tab:class}). 

\subsubsection{Spatial and flux distribution}
\begin{figure}
 \resizebox{\hsize}{!}{\includegraphics[clip,angle=0]{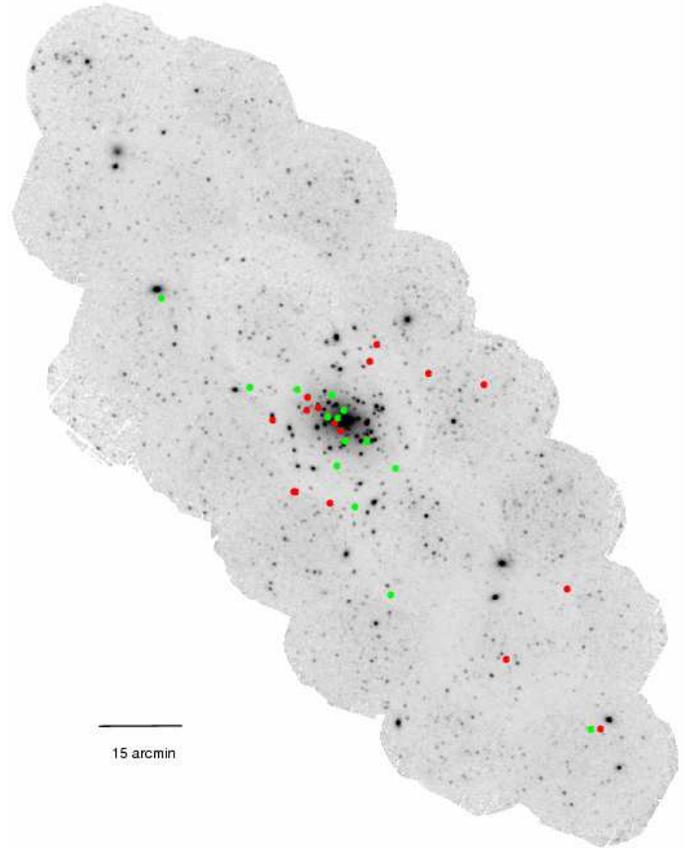}}
 \caption{The spatial distribution of SSSs classified in the \XLPt\ catalogue. The positions of the SSSs are marked with red and green dots. Sources that correlate with optical novae are given in green. An enhancement of sources in the central field is clearly visible.}
 \label{Fig:SSS_spdist} 
\end{figure}

Figure \ref{Fig:SSS_spdist} shows the spatial distribution of the SSSs. Clearly visible is a concentration of sources in the central field. There are two explanations for that central enhancement. The first is that the central region was observed more often than the remaining fields and therefore there is a higher chance of catching a transient SSS in outburst. The second reason is that the major class of SSSs in the centre of \m31\ are optical novae (PFF2005, PHS2007). Optical novae are part of the old stellar population which is much denser in the centre of \m31.

\begin{figure}
 \resizebox{\hsize}{!}{\includegraphics[clip,angle=0]{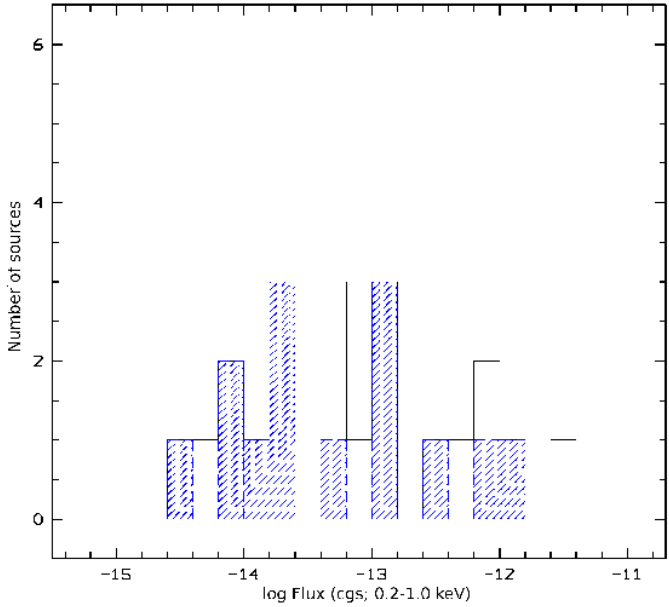}}
 \caption{Distribution of the source fluxes in the 0.2--1.0\,keV band. The diagram shows the number of SSSs per flux bin plotted versus the flux in logarithmic scale. 
The blue histogram gives the distribution of SSSs correlating with optical novae.}
 \label{Fig:SSS_fldist} 
\end{figure}

Figure \ref{Fig:SSS_fldist} gives the distribution of 0.2--1.0\,keV source fluxes for all SSSs (black) and for those correlating with optical novae (blue). The unabsorbed fluxes were determined assuming a 50\,eV blackbody model (PFF2005). The two brightest SSSs ($F_{\mr{X}}>$\oergcm{-12}) consist of  a persistent source with 217\,s pulsations \citep[\num\ 1\,061;][]{2008ApJ...676.1218T} and the nova M31N~2001-11a \citep[\num\ 1\,416;][]{2006IBVS.5737....1S}. A large fraction of SSSs are rather faint, with fluxes below 5\ergcm{-14}. Four sources have absorption-corrected luminosities below \oergs{36} (0.2--1.0\,keV), which was indicated as the limiting luminosity for SSSs. That does not necessarily imply that these sources are not SSSs, since it is possible that the blackbody fit chosen does not represent well the properties of these sources. A higher absorption or a lower temperature would lead to increased unabsorbed luminosities.  We also have to take into account that we might have observed the source during a phase of rising or decaying luminosity, \ie\ not at maximum luminosity.

\subsubsection{Correlations with optical novae}
\label{SubSec:opt_novae}
By cross-correlating with the nova catalogue\footnote{\url{http://www.mpe.mpg.de/~m31novae/opt/m31/M31_table.html}} indicated in Sect.\,\ref{Sec:CrossCorr_Tech}, 14 of the 30 SSSs can be classified as X-ray counterparts of optical novae. Of these 14 novae, eight (\num\ 748, \num\ 993, \num\ 1\,006, \num\ 1\,046, \num\ 1\,051, \num\ 1\,076, \num\ 1\,100, and \num\ 1\,236) are already discussed in PFF2005 and PHS2007.\@ Nova M31N~2001-11a was first detected as a supersoft X-ray source. Motivated by that SSS detection, \citet{2006IBVS.5737....1S} found an optical nova at the position of the SSS in archival optical plates which had been overlooked in previous nova searches. Nova M31N~2007-06b has been discussed in \citet[][]{2009A&A...500..769H}. The remaining four novae are discussed individually in more detail below. 

As was shown in the \xmm/\chandra\ \m31\ nova monitoring project\footnote{\url{http://www.mpe.mpg.de/~m31novae/xray/index.php}}, it is absolutely necessary to have a homogeneous and dense sample of deep optical and X-ray observations in order to study optical novae and their connections to supersoft X-ray sources. In the optical, the outer regions of \m31\ are regularly observed down to a limiting magnitude of $\sim$17\, mag (Texas Supernova Search (TSS); \citealt{Quimby2006}), while in X-rays only ``snapshots" are available. Hence, 
the correlations of optical novae with detected SSSs have to be regarded as lucky coincidences. That also means that the identified nova counterparts are detected at a random stage of their SSS evolution which does not allow us to constrain the exact start or end point of the SSS phase, nor the maximum luminosity of the SSS. We also cannot exclude the possibility that some of the SSSs observed in the outer parts of \m31\ correspond to the supersoft phase of optical novae for which the optical outburst was missed. In the outer regions of \m31, the samples of optical novae and X-ray SSSs are certainly incomplete, due to the rather high luminosity limit in the optical monitoring, and the lack of complete monitoring in X-rays, respectively. So one should be cautious in deriving properties of the disc nova population of \m31\ from the available data. 

\begin{figure}
 \resizebox{\hsize}{!}{\includegraphics[clip,angle=90]{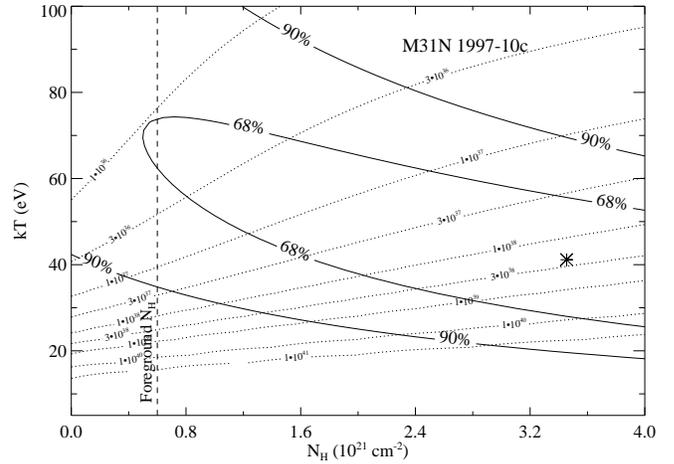}}
 \caption{Column density-temperature confidence contours inferred from the fit to the \xmm\ EPIC PN spectrum of M31N1997-10c. The formal best fit parameters are indicated by the star. Also drawn are lines of constant bolometric luminosity (in erg s$^{-1}$).  
The vertical dashed line indicates the Galactic foreground absorption in the direction of \m31.}
 \label{Fig:M31N1997-10c_ccont} 
\end{figure}

\paragraph{Nova M31N~1997-10c} was detected on 2 October 1997 at a B-band magnitude of $16.6$ \citep[ShA~58;][]{1998AstL...24..641S}.\@ An upper limit of 19 mag on 29 September 1997 was reported by the same authors. They classified this source as a very fast nova. In the \xmm\ observation c1 (25 June 2000), an SSS (\num\ 871), located within $\sim$1\,\farcs9 of the optical nova, was detected. The source was fitted with an absorbed blackbody model. The formal best fit parameters of the \xmm\ EPIC PN spectrum are: absorption $N_{\mr{H}}\approx3.45$\hcm{21} and $k_{\mr{B}}T\approx41$\,eV.\@ The unabsorbed luminosity in the 0.2--1\,keV band is $\approx5.9$\ergs{37}.\@ Confidence contours for absorption column density and blackbody temperature are shown in Fig.\,\ref{Fig:M31N1997-10c_ccont}. In the subsequent \xmm\ observation of that region taken about half a year later (c2; 27 December 2000) the source is not detected. 
Although the source position is covered in observations c3 (29 June 2001), c4 (6/7 January 2002) and b (16--19 July 2004) the source was not re-detected. Using the count rates derived for the variability study (see Sect.\,\ref{Sec:var}) and assuming the same spectrum for the source as in observation c1, upper limits of the source luminosity can be derived, which are given in Table~\ref{Tab:M31N1997-10c_uplim}.

\begin{table}[t]
\begin{center}
\caption{3$\sigma$ upper limits for the absorption-corrected luminosities for Nova M31N~1997-10c}
\begin{tabular}{cc}
\hline\noalign{\smallskip}
\hline\noalign{\smallskip}
\multicolumn{1}{l}{observation} & \multicolumn{1}{c}{$L_{\rm X}$/10$^{37}$\,erg\,s$^{-1}$ (0.2--1.0\,keV)}\\
\hline\noalign{\smallskip}
 c2 & 10.8$^{+}$\\
 c3 & 1.9\\
 c4 & 1.0\\
\noalign{\smallskip}
\hline
\noalign{\smallskip}
\end{tabular}
\label{Tab:M31N1997-10c_uplim}
\end{center}
Notes:\\
$^{ +~}$: The count rate detected in observation c2 gives a luminosity of 2.4$\pm$2.8\ergs{37}, which results in the upper limit given in the Table. The fact that this upper limit is higher than the luminosity detected in observation c1 is, at least in part, attributed to the very short effective observing time of less than 6\,000\,s.
\normalsize
\end{table}

\paragraph{Nova M31N~2005-01b} was discovered on 19 January 2005 at a white light magnitude of 16.3 by R.~Quimby.\footnote{\url{http://www.supernovae.net/sn2005/novae.html}} An SSS (\num\ 764) that correlates with the optical nova (distance: 4\,\farcs3; 3$\sigma$ error: 5\,\farcs5) was found in observation ss2 taken on 8 July 2006, which is 535 days after the discovery of the optical nova. Due to the severe background screening applied to observation ss2,  there is not enough statistics to obtain a spectrum of the X-ray source. To get an estimate of the spectral properties of that source we created a spectrum in the 0.2--0.8\,keV range of the \emph{unscreened} data. Although the spectrum was background corrected, we cannot totally exclude a contribution from background flares.
The spectrum is best fitted by an absorbed blackbody model with an absorption of $N_{\mr{H}}\approx1.03$\hcm{21} and a blackbody temperature of $k_{\mr{B}}T\approx45$\,eV.\@ The unabsorbed 0.2--1\,keV luminosity is $L_{\mr{X}}\sim$1.0\ergs{37}. In another \xmm\ observation taken 1\,073 days after the optical outburst (ss21; 28 December 2007) the X-ray source is no longer visible. The 3$\sigma$ upper limit of the unabsorbed source luminosity is $\sim3.3$\ergcm{35} in the 0.2--4.5\,keV band, assuming the spectral model used for source detection.

\begin{figure}
 \resizebox{\hsize}{!}{\includegraphics[clip,angle=-0]{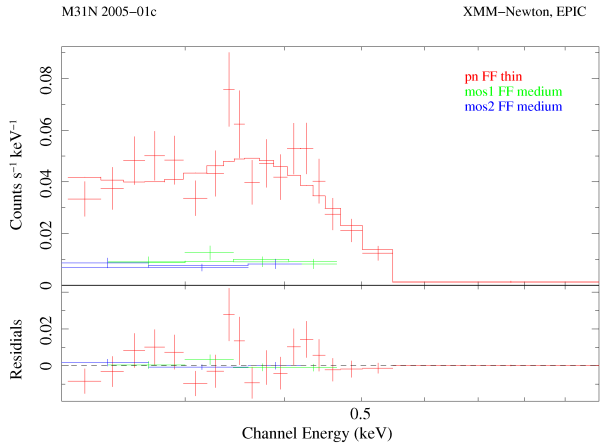}}
 \caption{\xmm\ EPIC spectrum of nova M31N~2005-01c. The absorbed black body fit to the data is shown in the upper panel.}
 \label{Fig:M31N2005-01c_spec} 
\end{figure}

\begin{figure}
 \resizebox{\hsize}{!}{\includegraphics[clip,angle=0]{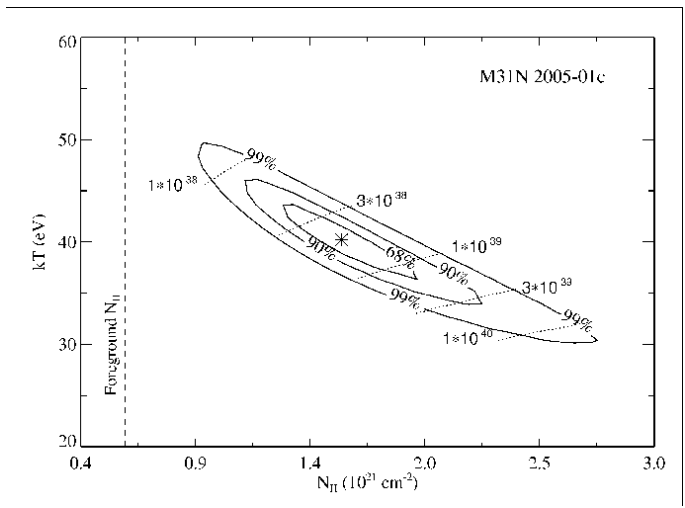}}
 \caption{Column density (\nh) - temperature ($k_{\mr{B}}T$) confidence contours inferred from the blackbody fit to the \xmm\ EPIC spectrum of M31N~2005-01c (see Fig.\,\ref{Fig:M31N2005-01c_spec}). The formal best fit parameters are indicated by the star. Also drawn are lines of constant bolometric luminosity and the vertical dashed line indicates the Galactic foreground absorption (see Fig~\ref{Fig:M31N1997-10c_ccont}).}
 \label{Fig:M31N2005-01c_ccont} 
\end{figure}

\paragraph{Nova M31N~2005-01c} was discovered on 29 January 2005 at a white light magnitude of 16.1 by R.~Quimby.\footnote{\url{http://www.supernovae.net/sn2005/novae.html}} In the \xmm\ observation from 02 January 2007 (ns2, 703 days after optical outburst) an SSS was detected (\num\ 1\,675) at a position consistent with that of the optical nova (distance: 0\,\farcs9). 
The X-ray spectrum (Fig.\,\ref{Fig:M31N2005-01c_spec}) can be well fitted by an absorbed blackbody model with the following best fit parameters: absorption $N_{\mr{H}}=1.58^{+0.65}_{-0.45}$\hcm{21} and $k_{\mr{B}}T=40\pm6$\,eV.\@ The unabsorbed 0.2--1\,keV luminosity is $L_{\mr{X}}\sim$1.2\ergs{38}. Confidence contours for absorption column density and blackbody temperature are shown in Fig.\,\ref{Fig:M31N2005-01c_ccont}.


\paragraph{Nova M31N~2005-09b} was discovered in optical images taken on 01 and 02 September 2005 at white light magnitudes of $\sim$18.0 and $\sim$16.5 respectively. From 31 August 2005, an upper limit of $\sim$18.7\,mag was reported \citep{2005ATel..600....1Q}.\@ The nova was spectroscopically confirmed \citep{2006ATel..850....1P} and classified as a possible Fe\,{\small II} or hybrid nova\footnote{\url{http://cfa-www.harvard.edu/iau/CBAT_M31.html}}. An X-ray counterpart (\num\ 92) was detected in the \xmm\ observation s3 (299 days after the optical outburst). Its position is consistent with that of the optical nova (distance: 0\,\farcs57). As observation s3 was heavily affected by background flares, we only could estimate the spectral parameters from the \emph{unscreened} data (see also paragraph about Nova M31N~2005-01b). A blackbody fit of the 0.2--0.8\,keV gives $N_{\mr{H}}\approx2.7$\hcm{21}, k$T\approx35$\,eV, and an unabsorbed 0.2--1\,keV luminosity of $L_{\mr{X}}\sim$5.4\ergs{38}. The X-ray source was no longer visible in observation s31, which was taken 391 days after observation s3.

\subsubsection{Comparing \textit{XMM-Newton}, \textit{Chandra} and \textit{ROSAT} catalogues}
\label{Sec:SSS_comp}
The results and a detailed discussion of a study of the long-term variability of the SSS population of \m31\ are presented in \citet{2010AN....331..212S}. In summary our comparative study of SSS candidates in \m31\ detected with \ros, \chandra\ and \xmm\ demonstrated that strict selection criteria have to be applied to securely select SSSs. It also underlined the high variability of the sources in this class and the connection between SSSs and optical novae.

\subsection{Supernova remnants}
\label{Sec:SNR_Diss}
After an supernova explosion the interaction between the ejected material and the ISM forms a supernova remnant (SNR).\@ The SNR X-ray luminosities typically vary between $10^{35}$ and \oergs{37} (0.2--10\,keV).\@ 

SNRs can be divided into two categories, (i) sources where the thermal components dominate the X-ray spectrum below 2\,keV, and (ii) the so-called ``plerions" or Crab-like SNRs with power law spectra. The former are located in areas of the X-ray colour/colour diagrams that overlap only with foreground star locii. If we assume that we have identified all foreground star candidates from the optical correlation and inspection of the optical images, the remaining sources can be classified as SNR candidates using the criteria given in Table~\ref{Tab:class}. Similar criteria were used to select supernova remnant candidates in \xmm\ observations of M~33 \citep{2004A&A...426...11P, 2006A&A...448.1247M}. \citet{2005AJ....130..539G}  and \citet{2010ApJS..187..495L}  confirmed the supernova remnant nature of many of these candidates based on optical and
radio follow-up observations. They also used a hardness ratio criterion to select supernova remnant candidates from \chandra\ data.

An X-ray source is classified as a SNR candidate if it either fulfils the hardness ratio criterion given in Table~\ref{Tab:class} (these are 25 such sources), or if it correlates with a known optical or radio SNR candidate (six sources). The sources assigned the classification of a SNR candidate based on the latter criterion alone, are marked in the comment column of Table~5 with the flag `\emph{only correlation}'. As these six SNR candidates would be classified as $<$hard$>$ on the basis of their hardness ratios, they are good candidates for being ``plerions". SNRs are taken as identified when they coincide with SNR candidates from the optical or radio and fulfil the hardness ratio criterion. For a discussion of detection of SNRs in different wavelength bands see \citet{2010ApJS..187..495L}. All together, we identified 25 SNRs and 31 SNR candidates in the \XLPt\ catalogue. 

This number is in the range expected from an extrapolation of the X-ray 
detected SNRs in the Milky Way as shown below. Assuming that our own Galaxy contains about 1\,440 X-ray sources which are brighter than $\sim$1\ergs{35} \citep{2006A&A...452..169R}, and that it contains $\sim$110 SNRs detected in X-rays \citep{2009BASI...37...45G}, we would expect to detect $\sim$50 SNRs in the \XLPt\ catalogue ($0.4\times\lb(1\,897 \mr{sources} - 263 \mr{fg Stars} \rb)$). This number is in good agreement with the number of identified and classified SNRs.

\begin{figure}
 \resizebox{\hsize}{!}{\includegraphics[clip]{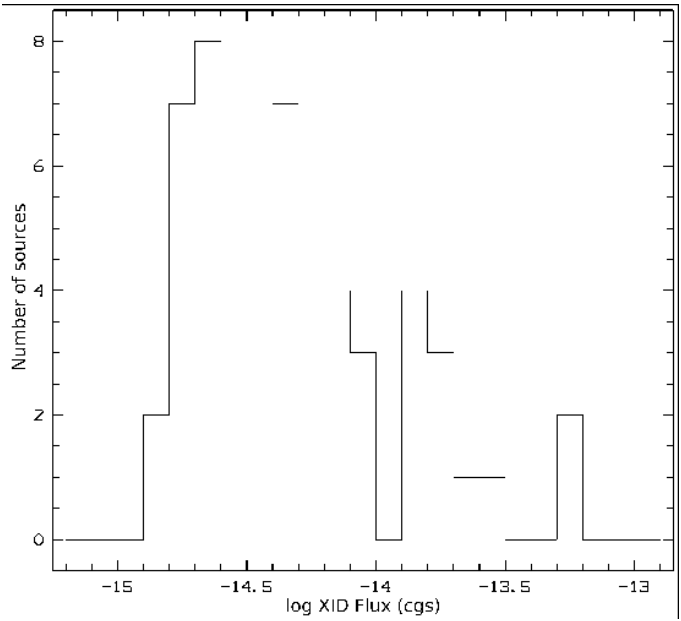}}
 \caption{Distribution of SNR fluxes in the 0.2--4.5\,keV (XID) band. The diagrams show the number of identified and classified SNRs at each flux bin, plotted versus the flux.} 
 \label{Fig:SNR_fldist} 
\end{figure}

The XID fluxes for SNRs range between 5.9\ergcm{-14} for source \num\  1\,234 and 1.5\ergcm{-15} for source \num\ 419. These fluxes correspond to luminosities of 4.3\ergs{36} to 1.1\ergs{35}. A diagram of the flux distribution of the detected SNRs and candidates is shown in Fig.\,\ref{Fig:SNR_fldist}. 

\begin{figure}
 \resizebox{\hsize}{!}{\includegraphics[clip,angle=-90]{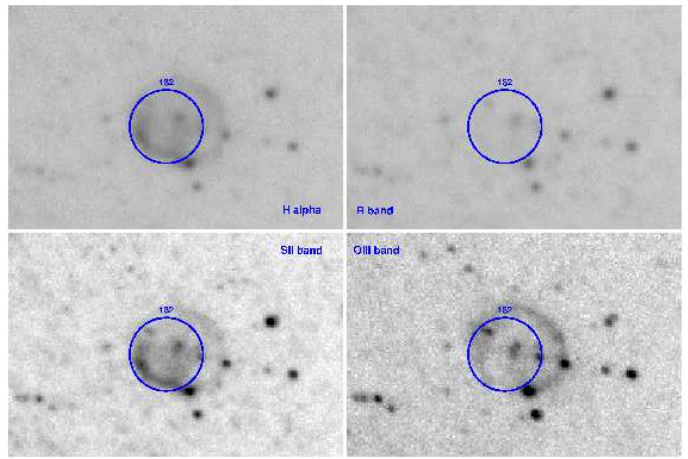}}
 \caption{H$\alpha$, R, S\,{\small II} and O\,{\small III} images, taken from the LGG Survey. Over-plotted is a circle at the position of source XMMM31~J003923.5+404419 with a radius of 5\farcs5 (3$\sigma$ positional error of the X-ray source).\@ The ring-like SNR is clearly visible in the H$\alpha$ and S\,{\small II} bands.}
 \label{Fig:src182_opt} 
\end{figure}

Among the 25 identified SNRs, there are 20 SNRs from the PFH2005 catalogue. Source [PFH2005]~146, which correlates with the radio source [B90]~11 and the SNR candidate BA146, was not found in the present study. 
Source [SPH2008]~858, which coincides with a source reported as a ring-like extended object in \chandra\ observations that was also detected in the optical and radio wavelength regimes and identified as a SNR \citep{2003ApJ...590L..21K}, was re-detected (\num\ 1\,050). Of the 31 SNR candidates ten have been reported by PFH2005.
In the following, we first discuss in more detail the remaining four identified SNRs, that appear in the new catalogue but were not included in PFH2005:

\paragraph{XMMM31~J003923.5+404419} (\num\ 182) was classified as a SNR candidate from its [S\,{\small II}]:H$\alpha$ ratio. It appears as an \textsl{`irregular ring with southerly projection'} \citep[][and Fig.\,\ref{Fig:src182_opt}]{1980A&AS...40...67D} and correlates with a radio source \citep{1969MNRAS.144..101P}. X-ray radiation of that source was first detected in the present study.

\begin{figure}
 \resizebox{\hsize}{!}{\includegraphics[clip,angle=-90]{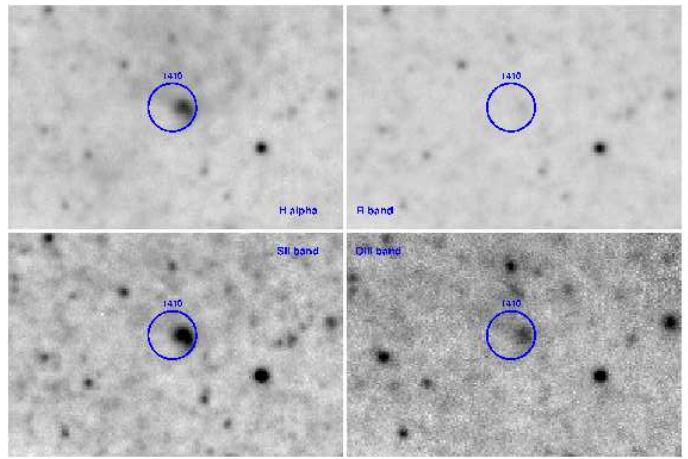}}
 \caption{H$\alpha$, R, S\,{\small II} and O\,{\small III} images, taken from the LGG Survey. Over-plotted is a circle at the position of source XMMM31~J004413.5+411954 with a radius of 3\farcs6 (3$\sigma$ positional error of the X-ray source).\@ The SNR `appears as a bright knot'.}

 \label{Fig:src1410_opt} 
\end{figure}

\paragraph{XMMM31~J004413.5+411954} (\num\ 1\,410) was classified as a SNR candidate from its [S\,{\small II}]:H$\alpha$ ratio \citep{1993A&AS...98..327B,1995A&AS..114..215M}. From Fig.\,\ref{Fig:src1410_opt} we can see that the source \textsl{`appears as a bright knot'}, as was already reported by \citet{1993A&AS...98..327B}. The source has counterparts in the radio \citep{1990ApJS...72..761B} and X-ray (SHP97) range. It was reported as a SNR by SHP97.

\paragraph{XMMM31~J004510.5+413251 and XMMM31~J004512.3+420029} 
(\num\ 1\,587 and \num\ 1\,593, respectively) are new X-ray detections and correlate with the radio sources: \#354 and \#365 in the list of \citet{1990ApJS...72..761B}. Source \num\ 1\,587 also correlates with source 37W209 from the catalogue of \citet{1985A&AS...61..451W}. No optical counterparts were reported in the literature. 

\begin{figure}
 \resizebox{\hsize}{!}{\includegraphics[clip,angle=-90]{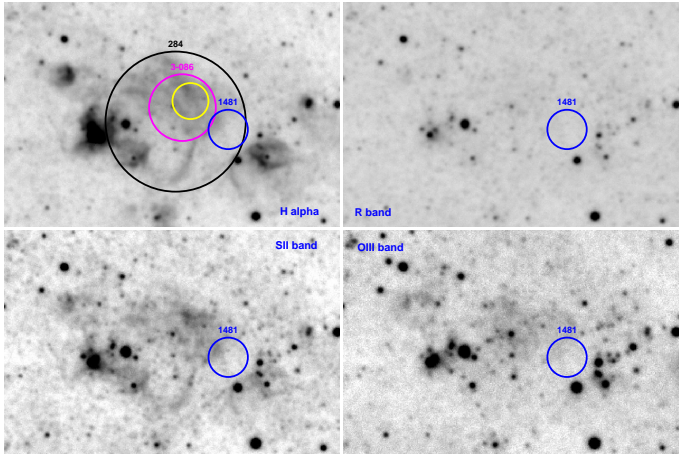}}
 \caption{H$\alpha$, R, S\,{\small II} and O\,{\small III} images, taken from the LGG Survey. Over-plotted is a blue circle at the position of source XMMM31~J004434.8+412512 with a radius of 5\farcs9 (3$\sigma$ positional error of the X-ray source).\@ Source [SPH97]~284 is indicated by a black circle with a radius of 21\arcsec\ (3$\sigma$ positional error), source 3-086 by the magenta circle with a radius of 10\arcsec; the position of the radio counterpart is marked by the yellow circle.}
 \label{Fig:src1481_opt} 
\end{figure}

\vspace{5mm}
In the following, we discuss two SNR candidates in more detail:

\paragraph{XMMM31~J004434.8+412512} (\num\ 1\,481) lies in the periphery of a super-shell with [S\,{\small II}]:H$\alpha\!>$0.5 \citep[][src 490]{1993A&AS...98..327B}. Located next to this source is a SNR candidate reported in \citet[][src 3-086]{1995A&AS..114..215M}, which has a radio counterpart from the NVSS catalogue. \num\ 1\,481 also correlates with \ros\ source [SPH97]~284, which was identified as a SNR in SPH97 due to its spatial correlation with source 3-086.\@ Figure~\ref{Fig:src1481_opt} shows the \xmm\ error circle over-plotted on LGGS images. From the \xmm\ source position it looks more likely that the X-rays are emitted from the HII region rather than  from the SNR candidate visible in the optical and radio wavelengths. Nevertheless the \xmm\ source detected is point-like and its hardness ratios lie in the range expected for SNRs. If the X-ray emission originated from the \HII-region, it should have been detected as spatially extended emission. Thus, \num\ 1\,481 is classified as SNR candidate. A puzzling fact, however, is the pronounced variability between \ros\ and \xmm\ observations of $F_{\mr{var}}\!=\!9.82$ with a significance of $S_{\mr{var}}\!\approx\!4$ (see Table~\ref{Tab:VarSNRs1}), which is not consistent with the long term behaviour of SNRs. There is still the possibility that the detected X-ray emission does not belong to either the \HII-region or a SNR at all.

\begin{figure}
 \resizebox{\hsize}{!}{\includegraphics[clip,angle=-90]{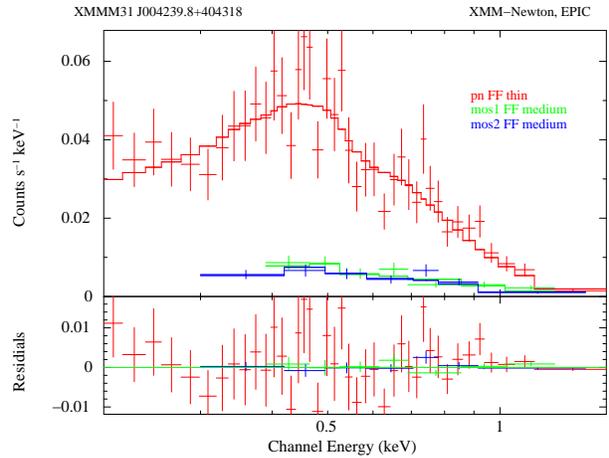}}
 \caption{0.2--3.0\,keV EPIC spectrum of source \num\ 969. The best fit absorbed non-equilibrium ionisation model is indicated by the solid lines.}
 \label{Fig:src969_sp} 
\end{figure}

\paragraph{XMMM31~J004239.8+404318} (\num\ 969) was already observed with \ros\ (SHP97, SHL2001) and \chandra\ \citep[][s1-84]{2004ApJ...609..735W}. No optical counterpart is visible on the LGGS images. The X-ray spectrum, which is shown in Fig.\,\ref{Fig:src969_sp}, is well fitted by an absorbed non-equilibrium ionisation model with the following best fit values: an absorption of $N_{\mr{H}}=1.76^{+0.46}_{-0.60}$\hcm{21}, a temperature of $k_{\mr{B}}T=219^{+32}_{-19}$\,eV, and an ionisation timescale of $\tau=1.75^{+0.82}_{-1.75}\times10^8$\,s\,cm$^{-3}$. The unabsorbed 0.2--5\,keV luminosity is $L_{\mr{X}}\sim6.5$\ergs{37}.\@ The soft spectrum with the temperature of $\sim$200\,eV is in good agreement with spectra of old SNRs \eg\ in the SMC \citep{2008A&A...485...63F}. Although the unabsorbed luminosity is rather high for an old SNR, it is still in the range found for other SNRs \citep[\cf\ ][]{2002ApJ...580L.125K,2007ApJ...663..234G}. Hence, XMMM31~J004239.9+404318 is classified as a SNR candidate.

\subsubsection{Comparing SNRs and candidates in \textit{XMM-Newton}, \textit{Chandra} and \textit{ROSAT} catalogues}
The second \ros\ PSPC catalogue (SHL2001) contains 16 sources classified as SNRs. The counterparts of 12 of these sources are also classified as SNRs or SNR candidates in the \XLPt\ catalogue. 

\begin{table*}
\scriptsize
\begin{center}
\caption{Flux comparison of SNRs and SNR candidates from the \XLPt\ catalogue with counterparts classified as SNRs in \ros\ and \chandra\ catalogues}
\begin{tabular}{rrccrcrrl}
\hline\noalign{\smallskip}
\hline\noalign{\smallskip}
\num\ & \multicolumn{1}{c}{XLPt} & \multicolumn{1}{c}{SHP97} & \multicolumn{1}{c}{SHL2001} & \multicolumn{1}{c}{KGP2002$^{+}$} & \multicolumn{1}{c}{WGK2004$^{+}$} & fvar & svar & reason why indicated vaiability is not reliable\\
& \multicolumn{5}{c}{XID Flux with error in \oergcm{-15}} & & &\\
\noalign{\smallskip}
\hline\noalign{\smallskip}
 474 &  5.27 $\pm$ 0.56 &  21.18 $\pm$ 4.46 &                    &                  &                    &  4.01 &  3.54  \\ 
 668 &  7.94 $\pm$ 1.36 &  26.30 $\pm$ 6.69 &                    &                  &                    &  3.31 &  2.69  \\ 
 883 &  2.83 $\pm$ 0.33 &                   &                    & 3.33 $\pm$ 0.83  &                    &  1.18 &  0.56  \\
1\,040 &  7.12 $\pm$ 0.47 &                   &                    & 12.49 $\pm$ 1.67  &                    &  1.75 &  3.11  \\
1\,050 &  8.25 $\pm$ 0.70 &                   &                    & 2.50 $\pm$ 0.83  &                    &  3.30 &  5.28  \\
1\,066 & 28.35 $\pm$ 1.16 &                   &  256.16$\pm$ 16.19 & 39.13 $\pm$ 3.33 & 25.29 $\pm$ 5.32    & 10.13 & 14.06 & \ros\ source is a blend of two \xmm\ sources\\ 
1\,234 & 59.12 $\pm$ 1.10 & 152.91 $\pm$13.82 &  268.98 $\pm$17.09 & 54.11 $\pm$ 3.33 & 109.13 $\pm$ 11.31 &  4.97 & 12.34  & embedded in diffuse emission in central area of \m31\\ 
1\,275 & 23.88 $\pm$ 1.08 &  53.50 $\pm$ 8.47 &   79.39 $\pm$ 9.90 &                  &                    &  3.32 &  5.58  \\ 
1\,328 &  9.25 $\pm$ 0.74 &                   &   26.99            &                  &                    &  1.00 &  0.00  \\ 
1\,351 &  4.96 $\pm$ 0.68 &  24.96 $\pm$ 8.92 &   17.77            &                  &                    &  1.00 &  0.00  \\ 
1\,372 &  2.12 $\pm$ 0.84 &                   &   29.91            &                  &                    &  1.00 &  0.00  \\ 
1\,410 &  7.40 $\pm$ 0.94 &  29.87 $\pm$ 7.13 &                    &                  &                    &  4.04 &  3.12  \\
1\,481 &  3.43 $\pm$ 0.97 &  33.66 $\pm$ 7.36 &                    &                  &                    &  9.82 &  4.07  & see Sect.\,\ref{Sec:SNR_Diss}\\
1\,535 & 14.73 $\pm$ 1.31 &  53.94 $\pm$ 9.14 &   34.41 $\pm$ 7.20 &                  &                    &  3.66 &  4.25  \\
1\,599 & 16.08 $\pm$ 0.92 &  54.39 $\pm$10.03 &   33.51 $\pm$ 6.97 &                  &                    &  3.38 &  3.80  \\
1\,637 & 12.72 $\pm$ 1.33 &                   &   27.21            &                  &                    &  1.00 &  0.00  \\
\noalign{\smallskip}
\hline
\noalign{\smallskip}
\end{tabular}
\label{Tab:VarSNRs1}
\end{center}
Notes:\\
$^{ +~}$: KGP2002: \citet{2002ApJ...577..738K},WGK2004: \citet{2004ApJ...609..735W}\\
\ros\ and \chandra\ count rates are converted to 0.2--4.5\,keV fluxes, using WebPIMMS and assuming a foreground absorption of \nh\,$=\!6.6$\hcm{20} and a photon index of $\Gamma\!=\!1.7$: ECF$_{\mr{SHP97}}\!=\!2.229\times$10$^{-14}$\,erg\,cm$^{-2}$\,cts$^{-1}$,  ECF$_{\mr{SHL2001}}\!=\!2.249\times$10$^{-14}$\,erg\,cm$^{-2}$\,cts$^{-1}$, and ECF$_{\mr{KGP2002}}\!=\!8.325\times$10$^{-14}$\,erg\,cm$^{-2}$\,cts$^{-1}$. For WGK2004 the luminosity given in Table~2 of WGK2004 was converted to XID flux using $F_{\mr{XID}}$[erg\,cm$^{-2}$\,s$^{-1}$]\,$=\!6.654\times$10$^{-15}\!\times\!L_{\mr{WGK2004}}$[10$^{36}$erg\,s$^{-1}$].
\normalsize
\end{table*}
Table~\ref{Tab:VarSNRs1} lists the \xmm, \ros, and \chandra\ fluxes of all SNRs and SNR candidates from the \XLPt\ catalogue that have counterparts classified as SNRs in \ros\ or \chandra\ source lists. In addition, the maximum flux variability and the maximum significance of the variability (following the variability calculation of Sect.\,\ref{Sec:DefVar}) are given. Three SNRs that have \ros\ counterparts show variability changing in flux by more than a factor of five. The most variable source (\num\ 1\,066) is discussed below, the second source was discussed in Sect.\,\ref{Sec:SNR_Diss} (XMMM31~J004434.8+412512, \num\ 1\,481), and the third source (\num\ 1\,234) is embedded in the diffuse emission of the central area of \m31. In this environment the larger PSF of \ros\ results in an overestimate of the source flux, since the contribution of the diffuse emission could not be totally separated from the emission of the point source.  

The remaining four \ros\ sources classified as SNRs and their \xmm\ counterparts are discussed in the following paragraph. 

\begin{table*}
\scriptsize
\begin{center}
\caption{Flux comparison of SNRs and SNR candidates from the \XLPt\ catalogue which have counterparts in \ros, and/or \chandra\ catalogues that are not classified as SNRs}
\begin{tabular}{lrccrcrrr}
\hline\noalign{\smallskip}
\hline\noalign{\smallskip}
\num\ & \multicolumn{1}{c}{XLPt} & \multicolumn{1}{c}{Chandra} & \multicolumn{1}{c}{PFJ93} & \multicolumn{1}{c}{SHP97} & \multicolumn{1}{c}{SHL2001} & fvar & svar & remark$^{\ddagger}$\\
& \multicolumn{5}{c}{XID Flux with error in \oergcm{-15}} & & & \\
\multicolumn{1}{c}{(1)} & \multicolumn{1}{c}{(2)} & \multicolumn{1}{c}{(3)} & \multicolumn{1}{c}{(4)} & \multicolumn{1}{c}{(5)} & \multicolumn{1}{c}{(6)} & \multicolumn{1}{c}{(7)} & \multicolumn{1}{c}{(8)} & \multicolumn{1}{c}{(9)} \\ 
\hline\noalign{\smallskip}
294    & $18.50 \pm 0.85$    &                  &         &$53.27 \pm 6.69$ & $46.78 \pm 7.87$ & 2.88 & 5.16 & \\
472    & $ 3.15 \pm 0.69$  &                  &         &                 & $26.09 \pm 6.07$  & 8.28 & 3.76 &  468 brt  \\
969    & $53.51 \pm 1.35$  & $84.51\pm 15.97^{+}$ &         & $34.55 \pm 6.91$ & $89.06 \pm 11.92$  & 2.58 & 3.96 &           \\
1\,079   & $ 4.19 \pm 0.59$  &                  &         &  $20.06 \pm 6.24 $&                   &  4.79 &  2.53 & brt \\ 
1\,291 & $14.55 \pm 0.75$  & $16.04^{*}         $ & $>$24.0 & $35.22 \pm 8.47$ & $40.93 \pm 7.87$  & 2.81 & 3.33 &           \\
1\,741 & $ 4.12 \pm 0.65$  & $4.17^{\dagger}          $ &         &                 &                  & 1.01 & ---  &      brt  \\
1\,793 & $ 3.70 \pm 0.52$  &                  &         & $26.08 \pm 6.46$ &                  & 7.06 & 3.46 & 1\,799 brt  \\
\noalign{\smallskip}
\hline
\noalign{\smallskip}
\end{tabular}
\label{Tab:VarSNRs}
\end{center}
Notes:\\
$^{ \ddagger~}$: Source number (from \XLPt\ catalogue) of another (brighter) \XLPt\ source which correlate with the same \ros\ source as the \XLPt\ source given in Col. 1; brt: \XLPt\ flux is below the \ros\ detection threshold (5.3\ergcm{-15}).\\
 \ros\ and \chandra\ count rates are converted to 0.2--4.5\,keV fluxes, using WebPIMMS and assuming a foreground absorption of \nh\,$=\!6.6$\hcm{20} and a photon index of $\Gamma\!=\!1.7$: ECF$_{\mr{SHP97}}\!=\!2.229\times$10$^{-14}$\,erg\,cm$^{-2}$\,cts$^{-1}$,  ECF$_{\mr{SHL2001}}\!=\!2.249\times$10$^{-14}$\,erg\,cm$^{-2}$\,cts$^{-1}$, ECF$_{\mr{HRI}}\!=\!6.001\times$10$^{-14}$\,erg\,cm$^{-2}$\,cts$^{-1}$, $^{ \dagger~}$: ECF$_{\mr{DKG2004}}\!=\!5.56\times$10$^{-12}$\,erg\,cm$^{-2}$\,cts$^{-1}$.
$^{ +~}$: For WGK2004 the luminosity given in Table~2 of WGK2004 was converted to XID flux using $F_{\mr{XID}}$[erg\,cm$^{-2}$\,s$^{-1}$]\,$=\!6.654\times$10$^{-15}\!\times\!L_{\mr{WGK2004}}$[10$^{36}$erg\,s$^{-1}$].
$^{ *~}$: For VG2007 the luminosity given in Table~2 of VG2007 was converted to XID flux using $F_{\mr{XID}}$[erg\,cm$^{-2}$\,s$^{-1}$]\,$=\!9.433\times$10$^{-15}\!\times\!L_{\mr{VG2007}}$[10$^{36}$erg\,s$^{-1}$].
\normalsize
\end{table*}

SHP97 report that [SHP97]~203 and [SHP97]~211 ($\hat{=}$[SHL2001]~206) correlate with the same SNR ([DDB80]~1.13), have the same spectral properties and have luminosities within the range of SNRs. A correlation with the \ros\ HRI catalogue (PFJ93) reveals that the true X-ray counterpart of [DDB80]~1.13 is located between the two \ros\ PSPC sources. Furthermore, PFJ93 report that this SNR is located `\textsl{within 19\,\arcsec of a brighter X-ray source}' which matches positionally with [SHP97]~211. These findings are confirmed by \xmm\ and \chandra\ observations. The X-ray counterpart of [DDB80]~1.13 is source \num\ 1\,066 in the \XLPt\ catalogue (or [PFH2005]~354 or r3-69 in \citealp{2002ApJ...577..738K}).\@ The second source, which correlates with [SHP97]~211, is the \xmm\ source \num\ 1\,077, which has a ``hard" spectrum and is $\sim\!6.7$ times brighter than \num\ 1\,066. Hence, [SHP97]~211 is a blend of the two \xmm\ sources \num\ 1\,066 and \num\ 1\,077. This also explains the pronounced variability between [SHL2001]~206 and \num\ 1\,066 given in Table~\ref{Tab:VarSNRs1}. Comparing the \chandra\ detections of the SNR counterpart with the \xmm\ flux gives a variability factor of $F_{\mr{var}}\!\approx\!1.12$. 

The distance between [SPH97]~203 and [DDB80]~1.13 is $\ga\!20$\arcsec. [SPH97]~203 was reported only in the first \ros\ PSPC catalogue. It was not detected in the observations of the second \ros\ PSPC catalogue or in any \xmm\ or \chandra\ observation of that region. Thus it seems very likely that [SPH97]~203 was either a transient source or a false detection. In both cases [SPH97]~203 cannot be a SNR. 
As the field of [DDB80]~1.13 was observed many times with \chandra, and as \chandra\ has detected weak SNRs in the central part of \m31\ \citep[][and \eg\ \protect{[DDB80]~1.13}]{2003ApJ...590L..21K}, \chandra\ should have detected X-ray emission corresponding to the \ros\ source [SPH97]~203, if it really belonged to a SNR. 

The remaining two \ros\ SNRs correlate with \xmm\ sources, which were not classified as SNRs or SNR candidates. Source [SHP97]~258 correlates with source \num\ 1\,337 and has a 3$\sigma$ positional error of 30\arcsec. From the improved spatial resolution of \xmm\ the total positional error reduces to 2\,\farcs3. Hence, we can see that the X-ray source belongs to a foreground star candidate (\cf\ Table~5) and not to the very nearby SNR. Source [SHL2001]~129 correlates with sources \num\ 743 and \num\ 761, which are classified as a GlC and a GlC candidate, respectively. The SNR candidate listed as the counterpart of [SHL2001]~129 is located between these two \xmm\ sources. In addition PFH2005 gives a third source which lies within the error circle of [SHL2001]~129 and which is classified as an AGN candidate. Thus it is very likely that [SHL2001]~129 is a blend of these three \xmm\ sources and that the correlation with the SNR candidate has to be considered as a chance coincidence. 

From the sources listed as SNRs in the different \chandra\ studies many are re-detected. Nevertheless two SNRs from \chandra\ were not detected in the \xmm\ observations. Source n1-85 has been reported as spatially correlated with an optical SNR by \citet{2004ApJ...609..735W}, but has also been classified as a repeating transient source in the same paper. An \xmm\ counterpart to n1-85 was detected neither in the study of PFH2005 nor in the \XLPt\ catalogue. The transient nature of this source is at odds with the SNR classification. 
Source CXOM31~J004247.8+411556 \citep{2003ApJ...590L..21K}, which correlates with the radio source [B90]~95, is located in the vicinity of two bright sources and close to the centre of \m31. Due to \xmm's larger point spread function this source cannot be resolved by \xmm\ in this environment. The larger PSF of \xmm\ is also the reason why source \num\ 1\,050 has a significant variability in Table~\ref{Tab:VarSNRs1}, since this source is located within the central diffuse emission of \m31.

Finally, we wanted to  determine whether any of the \XLPt\ SNRs and SNR candidates were previously observed, but not classified as SNRs. In total there are seven such sources.

 One of them (\num\ 1741) is classified as a SNR candidate based on its \xmm\ hardness ratios, and correlates with the \chandra\ source n1-48 (DKG2004). The fluxes obtained with \xmm\ and \chandra\ are in good agreement (see Table~\ref{Tab:VarSNRs}), but below the \ros\ detection threshold (5.3 \ergcm{-15}).  

For a further four sources, the corresponding sources were only detected previously with ROSAT. 
One of them (\num\ 1\,793$\hat{=}$[SHP97]~347) also correlates with a radio source \citep[source 472 of][]{1990ApJS...72..761B} and is therefore identified as a SNR. The rather high flux variability between the \ros\ and \xmm\ observations (see Table~\ref{Tab:VarSNRs}) can be attributed to source \num\ 1\,799, which is located within 19\,\farcs9 of \num\ 1\,793. This suggests that [SHP97]~347 is a combination of both \xmm\ sources, but as [SHP97]~347 was not detected in SHL2001, we cannot exclude a transient source or false detection as an explanation for the \ros\ source. Source \num\ 472 ($\hat{=}$[SHL2001]~84), source \num\ 294 ($\hat{=}$[SHP97]~53$\hat{=}$[SHL2001]~56), and source \num\ 1\,079 ($\hat{=}$[SHP97]~212) are SNR candidates based on their hardness ratios. The pronounced flux variability of source \num\ 472 is due to source \num\ 468, which is located within 18\farcs5 of \num\ 472 and is $\sim$8.6 times brighter than \num\ 472.\@ The observed flux for source \num\ 1\,079 was below the ROSAT threshold. Furthermore, the ROSAT source [SHP97]~212 was classified as a SNR, but did not appear in the SHL2001 catalogue. Hence ROSAT may have detected an unrelated transient instead.

Sources corresponding to the remaining two \xmm\ sources were detected with \ros\ and \chandra. Source \num\ 969 was detected in both \ros\ PSPC surveys ([SHP97]~185$\hat{=}$[SHL2001]~186) and correlates with \chandra\ source s1-84 \citep{2006ApJ...643..356W}. We classify it as a SNR candidate due to its hardness ratios and X-ray spectrum (see XMMM31~J004239.8+404318). Counterparts for source \num\ 1\,291 were reported in the literature as [PFJ93]~84, [SHP97]~251, [SHL2001]~255, [VG2007]~261 and source 4 in Table~5 of \citet{2006ApJ...643..844O}. Based on the \xmm\ hardness ratios and the correlation with radio source [B90]~166 \citep{1990ApJS...72..761B}, we identified the source as a SNR.
For sources \num\ 294, \num\ 969, and \num\ 1\,291 the variability between different observations may not be real because of systematic cross-calibration uncertainties. Therefore, we keep the $<$SNR$>$ and SNR classifications for these sources.

\subsubsection{The spatial distribution}
\begin{figure}
\resizebox{\hsize}{!}{\includegraphics[clip,angle=-90]{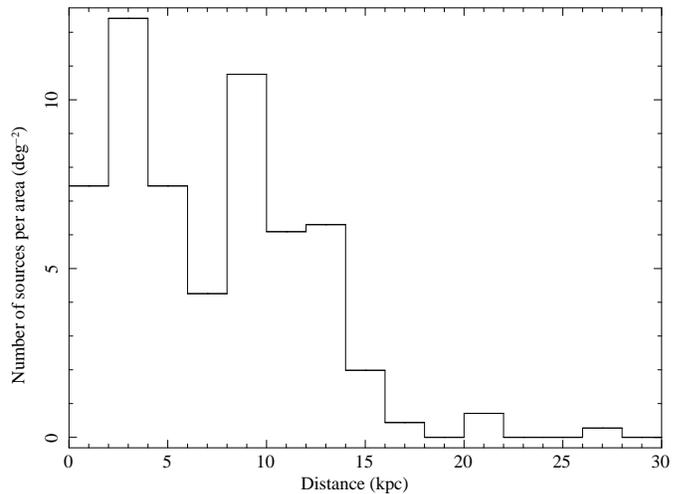}}
\caption[Projected radial distribution of SNRs and SNR candidates from the \XLPt\ catalogue.]{Projected radial distribution of SNRs and SNR candidates from the \XLPt\ catalogue. An enhancement in the source distribution corresponding to the 10\,kpc dust ring of \m31\ is visible.}
\label{Fig:SNRdepro_dist}
\end{figure}

\begin{figure}
 \resizebox{\hsize}{!}{\includegraphics[clip]{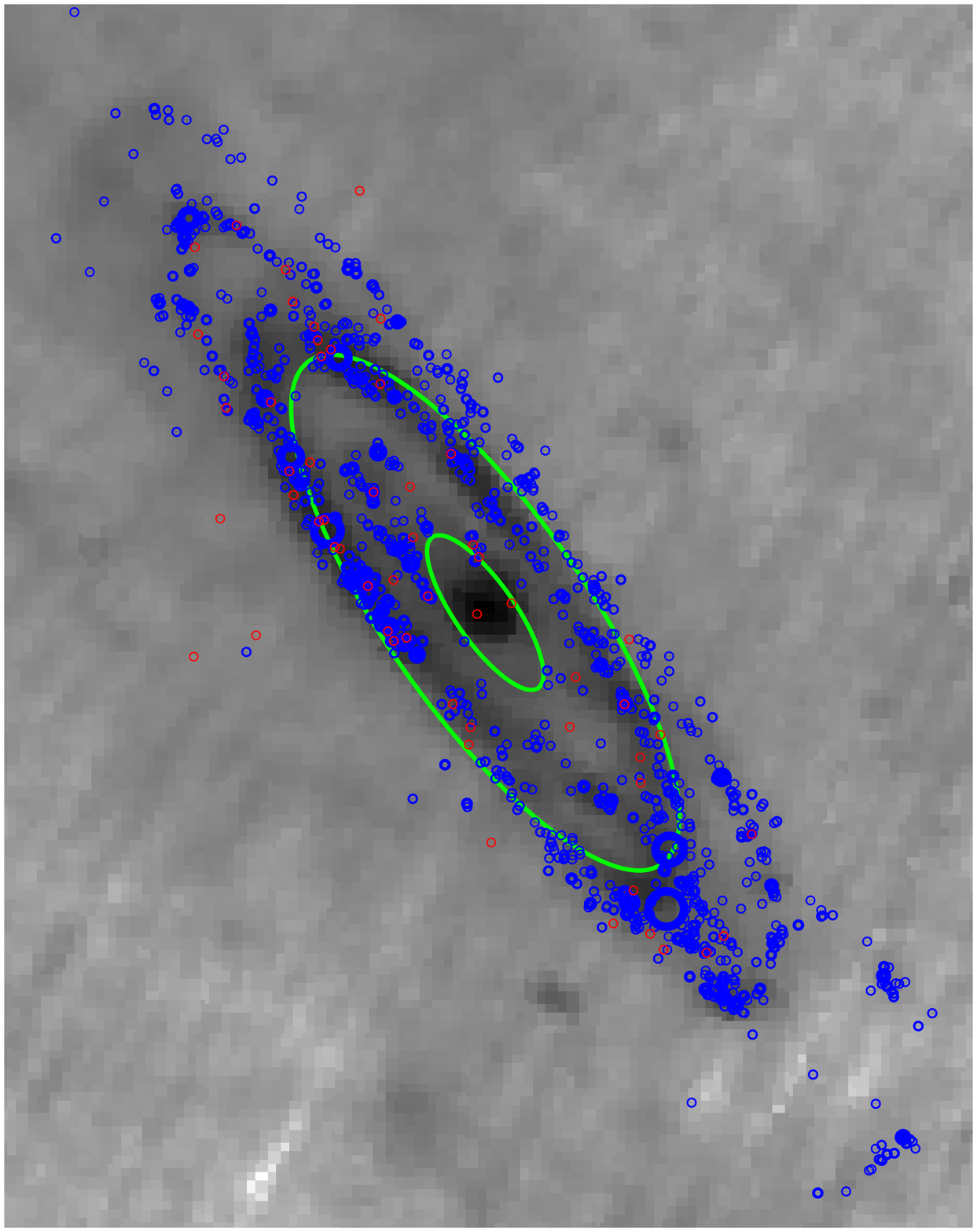}}
 \caption[An IRAS 60$\mu$m image, which clearly shows the dust ring located at $\sim\!10$\,kpc, over-plotted with the location of SNRs and candidates (red dots) from the \XLPt\ catalogue.]{An IRAS 60$\mu$m image \citep{1994STIN...9522539W}, which clearly shows the dust ring located at $\sim\!10$\,kpc, over-plotted with the location of SNRs and candidates (red dots) from the \XLPt\ catalogue. The coincidence between the SNRs and candidates and the structures of the image is visible. In addition the locations of star forming regions, which were obtained from GALEX data \citep{2009ApJ...703..614K}, are indicated by blue dots. Furthermore the two ellipses (green) at 3 and 10\,kpc from the centre correspond to the enhancemnets of sources from Fig.\,\ref{Fig:SNRdepro_dist}.}
 \label{Fig:SNR_dudist} 
\end{figure}

To examine the spatial distribution of the \xmm\ SNRs and SNR candidates, we determined projected distances from the centre of M~31. The  distribution of SNRs and SNR candidates (normalised per deg$^{2}$) is shown in Fig.\,\ref{Fig:SNRdepro_dist}. It shows an enhancement of sources around $\sim$3\,kpc, which corresponds to the SNR population in the 'inner spiral arms' of M~31. In addition, a second enhancement of sources around $\sim$10\,kpc is detected; this corresponds to the well known dust ring or star formation ring in the disc of \m31\ \citep{2006Natur.443..832B}. Only a few sources are located beyond this ring. Figure~\ref{Fig:SNR_dudist} shows the spatial distribution of the SNRs and SNR candidates from the \XLPt\ catalogue plotted over the IRAS 60$\mu$m image \citep{1994STIN...9522539W}. We see that most of the SNRs and SNR candidates are located on features that are visible in the IRAS image. This again demonstrates that SNRs and SNR candidates are coincident with the dust ring at $\sim\!10$\,kpc. In addition, the locations of star forming regions obtained from \textit{GALEX} data \citep[][and private communication]{2009ApJ...703..614K} are indicated in Fig.\,\ref{Fig:SNR_dudist}.\@ We see that many of the SNRs and SNR candidates are located within or next to star forming regions in M~31.

\subsection{X-ray binaries}
\label{SubSec:XRB}
X-ray binaries consist of a compact object plus a companion star. The compact object can either be a white dwarf (these systems are a subclass of CVs), a neutron star (NS), or a black hole (BH).\@ A common feature of all these systems is that a large amount of the emitted X-rays is produced due to the conversion of gravitational energy from the accreted matter into radiation by a mass-exchange from the companion star onto the compact object. 

X-ray binaries containing an NS or a BH are divided into two main classes, depending on the mass of the companion star:

\begin{itemize}
\item Low mass X-ray binaries (LMXBs) contain companion stars of low mass ($M\la$ 1\,M\subsun) and late type (type A or later), and have a typical lifetime of $\sim$\oexpo{8-9}~yr \citep{2006ARA&A..44..323F}. LMXBs can be located in globular clusters. Mass transfer from the companion star into an accretion disc around the compact object occurs via Roche-lobe overflow.

\item High mass X-ray binaries (HMXBs) contain a massive O or B star companion \citep[$M_{\mr{star}}\ga10$\,M\subsun,][]{Verbunt1994} and are short-lived with lifetimes of $\sim$\oexpo{6-7}\,yr \citep{2006ARA&A..44..323F}. One has to distinguish between two main groups of HMXBs: super-giant and the Be/X-ray binaries. In these systems wind-driven accretion onto the compact object powers the X-ray emission. Mass-accretion via Roche-lobe overflow is less frequent in HMXBs, but is still known to occur in several bright systems (\eg\ LMC\,X-4, SMC\,X-1, Cen\,X-3). HMXBs are expected to be located in areas of relatively recent star formation, between 25--60\,Myr ago \citep{2010ApJ...716L.140A}.
\end{itemize}

We should expect about 45 LMXBs in \m31, following a similar estimation as the one presented in Sect.\,\ref{Sec:SNR_Diss}. Here the number of LMXBs in the Galaxy was estimated from \citet{2002A&A...391..923G}.  In the \XLPt\ catalogue 88 sources are identified/classified as XRBs. This is not surprising as we may expect \m31\ to have a higher fraction of XRBs than the Galaxy since it is an earlier type galaxy composed of a higher fraction of old stars.

XRBs are the main contribution to the population of ``hard" X-ray sources in \m31. Despite some more or less reliable candidates, not a single, definitely detected HMXB is known in \m31. The results of a new search for HMXB candidates are presented in Sect.\,\ref{SubSec:XRB_HMXB}. The LMXBs can be separated into two sub-classes: the field LMXBs (discussed in this section) and those located in globular clusters. Sources belonging to the latter sub-class are discussed in Sect.\,\ref{SubSec:GlC}.\@ 

The sources presented here are classified as XRBs, because they have HRs indicating a $<$hard$>$ source and are either transient or show a variability factor larger than ten (see Sect.\,\ref{Sec:var}). 

\begin{figure}
 \resizebox{\hsize}{!}{\includegraphics[clip]{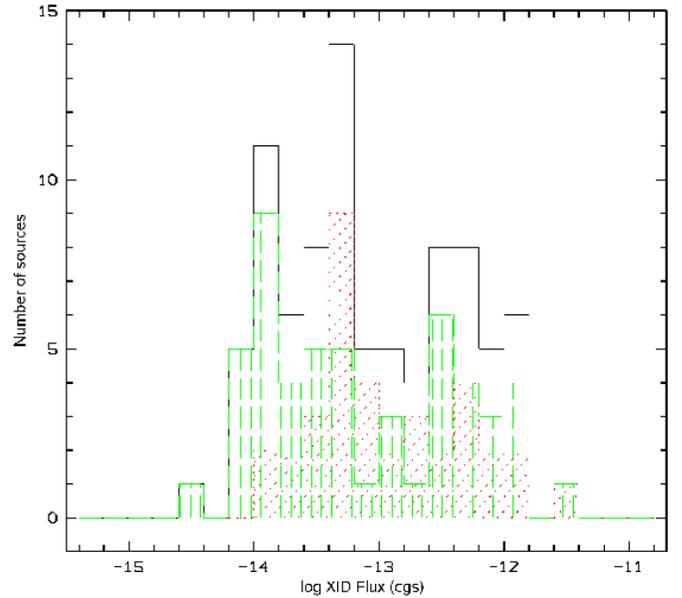}}
 \caption{Distribution of the source fluxes of XRBs and GlC sources in the 0.2--4.5\,keV (XID) band. The diagram shows the number of identified and classified XRBs and GlCs at each flux bin, plotted versus the flux. In addition, the individual distribution of (field)  XRBs (in red) as well as GlCs (in green) are given.
}
 \label{Fig:XRB_fldist} 
\end{figure}

In total 10 sources are identified and 26 are classified as XRBs by us, according to the classification criteria given in Table~\ref{Tab:class}. Apart from source \num\ 57 (XMMM31 J003833.2+402133, see below),  the identified XRBs had been reported as X-ray binaries in the literature (see comment column of Table~5).  Figure~\ref{Fig:XRB_fldist} (red histogram) shows the flux distribution of XRBs. We see that this class contains only rather bright sources. This is not surprising as the classification criterion for XRBs is based on their variability, which is more easily detected for brighter sources (\cf\  Sect.\,\ref{Sec:var}).
The XID fluxes range from 1.4\ergcm{-14} (\num\ 378) to 3.75\ergcm{-12} (\num\ 966), which correspond to luminosities from 1.0\ergs{36} to 2.7\ergs{38}. 

\begin{figure}
 \resizebox{\hsize}{!}{\includegraphics[clip]{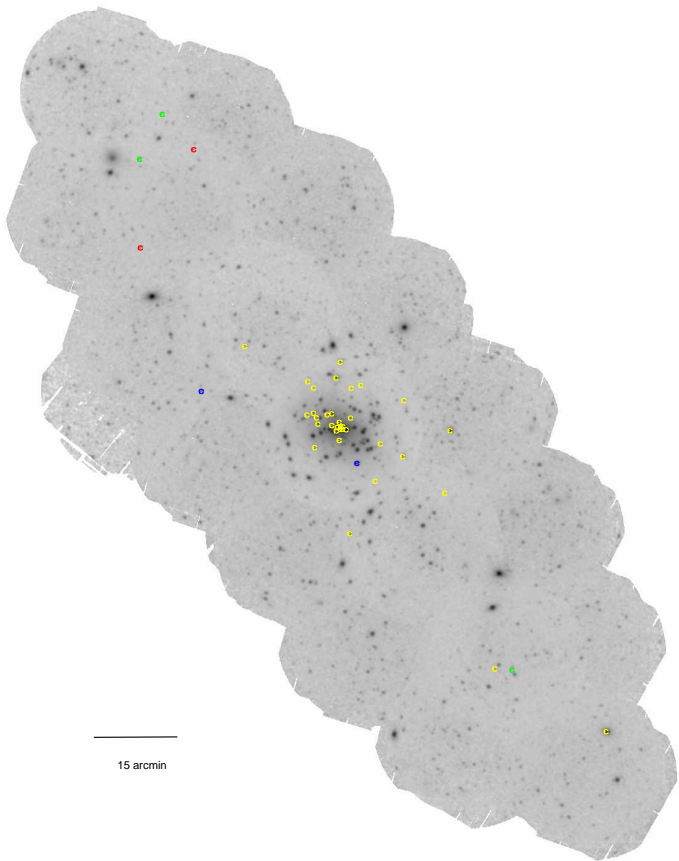}}
 \caption{The spatial distribution of XRBs and candidates from the \XLPt\ catalogue. The positions of the XRBs and candidates are marked with yellow dots; the two XRB candidates classified from their variability compared with \ros\ observations are marked with blue dots. An increase in the number density of sources in the central field is clearly visible. In addition the two new HMXB candidates presented in Sect.\,\ref{SubSec:XRB_HMXB} (red dots), and the three HMXB candidates of SBK2009 that satisfy our U-B/B-V selectrion criterion (green dots, see Sect.\,\ref{SubSec:XRB_HMXB}) are shown. XRBs which correlate with globular clusters are shown in Fig.\,\ref{Fig:GlC_spdist}.}
 \label{Fig:XRB_spdist} 
\end{figure}

\begin{figure}
 \resizebox{\hsize}{!}{\includegraphics[clip]{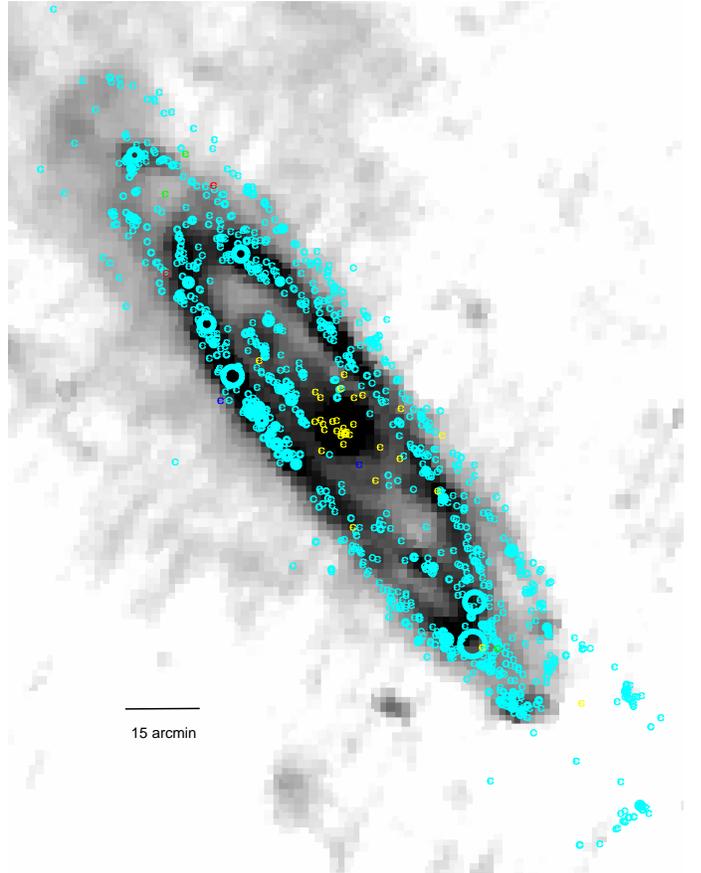}}
 \caption{The spatial distribution of XRBs and candidates from the \XLPt\ catalogue. Shown are the same sources as in Fig\,\ref{Fig:XRB_spdist}, but over-plotted on an IRAS 60\,$\mu$m image \citep{1994STIN...9522539W}, which shows the dusty star forming region in \m31. In addition the locations of star forming regions, which were obtained from GALEX data \citep{2009ApJ...703..614K}, are indicated by cyan dots.}
 \label{Fig:XRB_spdist_IRAS} 
\end{figure}

It is clear from Fig.\,\ref{Fig:XRB_spdist}, which shows the spatial distribution of the XRBs, that nearly all sources classified or identified as XRBs (yellow dots) are located in fields that were observed more than once (centre and southern part of the disc). This is partly a selection effect, caused by the fact that these particular fields were observed several times, thus allowing the determination of source variability. 
For sources located outside these fields, especially the northern part of the disc, the transient nature must have been reported in the literature to mark them as an XRBs. 
The source density of LMXBs, which follows the overall stellar density, is higher in the centre than in the disc of \m31. One would not expect HMXBs in the central region which is dominated by the bulge (old stellar population). 
From Fig.\,\ref{Fig:XRB_spdist_IRAS}, which shows the spatial distribution of the XRBs over-plotted on an IRAS 60\,$\mu$m image \citep{1994STIN...9522539W}, we see that only a few sources, classified or identified as XRBs, are located in the vicinity of star forming regions.

References for the sources, selected from their temporal variability, are given in Table~\ref{Tab:varlist}.\@ TPC06 report on four bright X-ray transients, which they detected in the observations of July 2004 and suggested to be XRB candidates. We also found these sources and classified source \num\ 705 and identified sources \num\ 985, \num\ 1\,153, \num\ 1\,177 as XRBs. One of the identified XRBs (\num\ 1\,177) shows a very soft spectrum. \citet{2005ApJ...632.1086W} observed source \num\ 1\,153 with \chandra\ and \textsl{HST}. From the location and X-ray spectrum they suggest it to be an LMXB. They propose that the optical counterpart of the X-ray source is a star within the X-ray error box , which shows an optical brightness change (in B) by $\simeq$1 mag. 
Source \num\ 985 was first detected in January 1979 by TF91 with the \ein\ observatory. WGM06 rediscovered it in \chandra\ observations from 2004. Their coordinated \textsl{HST} ACS imaging does not reveal any variable optical counterpart. From the X-ray spectrum and the lack of a bright star, WGM06 suggest that this source is an LMXB with a black hole primary.

In the following subsections we discuss three transient XRBs in more detail.

\paragraph{XMMM31~J003833.2+402133} (\num\ 57) was first detected in the \xmm\ observation from 02 January 2008 (s32) at an unabsorbed 0.2\,--\,10\,keV luminosity of $\sim\!2$\ergs{38}. 
From two observations, taken about 0.5\,yr (s31) and 1.5\,yr (s3) earlier, we derived upper limits for the fluxes, which were more than a factor of 100 below the values obtained in January 2008. 

\begin{figure}
\subfigure[XMMM31~J003833.2+402133]{\includegraphics[scale=0.3, angle=-90]{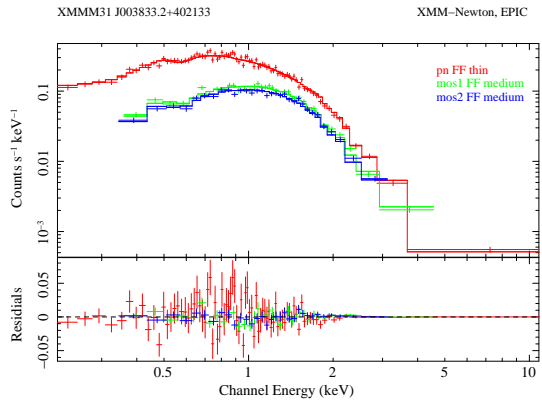}\label{SubFig:spec_1}}
\subfigure[CXOM31~J004059.2+411551]{\includegraphics[scale=0.3, angle=-90]{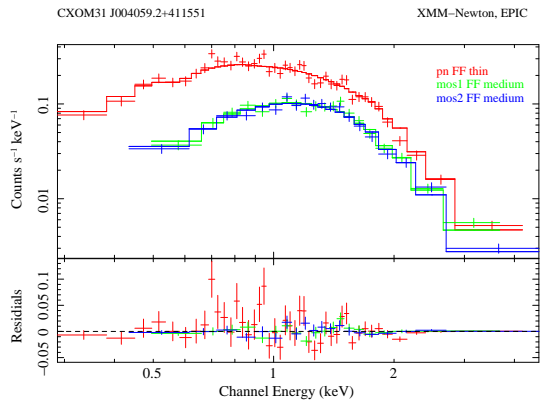}\label{SubFig:spec_2}}\\
\subfigure[XMMU~J004144.7+411110]{\includegraphics[scale=0.3, angle=-90]{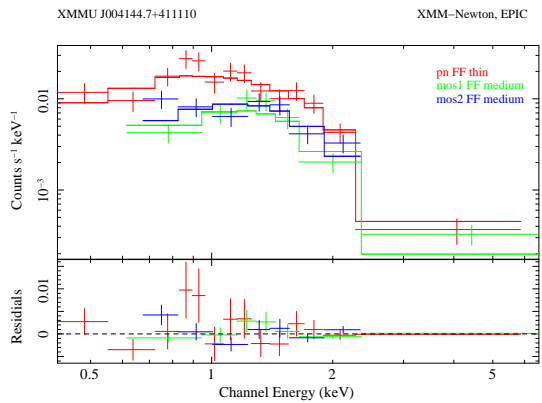}\label{SubFig:spec_3}}
\caption{EPIC spectra of the transient sources \subref{SubFig:spec_1} XMMM31~J003833.2+402133,  \subref{SubFig:spec_2} CXOM31~J004059.2+411551 and  \subref{SubFig:spec_3} XMMU~J004144.7+411110. The histograms show the best-fit model: PL+DISCBB in \subref{SubFig:spec_1}, DISCBB in \subref{SubFig:spec_2} and \subref{SubFig:spec_3}.}
\label{Fig:spec} 
\end{figure}

The combined EPIC spectrum from observation s32 (Fig.\,\ref{SubFig:spec_1}) is best fitted with an absorbed disc blackbody plus power-law model, with $N_{\mr{H}}\!=\!1.68^{+0.42}_{-0.48}$\hcm{21}, temperature at the inner edge of the disc $k_{\mr{B}}T_{\mr{in}}\!=\!0.462\pm0.013$\,keV and power-law index of $2.55^{+0.33}_{-1.05}$.\@ The contribution of the disc blackbody luminosity to the total luminosity is $\sim 59\,\%$. Formally acceptable fits are also obtained from an absorbed disc blackbody and an absorbed bremsstrahlung model (see Table~\ref{Tab:specprop}).

We did not find any significant feature in a fast Fourier transformation (FFT) periodicity search. The combined EPIC light curve during observation s32 was consistent with a constant value.

To identify possible optical counterparts we examined the LGGS images and the images taken with the \xmm\ optical monitor during the X-ray observation (UVW1 and UVW2 filters).
The absence of optical/UV counterparts and of variability on short timescales, as well as the spectral properties suggest that this source is a black hole LMXB in the steep power-law state \citep{2006csxs.book..157M}.

\paragraph{CXOM31~J004059.2+411551:}
\citet{2007ATel.1147....1G} reported on the detection of a previously unseen X-ray source in a 5\,ks \chandra\ ACIS-S observation from 05 July 2007. In an \xmm\ ToO observation \citep[sn11,][]{2007ATel.1191....1S} taken about 20 days after the \chandra\ detection, the source (\num\ 523) was still bright. 
The position agrees with that found by \chandra. We detected the source at an unabsorbed 0.2\,--\,10\,keV luminosity of $\sim\!1.1$\ergs{38}.

The combined EPIC spectrum (Fig.\,\ref{SubFig:spec_2}) can be well fitted with an absorbed disc blackbody model with $N_{\mr{H}}\!=\!\lb(2.00\pm{0.16}\rb)$\hcm{21} and with a temperature at the inner edge of the disc of  $k_{\mr{B}}T_{\mr{in}}\!=\!0.538\pm0.017$\,keV (Table~\ref{Tab:specprop}). The spectral parameters and luminosity did not change significantly compared to the \chandra\ values of \citet{2007ATel.1147....1G}.

We did not find any significant feature in an FFT periodicity search. The combined EPIC light curve was consistent with a constant value.

The examination of LGGS images and of images taken with the \xmm\ optical monitor (UVW1 and UVW2 filters) during the X-ray observation did not reveal any possible optical/UV counterparts. 

The lack of bright optical counterparts and the X-ray parameters (X-ray spectrum, lack of periodicity, transient nature, luminosity) are consistent with this source being a black hole X-ray transient, as already mentioned in \citet{2007ATel.1147....1G}.

\paragraph{XMMU~J004144.7+411110} (\num\ 705) was detected by \citet{2006ApJ...645..277T} in \xmm\ observations b1--b4 (July 2004) at an unabsorbed luminosity of 3.1--4.4\ergs{37} in the 0.3--7\,keV band, using a {\tt DISKBB} model. 
We detected the source in observation sn11 (25 July 2007) with an unabsorbed 0.2--10\,keV luminosity of $\sim$1.8\ergs{37}, using also a {\tt DISKBB} model.

In observation sn11, the source was bright enough to allow spectral analysis.
The spectra can be well fitted with an absorbed power-law, disc blackbody or bremsstrahlung model (Table~\ref{Tab:specprop}). 
The obtained spectral shapes (absorption and temperature as well as photon index) are in agreement with the values of \citet{2006ApJ...645..277T}.

An FFT periodicity search did not reveal any significant periodicities in the 0.3\,s to 2\,000\,s range.

No optical counterparts were evident in the images taken with the \xmm\ optical monitor UVW1 and UVW2 during the sn11 observation, nor in the LGGS images.
The lack of a bright optical counterpart and the X-ray parameters support that this source is a black hole X-ray transient, as classified by \citet{2006ApJ...645..277T}.

\begin{table*}
\scriptsize
\begin{center}
\caption{Spectral parameters of the transient sources.}
\begin{tabular}{ccccccccc}
\hline\noalign{\smallskip}
\hline\noalign{\smallskip}
\multicolumn{1}{c}{M 31 field} & \multicolumn{1}{c}{Model} &\multicolumn{1}{c}{$N_{\mr{H}}$} &
\multicolumn{1}{c}{$k_{\mr{B}}T$} & \multicolumn{1}{c}{$R_{in}\sqrt{\cos{i}}^*$} & \multicolumn{1}{c}{Photon} & \multicolumn{1}{c}{$\chi^2$} & \multicolumn{1}{c}{$L_X^{\dagger}$} & \multicolumn{1}{c}{Instrument} \\ 
\noalign{\smallskip}
& & \multicolumn{1}{c}{(\hcm{21})} & \multicolumn{1}{c}{(keV)} 
& \multicolumn{1}{c}{(km)}  & \multicolumn{1}{c}{Index} & \multicolumn{1}{c}{(d.o.f)}
& & \\
\noalign{\smallskip}\hline\noalign{\smallskip}\hline\noalign{\smallskip}
& & & \multicolumn{3}{c}{XMMM31~J003833.2+402133 (\num\ 57)} & &\\
\noalign{\smallskip}\hline\noalign{\smallskip}
s32 &PL+DISCBB&$1.68^{+0.42}_{-0.48}$&$0.462\pm0.013$&$106^{+9}_{-10}$ &$2.55^{+0.33}_{-1.05}$ &173.89(145)&2.04&PN+M1+M2\\{\smallskip}
s32 &DISCBB&$1.06\pm0.06$&$0.511\pm0.009$&$95\pm4$ & &270.01(147)&1.46&PN+M1+M2\\
s32 &BREMSS&$1.91\pm0.07$&$1.082^{+0.029}_{-0.030}$& & &208.65(147)&2.12&PN+M1+M2\\
\noalign{\smallskip}\hline\noalign{\smallskip}\hline\noalign{\smallskip}
& & & \multicolumn{3}{c}{CXOM31~J004059.2+411551 (\num\ 523)} &  &\\
\noalign{\smallskip}\hline\noalign{\smallskip}
sn11 &DISCBB&$2.00\pm0.16$&$0.538\pm0.017$&$75\pm6$ & &97.70(79)&1.12&PN+M1+M2\\
sn11 &BREMSS&$3.13\pm0.19$&$1.097^{+0.060}_{-0.056}$& & &93.17(79)&1.72&PN+M1+M2\\
\noalign{\smallskip}\hline\noalign{\smallskip}\hline\noalign{\smallskip}
& & & \multicolumn{3}{c}{XMMU~J004144.7+411110 (\num\ 705)} &  &\\
\noalign{\smallskip}\hline\noalign{\smallskip}
sn11 &DISCBB&$2.32^{+1.03}_{-0.87}$&$0.586^{+0.100}_{-0.087}$&$26^{+13}_{-8}$ & &29.74(23)&0.18&PN+M1+M2\\
sn11 &BREMSS&$3.72^{+1.14}_{-1.00}$&$1.216^{+0.373}_{-0.269}$& & &29.48(23)&0.29&PN+M1+M2\\
sn11 &PL&$6.17^{+1.72}_{-1.47}$&& &$3.23^{+0.46}_{-0.40}$& 31.57(23)&1.12&PN+M1+M2\\
\noalign{\smallskip}\hline\noalign{\smallskip}\hline\noalign{\smallskip}
\end{tabular}
\label{Tab:specprop}
\end{center}
Notes:\\
$^{ *~}$: effective inner disc radius, where $i$ is the inclination angle of the disc\\
$^{ {\dagger}~}$: unabsorbed luminosity in the $0.2$\,--\,$10.0$\,keV energy range in units of \oergs{38}\\ 
\normalsize
\end{table*}

\subsubsection{Sources from the XMM-LP total catalogue that were not detected by \textit{ROSAT}}
\begin{figure} 
\resizebox{\hsize}{!}{\includegraphics[clip,angle=0]{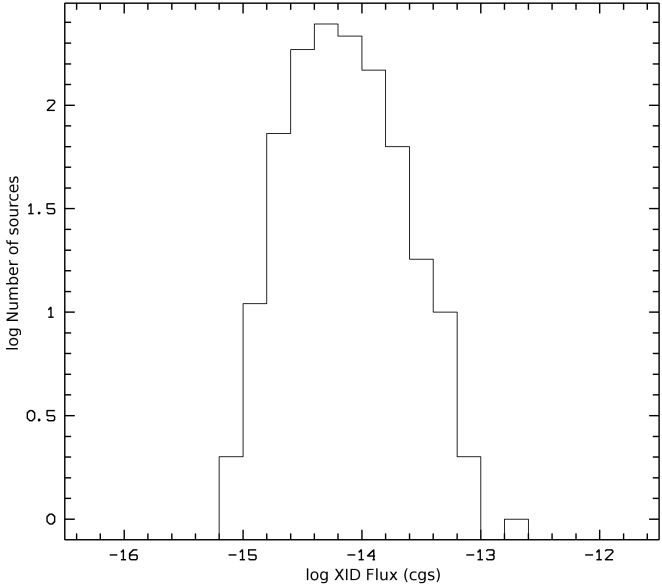}}
\caption{Distribution of the source fluxes in the 0.2\,--\,4.5\,keV (XID) band. The diagram shows the number of sources from the \XLPt\ catalogue that were classified as $<$hard$>$, and in addition do not correlate with a source listed in the \ros\ catalogues at each flux bin plotted versus the flux, using logarithmic scales.}
\label{Fig:noROSdist} 
\end{figure}

To search for additional XRB candidates, we selected all sources from the \XLPt\ catalogue, that were classified as $<$hard$>$ and which did not correlate with a source listed in the \ros\ catalogues (PFJ93, SHP97 and SHL2001). The flux distribution of the selected sources is shown in Fig.\,\ref{Fig:noROSdist}, and Table~\ref{Tab:noROSdist} gives the number of sources brighter than the indicated flux limit.

\begin{table}
\begin{center}
\caption{The cumulative number of sources from the \XLPt\ catalogue that were classified as $<$hard$>$, and in addition do not correlate with a source listed in the \ros\ catalogues. Four different limiting fluxes are indicated.}
\begin{tabular}{rr}
\hline\noalign{\smallskip}
\hline\noalign{\smallskip}
\multicolumn{1}{c}{XID flux limit} & \multicolumn{1}{c}{\# of sources}\\
\multicolumn{1}{c}{erg\,cm$^{-2}$\,s$^{-1}$} & \\
\hline\noalign{\smallskip}
5.5\,E-15 & 541 \\
1\,E-14 & 242 \\
5\,E-14 & 7 \\
1\,E-13 & 1 \\
\noalign{\smallskip}\hline\noalign{\smallskip}\hline\noalign{\smallskip}
\end{tabular}
\label{Tab:noROSdist}
\end{center}
\normalsize
\end{table}

Possible, new XRB candidates are sources that have an XID flux that lies at least a factor of ten above the \ros\ detection threshold (5.3\ergcm{-15}). These sources fulfil the variability criterion used to classify XRBs (\cf\ Sect.\,\ref{Sec:var}). The \XLPt\ catalogue lists five sources without \ros\ counterparts that have XID fluxes above 5.3\ergcm{-14}. These are: \num\ 239, \num\ 365, \num\ 910, \num\ 1\,164, and \num\ 1\,553. Between the \ros\ and \xmm\ observations more than ten years have elapsed. On this time scale AGN can also show strong variability. To estimate the number of AGN among the five sources listed above, we investigated how many sources of the identified and classified background objects from the \XLPt\ catalogue with an XID flux larger than 5.3\ergcm{-14} were not detected by \ros. The result is that \ros\ detected all background sources with an XID flux larger than 5.3\ergcm{-14} that are listed in the \XLPt\ catalogue. Thus, the probability that any of the five sources listed above is a background object is very small, in particular if the source is located within the D$_{25}$ ellipse of \m31. Therefore, the two sources located within the D$_{25}$ ellipse are listed in the \XLPt\ catalogue as XRB candidates, while the remaining three sources, which are located outside the D$_{25}$ ellipse, are classified as $<$hard$>$.\@ All five sources are marked in the comment column of Table~5 with `XRB cand.\ from \ros\ corr.'.

\subsubsection{Detection of high mass X-ray binaries}
\label{SubSec:XRB_HMXB}
As already mentioned, until now not a single secure HMXB in \m31\ has been confirmed. The reason for this is that the detection of HMXBs in \m31\ is difficult. \citet{2004ApJ...602..231C} showed that the hardness ratio method is very inefficient in selecting HMXBs in spiral galaxies. The selection process is complicated by the fact, that the spectral properties of BH HMXBs, which have power-law spectra with indices of $\sim$1-- $\sim$2 are similar to LMXBs and AGN. Therefore the region in the HR diagrams where BH HMXB are located is contaminated by other hard sources (LMXBs, AGN, and Crab like SNRs). For the NS HMXBs, which have power-law indices of $\sim$1, and thus should be easier to select, the uncertainties in the hardness ratios lead at best to an overlap -- in the worst case to a fusion -- with the area occupied by other hard sources \citep{2004ApJ...602..231C}.

Based on the spectral analysis of individual sources of \m31, SBK2009 identified 18 HMXB candidates with power-law indices between 0.8 and 1.2. One of these sources ([SBK2009]~123) correlates with a globular cluster, and hence it is rather an LMXB in a very hard state rather than an HMXB 
\citep[\cf\ ][]{2004ApJ...616..821T}. Four of their sources ([SBK2009]~34, 106, 149, and 295) do not have counterparts in the \XLPt\ catalogue.

\citet{Peter} developed a selection algorithm for HMXBs in the SMC, which also uses properties of the optical companion. X-ray sources were selected as HMXB candidates if they had HR2$+$EHR2$>$0.1 as well as an optical counterpart within 2\,\farcs5 of the X-ray source, with $-0.5\!<$B$-$V$<\!0.5$\,mag, $-1.5\!<$U$-$B$<\!-0.2$\,mag and V$<$17\,mag. 

We tried to transfer this SMC selection algorithm to \m31\ sources. In doing so, we encountered two problems: The first problem is that the region of the U-B/B-V diagram is also populated by globular clusters (LMXB candidates) in \m31. The second problem is that due to the much larger distance to \m31, the range of detected V magnitudes of HMXBs in the SMC of $\sim$13$<$V$<$17\,mag translates to a $\sim$19$<$V$<$23\,mag criterion for \m31. Thus the V magnitude of optical counterparts of possible HMXB candidates lies in the same range as the optical counterparts of AGN. Therefore the V mag criterion, which provided most of the discriminatory power in the case of the SMC, fails totally in the case of \m31. 

A few of the sources selected from the optical colour-colour diagram and HR diagrams are bright enough to allow the creation of X-ray spectra. That way two additional (\ie\ not given in SBK2009) HMXB candidates were found.

In addition, we determined the reddening free Q parameter: 
\begin{equation} 
Q = (\rm{U}-\rm{B})-0.72(\rm{B}-\rm{V})
\end{equation}
\citep[for definition see \eg\ ][]{cox2001allen}
which allowed us to keep only the intrinsically bluest stars, using Q $\le\!-0.4$ \citep[O-type stars typically have Q$<\!-0.9$, while -0.4 corresponds to a B5 dwarf or giant or an A0 supergiant, ][]{2007AJ....134.2474M}. U$-$B and B$-$V were taken from the LGGS catalogue. 

\paragraph{XMMM31~J004557.0+414830} (\num\ 1\,716) has an USNO-B1 (R2$=$18.72\,mag), a 2MASS and an LGGS (V$=$20.02\,mag; Q$=\!-0.44$) counterpart. The EPIC spectrum is best fitted ($\chi^2_{red}\!=\!0.93$) by an absorbed power-law with \nh$=\!7.4^{+6.0}_{-3.9}$\hcm{21} and photon index $\Gamma\!=\!1.2\pm0.4$. The absorption corrected X-ray luminosity in the 0.2--10\,keV band is $\sim$7.1\ergs{36}.

\paragraph{XMMM31~J004506.4+420615} (\num\ 1\,579) has an USNO-B1 (B2$=$20.87\,mag), a 2MASS and an LGGS (V$=$20.77\,mag; Q$=\!-1.04$) counterpart. The EPIC PN spectrum is best fitted ($\chi^2_{red}\!=\!1.6$) by an absorbed power-law with \nh$=\!0.48^{+2.4}_{-1.0}$\hcm{21} and photon index $\Gamma\!=\!1.0^{+0.7}_{-0.5}$. The absorption corrected X-ray luminosity in the 0.2--10\,keV band is $\sim$8.6\ergs{36}.\\

To strengthen these classifications spectroscopic optical follow-up observations of the optical counterparts are needed. An FFT periodicity search did not reveal any significant periodicities for either of the two sources and the light curves do not show eclipses.

From the sources reported as HMXB candidates in SBK2009, three sources ([SBK2009]~21, 236, and 256) are located in the region of the U-B/B-V diagram, that we used. Another three sources ([SBK2009]~123, 172, and 226) are located outside that region. The remaining sources of SBK2009 have either no counterparts with a U-B colour entry in the LGGS catalogue ([SBK2009]~99, 234, 294, and 302) or have no optical counterpart from the LGGS catalogue at all ([SBK2009]~9, 160, 197, and 305). The reddening free Q parameter for the SBK2009 sources that have counterparts in the LGGS catalogue are given in Table~\ref{Tab:SBK_Q}.

\begin{table}
\begin{center}
\caption{Reddening free Q parameter for HMXB candidates of SBK2009.}
\begin{tabular}{rrlr}
\hline\noalign{\smallskip}
\hline\noalign{\smallskip}
\multicolumn{1}{c}{\num} & \multicolumn{1}{c}{[SBK2009]} & \multicolumn{1}{c}{LGGS counterpart} & \multicolumn{1}{c}{$Q$} \\
\noalign{\smallskip}\hline\noalign{\smallskip}
 312 &  21 & J004001.50+403248.0 & -0.34\\
1668 & 236 & J004538.23+421236.0 & -0.77 \\
1724 & 256 & J004558.98+420426.5 & -0.81 \\
1109 & 123 & J004301.51+413017.5 & +1.77 \\
1436 & 172 & J004420.98+413546.7 & -0.65 \\
     &     & J004421.01+413544.3$^{*}$ & -0.29\\
1630 & 226 & J004526.68+415631.5 & -0.92\\
     &     & J004526.58+415633.1$^{*}$ & -0.72\\
\noalign{\smallskip}\hline\noalign{\smallskip}\hline\noalign{\smallskip}
\end{tabular}
\label{Tab:SBK_Q}
\end{center}
Notes: \\
$^{ *~}$: counterparts listed in SBK2009
\normalsize
\end{table}

\subsection{Globular cluster sources}
\label{SubSec:GlC}
A significant number of the luminous X-ray sources in the Galaxy and in \m31\ are found in globular clusters. X-ray sources corresponding to globular clusters are identified by cross-correlating with globular cluster catalogues (see Sect.\,\ref{Sec:CrossCorr_Tech}). Therefore changes between the \XLPt\ catalogue and the catalogue of PFH2005 in the classification of sources related to globular clusters are based on the availability of and modifications in recent globular cluster catalogues.

In total 52 sources of the \XLPt\ catalogue correlate with (possible) globular clusters. Of these sources 36 are identified as GlCs because their optical counterparts are listed as globular clusters in the catalogues given in Sect.\,\ref{Sec:CrossCorr_Tech}, while the remaining 16 sources are only listed as globular cluster candidates.

The range of source XID fluxes goes from 3.1\ergcm{-15} (\num\ 924) to 2.7\ergcm{-12} (\num\ 1\,057), or in luminosity from 2.3\ergs{35} to 2.0\ergs{38} (Fig.\,\ref{Fig:XRB_fldist}; green histogram). Compared to the fluxes found for the XRBs discussed in Sect.\,\ref{SubSec:XRB}, 14 sources that correlate with GlCs have fluxes below the lowest flux found for field XRBs. The reason for this finding is that the classification of field XRBs is based on the variable or transient nature of the sources, which can only be to detected for brighter sources (\cf\ Sect.\,\ref{Sec:var}) and not just by positional coincidence that is also possible for faint sources.

\begin{figure}
 \resizebox{\hsize}{!}{\includegraphics[clip]{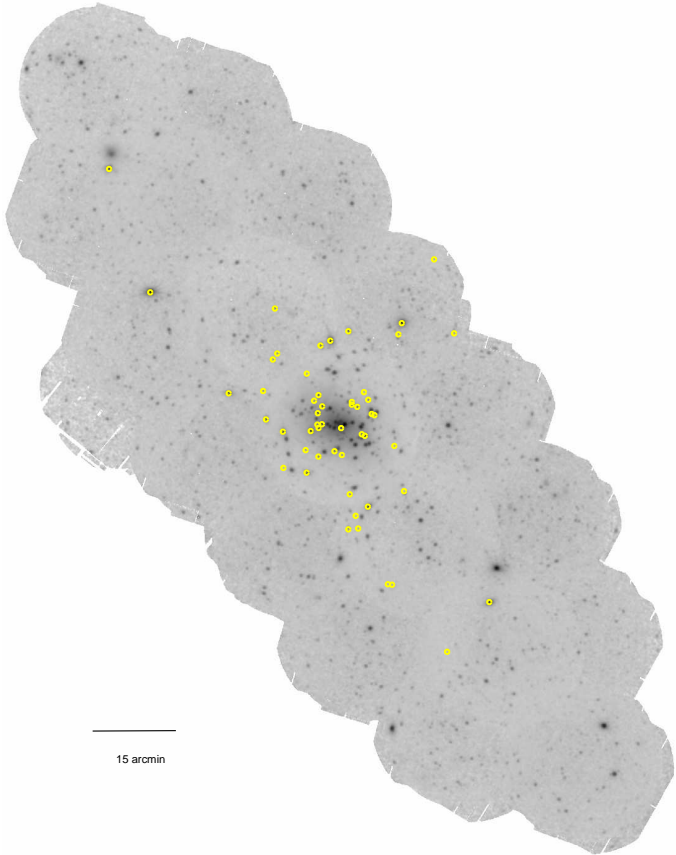}}
 \caption{The spatial distribution of X-ray sources correlating with GlCs and candidates from the \XLPt\ catalogue (yellow  dots). An enhancement of sources towards the central region of \m31\ is clearly visible.}
 \label{Fig:GlC_spdist} 
\end{figure}

Figure \ref{Fig:GlC_spdist} shows the spatial distribution of the GlC sources. X-ray sources correlating with GlCs follow the distribution of the optical GlCs, which are also concentrated towards the central region of \m31. 

The three brightest globular cluster sources, which are located in the northen disc of \m31, are \num\ 1\,057 (XMMM31~J004252.0+413109), \num\ 694 (XMMM31~J004143.1+413420), and \num\ 1\,692 (XMMM31~J004545.8+413941). They are all brighter than 8.4\ergs{37}. Source \num\ 694 was classified as a black hole candidate, due to its variability observed at such high luminosities. A detailed discussion of the three sources is given in \citet{2008ApJ...689.1215B}.

XMMM31~J004303.2+412121 (\num\ 1\,118) was identified as a foreground star in PFH2005, based on the classification in the ``Revised Bologna Catalogue" \citep{2004A&A...416..917G}. \citet{2004A&A...416..917G} took the classification from \citet{1997A&A...321..379D}, which is based on the velocity dispersion of that source. Recent `\textsl{HST images unambiguously reveal that this} [B147] \textsl{is a well resolved star cluster, as recently pointed out also by \citet{2007AJ....133.2764B}}' \citep{2007A&A...471..127G}. That is why source \num\ 1\,118 is now identified as an XRB located in globular cluster B147.

\subsubsection{Integrated optical properties of the globular clusters in which the X ray sources are located}

\begin{table}
\scriptsize
\begin{center}
\caption{Integrated V-I colours, dereddened V-I integrated colours, and age estimates of the globular clusters in which the X ray sources are located.}
\begin{tabular}{lcclllc}
\hline\noalign{\smallskip}
\hline\noalign{\smallskip}
\multicolumn{1}{c}{Name} & \multicolumn{1}{c}{Class$^{*}$} & \multicolumn{1}{c}{Age$^{+}$} & \multicolumn{1}{c}{V mag} & \multicolumn{1}{c}{VI} & \multicolumn{1}{c}{VI0} & \multicolumn{1}{c}{Age$^{\dagger}$} \\
\noalign{\smallskip}\hline\noalign{\smallskip}
B005   & confirmed & old & 15.69 & 1.29 & 1.15 & old \\
SK055B & candidate & - & 18.991 & 0.388 & 0.248 & -- \\
B024 & confirmed & old & 16.8 & 1.15 & 1.01 & old \\
SK100C & candidate & na & 18.218 & 1.181 & 1.041 & old \\
B045 & confirmed & old & 15.78 & 1.27 & 1.13 & old \\
B050 & confirmed & old & 16.84 & 1.18 & 1.04 & old \\
B055 & confirmed & old & 16.67 & 1.68 & 1.54 & old \\
B058 & confirmed & old:: & 14.97 & 1.1 & 0.96 & old-inter \\
MITA140 & confirmed & old & 17 & 9999 & - & -- \\
B078 & confirmed & old & 17.42 & 1.62 & 1.48 & old \\
B082 & confirmed & old & 15.54 & 1.91 & 1.77 & old \\
B086 & confirmed & old & 15.18 & 1.26 & 1.12 & old \\
SK050A & confirmed & - & 18.04 & 1.079 & 0.939 & old-inter \\
B094 & confirmed & old & 15.55 & 1.26 & 1.12 & old \\
B096 & confirmed & old & 16.61 & 1.48 & 1.34 & old \\
B098 & confirmed & old & 16.21 & 1.13 & 0.99 & old-inter \\
B107 & confirmed & old & 15.94 & 1.28 & 1.14 & old \\
B110 & confirmed & old & 15.28 & 1.28 & 1.14 & old \\
B117 & confirmed & old:: & 16.34 & 1 & 0.86 & inter \\
B116 & confirmed & old & 16.79 & 1.86 & 1.72 & old \\
B123 & confirmed & old & 17.416 & 1.29 & 1.15 & old \\
B124 & confirmed & old & 14.777 & 1.147 & 1.007 & old \\
B128 & confirmed & old:: & 16.88 & 1.12 & 0.98 & old-inter \\
B135 & confirmed & old & 16.04 & 1.22 & 1.08 & old \\
B143 & confirmed & old & 16 & 1.22 & 1.08 & old \\
B144 & confirmed & old:: & 15.88 & 0.59 & 0.45 & young \\
B091D & confirmed & old & 15.44 & 9999 & - & -- \\
B146 & confirmed & old:: & 16.95 & 1.09 & 0.95 & interm \\
B147 & confirmed & old & 15.8 & 1.27 & 1.13 & old \\
B148 & confirmed & old & 16.05 & 1.17 & 1.03 & old \\
B150 & confirmed & old & 16.8 & 1.28 & 1.14 & old \\
B153 & confirmed & old & 16.24 & 1.3 & 1.16 & old \\
B158 & confirmed & old & 14.7 & 1.15 & 1.01 & old \\
B159 & confirmed & old & 17.2 & 1.41 & 1.27 & old \\
B161 & confirmed & old & 16.33 & 1.1 & 0.96 & old-inter \\
B182 & confirmed & old & 15.43 & 1.29 & 1.15 & old \\
B185 & confirmed & old & 15.54 & 1.18 & 1.04 & old \\
B193 & confirmed & old & 15.33 & 1.28 & 1.14 & old \\
SK132C & candidate & - & 18.342 & 1.84 & 1.7 & old \\
B204 & confirmed & old & 15.75 & 1.17 & 1.03 & old \\
B225 & confirmed & old & 14.15 & 1.39 & 1.25 & old \\
B375 & confirmed & old & 17.61:: & 1.02 & 0.88 & interm \\
B386 & confirmed & old & 15.547 & 1.154 & 1.014 & old \\
\noalign{\smallskip}\hline\noalign{\smallskip}\hline\noalign{\smallskip}
\end{tabular}
\label{Tab:GlC_optprop}
\end{center}
Notes: \\
$^{ *~}$: classification as confirmed or otherwise comes from the revised Bologna catalogue (December 2009, Version 4) \url{http://www.bo.astro.it/M31/RBC_Phot07_V4.tab}\\
$^{ +~}$: age comes from \citet{2009AJ....137...94C}\\
V, and V$-$I are integrated colours that come from the revised Bologna catalogue (December 2009, Version 4)
           \url{http://www.bo.astro.it/M31/RBC_Phot07_V4.tab}\\
(V$-$I)$_{\rm{o}}$ is the dereddened V-I integrated colour, assuming E(B-V)=0.10+-0.03,  which is the average of the reddenings of all \m31\ clusters in \citet{2005AJ....129.2670R}.  (this E(B$-$V) corresponds to E(V$-$I)$=\!0.14$)\\
$^{ \dagger~}$: This dereddened colour (V$-$I)$_{\rm{o}}$ is used to estimate the age on the basis of the plots (V$-$I)$_{\rm{o}}$ versus log\,Age from \citet{2007AJ....133..290S}.
\normalsize
\end{table}

\begin{figure}
  \resizebox{\hsize}{!}{\includegraphics[clip,angle=-0]{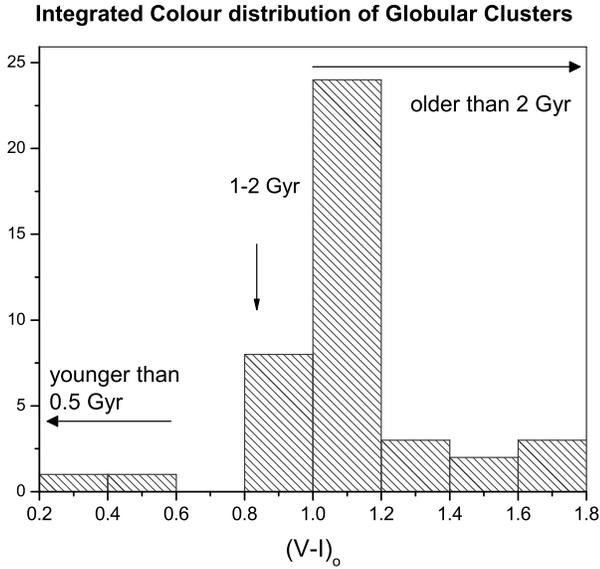}}
     \caption{The distribution of (V$-$I)$_{\rm{o}}$ for globular clusters (and candidates), with the approximate age-ranges marked, hosting \XLPt\ X-ray sources.}
    \label{Fig:GlC_agedist} 
\end{figure}

\begin{figure}
  \resizebox{\hsize}{!}{\includegraphics[clip,angle=-0]{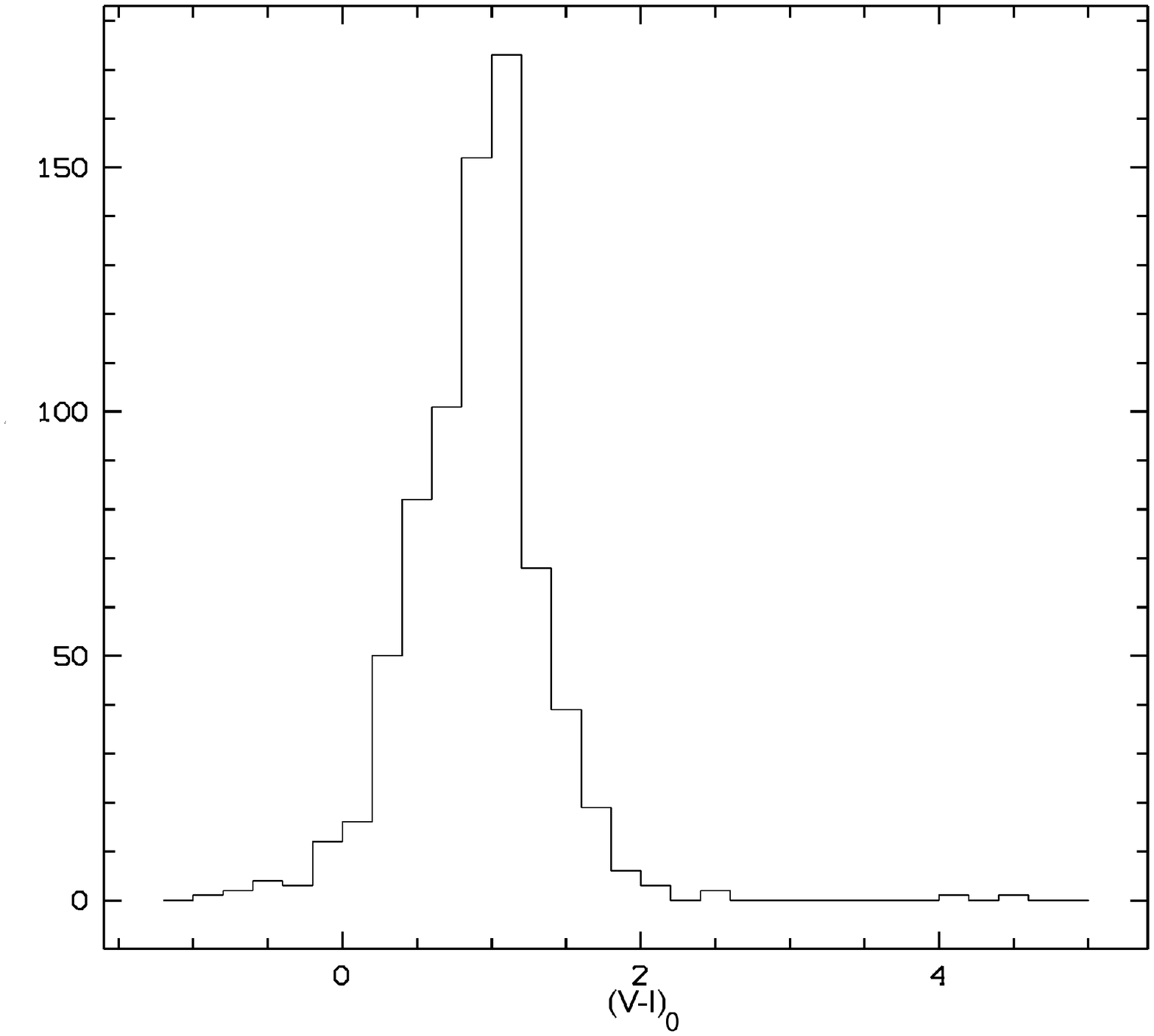}}
     \caption{The distribution of (V$-$I)$_{\rm{o}}$ for globular clusters (and candidates) from the RBC~V4, located in the \XLPt\ field.}
    \label{Fig:optGlC_agedist} 
\end{figure}

For each X-ray source which correlates with a globular cluster or globular cluster candidate in the optical, we investigated its integrated V-I colour and derived age estimates. Table~\ref{Tab:GlC_optprop} lists the name of the optical counterpart, its classification according to RBC V.4 \citep{2009A&A...508.1285G}, the age classification of \citet{2009AJ....137...94C}, the V magnitude and V-I colour given in RBC V.4, the dereddened V-I colour, and the age estimate derived by ourselves. 

The integrated V$-$I colours of the clusters can be found in RBC V.4 and can be used to provide estimates of the ages of the clusters, in conjunction with reddening values. We have adopted a reddening of E(B$-$V)$=\!0.10\pm0.03$, which is the average of the reddenings of all \m31\ clusters in \citet{2005AJ....129.2670R}. Using these values, we have derived (V$-$I)$_{\rm{o}}$ for our clusters. In most cases  (V$-$I)$_{\rm{o}}\!>\!1.0$ suggesting clusters older than $\simeq$2 Gyr according to \citet{2007AJ....133..290S}. 
The histogram in Fig.\,\ref{Fig:GlC_agedist} shows the distribution of (V$-$I)$_{\rm{o}}$ for our clusters, with the approximate age-ranges marked.

In general there is good agreement between the \citet{2009AJ....137...94C} and our age estimates. This result indicates that the great majority of the objects are indeed old globular clusters.

Figure~\ref{Fig:optGlC_agedist} shows the distribution of (V$-$I)$_{\rm{o}}$ for all confirmed and candidate globular clusters, listed in the RBC~V.4, which are located in the \XLPt\ field, and which have V as well as I magnitudes given. A comparison with Fig.\,\ref{Fig:GlC_agedist} again reveals that mainly counterparts of old globular clusters (age $\ga$2\,Gyr) are detected in X-rays.

\subsubsection{Comparing GlC and candidates in \textit{XMM-Newton}, \textit{Chandra} and \textit{ROSAT} catalogues}
\label{SubSub:comp_GlC}
The combined \ros\ PSPC catalogue (SHP97 and SHL2001) contains 33 sources classified as globular cluster counterparts. Of these sources one is located outside the field observed with \xmm.\@ Another two sources do not have counterparts in the \XLPt\ catalogue. The first one is [SHL2001]~232, which is not visible in any \xmm\ observation taken before December 2006 as was already reported in \citet{2004ApJ...616..821T}. The second source ([SHL2001]~231) correlates with B\,164 which is identified as a globular cluster in RBC~V3.5.\@ In addition [SHL2001]~231 is listed in PFH2005 as the counterpart of the source [PFH2005]~423. Due to the improved positional accuracy of the X-ray source in the \xmm\ observations, PFH2005 rejected the correlation with B\,164 and instead classified [PFH2005]~423 as a foreground star candidate.
  
Three \ros\ GlC candidates have more than one counterpart in the \XLPt\ catalogue. [SHL2001]~249 correlates with sources \num\ 1\,262 and \num\ 1\,267, where the latter is the X-ray counterpart of the globular cluster B\,185. [SHL2001]~254 correlates with sources \num\ 1\,289 and \num\ 1\,293, where the former is the X-ray counterpart of the globular cluster candidate mita\,311 \citep{1993PhDT........41M}. [SHL2001]~258 has a 1$\sigma$ positional error of 48\arcsec\ and thus correlates with sources \num\ 1\,297, \num\ 1\,305 and \num\ 1\,357.\footnote{In addition [SHL2001]~258 correlates with \num\ 1\,275, \num\ 1\,289 and \num\ 1\,293. However these sources have each an additional \ros\ counterpart.} The brightest of these three sources (\num\ 1\,305), which is actually located closest to the \ros\ position, correlates with the globular cluster candidate SK\,132C (RBC~V3.5).
  
Table \ref{Tab:ROSAT_GlC_tvar} gives the variability factors (Cols.~6, 8) and significance of variability (7, 9) for sources classified as GlC candidates in the \ros\ PSPC surveys. For most sources only low variability is detected. The two sources with the highest  variability factors found (\num\ 1262, \num\ 1293) belong to \ros\ sources with more than one \xmm\ counterpart. In these cases the \xmm\ sources that correlate with the same \ros\ source and the optical globular cluster source show much weaker variability. Interestingly, a few sources show low, but very significant variability. Among these sources is the Z-source identified in \citet[][\num\ 966]{2003A&A...411..553B} and two of the sources discussed in \citet[][\num\ 1\,057, \num\ 1\,692]{2008ApJ...689.1215B}.

\begin{figure*}
\sidecaption
\includegraphics[width=12cm]{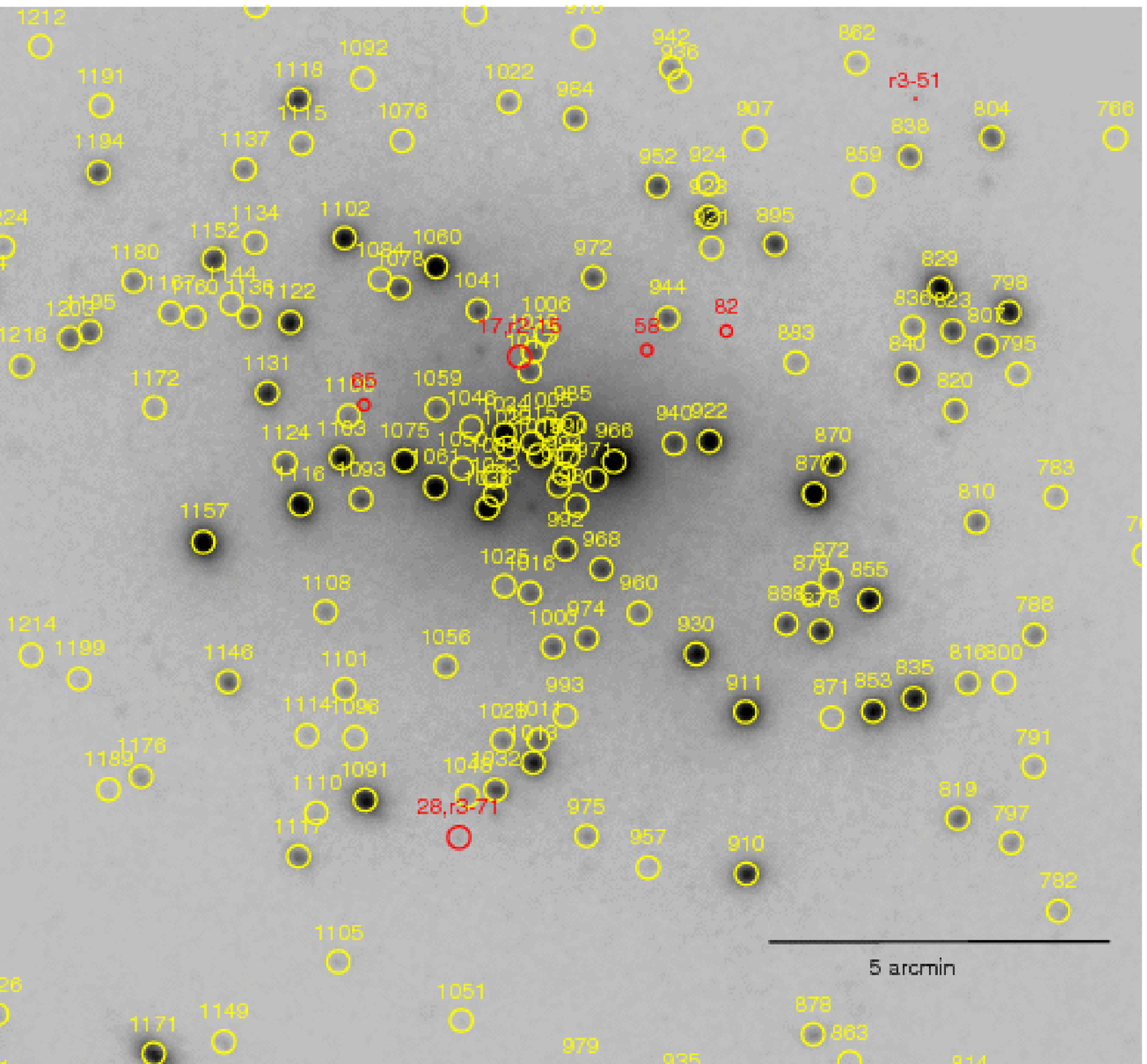}
\caption[Image of the central field of \m31\ over-plotted with the positions of six possible transient sources (red) and the sources of the \XLPt\ catalogue.]{Image of the central field of \m31\ over-plotted with the positions of six possible transient sources (red) and the sources of the \XLPt\ catalogue. Sources r2-15 and r3-71 are listed as sources \#17 and \#28, respectively, in \citet{2002ApJ...570..618D}. The three ``red" sources that are only marked with a number (\#58, \#65, \#82) are taken from \citet{2007A&A...468...49V}.}\label{Fig:posGlCtrans_pos}
\end{figure*}

\begin{table*}
\scriptsize
\begin{center}
\caption{Variability between \XLPt\ and \ros\ observations for sources classified as GlC candidates in the \ros\ PSPC surveys }
\begin{tabular}{rrrrrrrrrccc}
\hline\noalign{\smallskip}
\hline\noalign{\smallskip}
\multicolumn{1}{c}{SRC} & \multicolumn{1}{c}{SI$^{*}$} & \multicolumn{1}{c}{SII$^{*}$} & \multicolumn{1}{c}{XFLUX$^{+}$} & \multicolumn{1}{c}{EXFLUX$^{+}$} & \multicolumn{1}{c}{fv\_SI$^{\dagger}$} & \multicolumn{1}{c}{sv\_SI$^{\dagger}$} & \multicolumn{1}{c}{fv\_SII$^{\dagger}$} & \multicolumn{1}{c}{sv\_SII$^{\dagger}$} & \multicolumn{1}{c}{type} & \multicolumn{1}{c}{SIf$^{\ddagger}$}& \multicolumn{1}{c}{SIIf$^{\ddagger}$}\\
\multicolumn{1}{c}{(1)} & \multicolumn{1}{c}{(2)}& \multicolumn{1}{c}{(3)}& \multicolumn{1}{c}{(4)}& \multicolumn{1}{c}{(5)}& \multicolumn{1}{c}{(6)} & \multicolumn{1}{c}{(7)} & \multicolumn{1}{c}{(8)}& \multicolumn{1}{c}{(9)}& \multicolumn{1}{c}{(10)}& \multicolumn{1}{c}{(11)}& \multicolumn{1}{c}{(12)}\\
\noalign{\smallskip}\hline\noalign{\smallskip}
   383 &  73 &  68 & 1.45E-12 & 1.05E-14 &  1.26 & 11.58 &  1.23 &  9.04  & GlC        & *  & *  \\
   403 &  79 &  74 & 2.55E-14 & 2.36E-15 &  2.61 &  2.09 &  7.57 &  4.92  & Gal        &  &  \\
   422 &     &  76 & 2.01E-14 & 1.43E-15 &       &       &       &        & $<$hard$>$ &  &  \\
   694 & 122 & 113 & 1.52E-12 & 1.04E-14 &  1.38 & 10.56 &  1.26 &  7.46  & GlC        & * & *  \\
   793 & 138 & 136 & 4.76E-14 & 2.33E-15 &  1.10 &  0.55 &  1.02 &  0.14  & $<$Gal$>$  & *  &  \\
   841 & 150 & 147 & 1.40E-12 & 1.61E-14 &  1.77 & 19.81 &  2.11 &  26.37 & GlC        & * & --\\
   855 & 158 & 154 & 4.21E-13 & 3.70E-15 &  1.01 &  0.14 &  3.54 &  48.75 & GlC        & *  &  \\
   885 & 168 & 163 & 1.56E-14 & 1.60E-15 &  1.70 &  1.54 &       &        & GlC        &  &  \\
   923 & 175 & 175 & 1.47E-13 & 2.07E-15 &  1.87 &  7.68 &  1.30 &  3.44  & GlC        &  &  \\
   933 & 178 &     & 3.67E-14 & 1.64E-15 &  3.03 &  7.03 &       &        & GlC        &  &  \\
   947 & 180 & 179 & 3.24E-13 & 7.46E-15 &  2.43 & 12.78 &  2.29 &  12.16 & GlC        & * & --\\
   966 & 184 & 184 & 3.51E-12 & 9.21E-15 &  1.00 &  0.20 &  2.31 &  151.58 & XRB        &  &  \\
1\,057 & 205 & 199 & 2.67E-12 & 2.05E-14 &  1.72 & 26.26 &  1.79 &  28.48 & GlC       & * & *\\
1\,102 & 217 & 211 & 3.23E-13 & 2.93E-15 &  1.06 &  1.04 &  5.51 &  60.80 & GlC       & *  &  \\
1\,109 & 218 & 212 & 3.25E-13 & 9.08E-15 &  1.91 &  9.70 &  2.10 &   10.72 & GlC       & * & *\\
1\,118 & 222 & 216 & 1.16E-13 & 2.03E-15 &  1.46 &  3.63 &  1.89 &  7.46  & GlC       &   &  \\
1\,122 & 223 & 217 & 2.48E-13 & 2.72E-15 &  2.08 & 12.02 &  7.03 &  73.27 & GlC       &   &  \\
1\,157 & 228 & 223 & 7.59E-13 & 4.48E-15 &  1.06 &  1.68 &  1.08 &  2.75  & GlC       & *   & * \\
1\,171 & 229 & 227 & 4.68E-13 & 4.93E-15 &  1.82 & 12.92 &  1.75 &  12.19 & GlC       & * & *\\
1\,262 & 247 & 249 & 2.94E-14 & 3.10E-15 & 14.11 & 18.40 & 14.64 &  20.24  &           &   &  \\
1\,267 & 247 & 249 & 4.80E-13 & 4.57E-15 &  1.16 &  3.04 &  1.11 &  2.44  & GlC       & *   & *  \\
1\,289 & 250 & 254 & 2.88E-14 & 1.91E-15 &  1.16 &  0.57 &  1.98 &  3.15  & $<$GlC$>$ & *   &  \\
1\,293 & 250 & 254 & 6.70E-15 & 9.42E-16 &  3.73 &  2.80 &  8.53 &  5.72   & $<$AGN$>$ &   &  \\
1\,296 & 253 & 257 & 3.89E-14 & 1.58E-15 &  4.03 &  9.38 &  1.25 &  1.17  & GlC       &   &  \\
1\,297 & 252 & 258 & 5.59E-15 & 9.61E-16 &  4.46 &  2.69 &       &        & $<$hard$>$&   &  \\
1\,305 &     & 258 & 1.69E-14 & 9.87E-16 &       &       &       &        & $<$GlC$>$ &   &  \\
1\,340 & 261 & 266 & 6.07E-14 & 3.01E-15 &  1.77 &  4.20 &  1.30 &  1.64  & GlC       &   & *  \\
1\,357 &     & 258 & 7.04E-15 & 1.18E-15 &       &       &       &        & $<$hard$>$&   &  \\
1\,449 & 281 & 289 & 2.34E-14 & 1.00E-15 &  3.10 &  5.36 &  1.79 &  2.39  & fg Star   &   &  \\
1\,463 & 282 & 290 & 7.51E-13 & 8.38E-15 &  1.13 &  3.33 &  1.33 &  7.64  & GlC       & *  & *  \\
1\,634 & 302 & 316 & 7.70E-14 & 2.91E-15 &  3.14 &  5.60 &  1.89 &  4.33  & $<$hard$>$& * & *\\
1\,692 & 318 & 336 & 1.15E-12 & 2.00E-14 &  2.86 & 45.59 &  2.62 &  38.63 & GlC       & *  &  \\
1\,803 & 349 & 354 & 8.72E-13 & 9.17E-15 &  1.32 &  7.87 &  1.03 &  0.93  & GlC       & *  & *  \\
\noalign{\smallskip}\hline\noalign{\smallskip}\hline\noalign{\smallskip}
\end{tabular}
\label{Tab:ROSAT_GlC_tvar}
\end{center}
Notes:\\
$^{ *~}$: SI: SHP97, SII: SHL2001\\
$^{ +~}$: XID Flux and error in erg\,cm$^{-2}$\,s$^{-1}$\\
$^{ {\dagger}~}$: Variability factor and significance of variability, respectively, for comparisons of \xmm\ XID fluxes to \ros\ fluxes listed in SPH97 and SHL2001, respectively.\\ 
$^{ {\ddagger}~}$: An asterisk indicates that the XID flux is larger than the corresponding \ros\ flux. \ros\ count rates are converted to 0.2--4.5\,keV fluxes, using WebPIMMS and assuming a foreground absorption of \nh\,$=\!6.6$\hcm{20} and a photon index of $\Gamma\!=\!1.7$: ECF$_{\mr{SHP97}}\!=\!2.229\times$10$^{-14}$\,erg\,cm$^{-2}$\,cts$^{-1}$ and ECF$_{\mr{SHL2001}}\!=\!2.249\times$10$^{-14}$\,erg\,cm$^{-2}$\,cts$^{-1}$
\normalsize
\end{table*}

The 18 X-ray sources correlating with globular clusters which were found in the \ros\ HRI observations (PFJ93) were all re-detected in the \xmm\ data.
  
From the numerous studies of X-ray globular cluster counterparts in \m31\ based on \chandra\ observations \citep{2002ApJ...577..738K,2002ApJ...570..618D,2004ApJ...609..735W,2004ApJ...616..821T,2007A&A...468...49V}, only eight sources are undetected in the present study. One of them ([TP2004] 1) is located far outside the field of \m31\ covered by the Deep \xmm\ Survey. 
The transient nature of [TP2004] 35, and the fact that it is not observed in any \xmm\ observation taken before December 2006 was mentioned in Sect.\,\ref{SubSec:Chcat}. 
The six remaining sources (r2-15, r3-51, r3-71, [VG2007]~58, [VG2007]~65, [VG2007]~82) are located in the central area of \m31\ and are also not reported in PFH2005. Figure~\ref{Fig:posGlCtrans_pos} shows the position of these six sources (in red) and the sources of the \XLPt\ catalogue (in yellow). If the brightness of the six sources had not changed between the \chandra\ and \xmm\ observations, they would be in principle bright enough to be detected by \xmm\ in the merged observations of the central field, which have in total an exposure $\ga$ 100\,ks. Two sources (r2-15 and [VG2007]~65) are located next to sources detected by \xmm. Source r2-15 is located within 13\arcsec\ of \num\ 1\,012 and within 17\arcsec\ of \num\ 1\,017 and has -- in the \chandra\ observation -- a similar luminosity to both \xmm\ sources. The distance between \num\ 1\,012 and \num\ 1\,017 is 17\arcsec\, and within 20\arcsec\ of \num\ 1\,012, \xmm\ detected source \num\ 1\,006, which is about a factor 4.6 fainter than \num\ 1\,012.\@ Therefore, when in a bright state, source r2-15 should be detectable with \xmm. Source [VG2007]~65 is located within 17\arcsec of \num\ 1\,100, which is at least 3.5 times brighter than [VG2007]~65. This may complicate the detection of [VG2007]~65 with \xmm. The variability of [VG2007]~58, [VG2007]~65, and [VG2007]~82 is supported by the fact that these three sources were not detected in any \chandra\ study, prior to \citet{2007A&A...468...49V}. Hence, these six sources are likely to be at least highly variable or even transient.

Several sources identified with globular clusters in previous studies have counterparts in the \XLPt\ catalogue but are not classified as GlC sources by us.  Source \num\ 403 ([SHL2001]~74) correlates with B\,007, which is now identified as a background galaxy \citep[][RBC~V3.5]{2009AJ....137...94C,2007AJ....134..706K}. Sources \num\ 793 ([SHL2001]~136, s1-12) and \num\ 796 (s1-11) are the X-ray counterparts of B\,042D and  B\,044D, respectively, which are also suggested as background objects by \citet{2009AJ....137...94C}. Source \num\ 948 (s1-83) correlates with B\,063D, which is listed as a globular cluster candidate in RBC~V3.5, but might be a foreground star \citep{2009AJ....137...94C}. Due to this ambiguity in classification we classified the source as $<$hard$>$. Source \num\ 966 correlates with [SHL2001]~184, which was classified as the counterpart of the globular cluster NB\,21 (RBC~V3.5) in the \ros\ PSPC survey (SHL2001). In addition, source \num\ 966 also correlates with the \chandra\ source r2-26 \citep{2002ApJ...577..738K}. Due to the much better spatial resolution of \chandra\ compared to \ros, \citet{2002ApJ...577..738K} showed that source r2-26 does not correlate with the globular cluster NB\,21. \citet{2003A&A...411..553B} identified this source as the first Z-source in \m31. The nature of source \num\ 1\,078 is unclear as RBC~V3.5 reported that source to be a foreground star, while \citet{2009AJ....137...94C} classified it as an old globular cluster. Due to this ambiguity in the classification and due to the fact that source \num\ 1\,078 is resolved into two \chandra\ sources (r2-9, r2-10), we decided to classify the source as $<$hard$>$. Due to the transient nature \citep{2002ApJ...577..738K,2006ApJ...643..356W} and the ambiguous classifications reported by RBC~V3.5 (GlC) and \citet[][H{\small II} region]{2009AJ....137...94C}, we adopt the classification of PFH2005 ($<$XRB$>$) for source \num\ 1\,152.\@ SBK2009 classified the source correlating with source \num\ 1\,293 as a globular cluster candidate. We are not able to confirm this classification, as none of the globular cluster catalogues used, contains an entry at the position of source \num\ 1\,293. Instead we found a radio counterpart in the catalogues of \citet{2005ApJS..159..242G}, \citet{1990ApJS...72..761B} and NVSS.\@ We therefore classified the source as an AGN candidate, as was also done in PFH2005.

 For source \num\ 1\,449 ([SHL2001]~289) the situation is more complicated. SHL2001 report [MA94a]~380 as the globular cluster correlating with this X-ray source. Based on the same reference, \citet{2005PASP..117.1236F} included the optical source in their statistical study of globular cluster candidates. However, the paper with the acronym [MA94a] is not available. An intensive literature search of the papers by Magnier did not reveal any work relating to globular clusters in \m31, apart from \citet{1993PhDT........41M} which is cited in \citet{2005PASP..117.1236F} as ``MIT".\@ In addition the source is not included in any other globular cluster catalogues listed in Sect.\,\ref{Sec:CrossCorr_Tech}. In the X-ray studies of \citet{2004ApJ...609..735W} and PFH2005 and in \citet{1992A&AS...96..379M} the source is classified as a foreground star (candidate). Hence, we also classified source \num\ 1\,449 as a foreground star candidate, but suggest optical follow-up observations of the source to clarify its true nature.

 A similar case is source \num\ 422 ([SHL2001]~76), which is classified as a globular cluster by SHL2001, based on a correlation with [MA94a]~16. Here again the source is not listed in any of the globular cluster catalogues used. We found one correlation of source \num\ 422 with an object in the USNO-B1 catalogue, which has no B2 and R2 magnitude. Two faint sources (V$>\!22.5$\,mag) of the LGGS catalogue are located within the X-ray positional error circle. Thus source \num\ 422 is classified as $<$hard$>$.  While RBC~V3.5 classified the optical counterpart of source \num\ 1\,634 ([SHL2001]~316) as a globular cluster candidate, \citet{2009AJ....137...94C} regard SK\,182C as being a source of unknown nature. Therefore we decided to classify source \num\ 1634 as $<$hard$>$. 

\section{Conclusions}
\label{Sec:Concl}
This paper presents the analysis of a large and deep \xmm\ survey of the bright Local Group SA(s)b galaxy \m31. The survey observations were taken between June 2006 and February 2008. Together with re-analysed archival observations, they provide for the first time  full coverage of the M31 $\mr{D}_{25}$ ellipse down to a 0.2\,--\,4.5\,keV luminosity of $\sim$\oergs{35}.

The analysis of combined and individual observations allowed the study of faint persistent sources as well as brighter variable sources.

The source catalogue of the Large \xmm\ Survey of \m31\ contains 1\,897 sources in total, of which 914 sources were detected for the first time in X-rays. The XID source luminosities range from $\sim$4.4\ergs{34} to 2.7\ergs{38}. The previously found differences in the spatial distribution of bright ($\ga$\oergs{37}) sources between the northern and southern disc could not be confirmed.
The identification and classification of the sources was based on properties in the X-ray wavelength regime: hardness ratios, extent and temporal variability. In addition, information obtained from cross correlations with \m31\ catalogues in the radio, infra-red, optical and X-ray wavelength regimes were used.

The source catalogue contains 12 sources with spatial extent between 6\,\farcs2 and 23\,\farcs0. From spectral investigation and comparison with optical images, five sources were classified as galaxy cluster candidates. 

317 out of 1\,407 examined sources showed long term variability 
with a significance $>$3$\sigma$ between the \xmm\ observations. These include 173 sources in the disc that were not covered in the study of the central field (SPH2008). Three sources located in the outskirts of the central field could not have been detected as variable in the study presented in SPH2008, as they only showed variability with a significance $>$3$\sigma$ between the archival and the ``Large Project" observations. For 69 sources the flux varied by more than a factor of five between XMM-Newton observations; ten of these varied by a factor $>$100.

Discrepancies in source detection between the Large \xmm\ Survey catalogue and previous \xmm\ catalogues could be explained by different search strategies, and differences in the processing of the data, in the parameter settings of the detection runs and in the software versions used. Correlations with previous \chandra\ studies showed that those sources not detected in this study are strongly time variable, transient, or unresolved. This is particularly true for sources located close to the centre of \m31, 
where \chandra's higher spatial resolution resolves more sources. Some of the undetected sources from previous \ros\ studies were located outside the field covered with \xmm. However, there were several sources detected by \ros\ that had a \ros\ detection likelihood larger than 15. If these sources were still in a bright state they should have been detected with \xmm.\@ Thus, the fact that these sources are not detected with \xmm\ implies that they are transient or at least highly variable sources. On the other hand 242 $<$hard$>$ \xmm\ sources were found with XID fluxes larger than \oergcm{-14}, which were not detected with \ros. 

To study the properties of the different source populations of \m31, it was necessary to separate foreground stars (40 plus 223 candidates) and background sources (11 AGN and 49 candidates, 4 galaxies and 19 candidates, 1 galaxy cluster and 5 candidates) from the sources of \m31. 1\,247 sources could only be classified as $<$hard$>$, while 123 sources remained without identification or classification. The majority (about two-thirds, see Stiele et al. 2011 in preparation) of sources classified as $<$hard$>$ are expected to be background objects, especially AGN.

The catalogue of the Large \xmm\ survey of \m31\ contains 30 SSS candidates, with unabsorbed 0.2--1.0\,keV luminosities between 2.4\ergs{35} and 2.8\ergs{37}. SSSs are concentrated to the centre of \m31, which can be explained 
by their correlation with optical novae, and by the overall spatial distribution of \m31\ late type stars (\ie\ enhanced density towards the centre). Of the 14 identifications made of optical novae, four were presented in more detail. 

The 25 identified and 31 classified SNRs had XID luminosities between 1.1\ergs{35} and 4.3$\times$10$^{36}$ erg\,s$^{-1}$. Three of the 25 identified SNRs were detected for the first time in X-rays. For one SNR the \ros\ classification can be confirmed. Six of the SNR candidates were selected from correlations with sources in SNR catalogues from the literature. As these six sources had rather ``hard" hardness ratios they are good candidates for ``plerions". An investigation of the spatial distribution showed that most SNRs and candidates are located in regions of enhanced star formation, especially along the 10\,kpc dust ring in \m31. 
This connection between SNRs and star forming regions, implies that most of the remnants are from type II supernovae. Most of the SNR classifications from previous studies have been confirmed. However, in five cases these classifications are doubtful.

The population of ``hard" \m31\ sources mainly consists of XRBs. These rather bright sources (XID luminosity range: 1.0\ergs{36} to 2.7\ergs{38}) were selected from their transient nature or strong long term variability (variability factor $>$10; 10 identified, 26 classified sources). The spectral properties of three transient sources were presented in more detail.

A sub-class of LMXBs is located in globular clusters. They were selected from correlations with optical sources included in globular cluster catalogues (36 identified, 16 classified sources). The XID luminosity of GlCs ranges from 2.3\ergs{35} to 1.0\ergs{38}. The spatial distribution of this source class also showed an enhanced concentration to the centre of \m31.

From optical and X-ray colour-colour diagrams possible HMXB candidates were selected. If the sources were bright enough, an absorbed power-law model was fitted to the source spectra. Two of the candidates had a photon index consistent with the photon index range of NS HMXBs. Hence these two sources were suggested as new HMXB candidates. 

Follow-up studies in the optical as well as in radio are in progress or are planned. They will allow us to increase the number of identified sources and help us to classify or identify sources which can up to now only be classified as $<$hard$>$ or are without any classification.

This work focused on the overall properties of the source population of individual classes and gave us deeper insights into the long-term variability, spatial and flux distribution of the sources in the field of \m31\ and thus helped us to improve our understanding of the X-ray source population of \m31.

\begin{acknowledgements}
This publication makes use of the USNOFS Image and Catalogue Archive
operated by the United States Naval Observatory, Flagstaff Station
(http://www.nofs.navy.mil/data/fchpix/), 
of data products from the Two Micron All Sky Survey, 
which is a joint project of the University of Massachusetts and the Infrared 
Processing and Analysis Center/California Institute of Technology, funded by 
the National Aeronautics and Space Administration and the National Science 
Foundation, of the SIMBAD database,
operated at CDS, Strasbourg, France, 
and of the NASA/IPAC Extragalactic Database (NED) 
which is operated by the Jet Propulsion Laboratory, California 
Institute of Technology, under contract
with the National Aeronautics and Space Administration.
The XMM-Newton project is supported by the
Bundesministerium f\"ur Wirtschaft und Technologie/Deutsches Zentrum
f\"ur Luft- und Raumfahrt (BMWI/DLR, FKZ 50 OX 0001) and the Max-Planck
Society. HS acknowledges support by the
Bundesministerium f\"ur Wirtschaft und Technologie/Deutsches Zentrum
f\"ur Luft- und Raumfahrt (BMWI/DLR, FKZ 50 OR 0405).
\end{acknowledgements}

\bibliographystyle{aa}
\bibliography{papers2,/Users/apple/work/papers/my1990,/Users/apple/work/papers/my2000,/Users/apple/work/papers/my2001,/Users/apple/work/papers/catalog,/Users/apple/work/papers/my2007,/Users/apple/work/papers/my2008,/Users/apple/work/papers/my2010}

\longtab{9}{\begin{scriptsize}
\centering
\begin{longtable}{lrrrrcl}
\caption{Variable sources with flux variability larger than 5, ordered by variability.\label{Tab:varlist}}\\
\hline\noalign{\smallskip}
\hline\noalign{\smallskip}
\multicolumn{1}{c}{Source} & 
\multicolumn{1}{c}{fvar} &
\multicolumn{1}{c}{svar} &
\multicolumn{1}{c}{fmax$^{\ddagger}$} &
\multicolumn{1}{c}{efmax$^{\ddagger}$} &
\multicolumn{1}{c}{class$^{+}$} &
\multicolumn{1}{c}{Comment$^{\dagger}$}  \\ 
\noalign{\smallskip}\hline\noalign{\smallskip}
\endfirsthead
\caption[]{continued.}\\
\hline\noalign{\smallskip}
\hline\noalign{\smallskip}
\multicolumn{1}{c}{Source} & 
\multicolumn{1}{c}{fvar} &
\multicolumn{1}{c}{svar} &
\multicolumn{1}{c}{fmax$^{\ddagger}$} &
\multicolumn{1}{c}{efmax$^{\ddagger}$} &
\multicolumn{1}{c}{class$^{+}$} &
\multicolumn{1}{c}{Comment$^{\dagger}$}  \\ 
\noalign{\smallskip}\hline\noalign{\smallskip}
\endhead
\noalign{\smallskip}
\hline
\noalign{\smallskip}
\endfoot
523    &  692.64 & 63.33 & 106.61 & 1.68 & XRB      & \\
1\,032   &  660.61 & 54.21 & 33.89 & 0.62 & $<$GlC$>$    & 1(t, 92.2), 11, 16, 27(831.10)\\
57     &  644.03 & 96.796 & 147.76 & 1.52 & XRB      & \\
1\,131   &  558.48 & 80.18 & 38.93   & 0.48 & $<$XRB$>$    & 1(t, 954.2), 2(t, 2163), 19, 20(t), 22, 26, 27(624.05)\\
944    &  236.47  & 44.74 & 24.24 & 0.54 & XRB      &  1(t, 370.5), 14(t, BH-XRN), 20(t), 21(v,t), 26, 27(353.67)\\
705    &  185.02  & 39.85 & 32.52 & 0.81 & $<$XRB$>$    & 3(t), 27(178.79)\\
1\,007   &  161.84 & 42.33 & 44.30 & 1.04 & $<$XRB$>$    & 1, 12, 13, 26, 27(12.96)\\
1\,195   &  144.47 & 36.95 & 28.51 & 0.76 & $<$SSS$>$    & 2(t, 694), 14(v,t), 20(t), 26, 27(82.68)\\
1\,152   &  123.95 & 55.31 & 31.86 & 0.56   & $<$XRB$>$    & 1(t, 96.3), 2(t, 107), 9, 11(v), 12, 13(v), 19, 21(v,t), 26,\\
& & & & & &  27(131.73)\\
788    &  117.83 & 20.99 & 12.27 & 0.58 & $<$XRB$>$    & 1(t, 20.8), 2(t, 93), 11, 14(t), 20(t), 26, 27(97.89)\\
1\,136   &  87.84 & 25.00 & 9.79 & 0.39 & $<$XRB$>$    & 1(t, 46.1), 2(t, 155), 19, 20(t), 23, 26, 27(97.65)\\
934    &  84.94 & 28.35 & 21.26 & 0.90 & $<$SSS$>$    & \\
949    &  80.16 & 25.55 & 17.75 & 0.68 & $<$XRB$>$    & 2(t, 13), 20, 25, 26\\
1\,194   &  75.22 & 39.69 & 12.79 & 0.31 & $<$SSS$>$    & 1, 2(t, 96), 9, 12, 13(v), 14(v), 19(v), 21(v), 26, 27(85.93)\\
1\,416   &  72.25 & 21.00 & 10.51 & 0.49 & $<$SSS$>$    & 26, 28\\
1\,084   &  59.11 & 8.70 & 6.64 & 0.75 & $<$XRB$>$    & 1(t, 79.0), 2(t, 260), 14(t), 18, 20(t), 21(v,t), 26, 27(57.41)\\
985    &  51.66 & 80.00 & 80.46 & 0.97 & XRB         & 3(t), 5(t, LMXB), 9, 21(v), 27(76.27)\\
1\,017   &  49.85 & 20.73 & 9.64  & 0.45 & $<$XRB$>$       & 1(t, 99.5), 2(t, 158), 21(v,t), 26, 27(38.07)\\
990    &  48.90 & 61.77 & 82.16 & 1.28 & XRB         & 1(t, 468.8), 2(t, 285), 14(v,t, BH-XRT), 18(t), 21(v,t), 26, 27(201.71)\\
1\,059   &  48.10 & 23.64 & 9.16 & 0.37 & $<$XRB$>$       & 1(t, 64.6), 27(39.69)\\
698    &  41.20 & 18.73 & 9.33 & 1.37 & $<$XRB$>$       & 13, 23, 26, 27(44.67)\\
884    &  32.88 & 9.59 & 2.84 & 0.61 & $<$XRB$>$       & 1, 27(24.95)\\
1\,153   &  31.06 & 17.96 & 4.22 & 0.22 & XRB         & 1(t), 3(t), 6(t, LMXB), 27(30.50)\\
1\,000   &  26.86 & 20.17 & 5.75 & 0.48 &             & 1(t), 19, 21, 23, 26, 27(34.47)\\
1\,024   &  25.66 & 78.46 & 144.02 & 1.89  & $<$XRB$>$       & 1, 9, 11, 12, 13, 19, 21(v), 26, 27(34.02)\\
1\,177   &  24.00 & 19.31 & 4.59 & 0.22 & XRB         & 3(t), 27(27.06)\\
939    &  22.48  & 10.13 & 2.20 & 0.20 & $<$XRB$>$       & 1(t, 65.2), 27(28.02)\\
960    &  21.56 & 9.497 & 5.51 & 0.55 &  $<$fg Star$>$  & 1, 12, 13, 19, 21, 23, 26, 27(21.97)\\
1\,180   &  16.48 & 17.16 & 7.53 & 0.40 & $<$XRB$>$       & 1, 12, 19, 21(v), 26, 27(13.01)\\
378    &  16.21 & 9.39 & 6.83 & 0.98 & $<$XRB$>$       & 25, 26\\
714    &  14.89 & 13.77 & 10.35 & 0.68 & $<$fg Star$>$      & \\
748    &  14.45 & 13.86 & 3.62 & 0.23  & $<$SSS$>$       & 25, 26, 27(14.18)\\
92     &  14.23 & 5.73 & 1.86 & 0.30 & $<$SSS$>$       & \\
814    &  13.66 & 7.98 & 4.94 & 0.56 & $<$XRB$>$       & 1, 21(v), 25, 26, 27(15.76)\\
1\,006   &  11.30 & 14.34  & 1.44 & 0.08 & $<$SSS$>$       & 2(t, 51), 19, 23, 26, 27(10.70)\\
542    &  10.90 & 3.26  & 2.50 & 0.69 & $<$XRB$>$      & \\
1\,016   &  10.46  & 20.66 & 6.37 & 0.26 & $<$XRB$>$      & 1, 11 ,12, 19, 21, 26, 27(11.69)\\
1\,144   &  10.23 & 10.53 & 1.54 & 0.12 & $<$SSS$>$       & 2(t, 38), 26, 27(10.05)\\
1\,034   &  10.11  & 28.93 & 9.81 & 0.28 & $<$XRB$>$       & 1, 19, 21, 23, 27(10.24)\\
1\,099   &  9.95 & 4.50 & 5.12 & 1.01 & $<$hard$>$      & 25, 26\\
904    &  9.16 & 8.65 & 8.55 & 0.84 & $<$hard$>$      & 12, 26\\
872    &  9.13 & 17.16 & 7.31 & 0.38 & $<$GlC$>$       & 1, 7, 19 , 21(v), 26, 27(7.48)\\
1\,422   &  9.06  & 2.99 & 1.87 & 0.55 & $<$hard$>$      & \\
1\,183   &  7.66 & 11.93 & 8.43   & 0.57 & $<$hard$>$      & 12, 13, 26\\
422    &  7.23 & 9.65 & 2.10 & 0.16 & $<$hard$>$      & 13\\
823    &  7.16  & 28.29 & 20.25 & 1.30 & $<$GlC$>$       & 1, 4, 11, 14(v), 15, 19, 21(v), 26, 27(6.45)\\
1\,250   &  6.94 & 2.80 & 0.89 & 0.27  & $<$SSS$>$       & \\
398    &  6.79 & 2.93 & 1.79 & 0.52 & $<$hard$>$      & 13, 26\\
1\,266   &  6.67 & 7.54 & 2.37 & 0.24 & GlC        & 4, 26, 27(8.27)\\
975    &  6.58 & 7.98 & 1.76  & 0.35 & GlC        & 1, 4, 13, 15, 19, 20, 21(v), 26\\
522    &  6.47 & 6.62 & 2.21 & 0.26 & $<$hard$>$     & \\
805    &  6.22 & 3.66 & 0.96 & 0.21 & $<$hard$>$     & \\
916    &  6.21 & 3.73 & 2.00 & 0.44 & $<$hard$>$     & \\
862    &  6.17 & 5.33 & 1.38 & 0.20 & GlC        & 1, 4, 15, 21, 26, 27(5.35)\\
1\,124   &  6.04 & 12.93 & 5.73 & 0.23  & $<$GlC$>$      & 1, 19, 20, 21(v), 26\\
430    &  5.98 & 4.95 & 1.01 & 0.16 & $<$hard$>$     & \\
1\,825   &  5.57 & 6.11 & 2.97 & 0.39 &  fg Star   & 13, 26\\
1\,494   &  5.55 & 4.06 & 0.95 & 0.18 & $<$hard$>$     & 26\\
624    &  5.54  & 7.58 & 1.49 & 0.13 &  $<$fg Star$>$ & \\
1\,167   &  5.44 & 7.69 & 3.22 & 0.57 & $<$hard$>$     & 1, 19, 20, 21(v), 26\\
1\,450   &  5.44 & 6.51 & 2.24 & 0.24 & $<$hard$>$     & \\
66     &  5.39 & 3.80 & 2.31 & 0.49 & $<$hard$>$     & \\
1\,655   &  5.34 & 4.72  & 1.50 & 0.24 & $<$hard$>$     & \\
964    &  5.30 & 6.36 & 1.09 & 0.12 & $<$hard$>$     & \\
226    &  5.28 & 3.29 & 0.56 & 0.13 & $<$hard$>$     & \\
244    &  5.27 & 9.08 & 3.33 & 0.28 & $<$hard$>$     & 13, 25, 26\\
1\,361   &  5.21  & 3.12 & 1.36 & 0.34 & $<$hard$>$     & 27\\
933    &  5.19 & 4.99 & 8.73 & 1.30 & GlC        & 4, 11, 12, 25, 26\\
1\,366   &  5.13  & 3.02 & 1.05 & 0.27 & $<$hard$>$     & \\
\end{longtable}
\end{scriptsize}
\scriptsize
Notes:\\
$^{ {\ddagger}~}$: maximum flux and error in units of 1\ergcm{-14} or maximum luminosity and error in units of 7.3\ergs{35}\\
$^{ {+}~}$: class according to Table~\ref{Tab:class} \\
$^{ {\dagger}~}$: 1: \citet{2007A&A...468...49V}, 2: \citet{2006ApJ...643..356W}, 3: \citet{2006ApJ...645..277T}, 4: \citet{2004ApJ...616..821T}, 5: \citet{2006ApJ...637..479W}, 6: \citet{2005ApJ...632.1086W}, 7: \citet{2005A&A...430L..45P}, 8: \citet{2003ApJ...590L..21K}, 9: \citet{1991ApJ...382...82T}, 10: \citet{1990ApJ...356..119C}, 11: \citet{1993ApJ...410..615P}, 12: \citet{1997A&A...317..328S}, 13: \citet{2001A&A...373...63S}, 14: \citet{2001A&A...378..800O}, 15: \citet{2002ApJ...570..618D}, 16: \citet{2005PASP..117.1236F}, 17: \citet{2007A&A...465..375P}, 18: \citet{2000ApJ...537L..23G}, 19: \citet{2002ApJ...578..114K}, 20: \citet{2004ApJ...609..735W}, 21: \citet{2002ApJ...577..738K}, 22: \citet{2005ApJ...620..723W}, 23: \citet{2004ApJ...610..247D}, 24: \citet{2003A&A...411..553B}, 25: \citet{2009A&A...495..733S}, 26: \citet{2005A&A...434..483P}, 27: \citet{2008A&A...480..599S}, 28: \citet{2002IAUC.7798....2T} ; t: transient, v: variable, sv: spectrally variable, r: recurrent, d: dipping, z: Z-source candidate; BH: black hole, XRN: X-ray nova, XRT: X-ray transient, LMXB: low mass X-ray binary, NS: neutron star; numbers indicate the variability given by the corresponding paper 
}

\Online
\begin{appendix}
\label{App:XID_Images}
\begin{figure*}
\resizebox{\hsize}{!}{\includegraphics[clip]{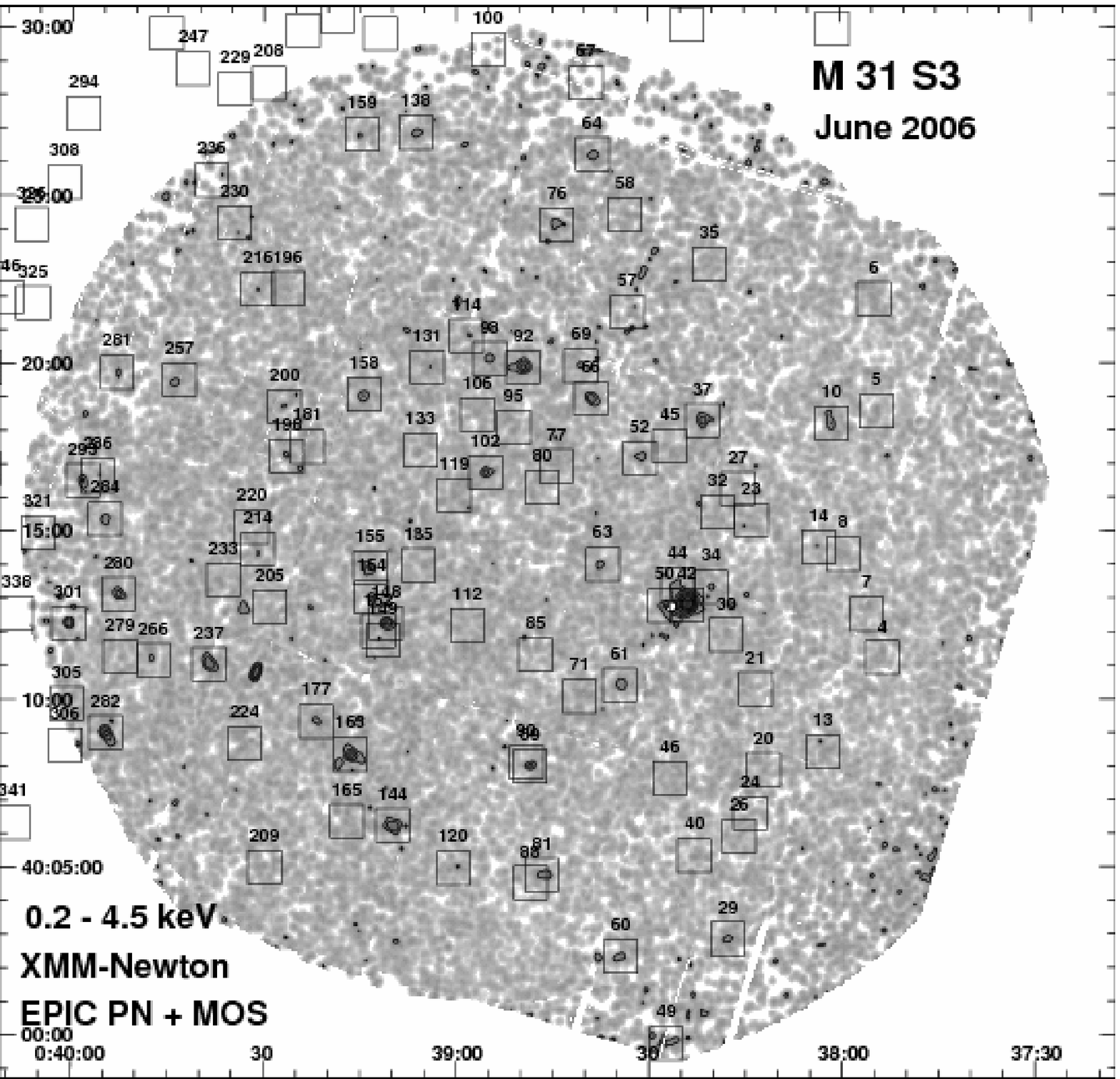}\hskip0.2cm\includegraphics[clip]{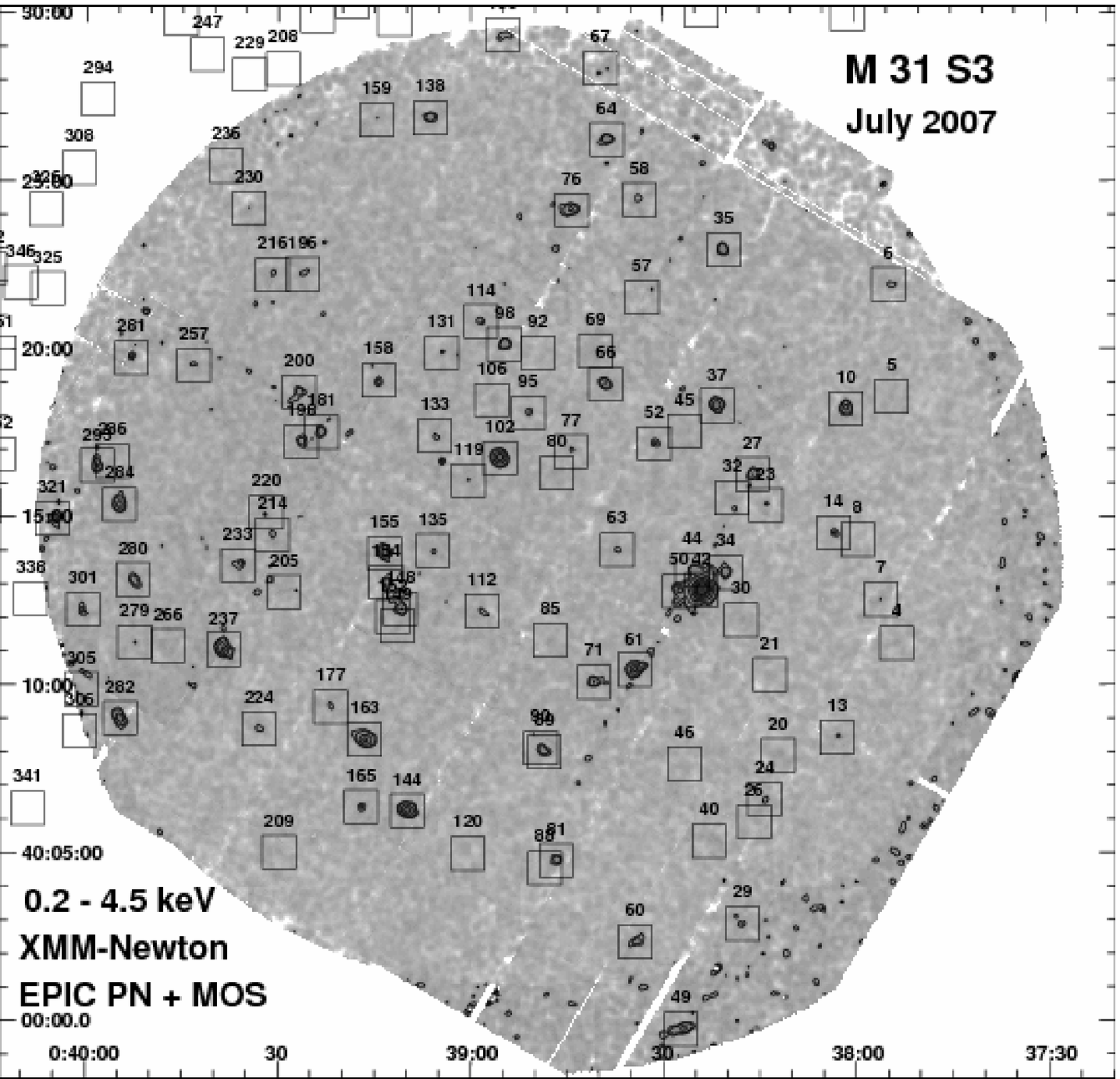}}
\resizebox{\hsize}{!}{\includegraphics[clip]{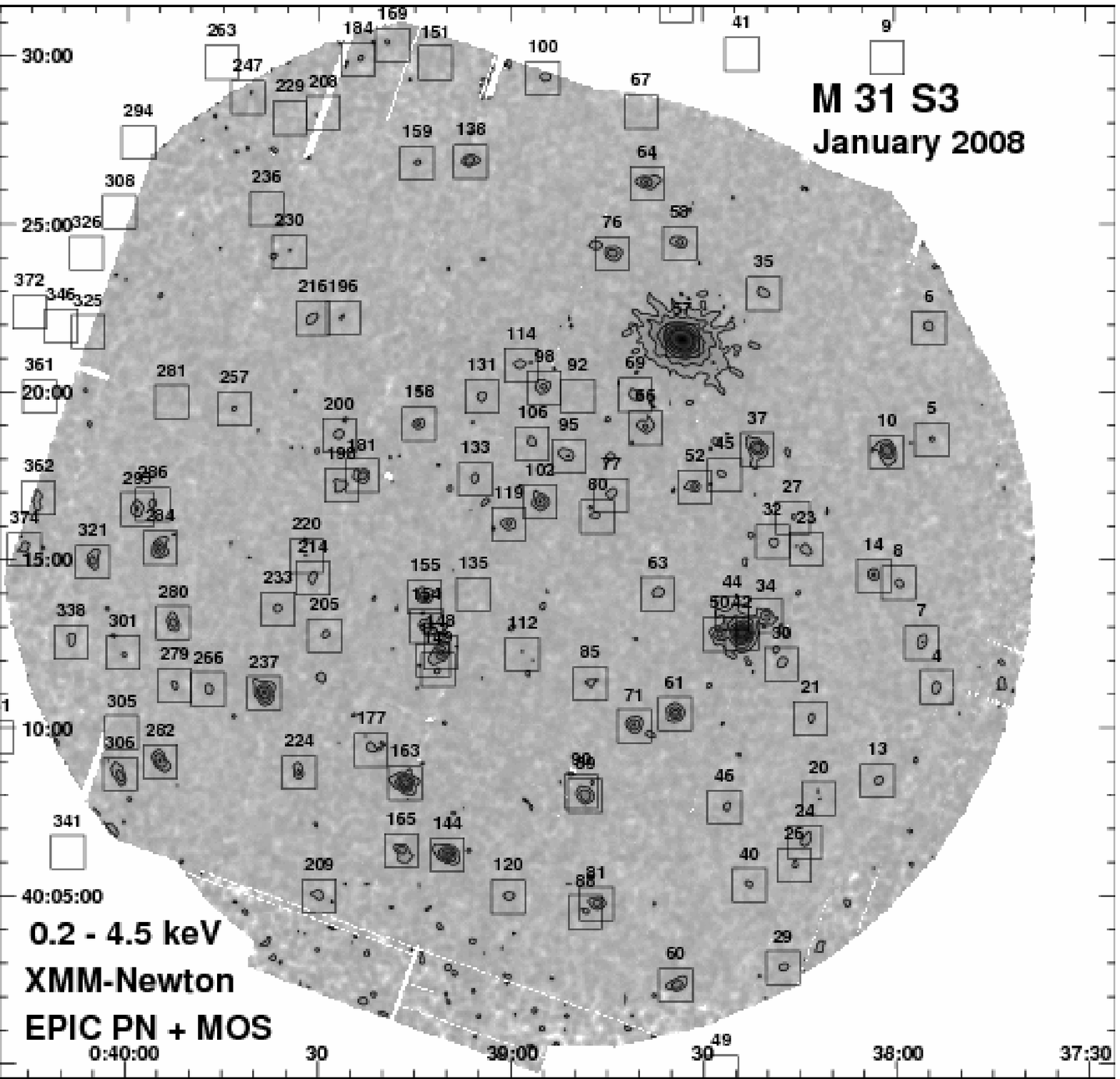}\hskip0.2cm\includegraphics[clip]{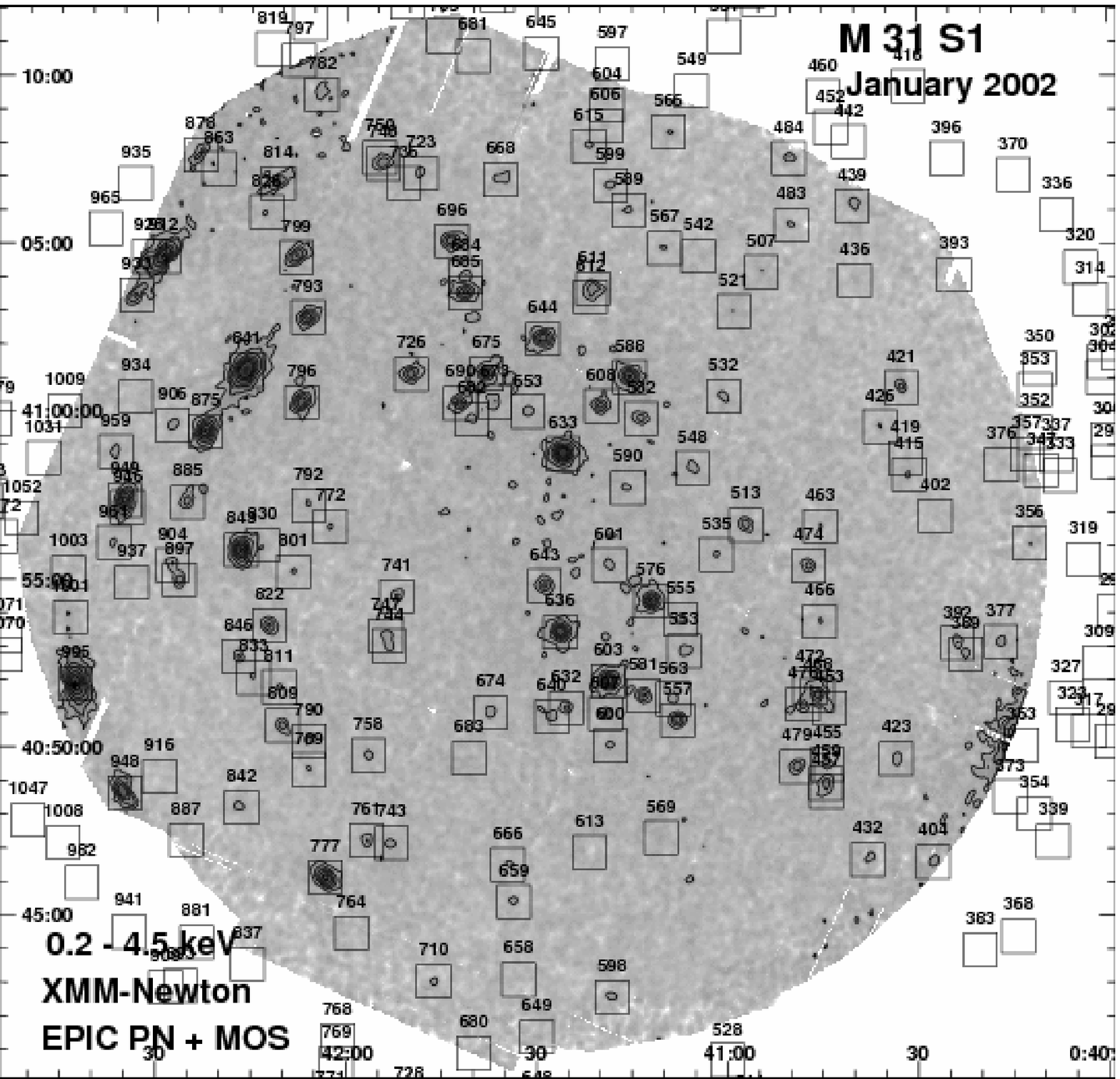}}
     \caption[Logarithmically scaled \xmm\ EPIC low background images
     integrated in 2\arcsec$\times$2\arcsec\ pixels of the \m31\ observations combining PN
     and MOS\,1 and MOS\,2 cameras in the (0.2--4.5) keV XID band.]{Logarithmically scaled \xmm\ EPIC low background images
     integrated in 2\arcsec$\times$2\arcsec\ pixels of the \m31\ observations combining PN
     and MOS\,1 and MOS\,2 cameras in the (0.2--4.5) keV XID band.  
     The data are
     smoothed with a 2D-Gaussian of FWHM 10\arcsec, which corresponds to the point spread function in the centre area. The images are corrected for unvignetted exposures. Contours in units of $10^{-6}$\,ct\,s$^{-1}$\,pix$^{-1}$ including a factor of two smoothing are at $(8, 16, 32, 64, 128)$ in the upper left panel, at $(6, 8, 16, 32, 64, 128)$ in the upper right panel, and at $(4, 8, 16, 32, 64, 128)$ in both lower panels. Sources from the \XLPt\ catalogue are marked as 60\arcsec$\times$60\arcsec\ squares. 
}
    \label{Ima:singleIma} 
\end{figure*}

\begin{figure*}
\addtocounter{figure}{-1}
\resizebox{\hsize}{!}{\includegraphics[clip]{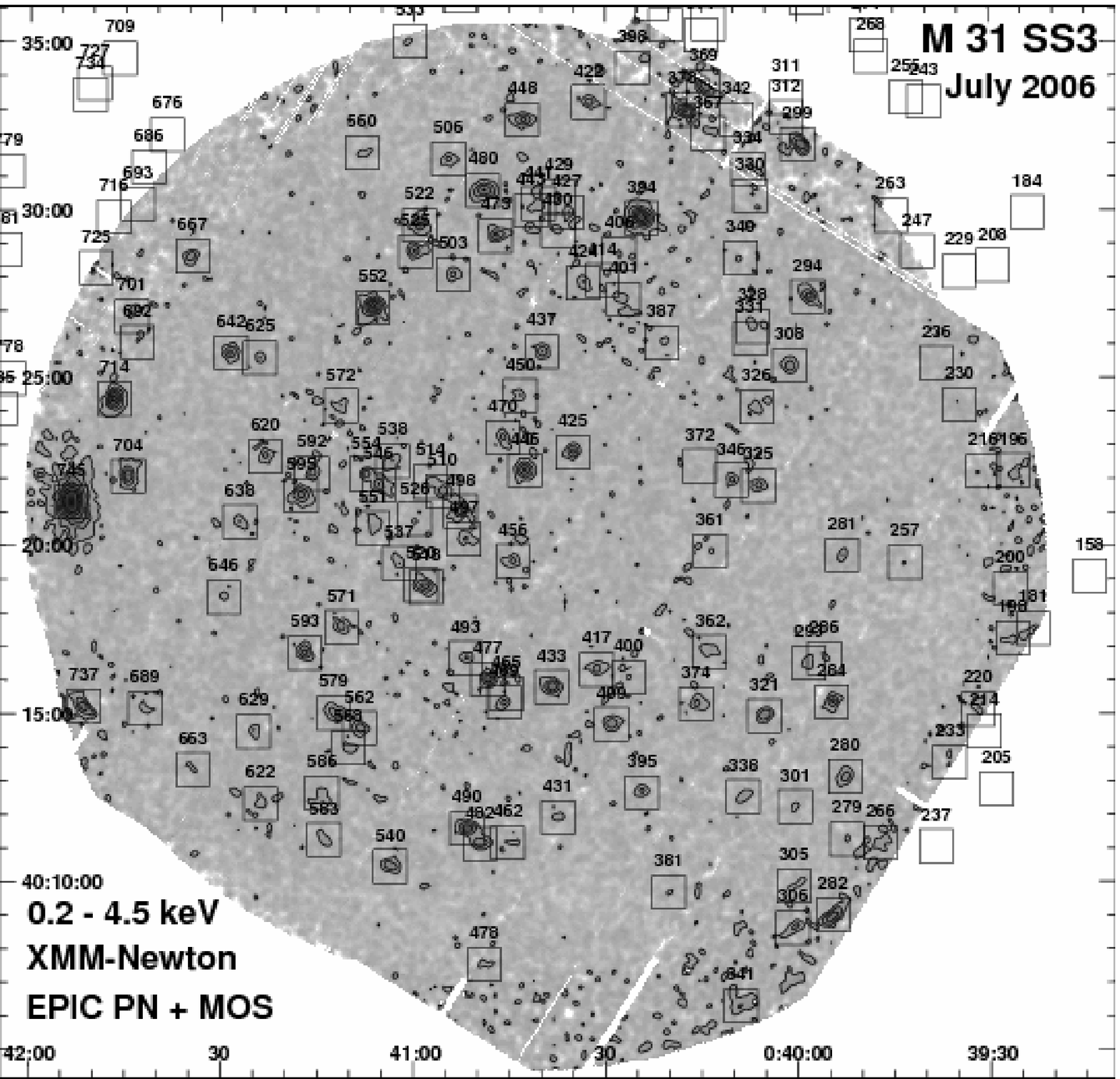}\hskip0.2cm\includegraphics[clip]{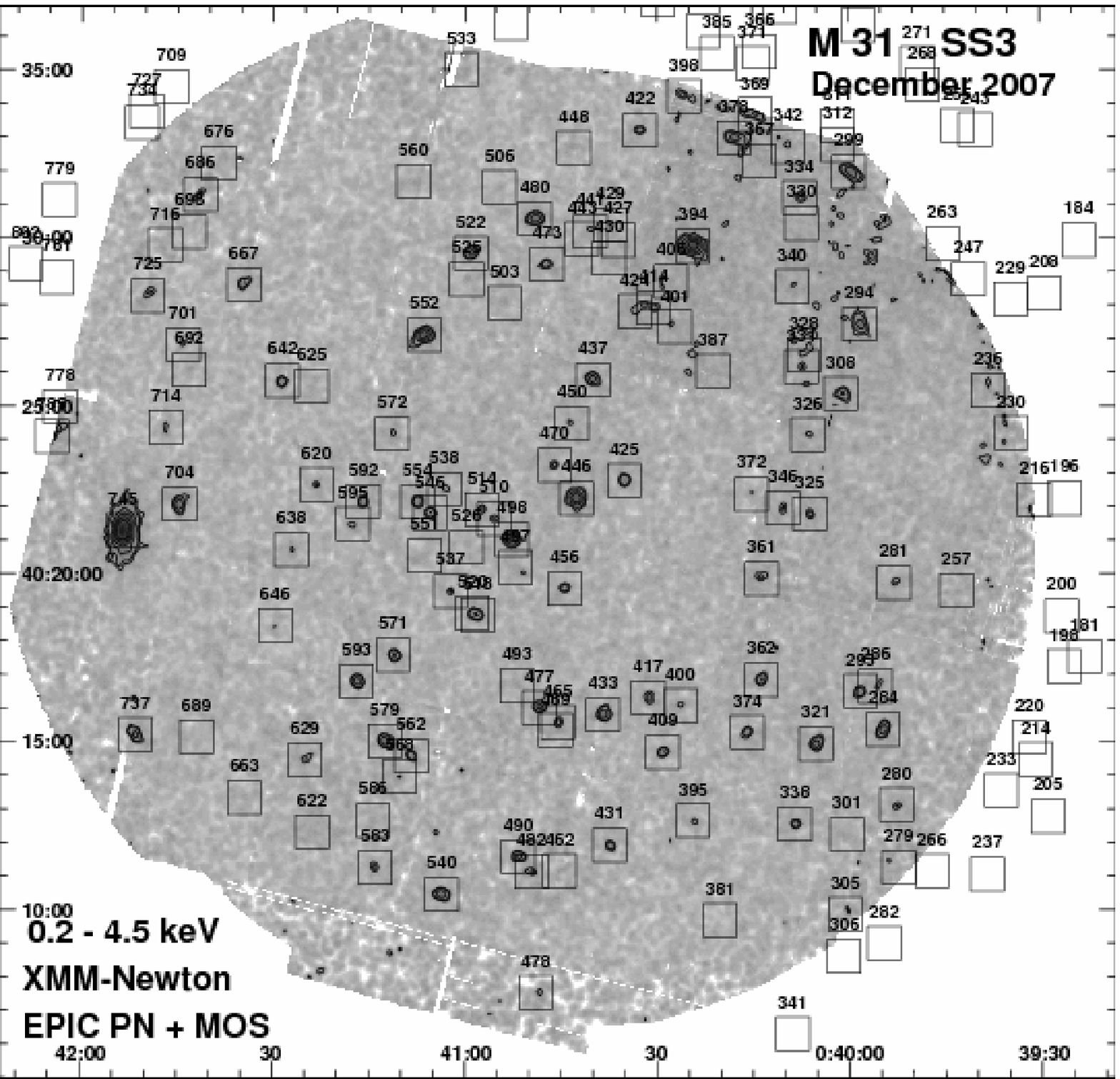}}
\resizebox{\hsize}{!}{\includegraphics[clip]{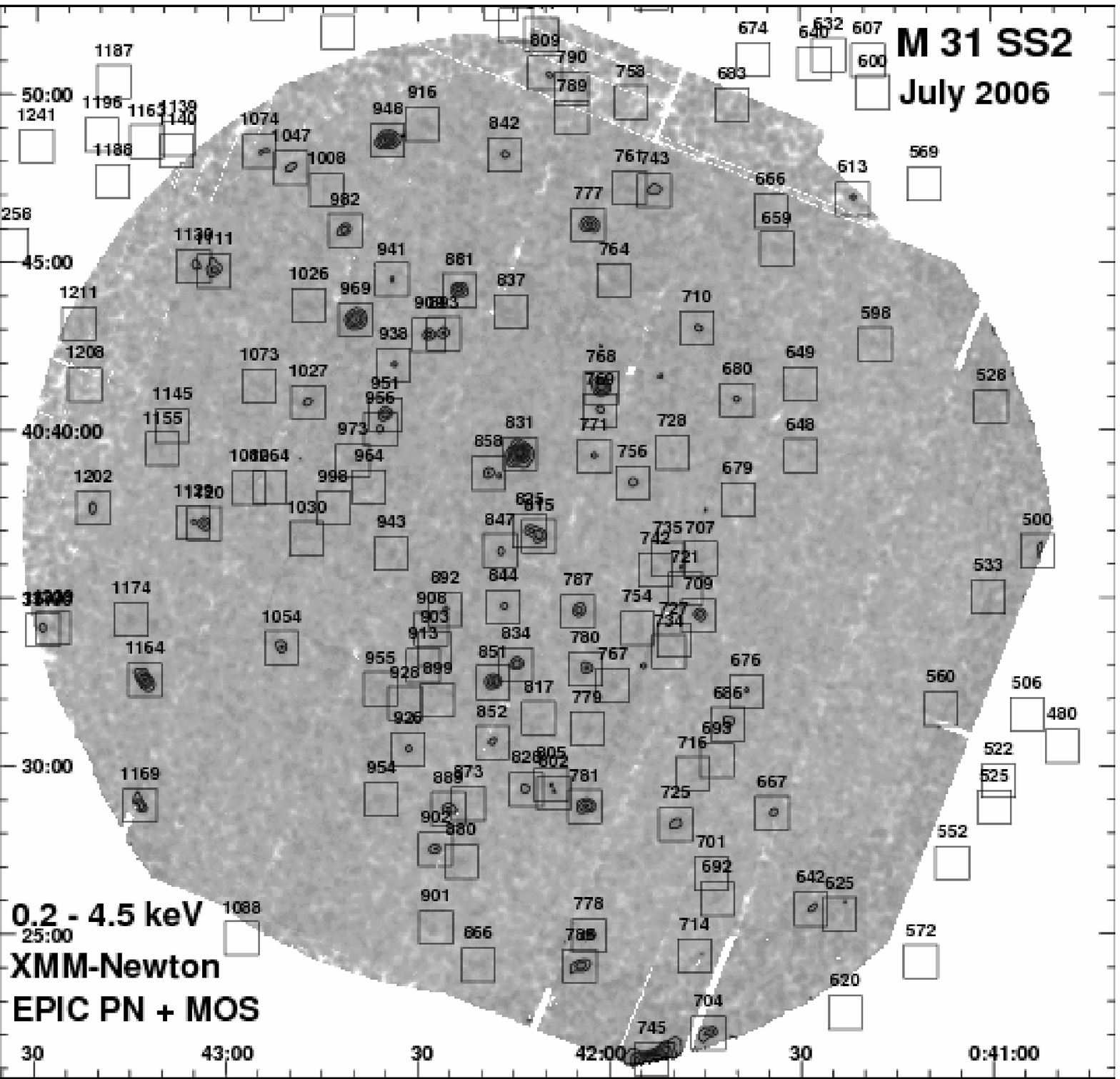}\hskip0.2cm\includegraphics[clip]{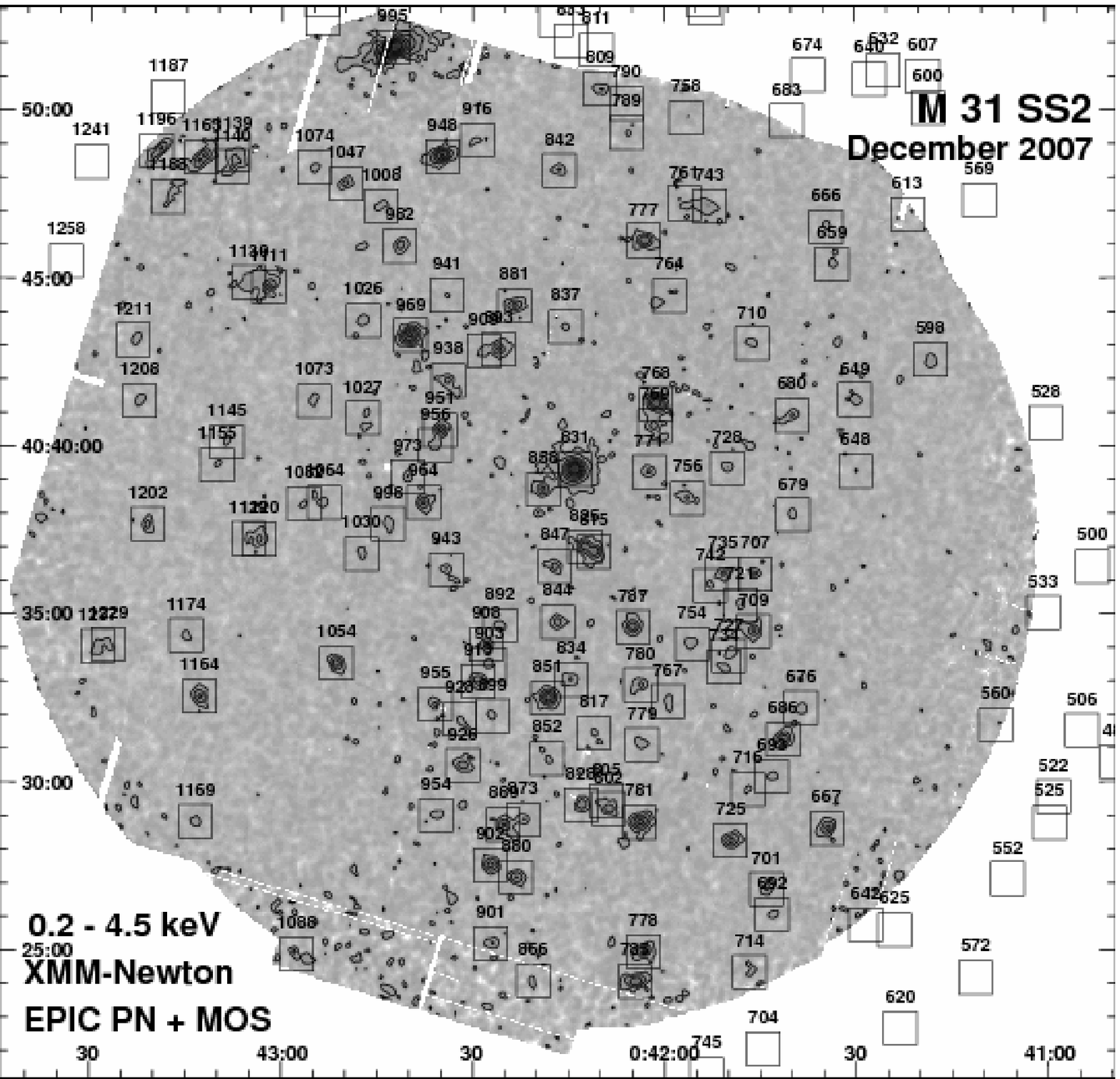}}
     \caption[]{(continued) Contours are at $(4, 8, 16, 32, 64, 128)$ in the upper left panel and lower right panel, at $(6, 8, 16, 32, 64, 128)$ in the upper right panel, and at $(8, 16, 32, 64, 128)$ in the lower left panel. }
\end{figure*}

\begin{figure*}
\addtocounter{figure}{-1}
\resizebox{\hsize}{!}{\includegraphics[clip]{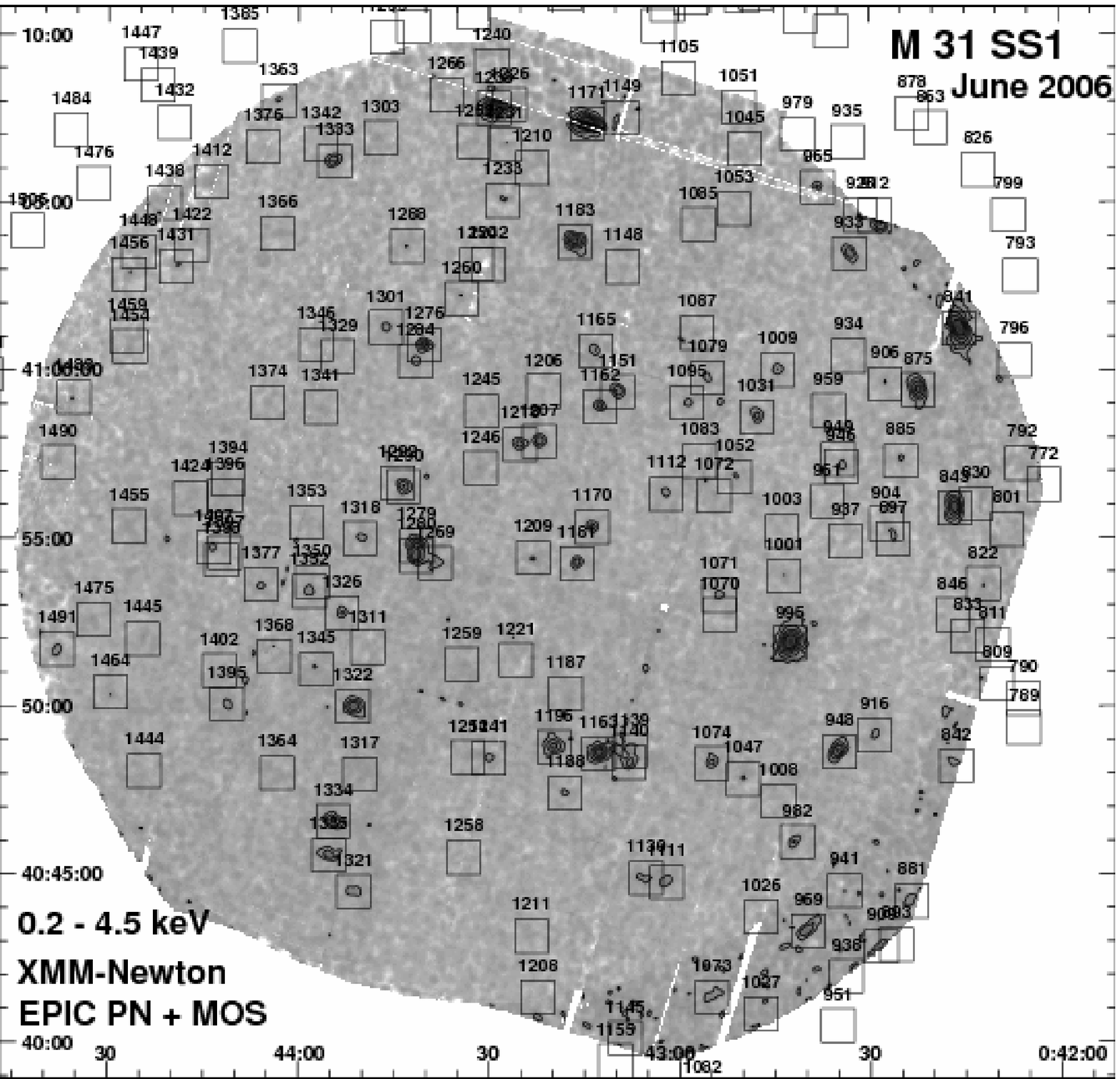}\hskip0.2cm\includegraphics[clip]{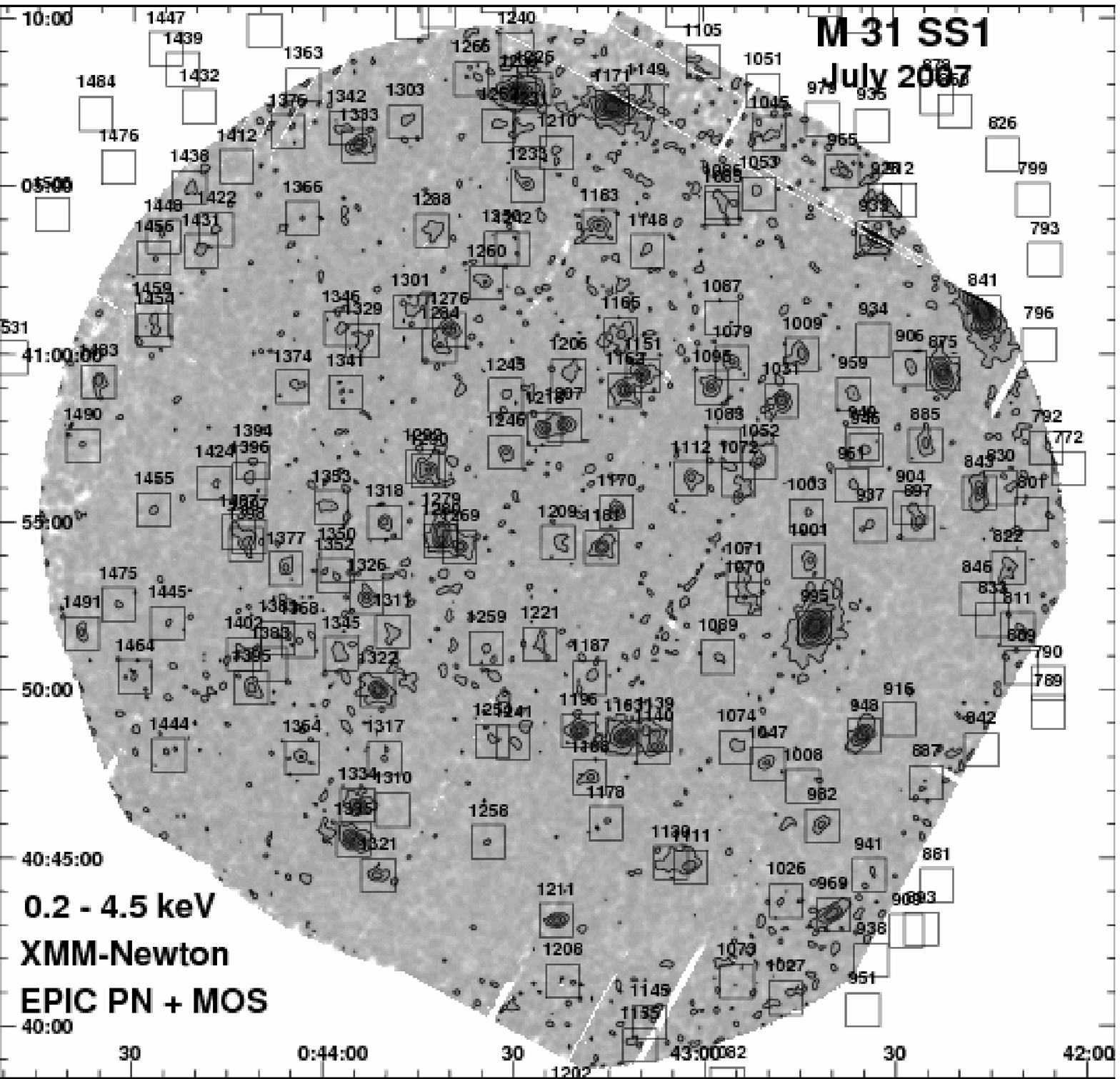}}
\resizebox{\hsize}{!}{\includegraphics[clip]{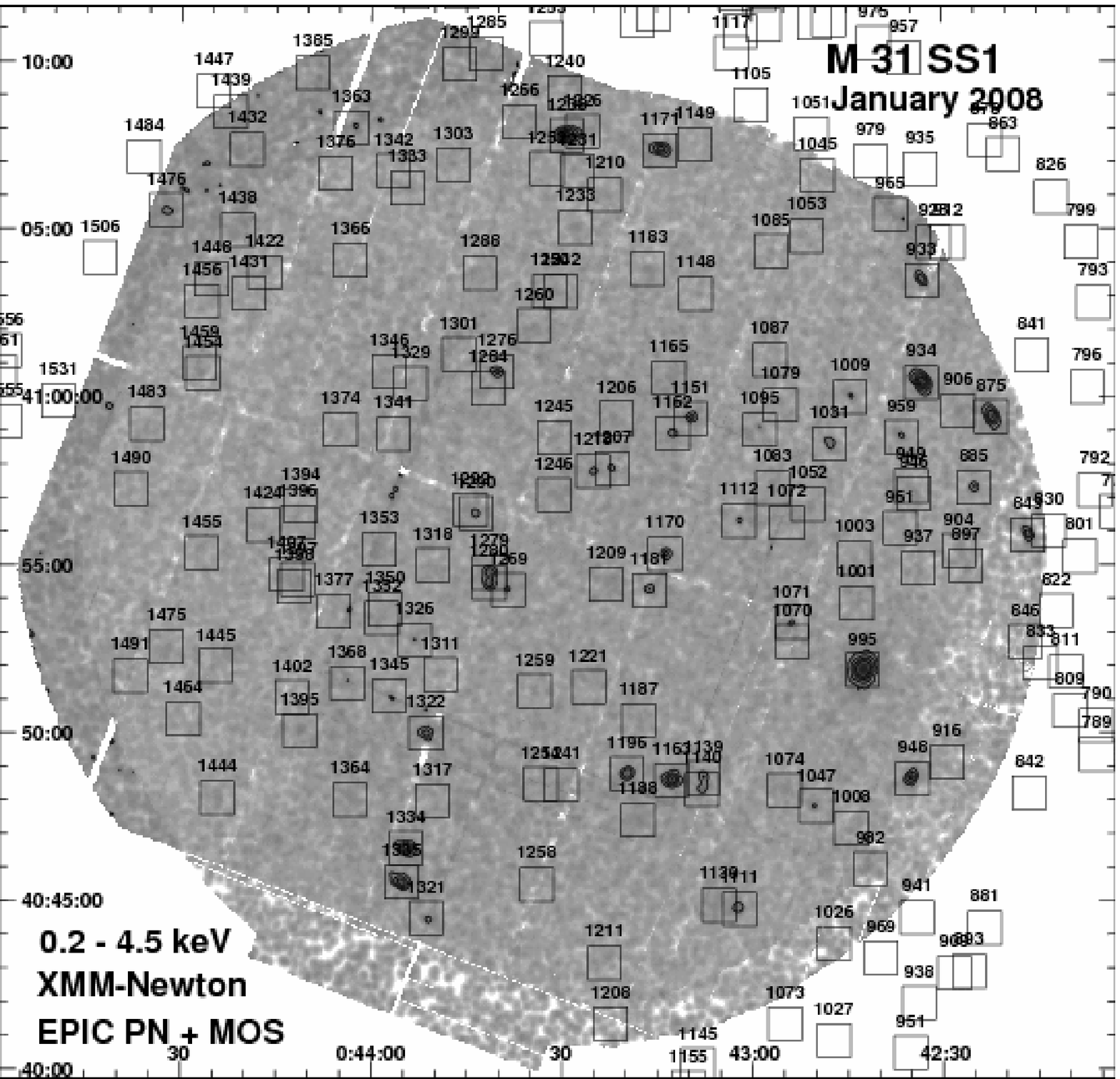}\hskip0.2cm\includegraphics[clip]{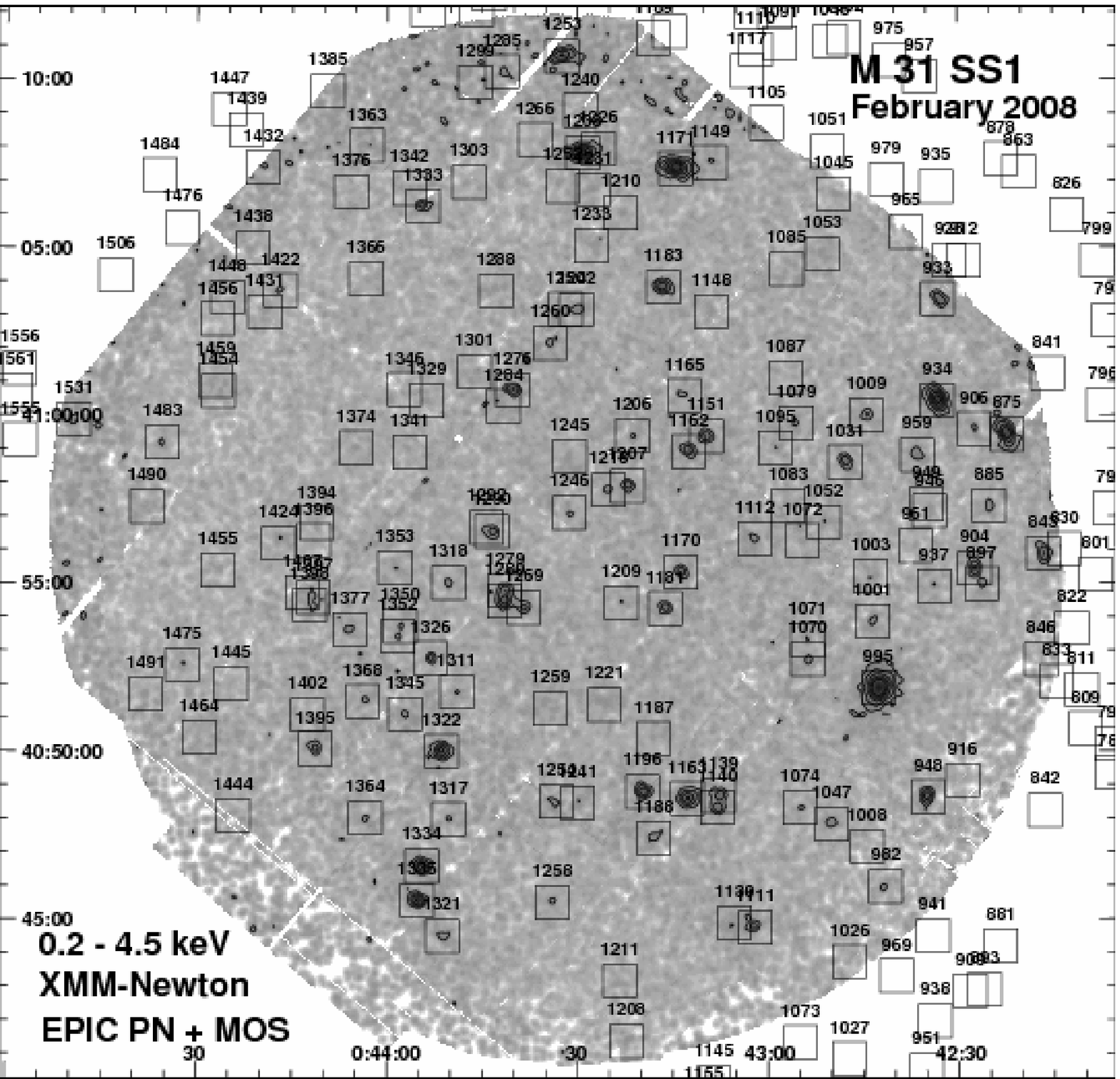}}
     \caption[]{(continued) Contours are at $(8, 16, 32, 64, 128)$ in the upper left panel and lower right panel, at $(4, 8, 16, 32, 64, 128)$ in the upper right panel, and at $(16, 32, 64, 128)$ in the lower left panel. 
}
\end{figure*}

\begin{figure*}
\addtocounter{figure}{-1}
\resizebox{\hsize}{!}{\includegraphics[clip]{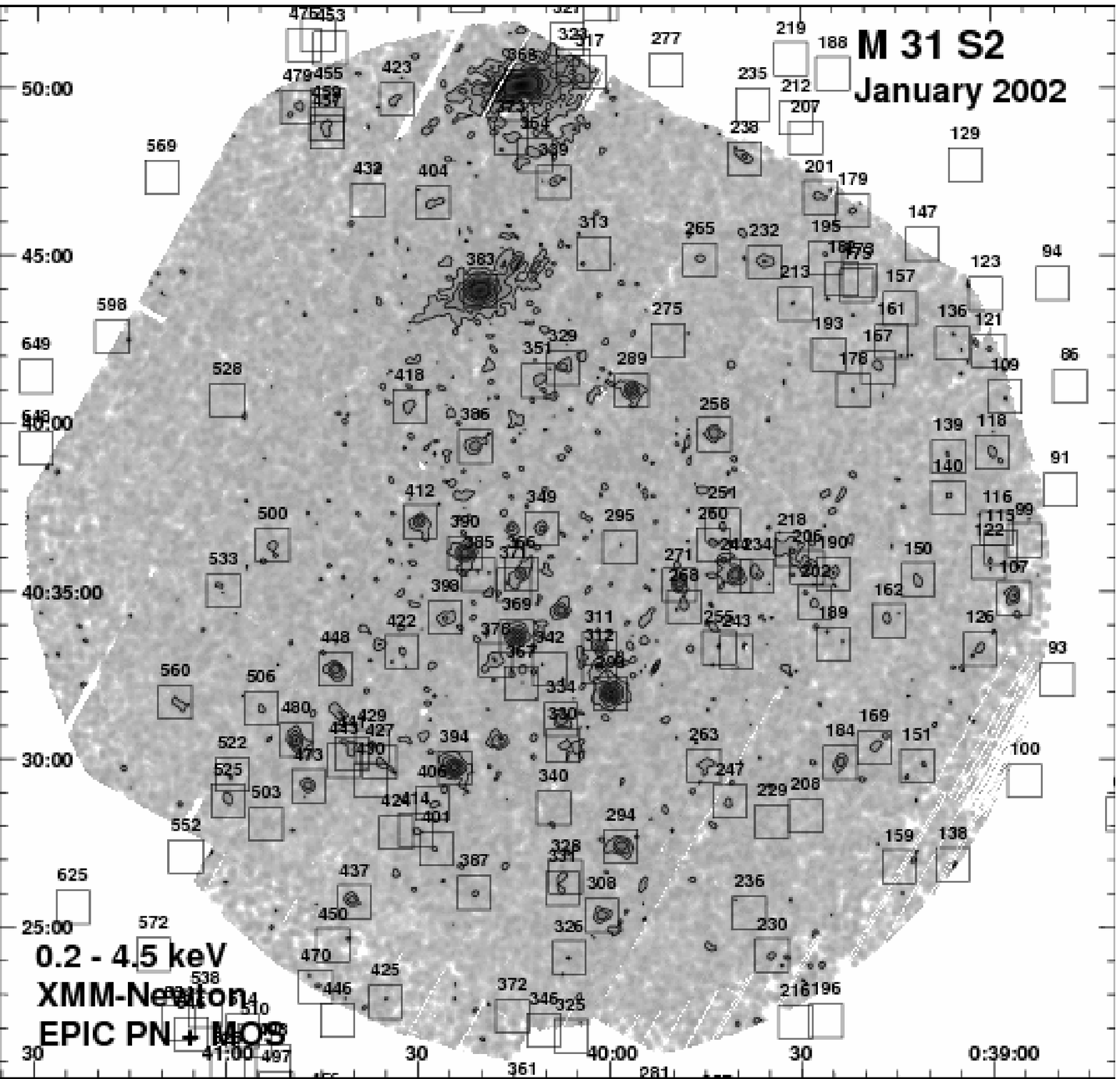}\hskip0.2cm\includegraphics[clip]{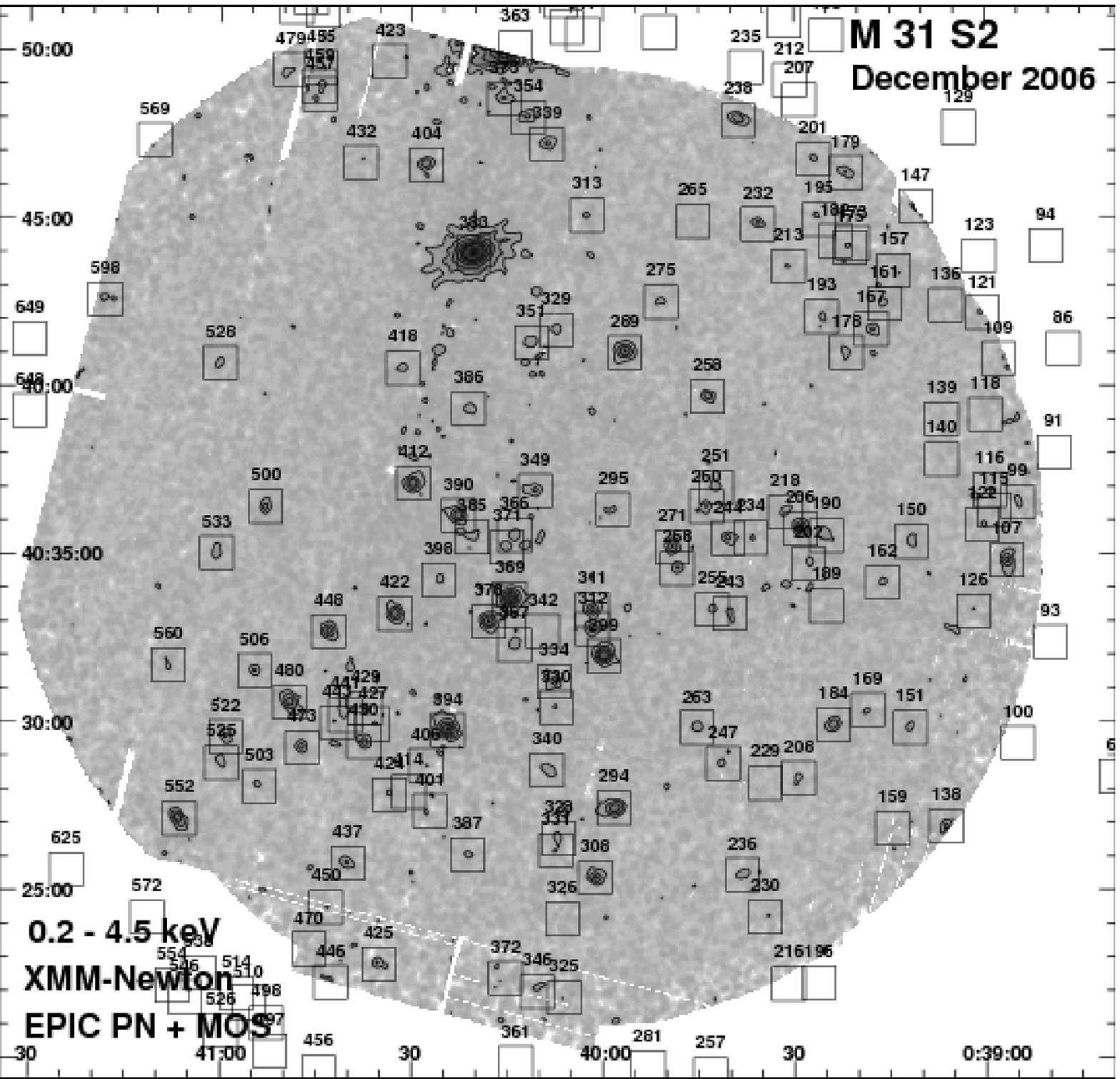}}
\resizebox{\hsize}{!}{\includegraphics[clip]{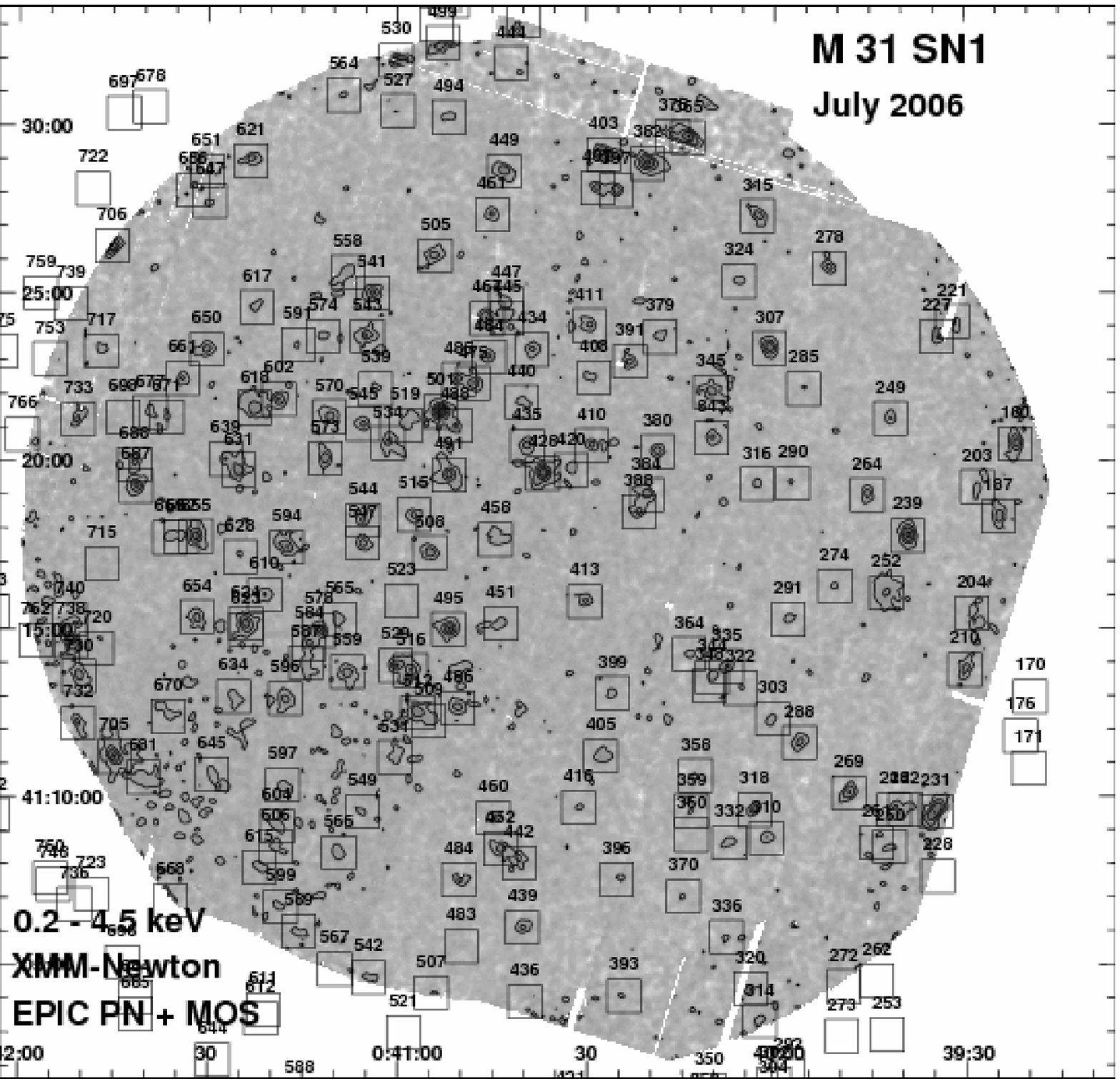}\hskip0.2cm\includegraphics[clip]{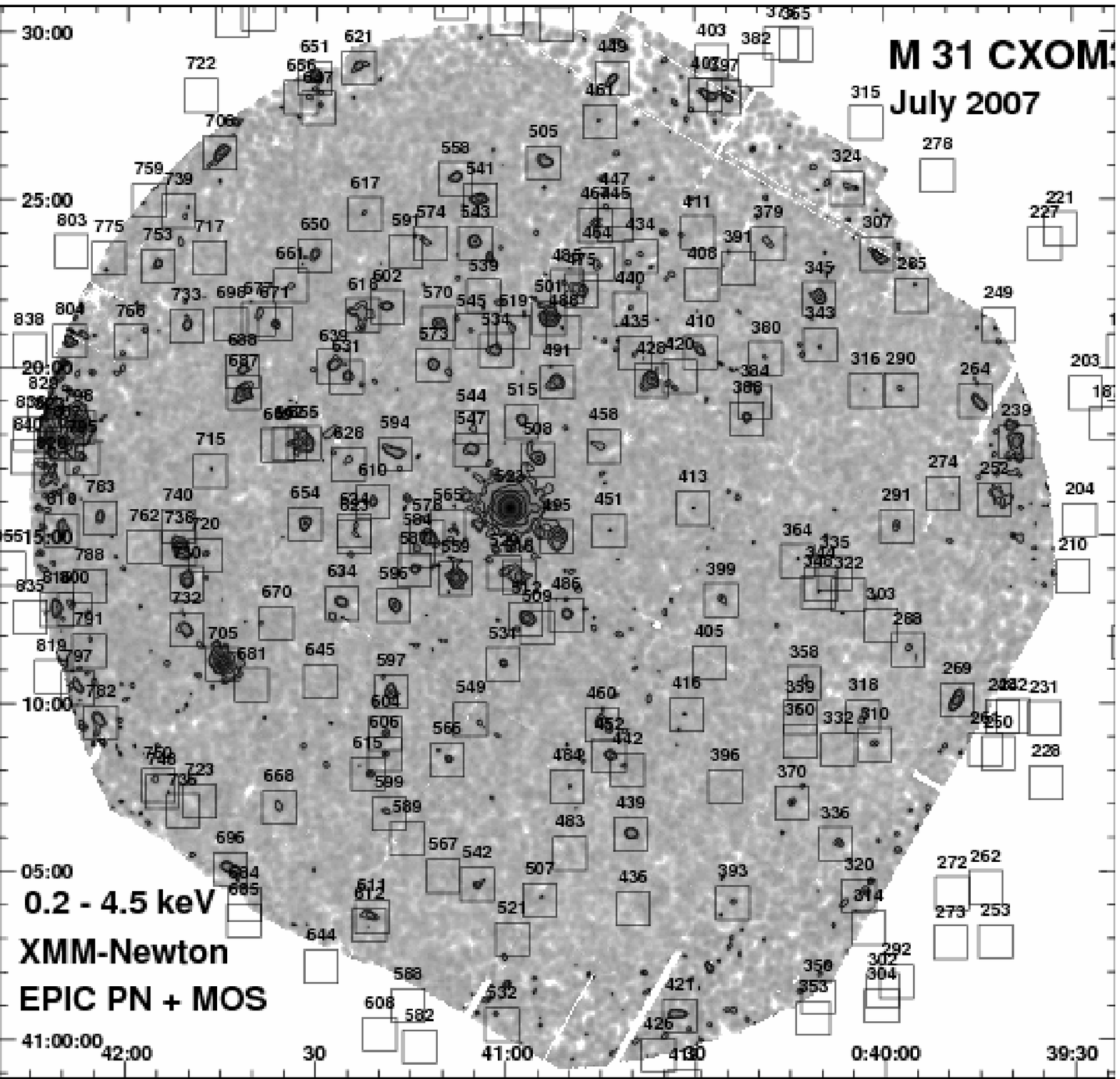}}
     \caption[]{(continued) Contours are at $(4, 8, 16, 32, 64, 128)$ in both upper panels and the lower left panel, and at $(6, 8, 16, 32, 64, 128)$ in the lower right panel. 
}
\end{figure*}

\begin{figure*}
\addtocounter{figure}{-1}
\resizebox{\hsize}{!}{\includegraphics[clip]{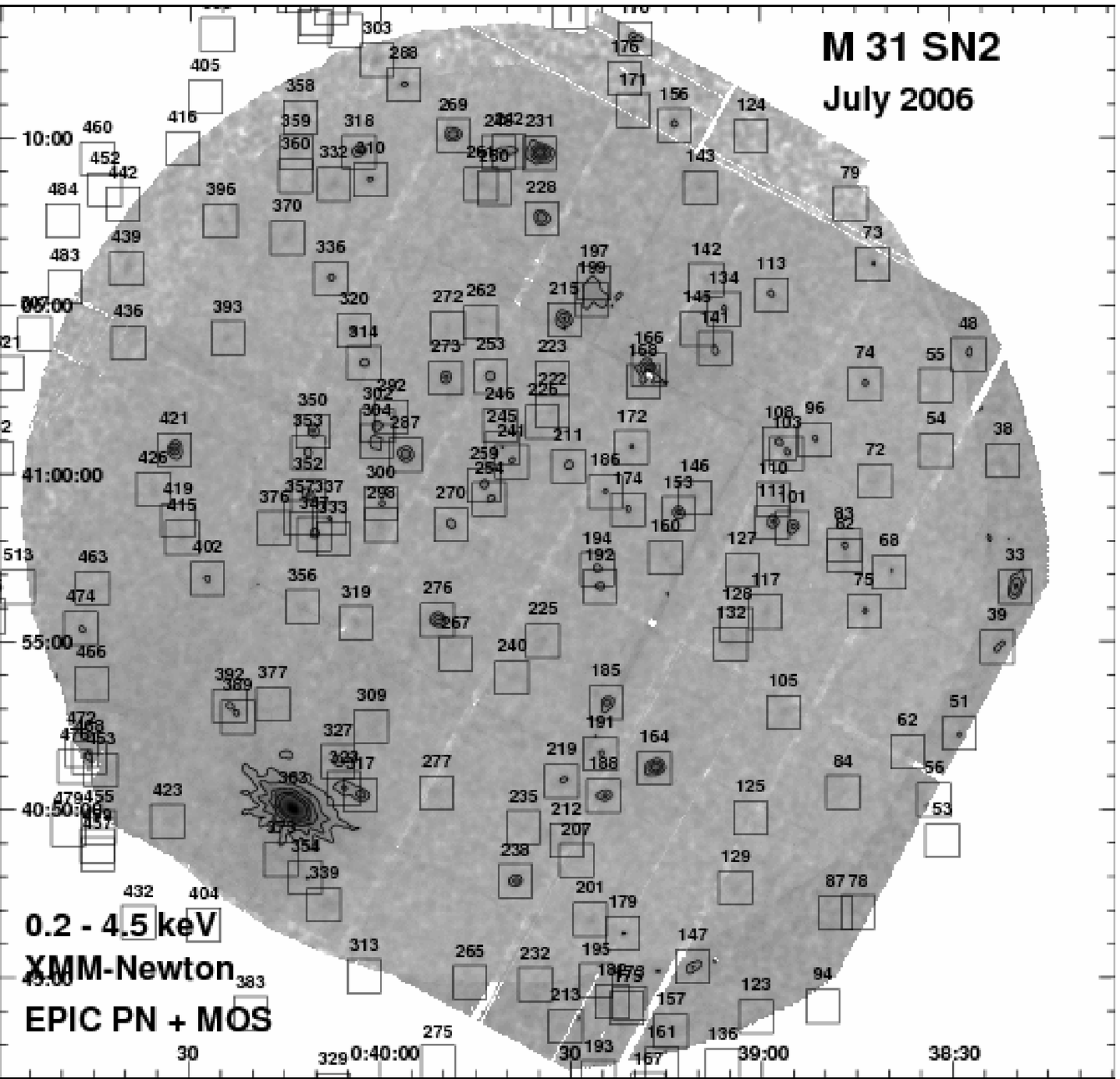}\hskip0.2cm\includegraphics[clip]{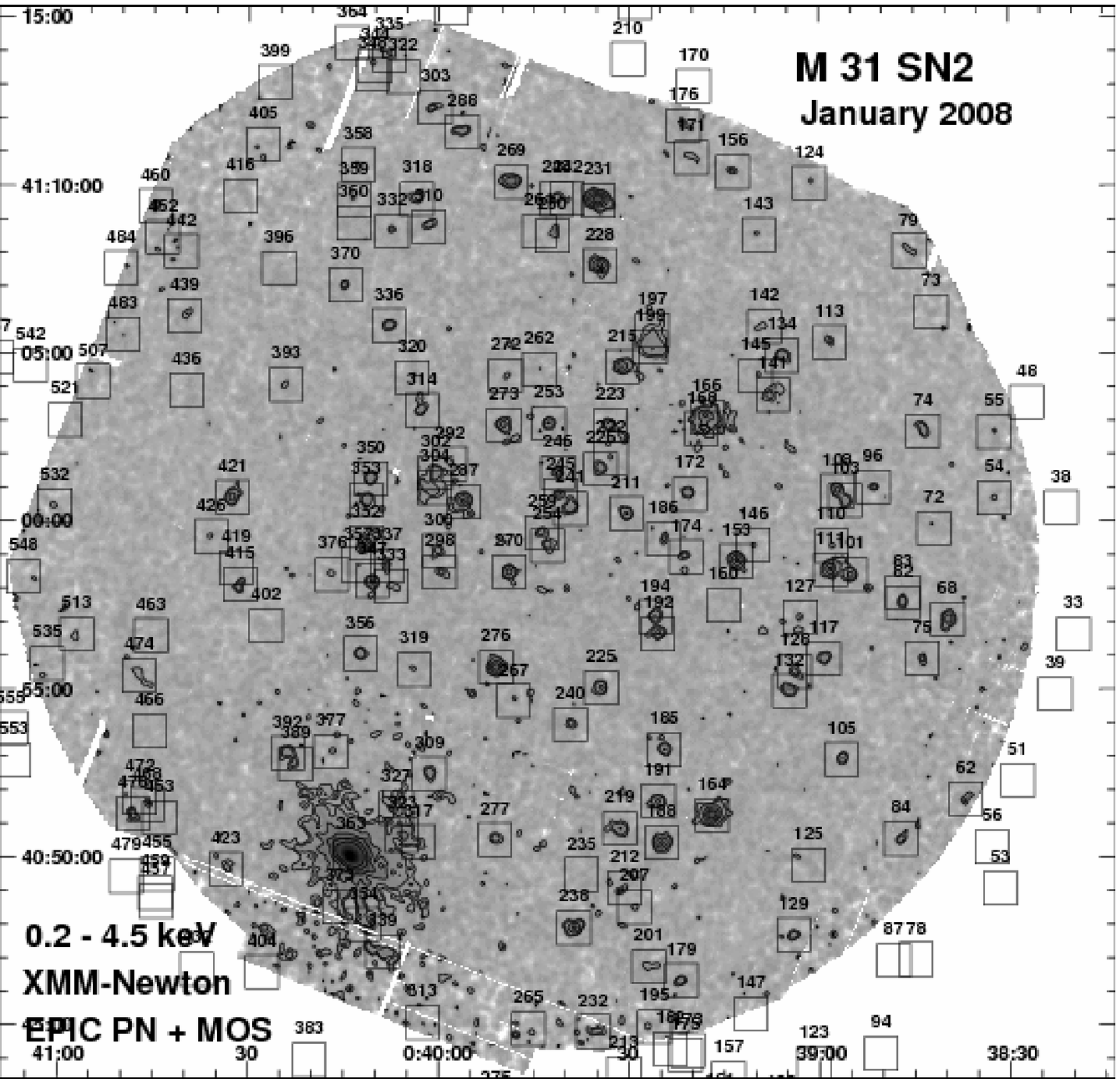}}
\resizebox{\hsize}{!}{\includegraphics[clip]{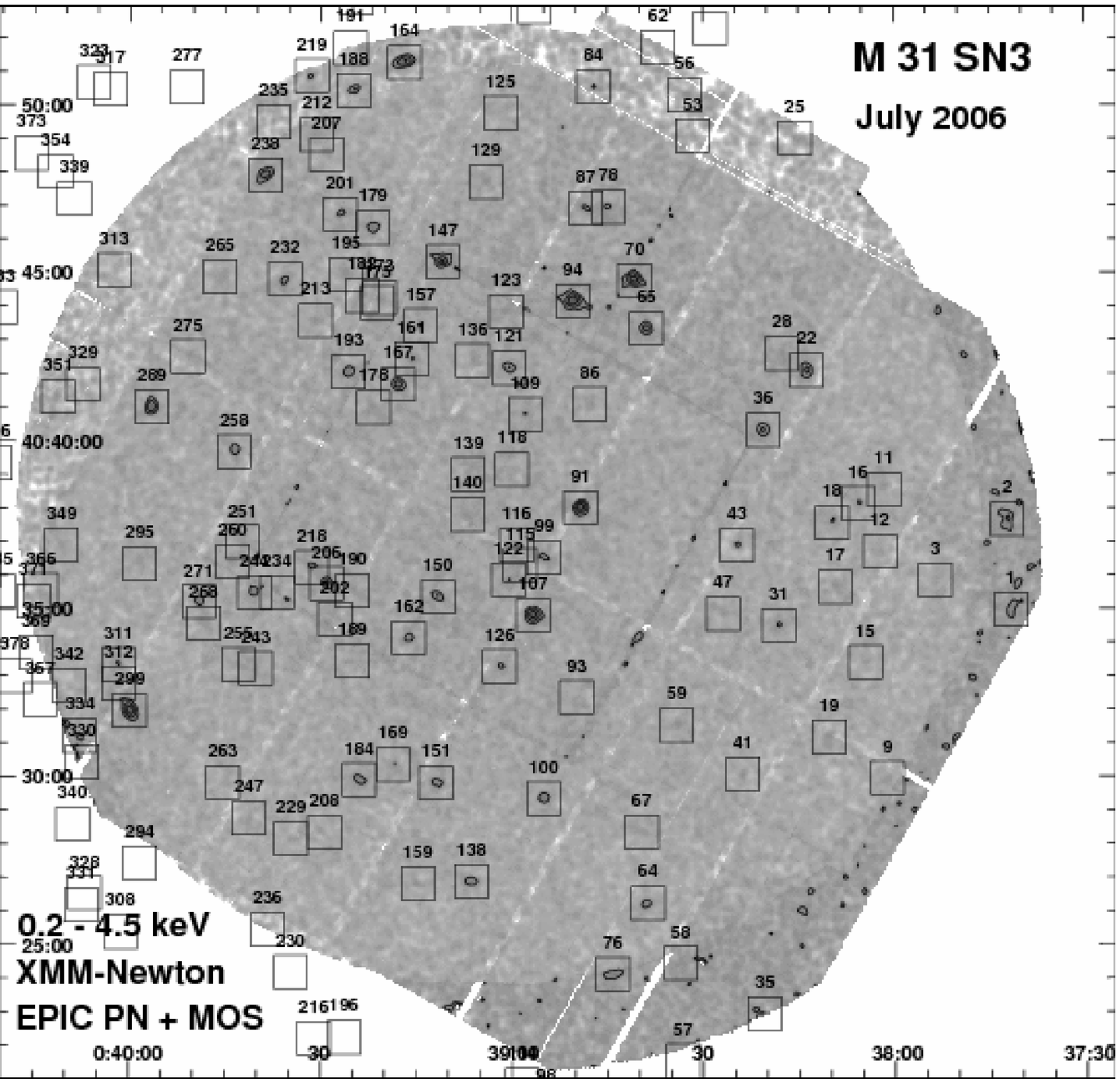}\hskip0.2cm\includegraphics[clip]{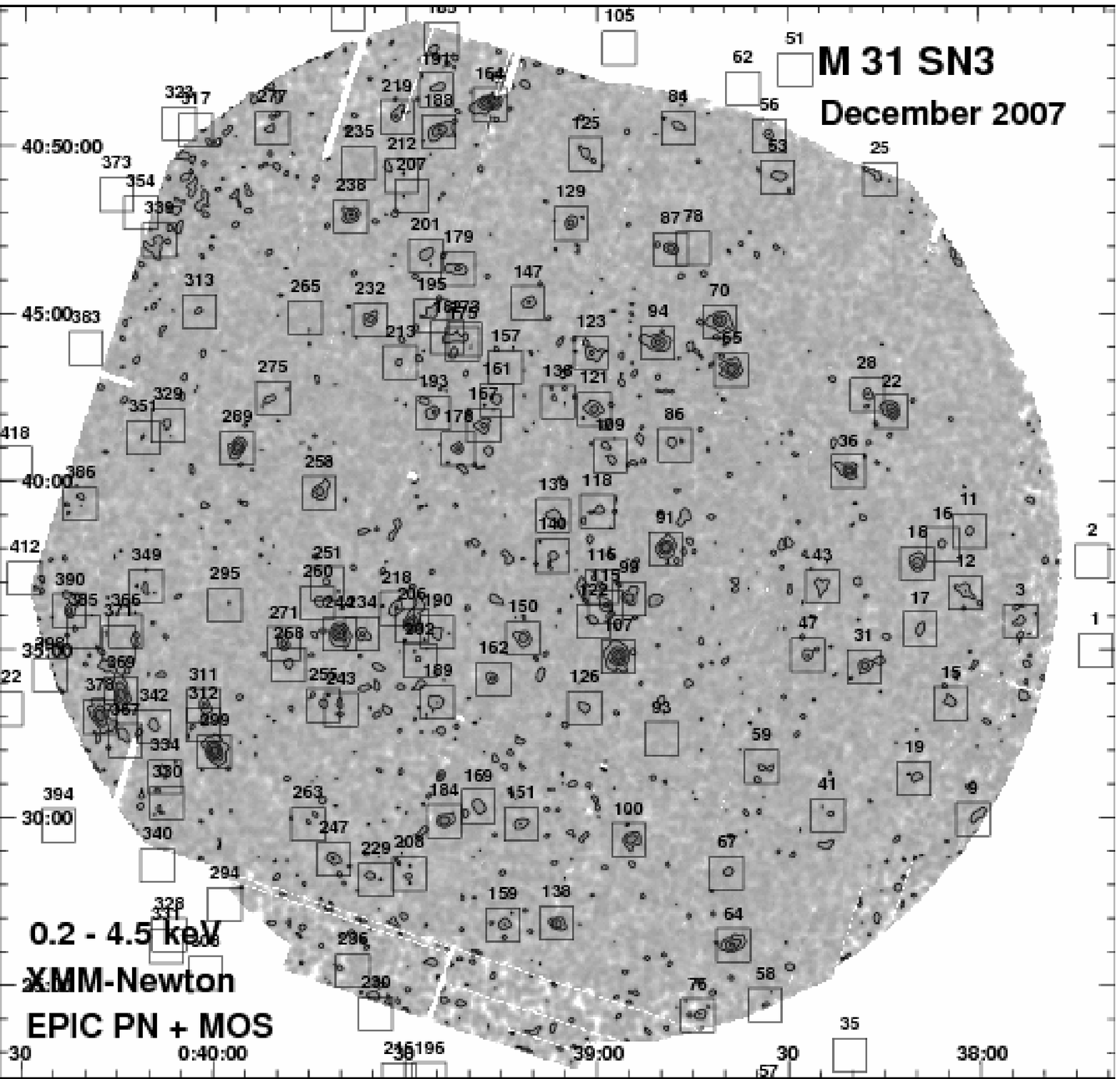}}
     \caption[]{(continued) Contours are at: $(8, 16, 32, 64, 128)$ in the upper left and lower left panels, $(6, 8, 16, 32, 64, 128)$  in the upper right panel, and at $(4, 8, 16, 32, 64, 128)$ in the lower right panel. 
}
\end{figure*}
\begin{figure*}
\addtocounter{figure}{-1}
\resizebox{\hsize}{!}{\includegraphics[clip]{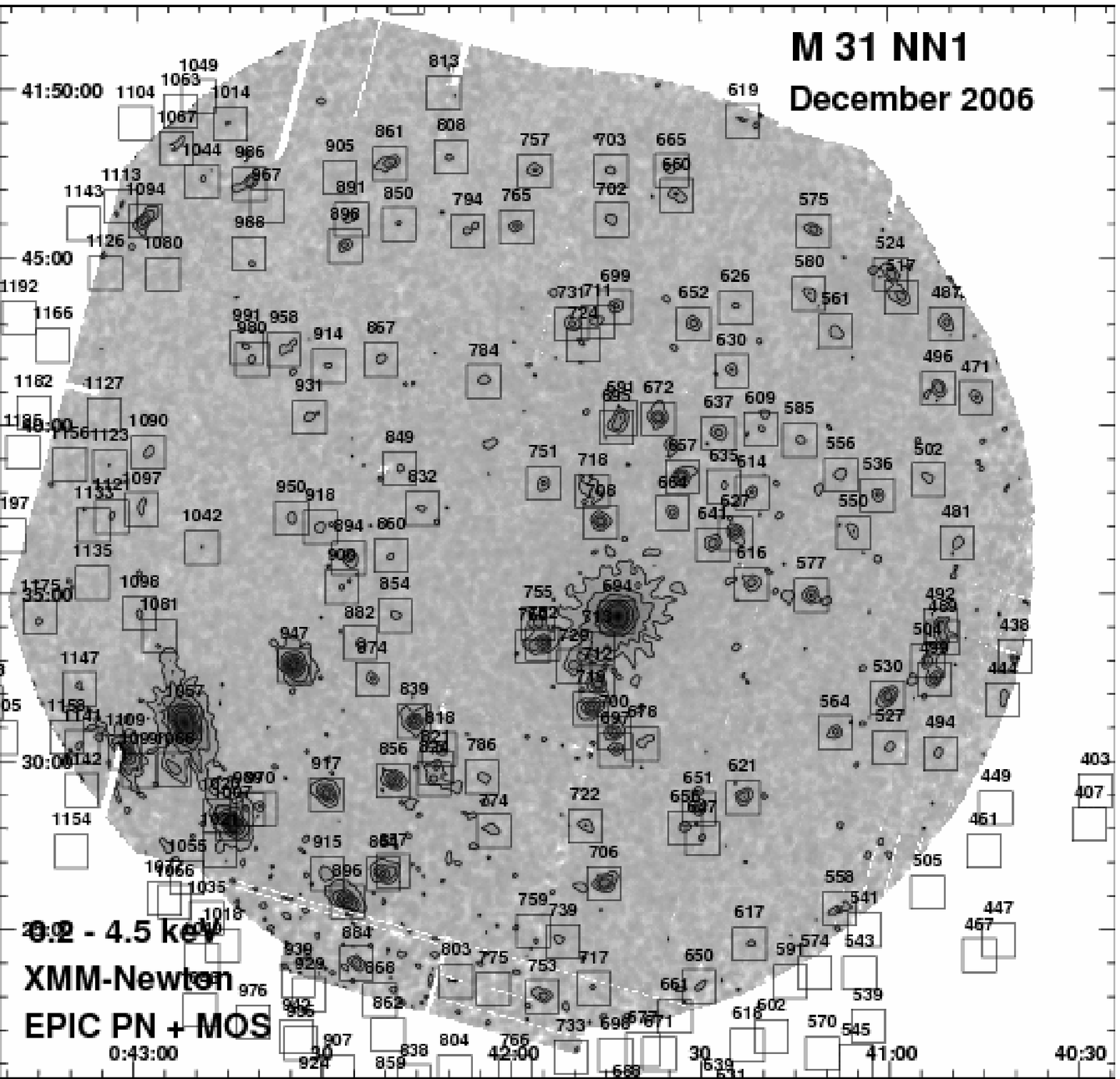}\hskip0.2cm\includegraphics[clip]{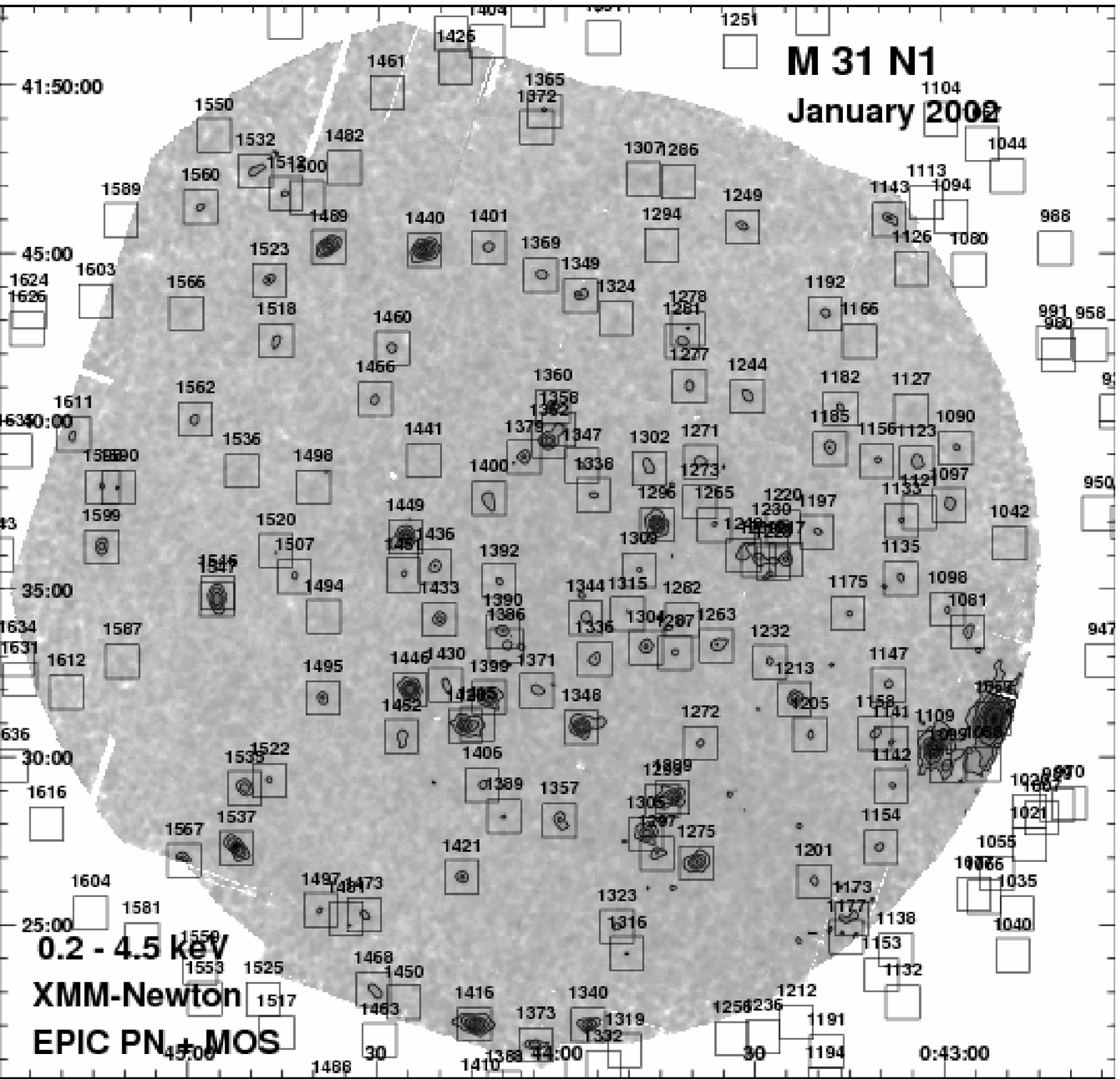}}
\resizebox{\hsize}{!}{\includegraphics[clip]{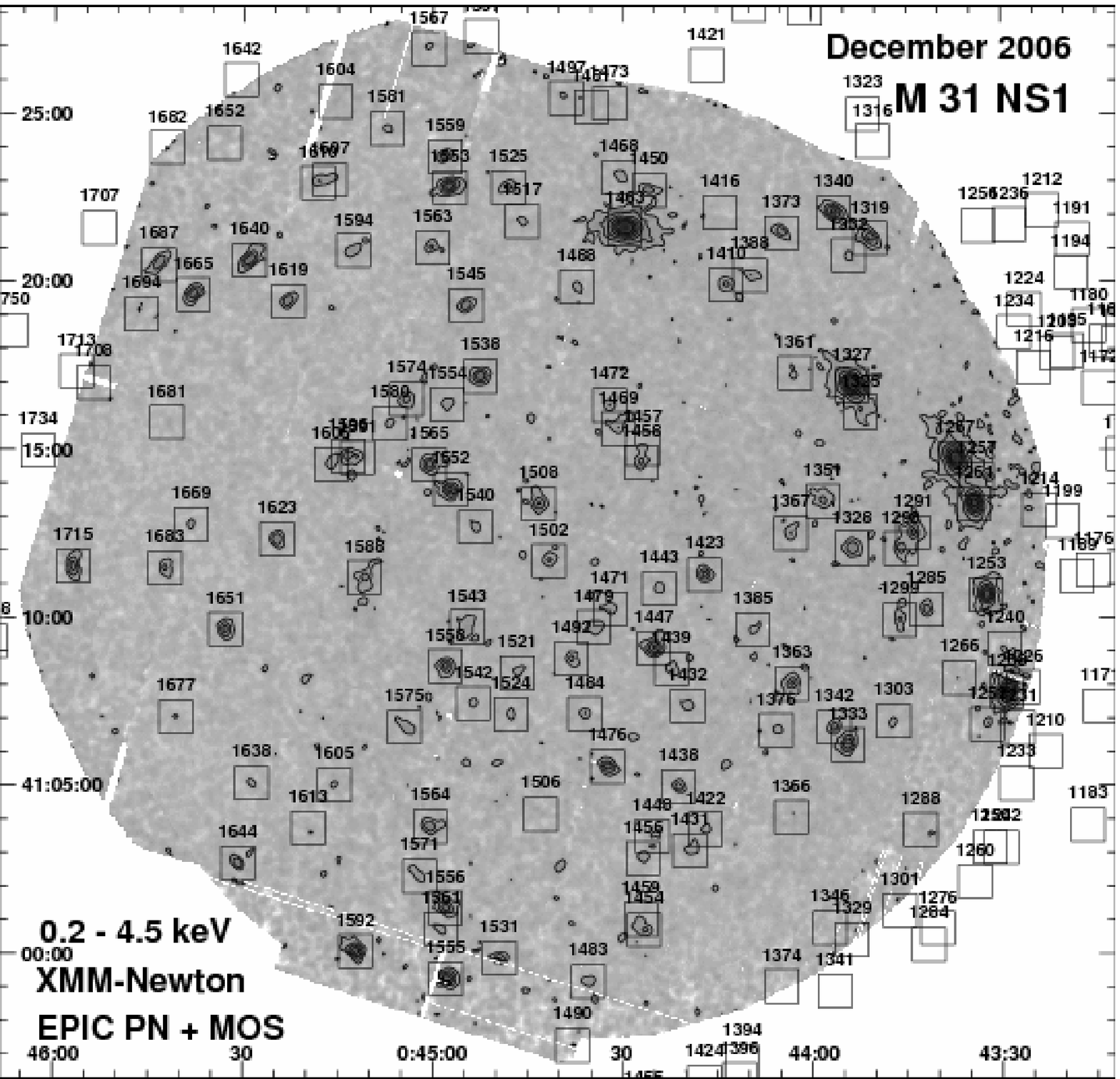}\hskip0.2cm\includegraphics[clip]{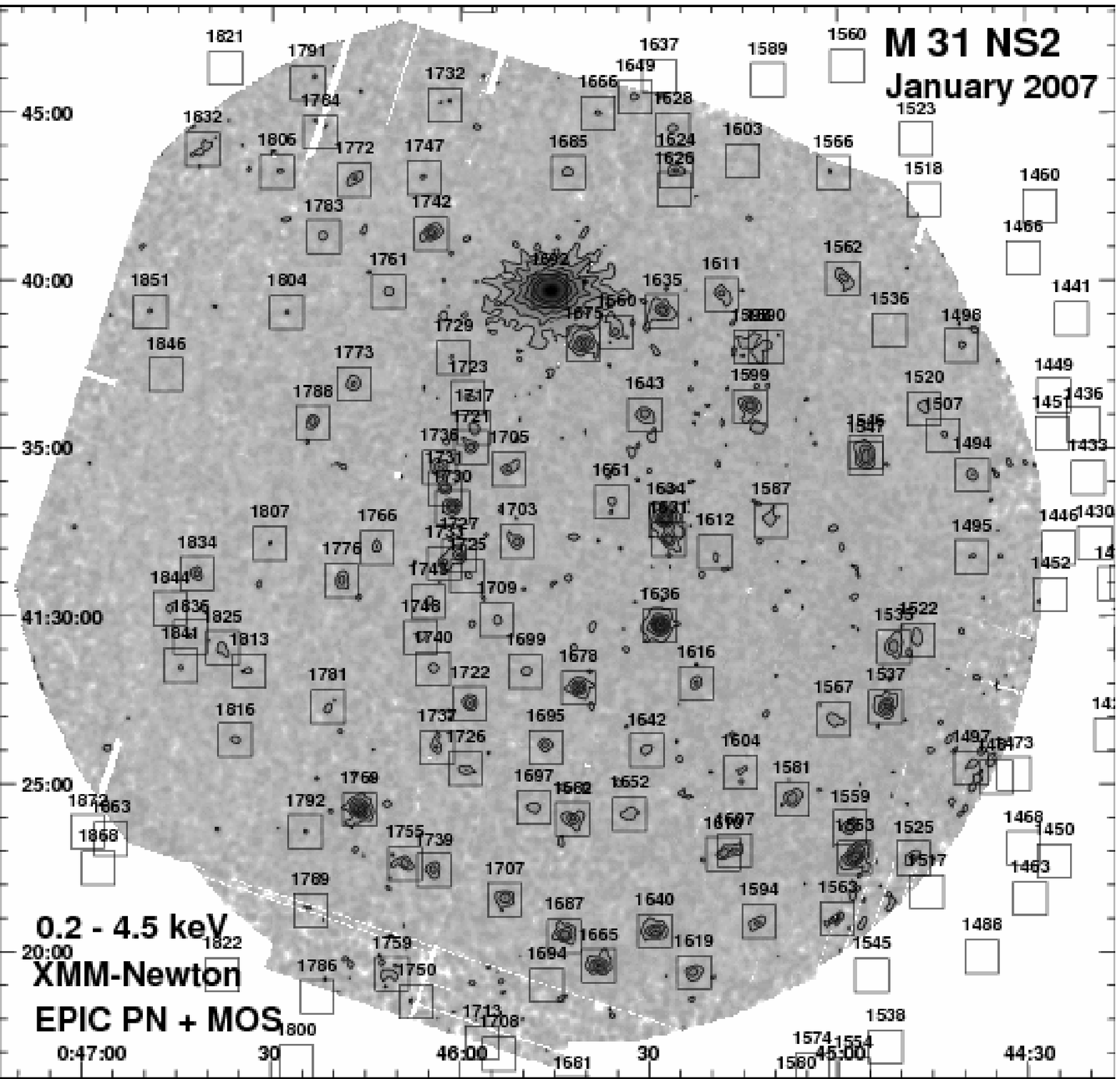}}
     \caption[]{(continued) Contours are at $(4, 8, 16, 32, 64, 128)$ in all panels. 
     }
\end{figure*}

\begin{figure*}
\addtocounter{figure}{-1}
\resizebox{\hsize}{!}{\includegraphics[clip]{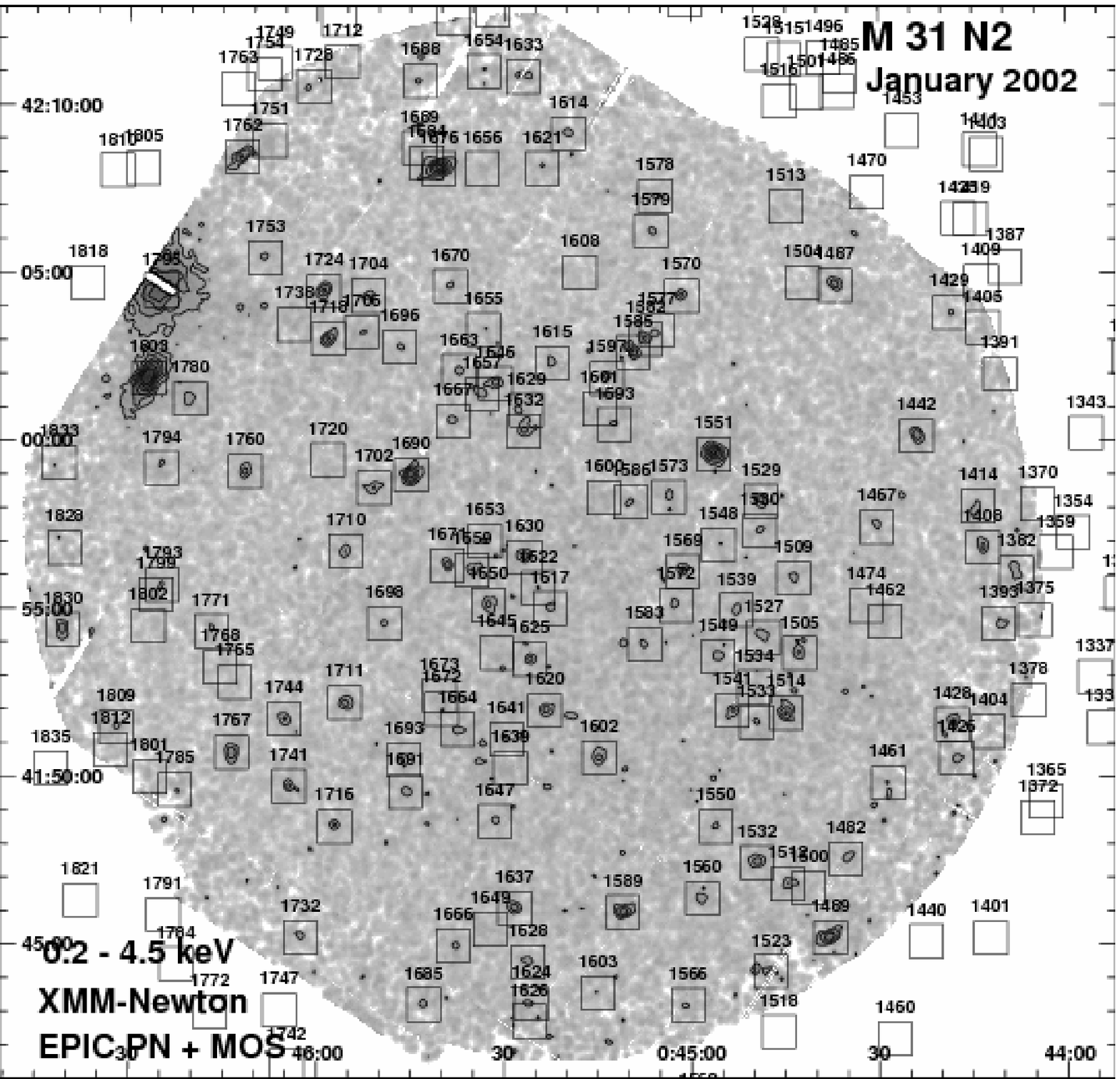}\hskip0.2cm\includegraphics[clip]{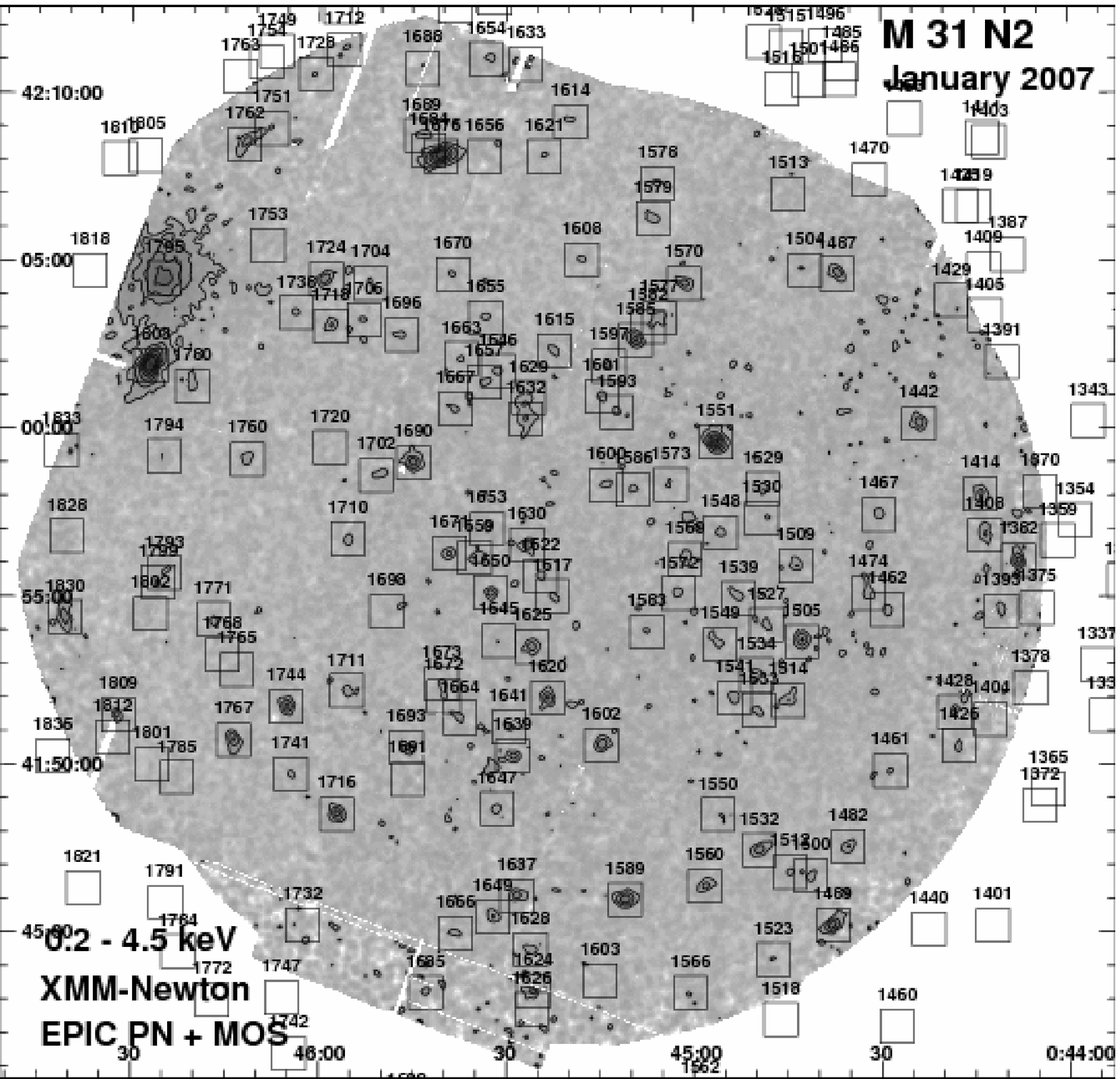}}
\resizebox{\hsize}{!}{\includegraphics[clip]{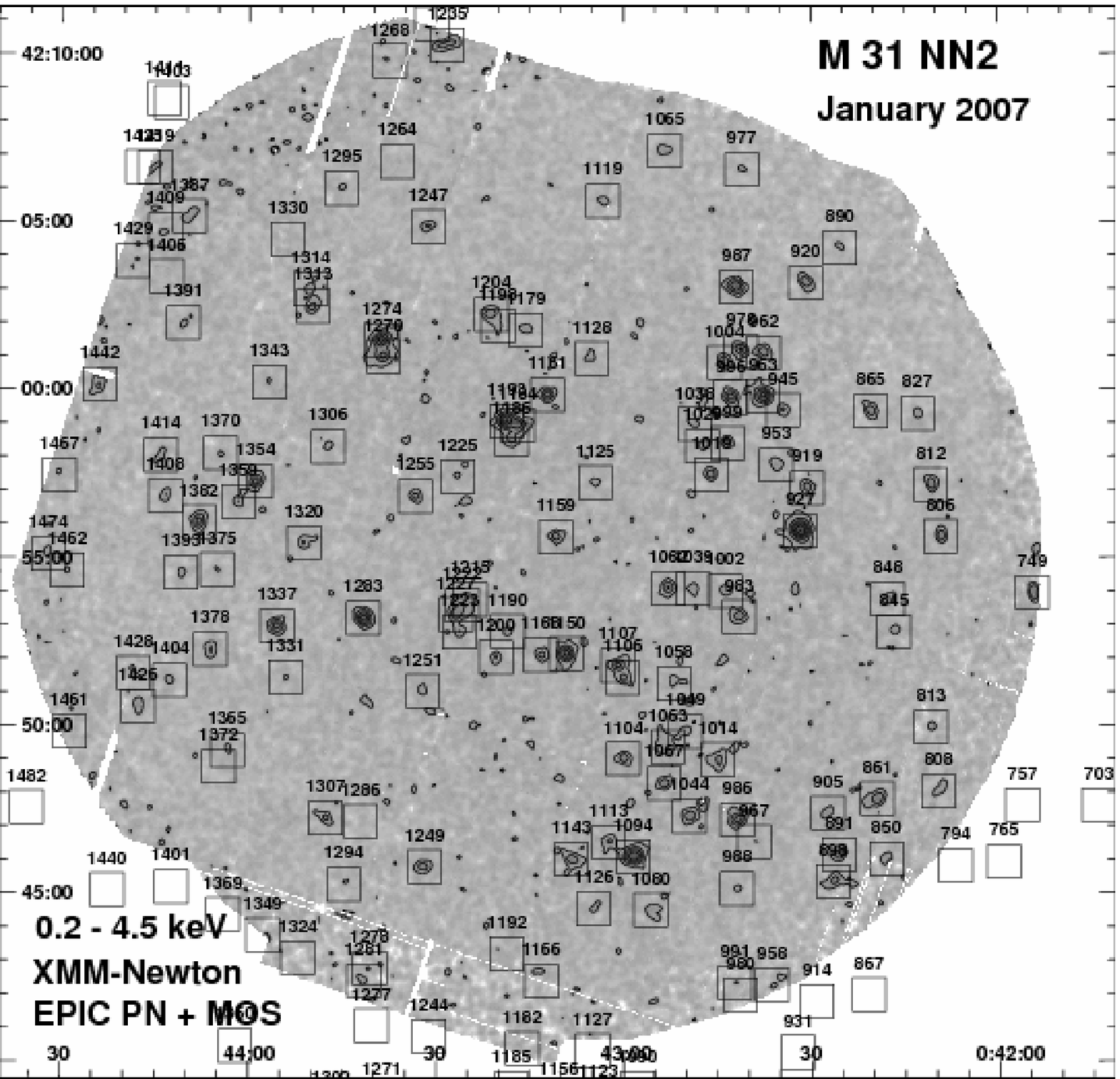}\hskip0.2cm\includegraphics[clip]{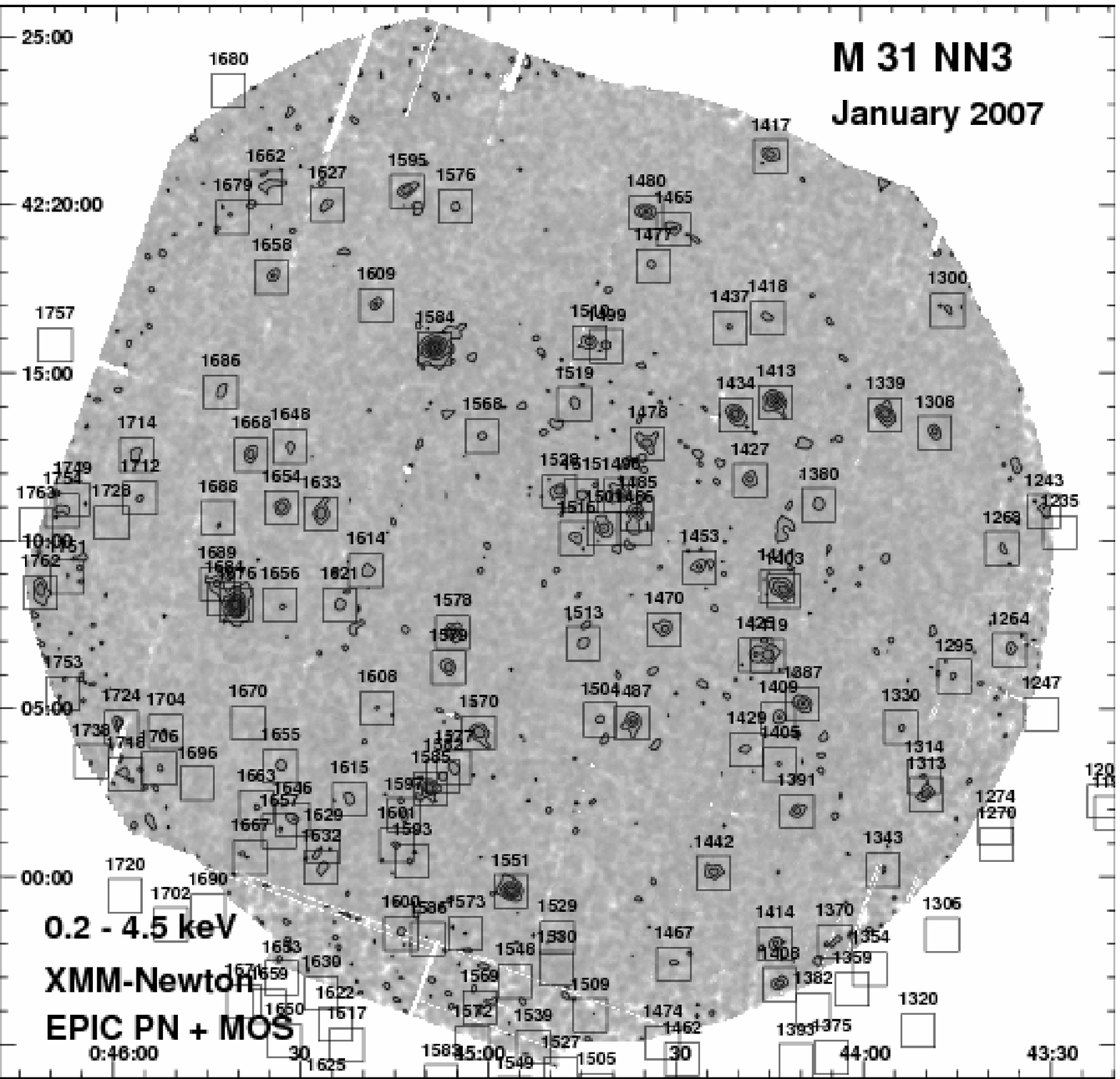}}
     \caption[]{(continued) Contours are at $(4, 8, 16, 32, 64, 128)$ in all panels. 
}
\end{figure*}

\begin{figure*}
\addtocounter{figure}{-1}
\resizebox{\hsize}{!}{\includegraphics[clip]{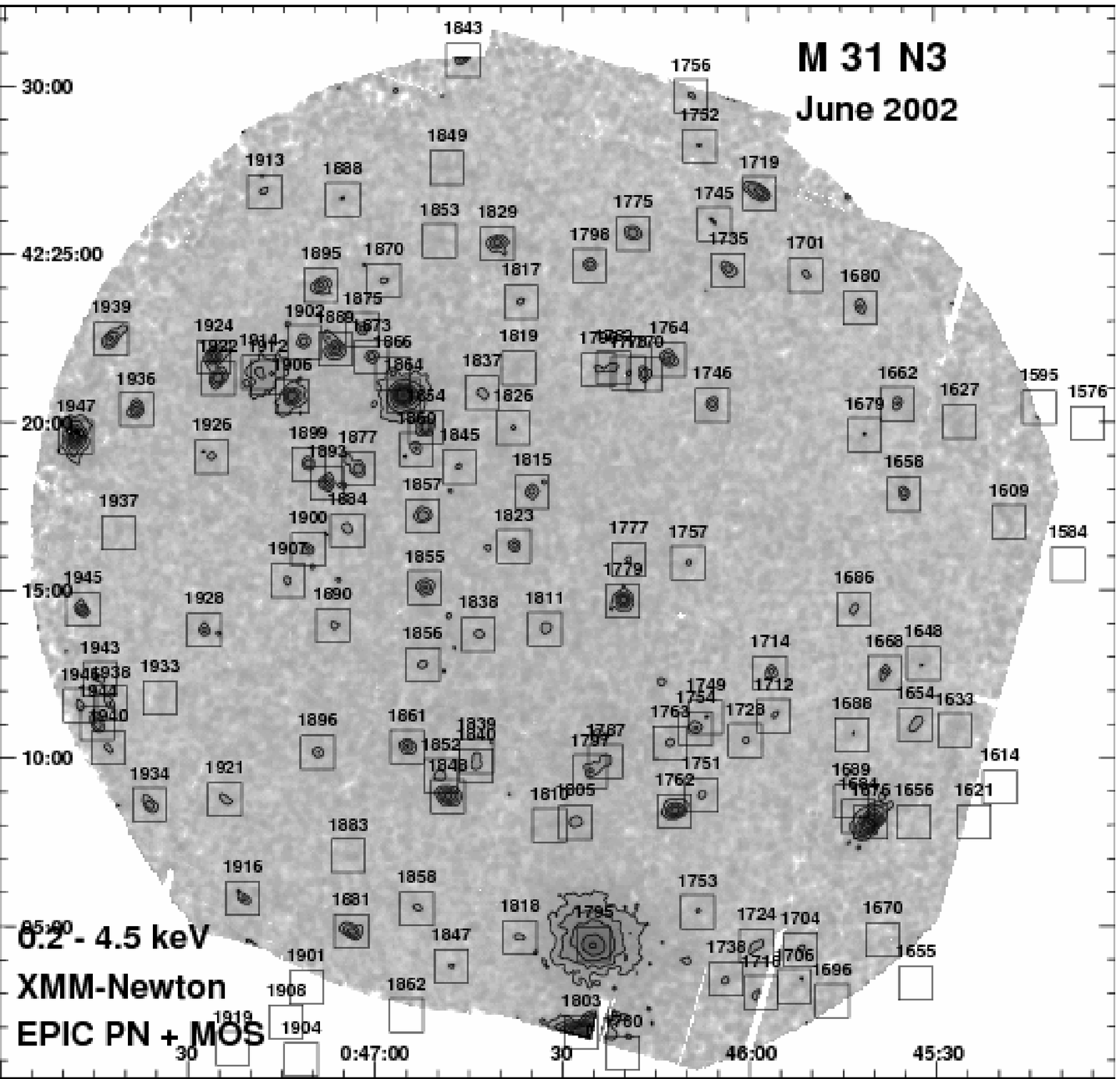}\hskip0.2cm\includegraphics[clip]{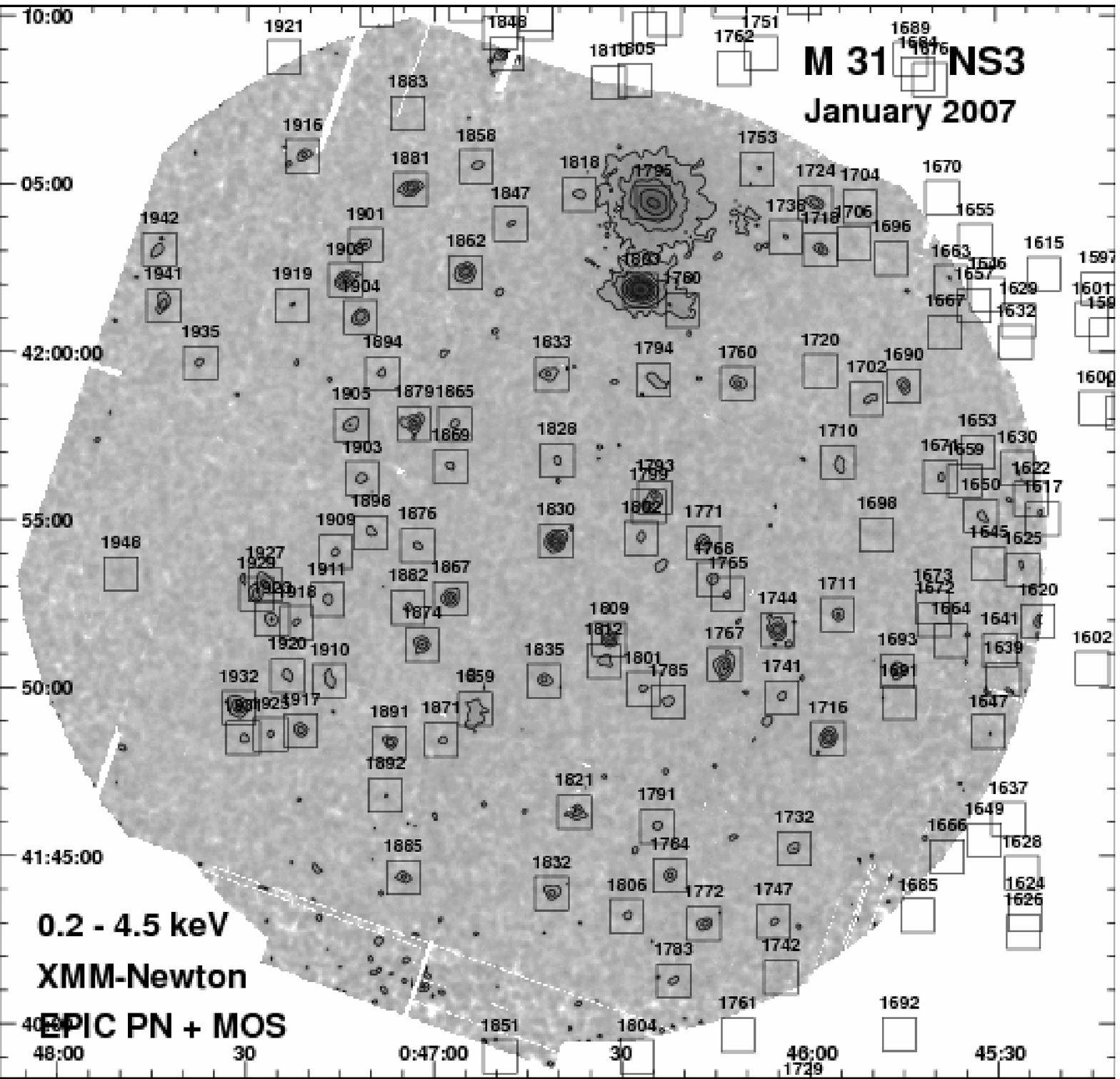}}
\resizebox{\hsize}{!}{\includegraphics[clip]{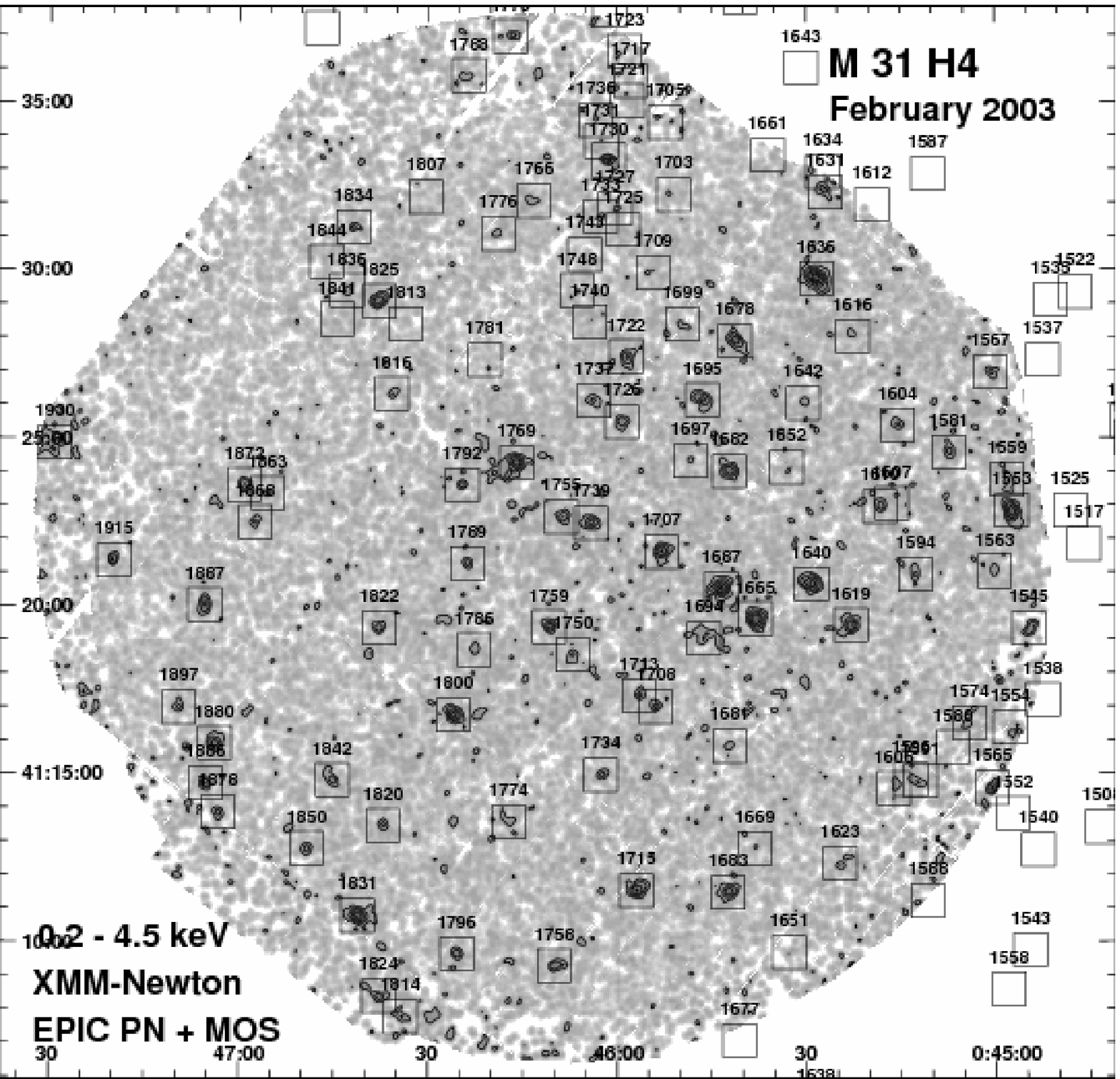}\hskip0.2cm\includegraphics[clip]{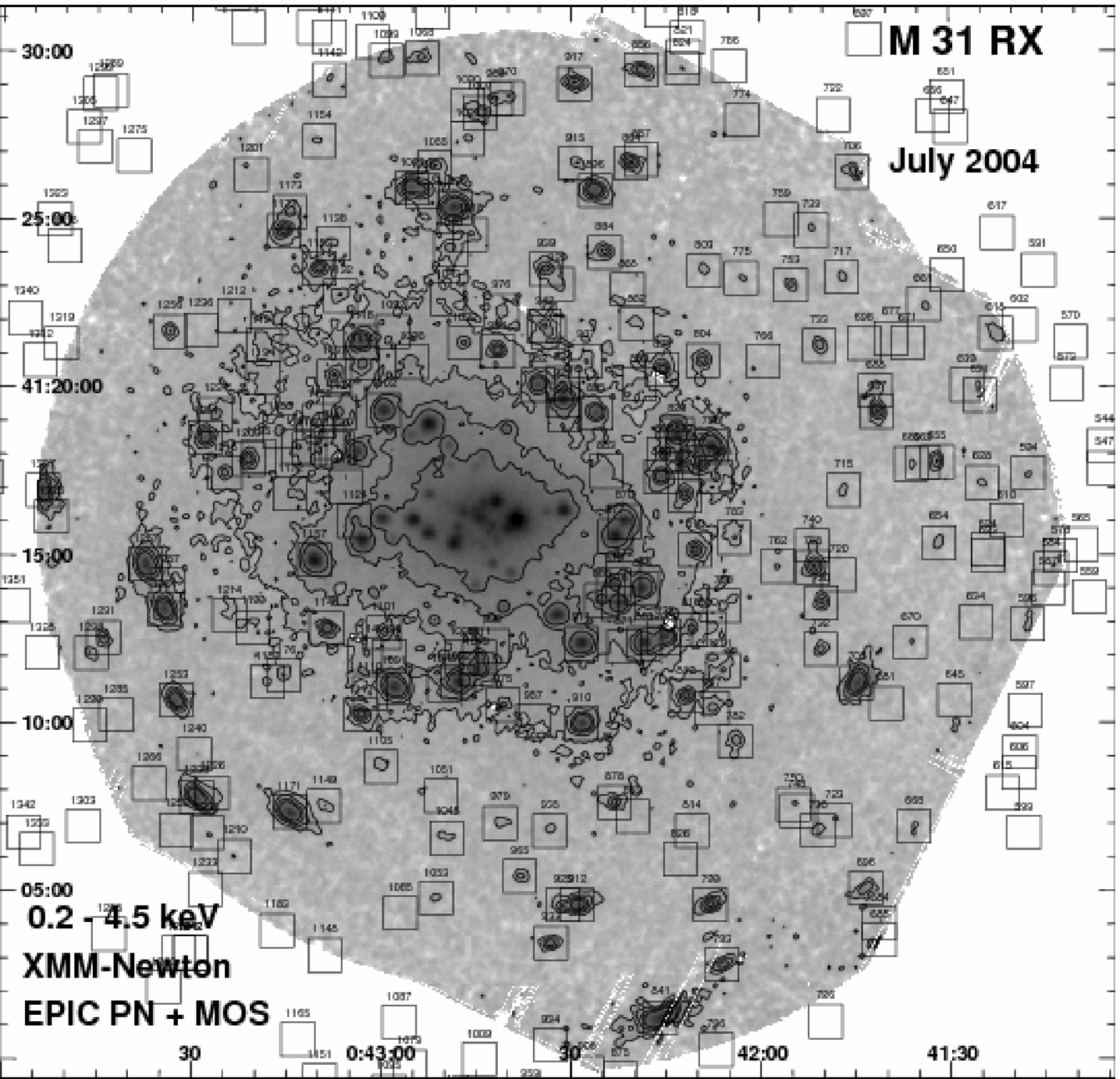}}
     \caption[]{(continued) Contours are at $(4, 8, 16, 32, 64, 128)$ in both upper panels and the lower left panel, and at $(4, 8, 16, 32)$ in the lower right panel. The inner area of the image shown in the lower right panel is shown in detail in Fig.\,\ref{Ima:singleIma_zoom}. 
}
\end{figure*}

\begin{figure*}
\addtocounter{figure}{-1}
\resizebox{\hsize}{!}{\includegraphics[clip]{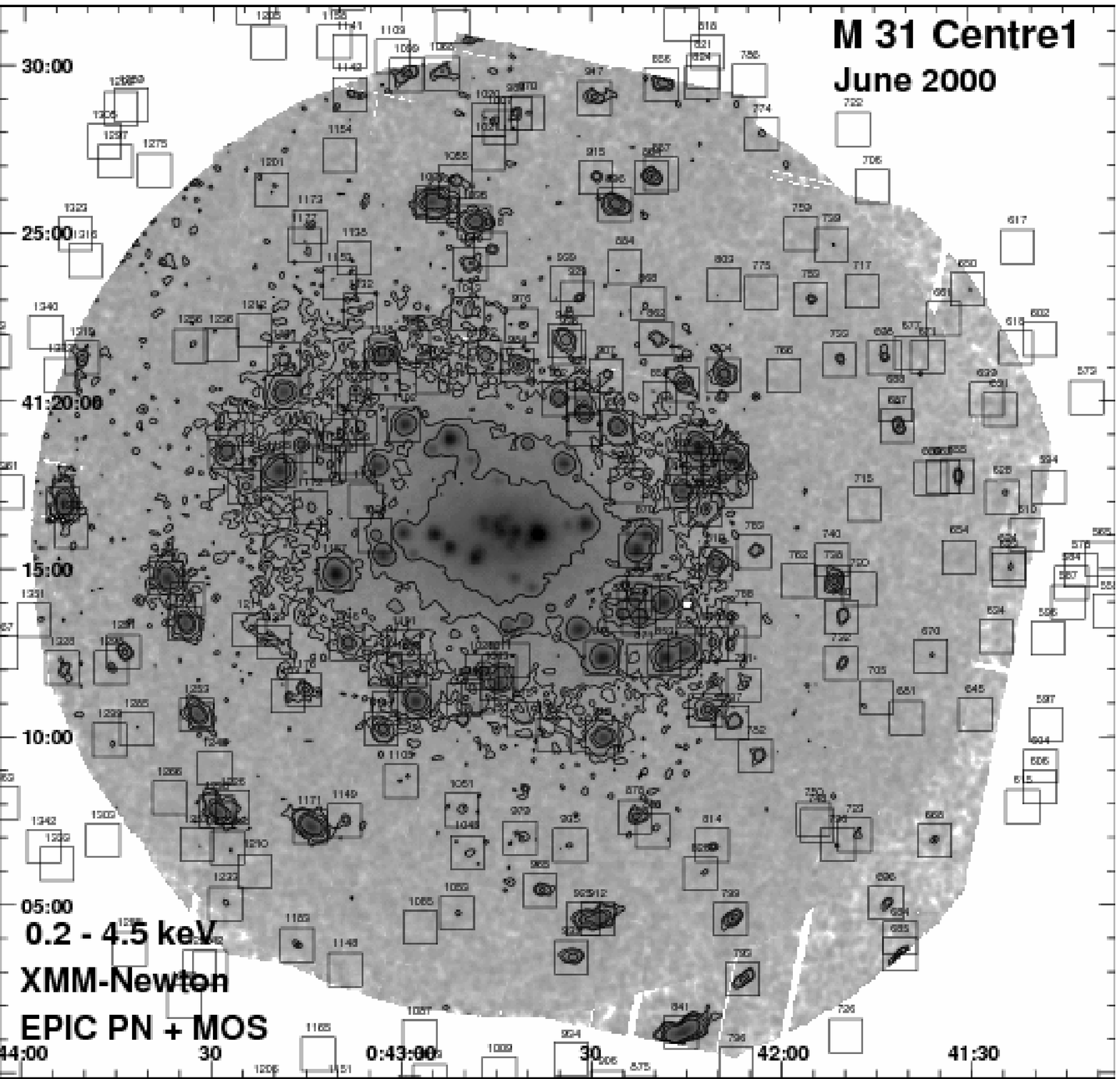}\hskip0.2cm\includegraphics[clip]{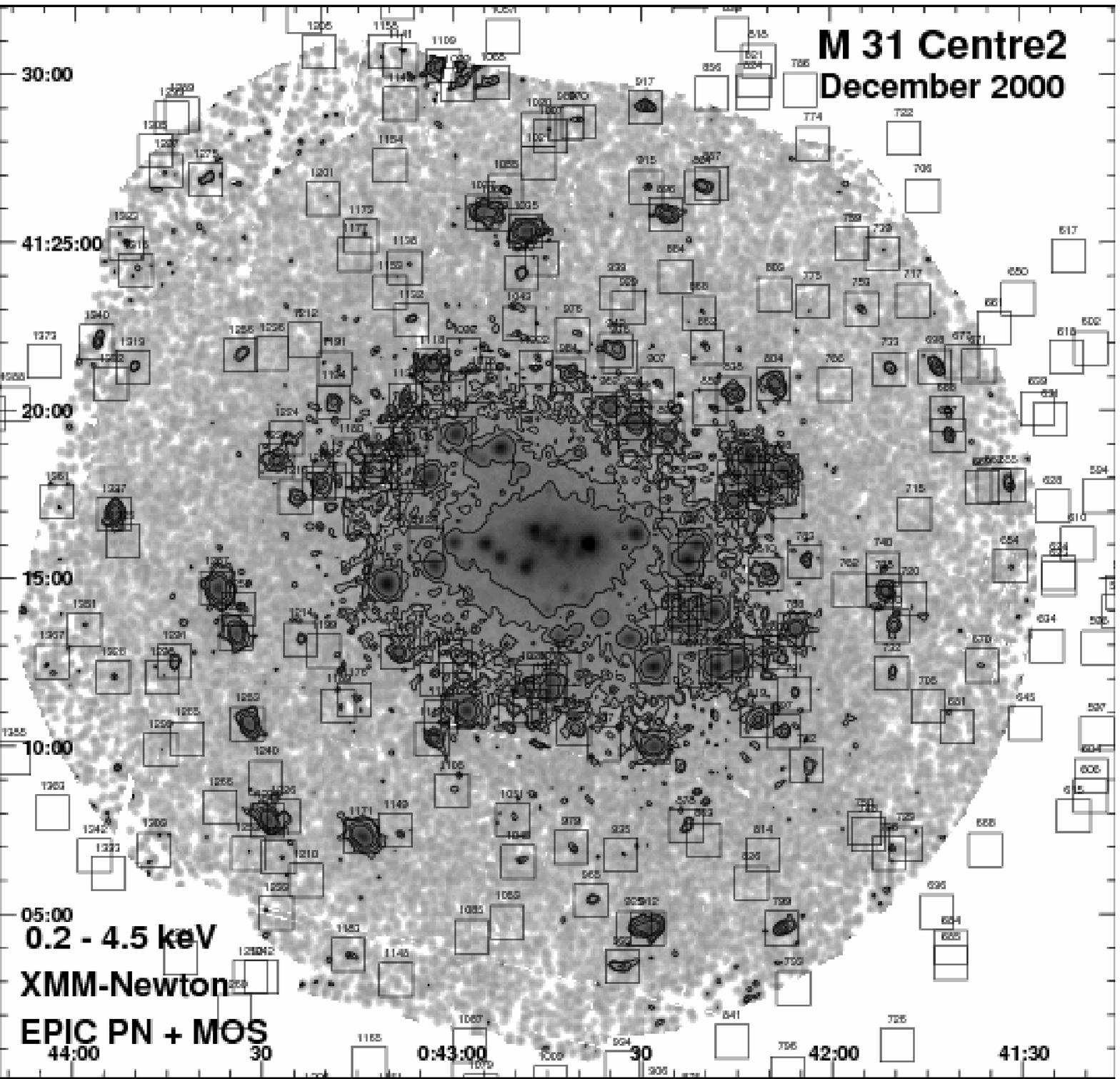}}
\resizebox{\hsize}{!}{\includegraphics[clip]{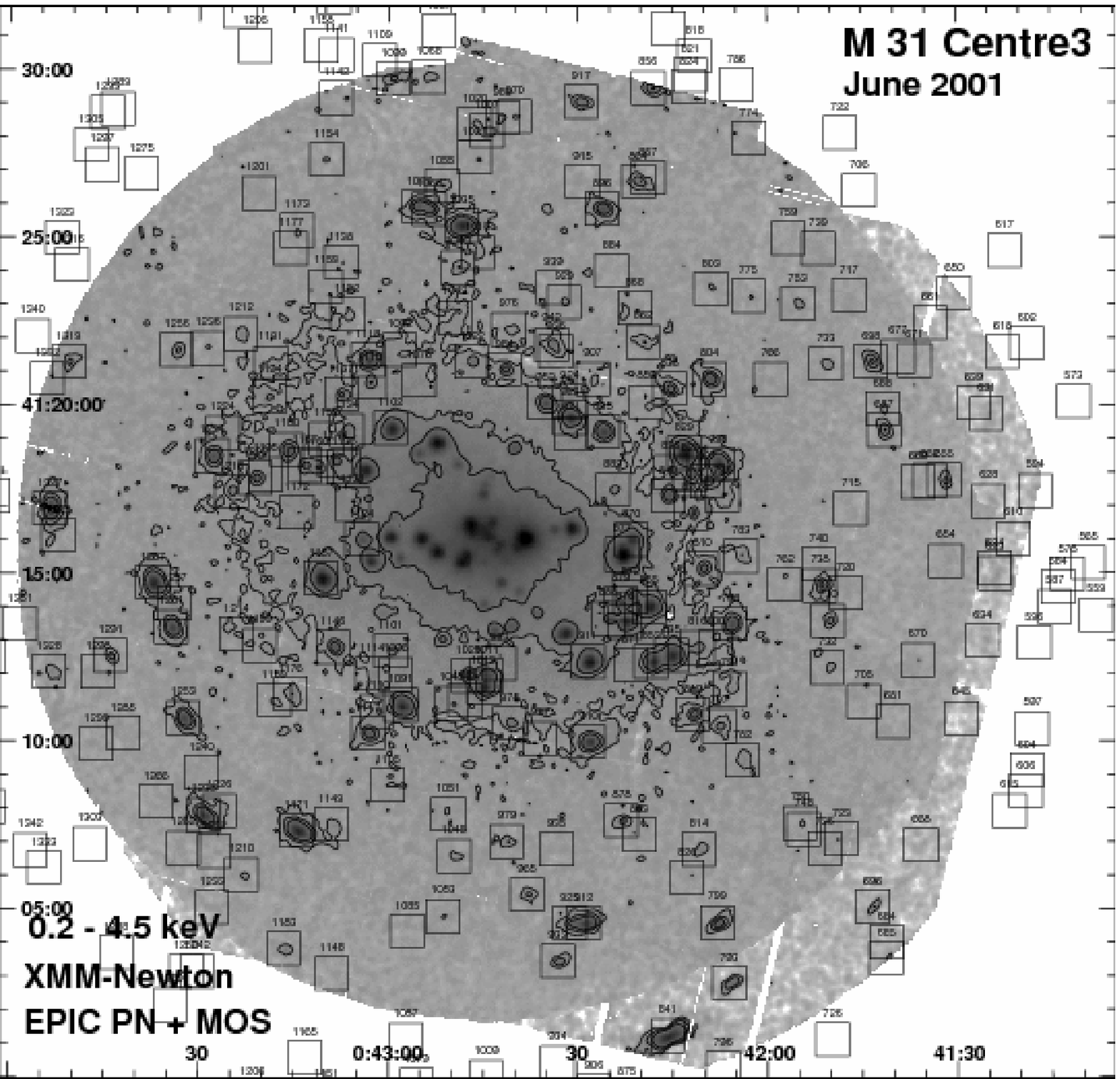}\hskip0.2cm\includegraphics[clip]{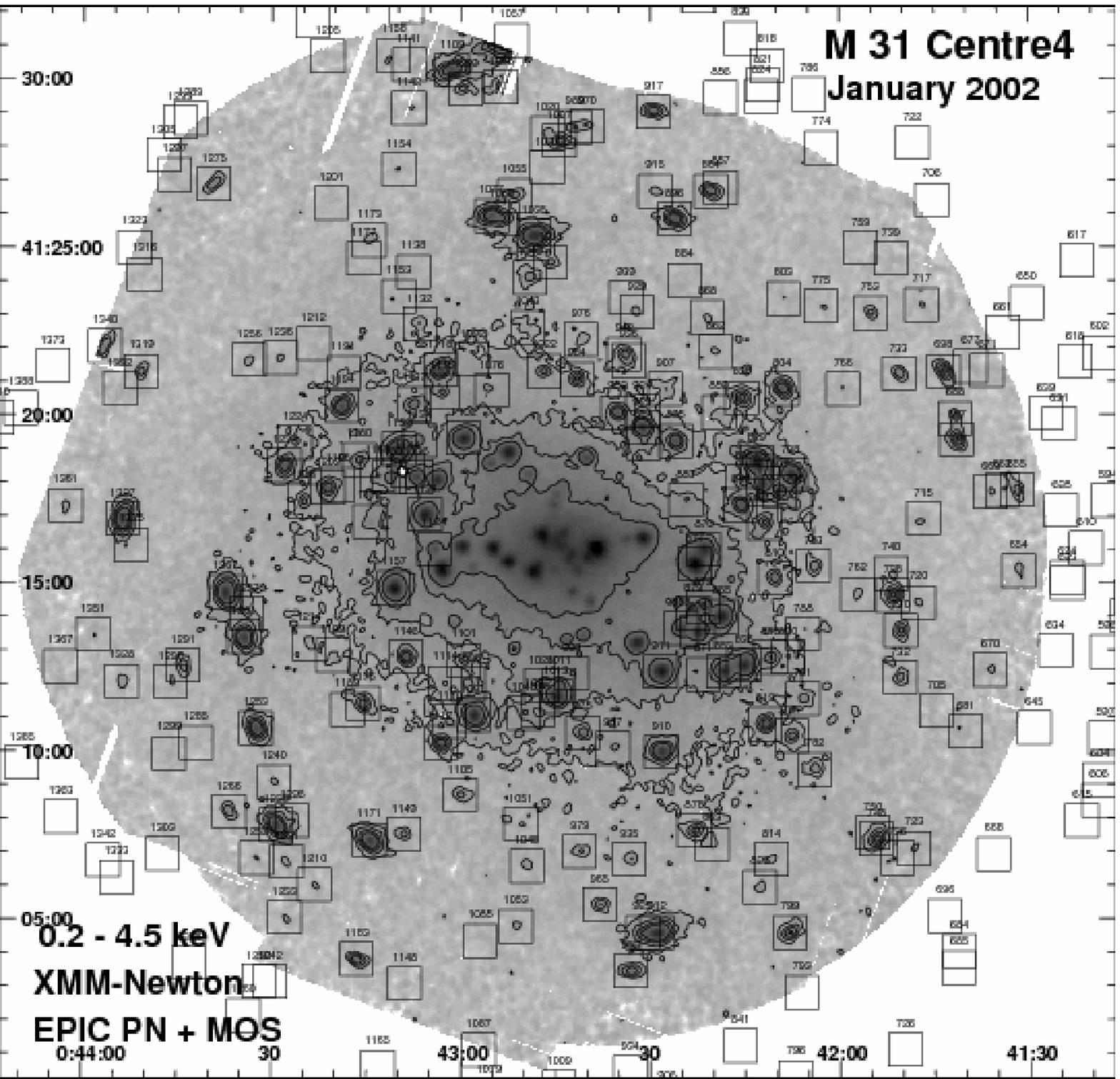}}
     \caption[]{(continued) Contours are at $(6, 8, 16, 32)$ in both upper panels, at $(8, 16, 32)$ in the lower left panel, and at $(4, 8, 16, 32)$ in the lower right panel. The inner area is shown in detail in Fig.\,\ref{Ima:singleIma_zoom}. 
}
\end{figure*}

\begin{figure*}
\resizebox{\hsize}{!}{\includegraphics[clip]{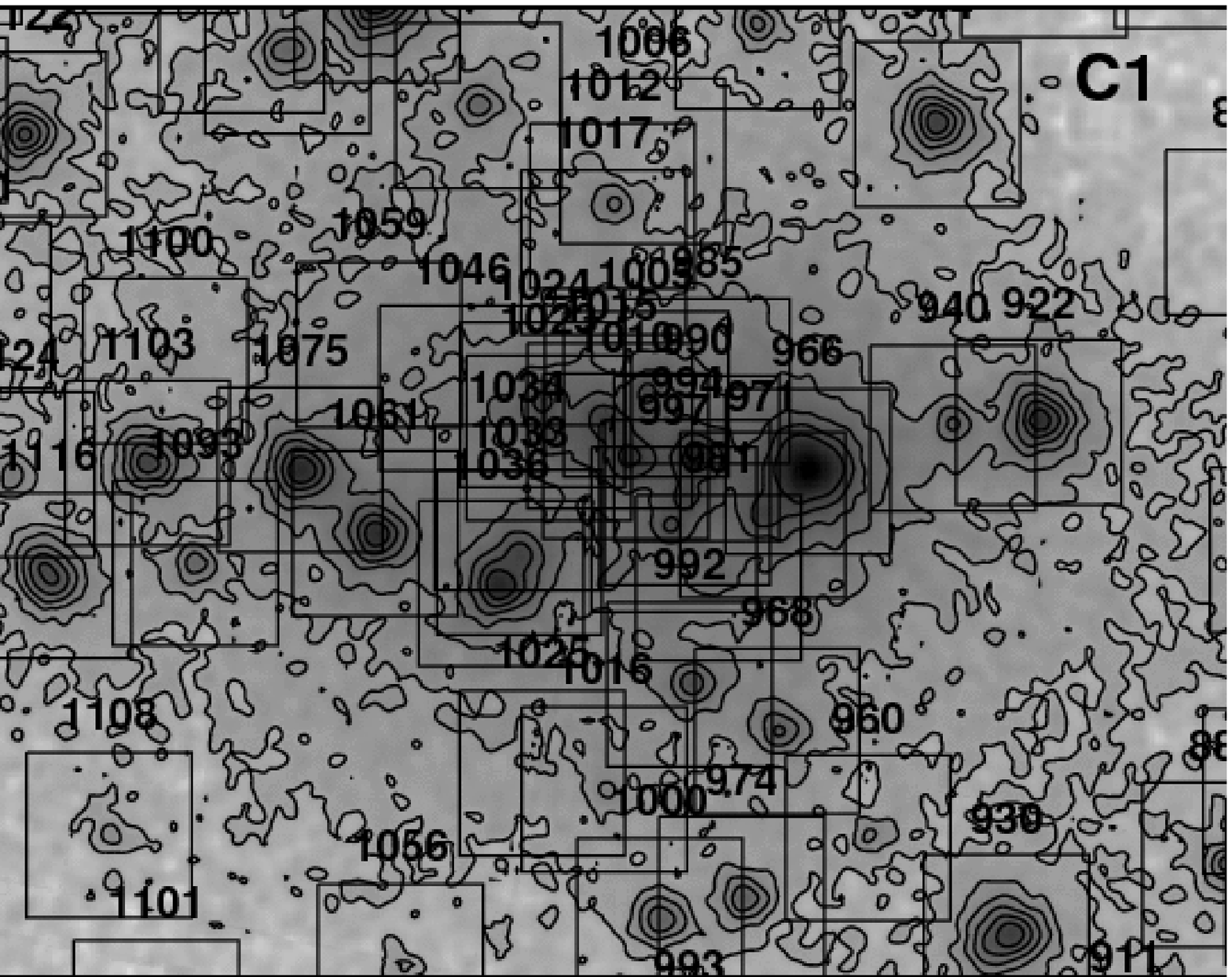}\hskip0.2cm\includegraphics[clip]{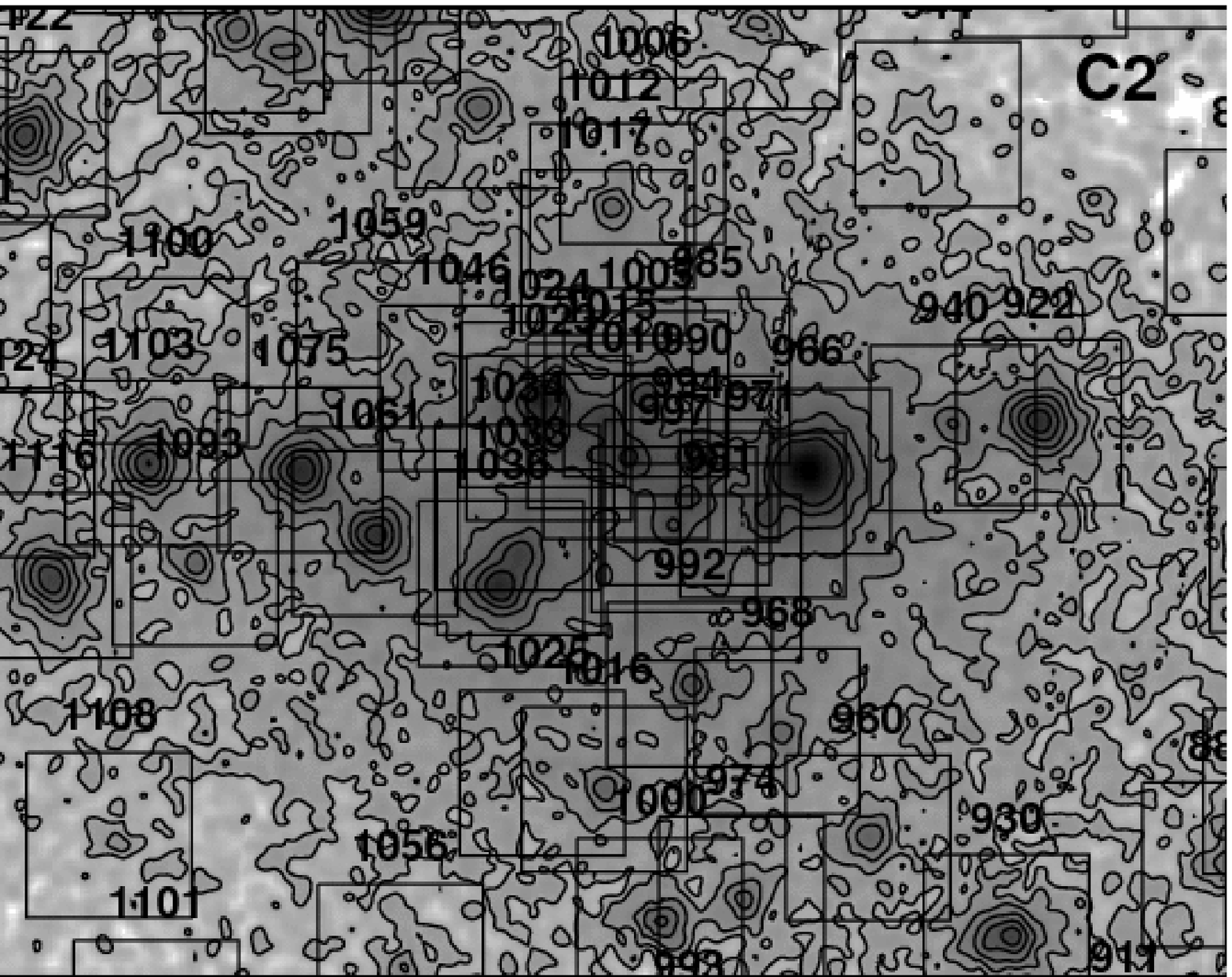}}
\resizebox{\hsize}{!}{\includegraphics[clip]{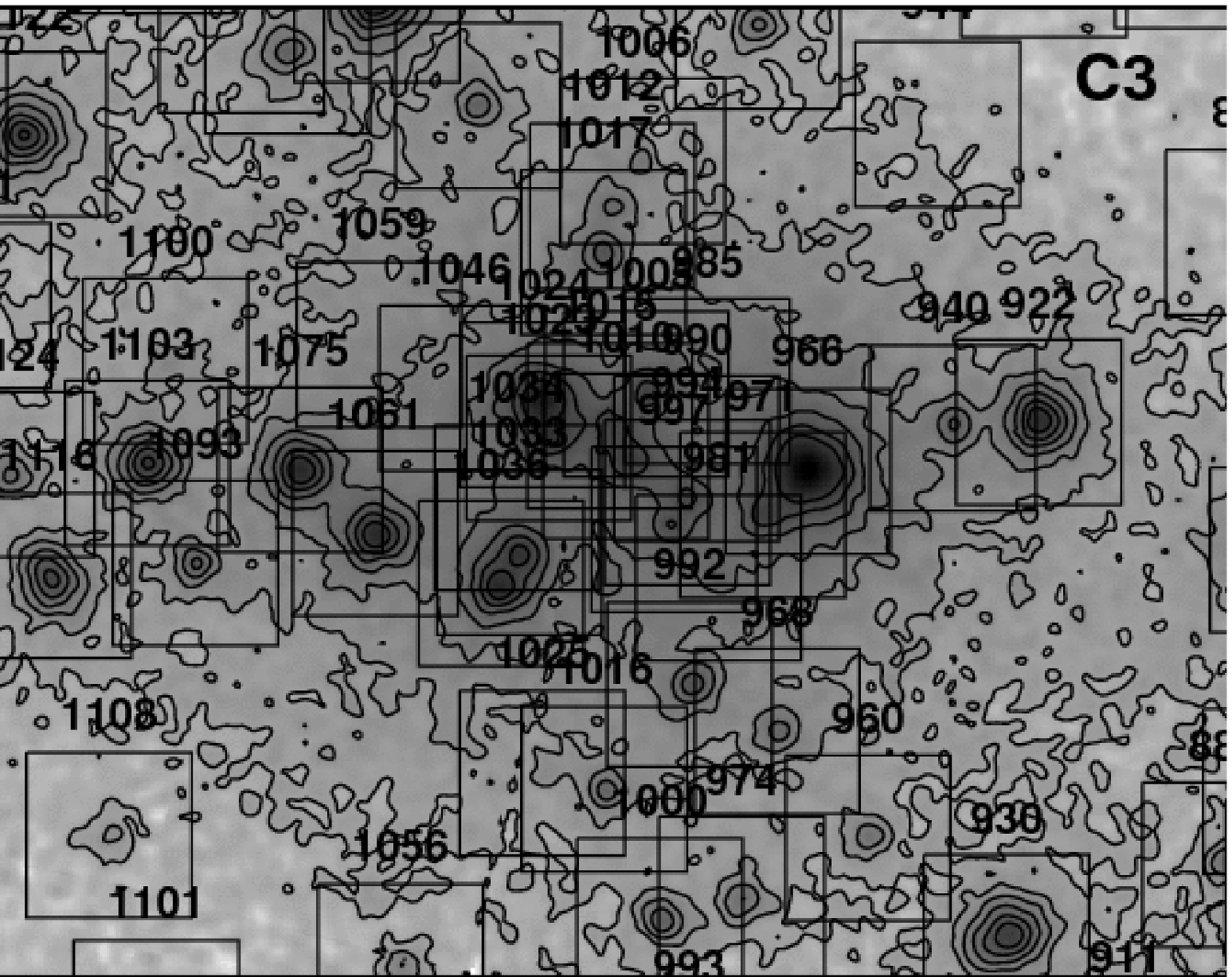}\hskip0.2cm\includegraphics[clip]{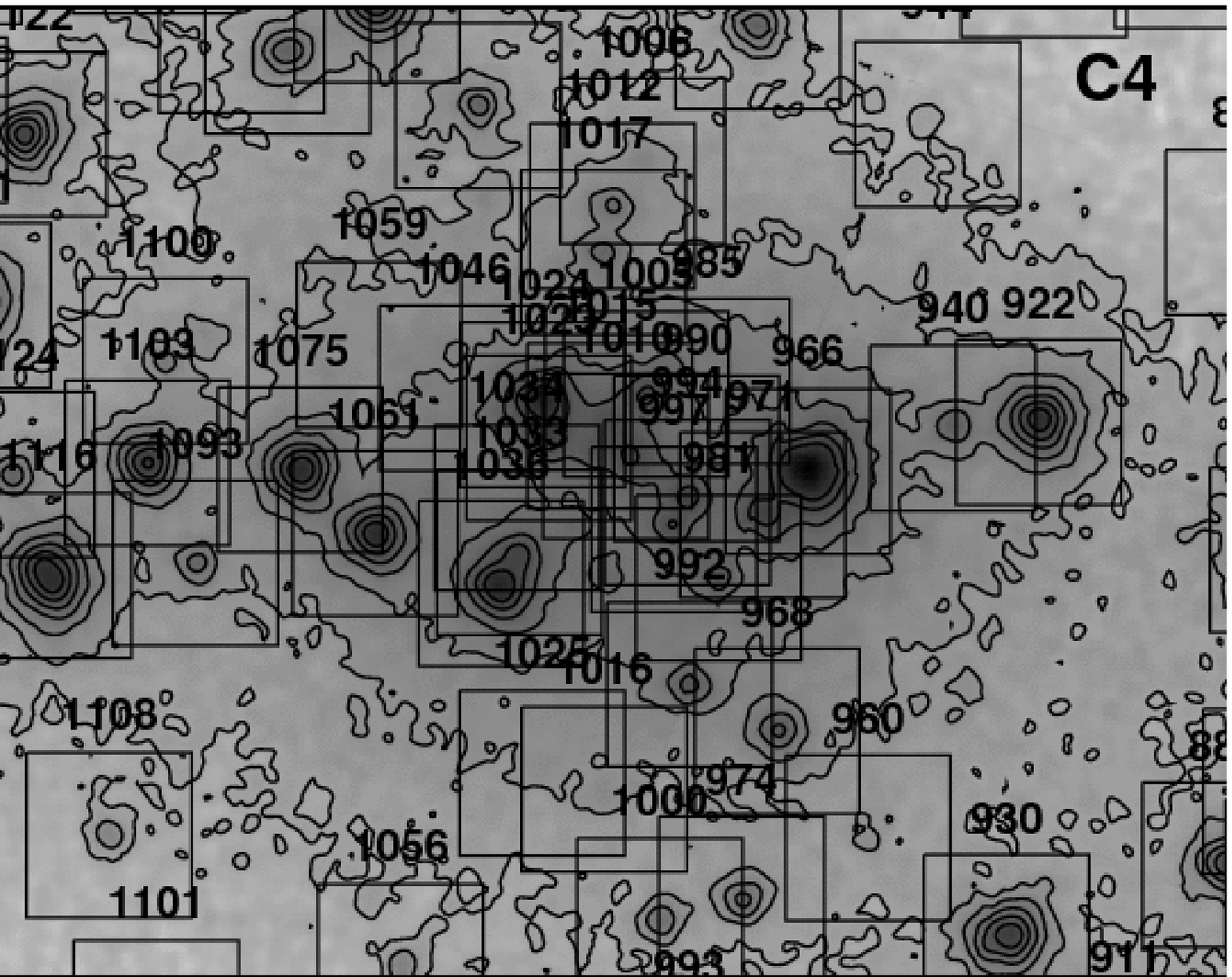}}
\resizebox{\hsize}{!}{\includegraphics[clip]{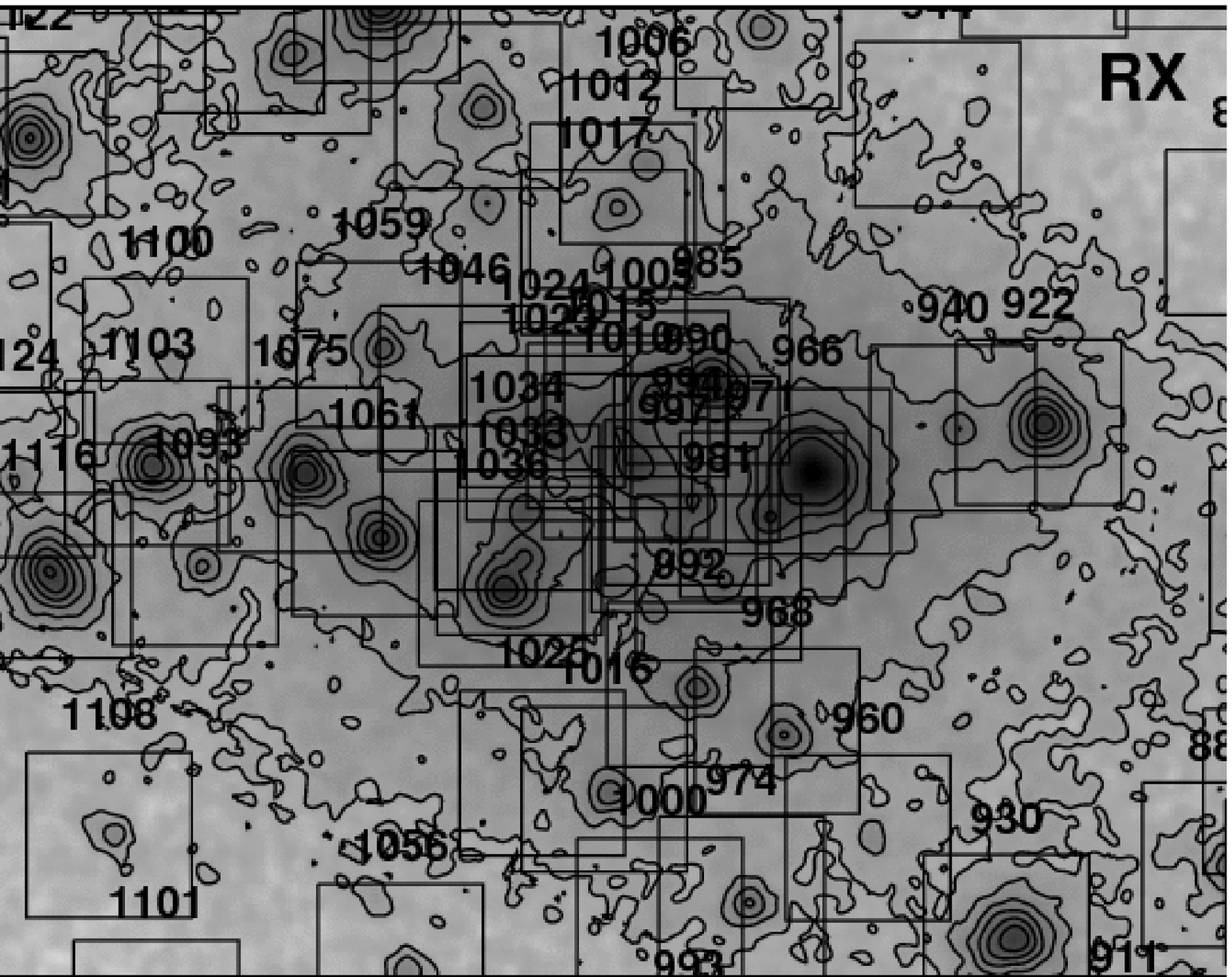}\hskip19.2cm}
     \caption[Inner area of \m31\ enlarged from Fig.\,\ref{Ima:singleIma}.]{Inner area of \m31\ enlarged from Fig.\,\ref{Ima:singleIma}. Contours are at $(4, 8, 16,
     32, 64, 128, 256)\times 10^{-6}$ ct s$^{-1}$ pix$^{-1}$ including a 
     factor of one smoothing. Sources from the large catalogue are marked as
     30\arcsec$\times$30\arcsec\ squares. The images are ordered as follows:
     Centre 1 ({upper left}),
     Centre 2 ({upper right}),
     Centre 3 ({middle left}),  
     Centre 4 ({middle right}) and
     Centre B ({lower left}).
}
    \label{Ima:singleIma_zoom} 
\end{figure*}

\end{appendix}

\end{document}